\documentclass{elsarticle}

\usepackage{graphicx,amsmath,amsfonts,amssymb}
\usepackage{bbm}
\usepackage{color}

\let\a=\alpha \let\b=\beta \let\g=\gamma \let\d=\delta
\let\e=\varepsilon  \let\h=\eta 
\let\l=\lambda   \let\x=\xi 
\let\s=\sigma \let\t=\tau  
   \let\G=\Gamma
\let\D=\Delta   
\let\Si=\Sigma   
\let\ee=\epsilon   \let\io=\infty

\def\EE{{\cal E}} 
 \def\HH{{\cal H}}
\def\TT{{\cal T}}\def\NN{{\cal N}}

 \def\SS{{\cal S}}
  
\def\ZZ{{\cal Z}}

\def\to{\rightarrow} 
 
\def\us{\underline\sigma}

\newcommand{\beq}{\begin{equation}} \newcommand{\eeq}{\end{equation}}
\newcommand{\wh}{\widehat} 
\newcommand{\Tr}{\text{Tr}}

\def\tb{{\widetilde{\beta}}}
\def\Next{{{\cal N}_{\rm ext}}}
\def\Nint{{{\cal N}_{\rm int}}}
\def\Ntraj{{{\cal N}_{\rm traj}}}
\def\hsix{{\widehat{\sigma}_i^x}}
\def\hsiz{{\widehat{\sigma}_i^z}}
\def\la{\langle} 
\def\ra{\rangle}
\def\da{{\partial a}}
\def\di{{\partial i}}
\def\dami{{\partial a \setminus i}}
\def\dima{{\partial i \setminus a}}
\def\dbmi{{\partial b \setminus i}}
\def\db{{\partial b}}
\newcommand{\bea}{\begin{eqnarray}} \newcommand{\eea}{\end{eqnarray}}
\def\tq{\widetilde{q}}
\def\tH{\widetilde{H}}
\def\tW{\widetilde{W}}
\def\eqd{\overset{\rm d}{=}}
\def\cP{{\cal P}}
\def\E{\mathbb{E}}
\def\dd{{\rm d}}

\def\hX{\widehat{X}}
\def\hH{\widehat{H}}
\def\hU{\widehat{U}}
\def\hHi{\widehat{H}_{\rm i}}
\def\hHf{\widehat{H}_{\rm f}}

\def\hT{\widehat{T}}

\def\hs{\widehat{\sigma}}

\def\Tr{{\rm Tr} \,}
\def\bs{{\boldsymbol{\sigma}}}
\def\ubs{\underline{{\boldsymbol{\sigma}}}}
\def\Ns{{N_{\rm s}}}
\def\bvh{\vec{\boldsymbol{h}}}
\def\vh{\vec{h}}
\def\uzero{\underline{0}}

\def\hP{\widehat{P}}
\def\hA{\widehat{A}}
\def\ut{{\underline{\tau}}}

\newcommand{\ket}[1]{\left\vert#1\right\rangle}
\newcommand{\bra}[1]{\left\langle#1\right\vert}

\journal{Physics Reports}

\begin{document}

\begin{frontmatter}

\title{The Quantum Adiabatic Algorithm \\ applied to random optimization problems: \\
the quantum spin glass perspective}

\author{V.~Bapst$^1$, L.~Foini$^{2}$, F.~Krzakala$^3$, G.~Semerjian$^1$ 
and F.~Zamponi$^1$}

\address{
$^1$ LPT,
\'Ecole Normale Sup\'erieure, UMR 8549 CNRS, 24 Rue Lhomond, 75005 France \\
$^2$ LPTHE, UPMC Paris 06, 4 Place Jussieu, 75252 Paris Cedex 05, France \\
$^3$ ESPCI ParisTech, CNRS UMR 7083 Gulliver, 10 rue Vauquelin, 75005 Paris, 
France
}

\begin{abstract}
Among various algorithms designed to exploit the specific properties of
quantum computers with respect to classical ones, the quantum adiabatic
algorithm is a versatile proposition to find the minimal value of an
arbitrary cost function (ground state energy). Random optimization problems
provide a natural testbed to compare its efficiency with that of classical
algorithms. These problems correspond to mean field spin glasses that have
been extensively studied in the classical case. This paper reviews recent
analytical works that extended these studies to incorporate the effect of
quantum fluctuations, and presents also some original results in this 
direction. 
\end{abstract}


\end{frontmatter}

\tableofcontents

\section{Introduction}

A central issue in computer science is the classification of the difficulty
of computational tasks, i.e. the existence or not of algorithms with small 
requirements (in terms of the time of execution and the necessary memory)
that perform a given task~\cite{GareyJohnson,Papadimitriou94}. This classification
is called computational complexity theory. 
A rough distinction between easy and hard tasks is made by distinguishing
algorithms that need to perform a number of elementary operations growing
either polynomially or exponentially with the size of their input.
One of the central tasks analyzed in this context concerns combinatorial
optimization problems~\cite{Pa83}: given a cost function defined
on $N$ variables, each taking a finite number of values, the question is
to classify families of cost functions such that algorithms can, or cannot,
find their global minimum by executing a number of operations smaller than
some polynomial of $N$. The current consensus is that there exist families
of cost functions such that no algorithm can achieve this goal (this is the
famous P$\neq$NP conjecture). One example of such difficult problems is
the graph $q$-coloring (for $q\ge 3$): 
given a graph, i.e. a collection of $N$ vertices and 
$M$ edges linking some of the pairs of vertices, the cost function associates 
to each of its $q$-coloring 
(one out of $q$ colors is chosen for each vertex) the number of monochromatic
edges, linking two vertices of the same color.

It was understood above that the term ``elementary operation'' meant
some simple process like adding two numbers, or other arithmetic tasks, 
which can all be reduced to logical operations on boolean variables. In 
that context the basic elements of a computer are bits that behave 
``classically'', i.e. they are in a well defined state 0 or 1, that is altered
deterministically by logical operations involving one or a few of them.
But what happens if the basic elements of a computer behave ``quantumly'',
i.e. if instead of bits one deals with ``qubits'' that can not only be in
the states 0 or 1 but in any linear combination of the two? Will such a
quantum computer be able to solve efficiently some of the tasks on which
classical computers get stuck? These questions were first raised in the 
eighties by Feynman~\cite{Fe82} and Deutsch~\cite{Deutsch85} and opened
the way to a new branch of science at the interface between computer
science and physics, known today under the names of quantum computing and 
quantum information theory~\cite{NielsenChuang,Mermin,lectures_QI}.

Several specific quantum algorithms have been discovered since 
then~\cite{Deutsch85,DeutschJozsa92,Simon94,Shor94,Grover97}, providing 
``quantum speedup'' with respect to their fastest classical counterparts.
Most of them concern arithmetic problems, notably Shor's algorithm for
factoring integers~\cite{Shor94}, yet none of them solves efficiently
a representant of the classically hardest problems (the so-called NP-complete 
ones). A quantum analog of the computational complexity theory has been
developed~\cite{BeVa97,Wa00,KiShVy02}, with the introduction of complexity 
classes
of easy and hard problems, the notion of difficulty being now with respect
to the number of required operations on a quantum, instead of classical,
computer.

A generic strategy to solve optimization problems with a quantum computer,
called quantum annealing or quantum adiabatic 
algorithm~\cite{qa_first,qa_second,KaNi98,Aeppli99,Fa01}, proceeds in the 
following way 
(see~\cite{qa_review_santoro,qa_book_das_chakrabarti,qa_rmp_das_chakrabarti,review_Nishimori} 
for reviews of this procedure).
The evolution of the state of a quantum computer obeys Schr\"odinger
equation, with a time-evolving Hamiltonian $\hH(t)$ that is controlled by the 
programmer. If in an initialization step the system is prepared in the
ground state of a simple Hamiltonian $\hHi$, and if the time evolution of the 
Hamiltonian is slow enough, the adiabatic theorem~\cite{Messiah} ensures
that the system remains, with high probability, in the instantaneous ground state
at all subsequent times. This property can be exploited by driving the
Hamiltonian towards one that corresponds to the cost function of the 
optimization problem to be solved, let us call it $\hHf$. 
Indeed its ground state, which is the final
state of the system according to the adiabatic theorem, provides precisely the
answer to the combinatorial optimization problem. The crucial question
of the efficiency of such an algorithm reduces thus to a criterion for the
validity of the adiabatic approximation. Roughly speaking the adiabatic
theorem states that the evolution time of the Hamiltonian (hence the running
time of the algorithm) has to be larger than the inverse square of the minimal
energy gap between the ground state and the first excited state encountered
during the process. Instances of optimization problems $\hHf$ such that this gap
is exponentially small in the size of the problem (and thus require an 
exponentially large time to be solved adiabatically) were exhibited early
on~\cite{DaMoVa01,DaVa01}. It was also realized that some choices of the
initial Hamiltonian $\hHi$ led ineluctably to exponentially small 
gaps~\cite{farhi08,ZnHo06}. 

One can however wonder if, for ``reasonable'' choices of $\hHi$, and for 
``most instances'' $\hHf$ belonging to hard optimization problem classes,
this annealing procedure leads to a quantum speedup with respect to classical
algorithms. A precise meaning can be given to the expression 
``most instances'' by considering ensemble of random instances. Continuing
with the example of the graph coloring problem defined above, one can for
instance define probability laws on the set of all graphs of $N$ vertices,
the most famous one being the Erd\H{o}s-R\'enyi random graph~\cite{Janson}
in which the $M$ edges are chosen uniformly at random. Then a property
holds for ``most instances'' if its probability with respect to the choice
of the random graph goes to 1 in the large size (thermodynamic) limit 
$N\to \infty$. Such ensembles of random optimization problems were actually
introduced in computer science~\cite{MitchellSelman92} as generators
of hard problems on which to benchmark classical algorithms. Since then an
intense research effort was devoted to their study, in theoretical computer
science and discrete mathematics of course, but also in statistical mechanics.
Random optimization problems can indeed be handled by methods first devised
for the study of disordered physical systems, spin glasses in 
particular~\cite{Beyond}: renaming energy the cost function, optimization
amounts to low temperature statistical mechanics (one can view for instance
the graph coloring problem as an antiferromagnetic $q$-states Potts model),
an optimal configuration becomes a ground state, 
and the randomness in the instance corresponds to the quenched disorder
of spin glasses. This interdisciplinary approach turned out to be very
fruitful, and in the last decade a detailed understanding of the shape of
the configuration space of random optimization problems was reached thanks
to the non-rigorous methods of statistical mechanics (some of these predictions
were later on put on a rigorous mathematical basis). In the thermodynamic
limit these problems undergo several phase transitions when some control
parameter of the random ensemble (for instance the finite ratio $M/N$ in the
case of the graph coloring) is varied, in particular the set of ground states
gets split into a large number of clusters of close-by configurations,
the clusters being well separated one from the other in the configuration
space. 
This understanding also allowed to devise specific ensembles of random
instances where one can ``hide'' an arbitrarily chosen unique ground state,
that remains hard to find if no direct information is available on the
hidden configuration. This is particularly useful in the context of the
quantum adiabatic algorithm, which has most often been studied on instances
with a Unique Satisfying Assignment (USA).

As explained above the random optimization problems provided useful
benchmarks for classical algorithms, it is thus natural to test the
efficiency of the quantum adiabatic algorithm on them, and indeed one
of the first proposals~\cite{Fa01} studied such random instances.
Unfortunately simulating quantum computers on classical ones is a hard
computational task because the dimension of the Hilbert space grows
exponentially with the number of qubits, hence the numerical integration
of Schr\"odinger equation, or the exact diagonalization of the time-varying
Hamiltonian is restricted to rather small system sizes, whereas computational
complexity theory classifies the difficulty of problems in the infinite size
limit. However random optimization problems, viewed from the perspective
of statistical mechanics of disordered systems, are mean field systems (there
is no finite-dimensional lattice underlying their definitions) and are as such
amenable to an analytic resolution. It is thus possible to build upon their 
classical statistical mechanics studies in order to include the quantum effects
induced by the interpolation procedure at the core of the quantum annealing
procedure. In particular one can investigate the fate of the classical phase
transitions mentioned above when quantum effects are added; when these
become quantum phase transitions~\cite{sachdev2001} as a function of the time 
parameter of
the adiabatic interpolation, energy gaps close in the thermodynamic limit
and this sets a lower bound on the running time of the algorithm, as the
adiabatic criterion has to be fulfilled. It is thus of a crucial importance
to understand the quantum phase transitions of random constraint satisfaction
problems in presence of quantum fluctuations, and in particular their order:
generically second order phase transitions are associated to polynomially
small gaps, while first order transitions (which are commonly found in
quantum mean field spin glasses~\cite{Go90,NR98,BC01,CGS01,JKKM08}) cause
exponentially small gaps, and in consequence an exponentially long evolution 
time is required for the adiabatic criterion to hold. Another, distinct, 
mechanism for the appearance of small gaps was pointed out 
in~\cite{AC09,AKR10}, based on an analysis of the perturbative 
effects of quantum fluctuations on the classical energy levels of optimization
problems. Some variants of the quantum adiabatic algorithm were claimed to
circumvent the effects of these ``perturbative crossings'' 
in~\cite{FGGGS10,Ch11,DicAm11}.

In this brief presentation, as in most of the literature, the quantum adiabatic
algorithm is viewed as an algorithm to find the ground state of an Hamiltonian, 
or in computer science terms to solve exactly an optimization problem. One
can however think of it more generically as an approximation 
algorithm~\cite{vazirani2001}: if the allowed evolution time is smaller
than required by the adiabaticity criterion then the system ends up in an
excited state, corresponding to energies higher than the global minimum of
the Hamiltonian (this is, in physical terms, reminiscent of the Kibble-Zurek
problem, see~\cite{review_kz} for a recent review). But it might be that a 
good compromise can be found between
short execution times on the one hand, and small excitation energies on the 
other hand. This would be as important from a complexity point of view as being
able to find the exact ground state. Indeed for several optimization problems
it is computationally hard to find an approximate value of the ground state 
energy, and for some of them it is hard even to make an estimate better than
the energy of a configuration chosen uniformly at random~\cite{hastad01}. 

In this paper we shall review and extend recent works on the behaviour
of the quantum adiabatic algorithm on random optimization problems. As we
explained above this is strongly related to the understanding of the 
low temperature phases of quantum spin glasses. The paper is organized as
follows.
In Sec.~\ref{sec:quantum_computers} we shall make a brief introduction to 
classical and quantum computational complexity theory, and define more 
precisely the quantum adiabatic algorithm.
Sec.~\ref{sec:classical_mean_field} contains a review on classical random
optimization problems, their phase transitions and their relations to
spin glasses.  
Special attention will be given in Sec.~\ref{sec:generating_USA}
to the problem of generating random instances with prescribed properties, in 
particular to ensure the non-degeneracy of the ground state (USA instance).
In Sec.~\ref{sec:low_energy} we discuss the thermodynamic properties of
quantum spin glasses, concentrating in particular on their low energy
properties and quantum phase transitions, without entering into technical
details. 
The latter are touched upon in Sec.~\ref{sec:methods}, where we present 
several methods, both analytical and numerical, for the study of quantum
disordered systems. Some of these methods are then applied to a few
representative examples of random optimization problems subject
to quantum fluctuations in Sec.~\ref{sec:results}. We finally draw our
conclusions in Sec.~\ref{sec:conclusions}.

In addition to its review character this paper contains original material:
in Sec.~\ref{sec:qsubcubes} and \ref{sec:XORSAT} we present some details and
additional results of two works that previously appeared as 
letters~\cite{FSZ10,JKSZ10}; among the results of these sections
the study of the gap in presence of an exponential degeneracy of the
ground state in Sec.~\ref{sec:res_xor_expdegen} should have a general relevance. 
The discussion of the quantum $q$-coloring (or
antiferromagnetic Potts model) in Sec.~\ref{sec:results_coloring} was not
published before and will be further developed in a forthcoming 
publication~\cite{BSZ12}.
In Sec.~\ref{sec:qmc} we propose a method
to extract the gap from Quantum Monte Carlo (QMC) numerical simulations, that,
to the best of our knowledge, was not discussed previously.
The discussion on the generation of USA instances of 
Sec.~\ref{sec:generating_USA} also bears some originality in the quantum
context.
Finally, in Sec.~\ref{sec:qmc_ann} and~\ref{sec:XORSAT_USA} we show how one can use
QMC simulations to detect the clustering transition (to be introduced in Sec.~\ref{sec:classical_mean_field})
of quantum models.

Despite its length this review has no pretension of exhaustivity; complementary
point of views on the quantum adiabatic algorithm can be found in the 
reviews~\cite{qa_review_santoro,qa_book_das_chakrabarti,qa_rmp_das_chakrabarti,review_Nishimori,review_Knysh,review_Tanaka,review_Ohzeki} and references therein.

\section{Classical and quantum computations}
\label{sec:quantum_computers}

\subsection{Classical computation theory}
\label{sec:classical_complexity}

\subsubsection{Examples of optimization problems}
\label{sec:examples_optimization}

We shall give in this section a brief introduction to the classical theory
of computational complexity~\cite{GareyJohnson,Pa83,Papadimitriou94}, and set up some notations
that we shall use in the rest of the paper. For concreteness we will
concentrate on computational tasks related to combinatorial optimization
problems. We shall thus consider a discrete configuration space of $N$
variables denoted $\s_1,\dots,\s_N$, each of them taking values in a
finite set $\chi$, and denote a global configuration 
$\us=(\s_1,\dots,\s_N) \in \chi^N$. In most computer science applications
the variables considered are boolean, and one usually takes 
$\chi=\{\text{True},\text{False}\}$ or $\chi=\{0,1\}$. For consistency with
the conventions in vigor in physics we shall also use 
$\chi=\{+1,-1\}$,
the translations between the various conventions being straightforward.
The computational tasks we are interested in are defined in terms of a
cost function that assigns to each configuration $\us$ a real number. We
will call this cost the energy of the configuration, or the value of its
Hamiltonian, and denote it $E(\us)$. Let us give some examples:
\begin{itemize}
\item The graph $q$-coloring problem, in short $q$-COL, was mentioned in
the introduction and is formalized as follows. Given a graph $G=(V,L)$
with $V$ a set of $N$ vertices and $L$ a set of $M$ edges between pairs of vertices,
one takes $\chi=\{1,\dots,q\}$, with $q \ge 2$ an arbitrary integer, so 
that each configuration $\us$ corresponds to the coloring where vertex $i$ is 
given the color $\s_i$. The cost function is
\beq
E(\us) = \sum_{\la i,j\ra \in L} \delta_{\sigma_i,\sigma_j} \ ,
\label{def:COL}
\eeq 
where the sum runs over all edges of the graph, and $\delta$ denotes the
Kronecker symbol. The cost function thus counts the number of monochromatic
edges in the configuration $\us$.
\end{itemize}
The following examples involve binary variables that, as explained above,
we encode with Ising spins, $\chi=\{+1,-1\}$.
\begin{itemize}
\item The $k$-XORSAT problem is defined on a $k$-hypergraph $G=(V,L)$:
each of the $M$ hyper-edges $L$ involves a $k$-uplet of variables, with 
$k\ge 2$, thus generalizing the notion of usual graphs that corresponds
to $k=2$. We label the hyper-edges with an index $a=1,\dots,M$, 
and denote $i_a^1,\dots,i_a^k$ the indices of the vertices linked by the
$a$-th hyper-edge. In addition to the hyper-graph the problem is defined
by $M$ constants $J_a \in \{+1,-1\}$, and the cost function reads
\beq
E(\us) =  \sum_{a=1}^M \frac{1- J_a \underset{j=1}{\overset{k}{\prod}}
\s_{i_a^j}}{2} \ .
\label{def:XORSAT}
\eeq
It is easily seen that this sum equals the number of hyper-edges $a$
for which the condition $\s_{i_a^1} \dots \s_{i_a^k} = J_a$ is violated.
There are various equivalent interpretations of this condition;
in the language of coding theory~\cite{RichardsonUrbanke} this is a parity 
check rule. By associating Ising spins to $\{0,1\}$ variables according
to $\s_i=(-1)^{x_i}$ it is also equivalent to a linear equation of the
form $x_{i_a^1}+ \dots +x_{i_a^k} = y_a$, where $J_a=(-1)^{y_a}$ and the 
additions are interpreted modulo 2. Finally it can be seen as a condition on 
the eXclusive OR of $k$ boolean variables $\{\text{True},\text{False}\}$,
hence the name of the problem.

\item In the $k$-SAT problem one is given an hypergraph and, for each
hyper-edge, $k$ constants $J_a^1,\dots,J_a^k \in \{+1,-1\}$; the cost
of a configuration is then defined as
\beq
E(\us) = \sum_{a=1}^M
 \prod_{j=1}^k \frac{1- J_a^j \s_{i_a^j}}{2} \ .
\label{def:SAT}
\eeq
Each term of the sum is equal to 1 if, for all the $k$ vertices involved
in that hyper-edge, one has $\s_{i_a^j} \neq J_a^j$; on the contrary it vanishes
as soon as one of the $k$ vertices fulfill $\s_{i_a^j} = J_a^j$. In terms of
boolean variables this is the disjunction (logical OR) of $k$ literals,
that are equal to a variable or its logical negation depending on the sign
of $J_a^j$.
\item Another example is the so-called $1$-in-$3$ SAT (or Exact Cover) 
problem, defined on a hypergraph of triplet of vertices with the cost function 
\beq
E(\us) = \sum_{a=1}^{M} \frac{5 - \s_{i_a^1} - \s_{i_a^2} - \s_{i_a^3}+
\s_{i_a^1} \s_{i_a^2} + \s_{i_a^1} \s_{i_a^3} + \s_{i_a^2} \s_{i_a^2} 
+ 3 \s_{i_a^1} \s_{i_a^2} \s_{i_a^3} }{8} \ .
\label{def:COVER}
\eeq
Each term of the sum is equal to 0 or 1, the former case being realized
if exactly one out of the three variables involved is equal to -$1$, the
two others being equal to 1.
\end{itemize}

Note that in all the examples above the energy function is constructed
as a sum of $M$ indicator functions that take the value 0 (resp. 1) if some
constraint involving $k$ variables is satisfied (resp. unsatisfied). These
examples thus belong to the class of Constraint Satisfaction Problems (CSP). In
this context one often calls a clause each of the individual constraint,
and formula the conjunction of all the constraints. A formula is said to be
satisfied by an assignment $\us$ of the variables if and only if all the 
individual constraints are satisfied. A formula is satisfiable if and only if
there exists at least one configuration that satisfies it; in physical terms
this correspond to the ground state energy being equal to 0.

\subsubsection{Classical complexity classes}
\label{sec:classical_complexity_classes}

Given an arbitrary cost function $E(\us)$ on a discrete configuration space,
one can define various computational tasks:
\begin{itemize}
\item The decision task is to answer yes or no to the question ``is there
a configuration $\us$ whose cost is smaller or equal than a given constant 
$C$?''. In the context of CSP one can further 
specialize this question by taking $C=0$; the question thus becomes ``is the
formula satisfiable?''

\item The optimization task is to compute the minimal value of $E$ over
the configuration space; the output is thus a real number instead of the yes/no
answer of the decision task.

\item One can also ask to compute the number of configurations of minimal 
energy (a counting task).

\item Another task is to output explicitly one configuration of minimal 
energy; this could either be any such configurations, or one could require
in addition that the output configuration is a random configuration, with
for instance the uniform distribution over all configurations of minimal
energy (this is a sampling task).
\end{itemize}

The goal of computational complexity theory is to classify the difficulty of 
these tasks, in terms of the time and space (memory) requirements of
algorithms that perform them. Let us emphasize some subtleties in the 
vocabulary to be used: a \emph{problem}, is, in its loose sense, a set
of cost functions. For instance the $q$-coloring problem means all the 
functions $E(\us)$ defined in Eq.~(\ref{def:COL}), for all possible graphs. 
To be more precise one has to indicate, along with the set of cost functions, 
the \emph{version} of the problem, among the decision, optimization, 
and counting variants defined above. Finally an \emph{instance} of a
problem means one representant of the class of cost functions it includes.
An instance of the $q$-coloring problem is thus defined by a graph.

Let us concentrate first on the decision problems. The NP (standing for 
Non-deterministic Polynomial) complexity class 
contains the problems for which it is easy, for every instance, to check the 
correctness of the ``yes'' answer, if the algorithm
provides as a certificate a configuration $\us$ with $E(\us) \le C$. In
other words for NP problems computing the value of $E(\us)$, given $\us$,
is by itself an easy task, which means a task that can be performed with
a number of operations growing only polynomially in the size of the input.
This is indeed the case for all the examples we have given above (the size
of the input being here controlled by $N$ and $M$). Of course, even if the
answer is easy to check a posteriori, this does not mean that the certificate
is easy to find a priori. This is true only for a subset of the problems in
NP, the so-called P (for Polynomial) problems. An example of a problem in P is
deciding the satisfiability of XORSAT formulas: 
thanks to the mapping onto a problem of linear equations, 
for any choice of the hyper-graph and of the constants $J_a$
one can use Gaussian elimination and check in a number of operations growing
as $N^3$ whether or not there exists a configuration satisfying all the 
constraints, thus answering the decision question with $C=0$. The decision
versions of 2-SAT and 2-COL, with $C=0$, are also in P.
On the contrary
for $k$-SAT or $q$-COL with $k,q\ge 3$, no algorithm is known to answer the 
decision question for all possible instances in polynomial time (this can
of course be done in an exponential time just by inspecting all the $2^N$
configurations one by one). In fact it is strongly believed that no such
algorithm can exist: $k$-SAT and $q$-COL (with $k,q\ge 3$) belong to the
NP-complete subset of problems in NP, that are the hardest problems of NP
in the sense that any instance of any problem in NP can be translated (with
a polynomial overhead) to an instance of a NP-complete problem. Exhibiting a 
polynomial algorithm for a single NP-complete problem would imply that P=NP, 
such a collapse of the complexity class being held as rather improbable.

The NP class and its subsets we have briefly discussed is only one example
among a very large number of complexity classes; let us emphasize that
NP concerns only the decision tasks whose answer is a yes or no. In general
the other versions of the problem, which must return a number or a 
configuration, may fall into other complexity classes. In some cases
the optimization can be essentially reduced to the decision version: if 
the possible costs are bounded and discrete, one can simply make a dichotomy
on the value of $C$ and finds the optimal value of the cost function by
calling the decision problem a number of times growing logarithmically
with the number of possible values of the cost. But counting problems
are not reducible in this way, and belong to other complexity classes,
known as $\#$P and its variants. 

One should also keep in mind that the easiness of a
class of problems for the decision task does not imply that the other tasks,
on the same problem, are easy as well. The XORSAT problem is very illuminating
in this respect: even if its decision version is in P when $C=0$, thanks to
the Gaussian elimination algorithm, solving the decision problem with $C>0$
or finding the optimal 
cost of an arbitrary instance takes in the worst-case an exponential time
(in the sense that no polynomial algorithm is known that performs this task).
Indeed, if the Gaussian elimination shows that there are no configurations
satisfying simultaneously all the constraints, it gives no clue on how to
find optimal assignments of the variables. The situation is actually even
worse: not only it is hard to find exactly the optimal cost, but even finding
a good approximation for it is also hard. This question of approximate 
resolution of optimization problems is an important issue in computer
science~\cite{vazirani2001}. In the case of $k$-SAT and $k$-XORSAT
it was shown in~\cite{hastad01} that finding an approximation of the optimal
energy which is more accurate than the one obtained by taking uniformly at 
random a configuration of the variables is harder than any NP-complete problem.

Let us finally comment on the notion of execution time of classical algorithms.
In the formal studies of computational complexity theory this time is measured
in the units of the number of steps performed by a so-called Turing machine to
execute the algorithm. The Turing machine is a very simplified model of a
``computing machine'', very far from the complexity of today's computer.
Its power as a formal tool for the analysis of algorithms relies in its 
universality, formulated in the Church-Turing hypothesis: all classical
computers can be emulated by a Turing machine with an overhead that grows
only polynomially with the size of the input to the algorithm, hence the
distinction between polynomial and exponential run-time is independent of
the precise computational model.

\subsection{Quantum computation theory}

We shall now describe a few aspects of the quantum computation theory,
and contrast it with the classical one. This review being theoretical
in nature we shall not address the experimental challenges for building
quantum computers (known as the DiVincenzo criteria~\cite{DiVicenzo_criteria})
and only provide a few references to experimental works.

\subsubsection{Quantum circuits model and examples of quantum 
algorithms}
\label{sec:quantum_circuits}

We shall now give a brief presentation of the basics of quantum computer
science~\cite{NielsenChuang,Mermin,lectures_QI}, assuming knowledge of the 
laws and notations of quantum mechanics. An introduction to quantum computer
science written by and for physicists can be found in~\cite{LaMoScSo10}.
The paradigmatic shift
from classical computer science is the assumption that the elementary
variables at the core of the computer behave quantumly: instead of bits
which can take either the value 0 or the value 1, one deals with qubits
which can be in a coherent superposition of the two values. Let us introduce
some notations: for a system of $N$ qubits we denote ${\mathcal H}$ the
Hilbert space spanned by the orthonormal 
basis $\{|\us \ra : \us \in \chi^N \}$. This
basis, indexed by the classical configurations, is called the 
computational basis. If $\chi$ has $d>2$ elements one often speaks
of qudits, to emphasize the $d$-dimensionality of the Hilbert space of a
single element. According to the laws of quantum mechanics the state
of the system is described by a vector $|\psi \ra$ of this Hilbert space,
i.e. a (complex) linear combination of the vectors of the computational
basis, which has norm 1. The state of the computer evolves during the
execution of an algorithm; according to the laws of quantum mechanics
this evolution is represented by the action of a linear operator on the
Hilbert space, $|\psi \ra \to \hU |\psi \ra$, where the linear operator
$\hU$ must be unitary in order to conserve the norm of $|\psi \ra$.
Every quantum algorithm thus corresponds to an unitary operator; in principle
this operator acts on all the qubits of the system, making a practical
implementation of non-trivial algorithms a seemingly impossible task.
Fortunately it has been shown~\cite{DiVi95,DeBaEk95,Ll95,BaBeChCletal95}
that any unitary operator can be factorized (with arbitrary precision)
as a product of simple operators, called gates in this context, that act only 
on one or two qubits (this is similar, in the classical case, to the 
reducibility of any Boolean function as a combination of NotAND gates).
Moreover there exist universal sets of gates that contain only
a finite number of operators. For example, in the case of binary qubits,
it is enough to take as one-qubit gates the operators $\hP$ and $\hA$, 
defined by their matrix representation
\beq
\hP = \begin{pmatrix} 1 & 0 \\ 0 & e^{i \frac{\pi}{4}} \end{pmatrix} \ , \quad
\hA =\frac{1}{\sqrt{2}} \begin{pmatrix} 1 & 1 \\ 1 & -1 \end{pmatrix} \ ,
\eeq
i.e. 
$\hP$ adds a phase of $\pi/4$ between the two states of the qubit, while the
Hadamard gate $\hA$ converts the two vectors of the computational basis
into their symmetric and antisymmetric linear combinations. The only
two-qubit gate that completes this universal set is called the
controlled NOT (cNOT) gate, that acts on the computational basis of the two 
qubits as $| \s_1, \s_2 \ra \to |\s_1, \s_1 \oplus \s_2 \ra$, where we
use in this section the convention $\chi=\{0,1\}$, and $\oplus$ denotes
the addition modulo 2. In other words the cNOT gate leaves the second 
(controlled) bit constant if and only if the first (controlling) bit is 0.

\begin{figure}
\centerline{\includegraphics{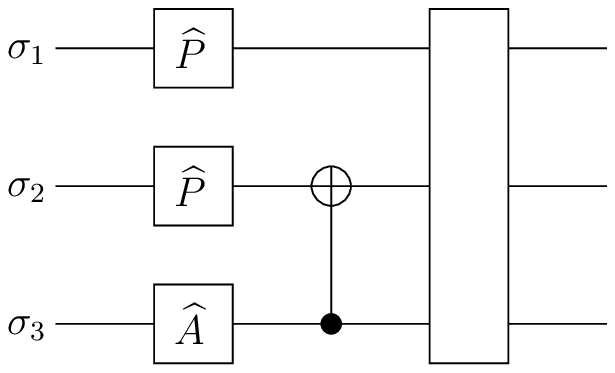}}
\caption{An example of a quantum circuit on three qubits. In the first step
the two first qubits are submitted to the phase operator while the third one
is acted upon by the Hadamard gate. Then $\s_2$ is submitted to a NOT 
controlled by $\s_3$, and finally the system is acted on by a
(here unspecified) three qubit gate.
}
\label{fig:example_qc}
\end{figure}

Quantum algorithms can be conveniently represented graphically by quantum
circuits: the unitary operator $\hU$ that encodes the algorithm can
be factorized as the consecutive applications of simpler unitary operators, acting
possibly on a subset of the whole qubits. These elementary operators can be
written as products of the universal gates displayed above, but this is not
compulsory and it is often simpler to describe the action of a gate on a
large number of qubits than its decomposition on one and two qubit gates.
An example of quantum circuit is shown on Fig.~\ref{fig:example_qc}.

Let us now describe some quantum algorithms that exhibit a velocity gain
with respect to classical computations. The simplest ones shall deal
with binary functions $f(\us)$ from $\{0,1\}^N$ to $\{0,1\}^M$. In the
quantum setting these functions are implemented as unitary linear
operators $\hU_f$; note that an unitary transformation is invertible,
hence $\hU_f$ must somehow keep trace both of the input $\us$ and
of the output $f(\us)$ of the function $f$. A convenient way to fulfill
this request is to let $\hU_f$ act on the Hilbert space of $N+M$ qubits,
its action being defined on the computational basis as
\beq
\hU_f |\us , \us' \ra = |\us , \us' \oplus f(\us) \ra \ ,
\eeq
where $\oplus$ is here the bitwise addition modulo 2.
We shall use equivalently the notations $|\us , \us' \ra$ and
$|\us \ra | \us' \ra$, with $\us=(\s_1,\dots,\s_N)$
and $\us'=(\s'_1,\dots,\s'_M)$, for the computational basis vectors, the
second notation emphasizing the tensorial product between the input and
output qubits. At a first (too optimistic) look, the laws of quantum 
mechanics allow to treat in a ``parallel'' way the $2^N$ possible
inputs of the function $f$; suppose indeed that the quantum computer is
prepared in the state
\beq
\frac{1}{2^{N/2}} \sum_{\us \in \chi^N} |\us \ra |\uzero \ra \ ,
\eeq
where $\uzero=(0,\dots,0)$, which can be reached by the application of
Hadamard gates on the first $N$ qubits to the initial state 
$|\uzero,\uzero \ra$. Then $\hU_f$ transforms this state into
\beq
| \psi \ra = \frac{1}{2^{N/2}} \sum_{\us \in \chi^N} |\us , f(\us) \ra \ ,
\eeq
which seems indeed to contain all the information about the behaviour
of $f$ on its $2^N$ possible inputs. However this information is not
reachable by an observer, because of the measurement axioms of quantum
mechanics. A measurement of the qubits in the state $| \psi \ra$ written
above leads to nothing but a random choice of a single configuration
$\us$ among the $2^N$ possible ones, and to the associated value
$f(\us)$. This trivial observation explains why some thought has to be put
in devising quantum algorithms that outperform classical ones; the 
availability of linear superpositions is not enough for that, one has to
use in a more clever way the possibility of interferences between states.

The simplest example of this strategy is Deutsch's algorithm~\cite{Deutsch85}.
Given a function $f$ from $\{0,1\}$ to $\{0,1\}$ (i.e. $N=M=1$), the task
is to determine whether $f$ is constant or not. Classically one cannot avoid
the computation of both $f(0)$ and $f(1)$ to answer this question. On a 
quantum computer this task can however be performed with a single application
of $\hU_f$. Indeed, starting from the state $|0,1\ra$ and applying Hadamard
gates to both qubits leads to
\beq
\frac{1}{2} (|0\ra + |1\ra ) (|0\ra - |1\ra ) \ . 
\eeq
Applying the operator $\hU_f$, followed by the Hadamard gate on the first
qubit, produces the state
\beq
|0\ra \frac{|f(0)\ra - |\overline{f(1)}\ra + |f(1)\ra - |\overline{f(0)}\ra }
{2 \sqrt{2}}
+ 
|1\ra \frac{|f(0)\ra - |f(1)\ra + |\overline{f(1)}\ra  - |\overline{f(0)}\ra }
{2 \sqrt{2}} \ ,
\nonumber
\eeq
where we denoted $\overline{\bullet}$ the logical negation 
($\overline{0}=1$, $\overline{1}=0$). If $f$ is constant the second term
vanishes, otherwise it is the first term that cancels out; measuring the
state of the first qubit thus yields, without any probability of error,
the answer to the question.

In the previous example the ``quantum speedup'' was rather modest, reducing
the computational cost from two classical evaluations of $f$ to one 
application of $\hU_f$. However, it illustrated the essential ideas behind the
much more impressive gains of quantum algorithms that have been developed
later on, and that we shall only sketch here. 

One of the problems solved by Deutsch and Jozsa in~\cite{DeutschJozsa92} 
concerns functions $f$ from $\{0,1\}^N$ to $\{0,1\}$, with $N$ arbitrary but 
with the promise that $f$ is either constant or balanced (i.e. takes the value
0 on exactly $2^{N-1}$ distinct inputs). Their quantum algorithm 
(in the refined formulation of~\cite{ClEkMaMo98}) decides
between these two alternatives with a single application of $\hU_f$, and
without possibility of error, whereas a classical algorithm needs $2^{N-1}+1$
evaluations of $f$ before a definite answer can be given; however a classical
randomized algorithm can answer after a finite (with respect to $N$) number of 
evaluations with a probability of error arbitrarily small.

Simon's algorithm~\cite{Simon94} is given as an input a function 
$f$ from $\{0,1\}^N$ to $\{0,1\}^N$, with the promise that either $f$ is
bijective, or that it is ``periodic'' in the (unusual) sense that there exists
a binary string $\ut \neq \uzero$ such that $f(\us)=f(\us') \Leftrightarrow
\us = \us' \oplus \ut$, with again $\oplus$ the bitwise addition modulo 2.
This quantum algorithm decides between these two alternatives, and allows to
determine $\ut$ in the second case, with an expected number of applications
of $\hU_f$ growing linearly with $N$. This has to be contrasted with the
exponential number of evaluations of $f$ that are necessary for a classical
algorithm to solve the same problem.

A crucial step in Simon's algorithm is a unitary transform known as a
Quantum Fourier Transform. This idea was also exploited by Shor 
in~\cite{Shor94} to devise a quantum algorithm for finding the period $\tau$
of a function $f$ from $\mathbb{Z}$ to $\mathbb{Z}$, where now the term period
is used in a more usual sense: $f(x)=f(y) \Leftrightarrow x = y $ mod $\tau$
(with the promise that $\tau$ is smaller than some given integer). The
importance of this result stems from its consequences in the context of 
number theory, and, in a more applied way, to cryptography. As a matter of
fact the problem of factorizing integers is reducible, via arithmetic
theorems, to the period finding problem that Shor's algorithm solves in an
efficient (polynomial) way. Moreover the security of the famous 
RSA~\cite{RSA} public-key protocol of cryptography is based on the inexistence
of an efficient classical algorithm for integer factoring. Hence the
construction of a large quantum computer would have drastic consequences
for the security of encrypted communications 
(see~\cite{Shor_exp1,Shor_exp2,Shor_exp3,Shor_exp4} 
for small-scale experimental demonstrations). From the more theoretical
point of view of computational complexity the factoring problem
(in its decision version, i.e. given $N,M$ two integers, is there $p$ with
$1<p\le M$ such that $p$ divides $N$, which can be used to exhibit 
a factor of $N$ via a dichotomy on $M$) is most likely in an intermediate
difficulty class, namely in NP but outside P and NP-complete 
(note that the primality decision problem, i.e. the existence of a 
factor $p$ of $N$ was relatively recently shown to be in P~\cite{AKS}, yet 
without the condition $p\le M$ this does not solve the factoring problem). 
Hence the efficient quantum algorithm devised for solving the factoring problem 
cannot be used, via reductions, to solve all the NP problems.

Let us also mention another quantum algorithm that is unrelated to
those mentioned above, and that can be described as follows. Let 
$f_\ut$ be the function from $\{0,1\}^N$ to $\{0,1\}$ that maps all
its $2^N$ inputs to 0, except one fixed string $\ut$ that is mapped to
1. This can be interpreted as an unsorted database with one single marked
element. In order to discover the value of $\ut$ a classical algorithm
(even randomized) cannot do better than computing the value of the function on 
$O(2^N)$ inputs. On the contrary Grover exhibited in~\cite{Grover97} a quantum
algorithm that solves this problem in a number of steps of order $2^{N/2}$,
i.e. with a quadratic speedup with respect to the classical execution time. In fact this problem can be reinterpreted as the search for the ground state of the operator $\hHf = \widehat{I}- | \ut \ra \la \ut |$ 
(where $\widehat{I}$ is the identity operator) 
on the Hilbert space of $N$ qubits. 
Grover's algorithm works by successive applications of the operators 
$(-1)^{| \ut \ra \la \ut|}$ and \hbox{$\hHi = \widehat{I} - | \Psi_0 \ra \la \Psi_0|$} 
(these notations will be useful in Sec.~\ref{sec:Def-Imp} where we shall 
come back on this problem), where $|\Psi_0\ra =2^{-N/2} \sum_{\us} |\us \ra$ 
is the uniform superposition of all the states of the Hilbert space. $\hHi$ 
connects any two vectors $|\us\ra,|\us'\ra$, allowing for a ``quantum diffusion'' 
between states. The important point is that the convergence can be guaranteed 
within $2^{N/2}$ applications of each operator~\cite{Grover97}, allowing for the 
quantum quadratic speedup. Grover's algorithm is known to be optimal \cite{Bennett97} 
and has been experimentally tested,
see~\cite{Grover_exp} and references therein.

Finally, other quantum algorithms have been developed to solve systems
of linear equations, see in particular~\cite{qlineq_1,qlineq_2}.

\subsubsection{Quantum complexity classes}
\label{sec:QCC}

The classification of problems according to their computational complexity
presented in Sec.~\ref{sec:classical_complexity_classes} relied on the
Church-Turing hypothesis, namely the equivalence (within polynomial 
reductions) of all classical computing devices. We shall now briefly
sketch the analogous classification that has been 
developed~\cite{BeVa97,Wa00,KiShVy02}, taking as a computing model a
quantum computer operating  algorithms described by quantum circuits.

The class BQP contains the problems that can be solved with a quantum circuit
containing a polynomial number of gates; for instance the existence of
Shor's algorithm~\cite{Shor94} demonstrates that the factoring problem
belongs to the BQP class. This is the quantum analog of the P class, or 
more precisely of the BPP class; indeed the measurement process at the end
of a quantum computation induces in general some probability of error, that
is required to be Bounded with respect to the size of the input in the BQP
class.

The closest quantum analog of NP is known as the Quantum Merlin Arthur (QMA) 
class of problems. Let us recall that the (rough) definition of NP we gave
was the class of decision problems for which a yes answer has certificates that
can be efficiently checked; for the examples of 
Sec.~\ref{sec:examples_optimization} a certificate could
be provided by a classical configuration $\us$, for which the computation
of the energy $E(\us)$ was an easy task. In an interactive definition
Merlin is the provider of the answer and its certificate (using for instance
a non-deterministic Turing machine) while Arthur is the checker of the 
certificate. In the quantum transposition of this definition Merlin is
allowed to give as a certificate of his answer an element of the Hilbert
space of a quantum computer with a polynomial number of qudits, and Arthur can apply a quantum circuit with
polynomially many gates to this vector in order to verify its validity. 
Because of the inherent stochasticity in the quantum measurement processes
some error tolerance has to be included in the precise definition of
QMA~\cite{KiShVy02}, namely the yes instances must have at least one 
certificate that will be accepted by Arthur with a probability close to 1,
while for the no instances Arthur should be able to reject all the 
certificates that Merlin could try with again a probability close to 1.

In computational complexity theory the notion of completeness
plays a central role: for instance the NP-complete problems are the
hardest of the NP ones, and as such contains the essence of the difficulty
of this class. Similarly in the quantum context the QMA-complete problems
are those problems to which any member of QMA can be reduced (within a
polynomial overhead). In order to describe some of the known
QMA-complete problems, let us first define the notion of local Hamiltonians.
Consider the Hilbert space of $N$ qudits spanned by 
$\{|\us \ra : \us \in \chi^N \}$ (with $|\chi|$ finite but possibly $>2$).
A $k$-local Hamiltonian is a Hermitian operator $\hH$ acting on this space, 
that can be written as $\hH=\sum_{a=1}^M \hH_a$ where each $\hH_a$ acts on at 
most $k$ qudits among the $N$. The decision problem associated to $\hH$ is
to determine whether its smallest eigenvalue (ground state energy) is either
$<a$ or $>b$, where $a<b$ are two given reals and with the promise that one of
the two alternatives is true (i.e. the smallest eigenvalue of $\hH$ is not in 
the interval $[a,b]$). The first QMA-completeness result was obtained
in~\cite{KiShVy02}, where it was shown that the 5-local Hamiltonian problem
is indeed QMA-complete (for simplicity we keep understood some necessary 
hypothesis on the norms of the $\hH_a$ and on the size of the promise gap
$b-a$). This first result was then strengthened in a series
of works, that showed the completeness of the 3-local Hamiltonian 
problem~\cite{KeRe03}, then of the 2-local Hamiltonian 
problem~\cite{KeKiRe06}, and finally of the 2-local Hamiltonian 
problem with the further restriction that the local interactions $\hH_a$
only couple nearest neighbor qudits on an unidimensional 
lattice~\cite{AhGoIrKe09}; this last result only holds if the internal 
dimension $|\chi|$ of the qudits is at least 12.

The locality condition is reminiscent of the form of the classical cost 
functions (\ref{def:COL}), (\ref{def:XORSAT}), (\ref{def:SAT}), (\ref{def:COVER}) for
CSP, that also takes the form of a sum of
terms acting on a small subset of variables. Consider in particular
the $k$-SAT problem defined in Eq.~(\ref{def:SAT}): this is precisely
a $k$-local Hamiltonian, diagonal in the computational basis of $N$
qubits, with each term $\hH_a$ a projector onto the state 
$(-J_a^1,\dots,-J_a^k)$ of the $k$ qubits $i_a^1,\dots,i_a^k$. This observation
triggered the study of a quantum generalization of the $k$-SAT problem,
known as $k$-QSAT, where the $\hH_a$ are arbitrary projectors acting
on $k$ qubits. This problem was first introduced in~\cite{Br06}, where it
was shown that the case $k=2$ is easy (even on a classical computer) 
while for $k\ge 4$ it falls in the QMA-complete class.
Several results on this problem, and in particular its random version, can
be found in the original 
papers~\cite{Br06,BrMoRu09,AmKeSa10,LaMoScSo10_qsat,LaLaMoScSo10} 
and are reviewed in~\cite{LaMoScSo10}.

\subsubsection{Other approaches to quantum computation}

In the above presentation we have described quantum computations in terms
of quantum circuits, i.e. the successive action of unitary operators (gates)
acting on a few qubits. Several alternatives to the quantum circuit strategy
have been proposed, and the main focus of this review is one of them, the
quantum adiabatic algorithm. Before presenting it in more details let
us just name a few of the other perspectives on quantum computation, some
of which having been proven to be as universal as the quantum circuit model.

In the topological quantum computation scheme (see~\cite{review_topological_qc}
for a review) the quantum states that shall be used as qubits are encoded in
non-local (topological) degrees of freedom, that increase their tolerance
to local decoherence caused by an imperfect decoupling from the environment.
Experimental realizations of this scheme have been proposed, exploiting the
non-abelian anyonic statistics of excitations in fractional quantum Hall 
states. 

Another proposal, named ``one-way quantum computation''~\cite{oneway_qc},
relies crucially on two exquisitely quantum properties, i.e. entanglement
and projective measurement. In this scheme the system is initially
prepared in an highly entangled state, and this entanglement is used
as a resource for computation, that proceeds by a succession of projective
measurements on subparts of the system.

Let us finally mention the quantum walk strategy 
(see~\cite{qwalk_Farhi,qwalk_Kempe,qwalk_Ambainis,qwalk_Childs,qwalk_Reitzner} 
and references therein) 
that promotes the classical random walk procedure
to explore some configuration space to the quantum level, allowing to
exploit interference effects between the paths followed by the walk.

\subsection{Quantum annealing, or quantum adiabatic algorithm}

\subsubsection{Definitions}
\label{sec:QAA_def}

In contrast with the generic quantum computation considerations presented
above, the main focus of this review is a specific quantum algorithm
to solve optimization problems, namely the quantum annealing or quantum
adiabatic algorithm. Let us first emphasize that
optimization problems are intimately related to low temperature
statistical mechanics. Considering the cost function $E(\us)$ as
an energy, the Gibbs-Boltzmann probability law at inverse temperature
$\beta$ reads
\beq
 \mu(\us) = \frac{ e^{-\beta E(\us)}}{Z(\beta)} \ , \qquad
Z(\beta)=\sum_{\us} e^{-\beta E(\us)} \ ,
\eeq
where the partition function $Z(\beta)$ ensures the normalization of
the probability law. The latter concentrates on the minima of $E(\us)$ in 
the zero temperature limit ($\beta \to \infty$). One can set up a short 
dictionary translating between the optimization and the
statistical physics vocabulary:
\vspace{0.5cm} \begin{center}
  \begin{tabular}{|c|c|}
  \hline
  Optimization & Statistical Physics \\
 \hline
 cost function & energy or Hamiltonian \\
 optimal configuration & ground state \\
 minimal cost & ground state energy \\
 boolean variables & spins \\
\hline
\end{tabular}
\end{center}
\vspace{0.5cm} 
In the classical setting this analogy suggested the so-called simulated 
annealing algorithm~\cite{KirkpatrickGelatt83}: in order to find the
minima of the cost function $E$ one can perform a random walk in the
configuration space, with transition probabilities respecting the detailed
balance condition (reversibility in the mathematical language) with respect to 
the Gibbs-Boltzmann distribution, with a time-varying temperature that is
slowly decreased towards zero. If this decrease is slow enough thermal 
equilibrium is ensured at all times, and at the end of the annealing the
system is found in one of the minima of $E$: thermal fluctuations
allow to explore the configuration space and to overcome energy barriers
between local minima.

\begin{figure}
\centering
\includegraphics[width=.6\textwidth]{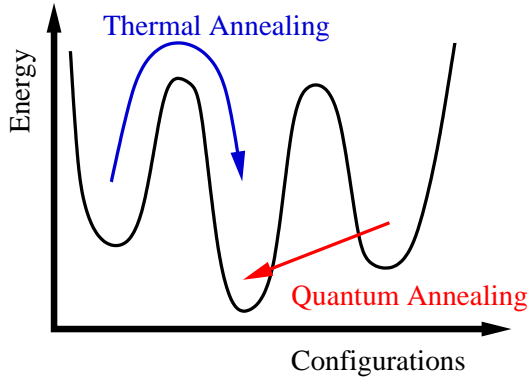}
\caption{Schematic picture of the thermal and quantum annealing processes.}
\label{Fig:Annealing}
\end{figure}

The quantum annealing, or Quantum Adiabatic 
Algorithm (QAA)~\cite{qa_first,qa_second,KaNi98,Aeppli99,Fa01}, exploits a similar
idea but with quantum fluctuations (and barrier penetration via tunnel effect) 
replacing thermal ones (see Fig.~\ref{Fig:Annealing} for a schematic
representation of this idea). 
To define it more precisely let us introduce
the operator $\hHf$, which acts in the Hilbert space spanned by the
classical configurations $\{|\us \ra : \us \in \chi^N \}$. For any
cost-function $E$ we define the associated operator $\hHf$, diagonal in
the computational basis, with $\hHf | \us \ra = E(\us)  | \us \ra $.
The state $|\psi(t) \ra$ of the quantum computer evolves according to
Schr\"odinger equation,
\beq
i \frac{\dd}{\dd t} |\psi(t) \ra = \hH(t) |\psi(t) \ra \ ,
\eeq
where we used a system of units where Planck's constant $\hbar$ is equal
to 1, and $\hH(t)$ is the time-varying Hamiltonian of the system. The
algorithm shall be run during an interval $\TT$ of physical time, 
it will thus be more convenient in the
following to trade the time $t$ with a reduced time $s=t/\TT \in [0,1]$.
To perform a quantum annealing one has to choose another operator
$\hHi$ and control the system in order to implement an interpolation
between the initial and final Hamiltonians $\hHi$ and $\hHf$, for instance
linearly (more general interpolations will be discussed below). In this
way the state of the system evolves according to
\beq
\frac{i}{\TT} \frac{\dd}{\dd s} |\psi(s) \ra = \hH(s) |\psi(s) \ra \ , \qquad
\hH(s) = (1-s) \hHi + s \hHf \ .
\label{Hquantum}
\eeq
If the initial condition $|\psi(0) \ra$ is the ground state of $\hHi=\hH(0)$,
and if $\TT$ is sufficiently large for the adiabatic condition~\cite{Messiah} 
to hold, then for all $s$ the state $|\psi(s) \ra$ is close to the 
instantaneous ground state of $\hH(s)$. In particular at the end of the
annealing $|\psi(1) \ra$ is nearly the ground state of $\hHf$, and a measure 
of the $N$ qubits returns an optimal configuration for the cost function
$E(\us)$.

This definition leaves a large variety of possible implementations of the
quantum adiabatic idea. In particular the initial Hamiltonian $\hHi$ 
can be chosen arbitrarily a priori, with a few conditions:
\begin{itemize}
\item it should not commute with $\hHf$, otherwise the dynamics is trivial.
\item its ground state should be easy to prepare.
\item its construction should not rely on a detailed knowledge of the
ground state of $\hHf$, that is precisely the problem one tries to solve.
\end{itemize}
When the classical variables are Ising spins ($\chi=\{+1,-1\}$) the 
computational basis can be viewed as the basis of common eigenstates
of the Pauli matrices along the axis $z$: $\hsiz |\us \ra = \s_i |\us\ra$.
In this way $\hHf$ is obtained very simply from $E$ with the replacement
$\s_i \to \hsiz$. Then a natural choice for $\hHi$, that fulfills the
conditions above, is the action of a transverse field in one direction
perpendicular to $z$, say $x$ for instance: $\hHi = - \sum_{i=1}^N \hsix$. Let
us recall that $\hsix$ acts on the computational basis by flipping the
$i$-th spin, $\hsix | \us \ra = | \us^{(i)}\ra$, with 
$\us^{(i)}=(\s_1,\dots,\s_{i-1},-\s_i,\s_{i+1},\dots,\s_N)$.
Note also that if one is interested in the decision problem of the existence of zero
energy ground states, 
then one
can also consider different cost functions that vanish on the same set 
of configurations. This allows to change $\hH_{\rm f}$ in order to improve
the efficiency of the algorithm, see e.g.~\cite{Ch11}.

An experimental realization of quantum annealing, and a comparison of
its efficiency with respect to thermal annealing, for
a disordered Ising system in a transverse field can be found 
in~\cite{Aeppli99}. This study was performed on a macroscopic sample
that allowed little control on the final Hamiltonian $\hHf$. The experiment
of~\cite{StDaHoBrCh03} concerned a 3 qubit NMR implementation of a
quantum adiabatic algorithm; more recently~\cite{DWave12} claimed to have
controlled an 83 qubit quantum computer based on superconducting loops.

\subsubsection{The adiabatic condition}
\label{sec:QAA-gap}

The first appearance of an adiabatic theorem in the context of quantum mechanics can be traced back to early works of Born and Fock~\cite{BoFo28}, later rephrased in more mathematical terms in~\cite{Kato50} (see~\cite{ChHoWi11,ElHa12} and references therein for more recent discussions); its common formulation in~\cite{Messiah} states that, for a system evolving according to the time-dependent Schr\"odinger equation (\ref{Hquantum}), in the absence of eigenvalue crossings, the system will follow the instantaneous ground state in the limit where the total evolution time $\TT$ tends to infinity. A more precise condition can be found in~\cite{Messiah}: let us define $\Delta(s) = E_1(s)   - E_0(s) $ the 
instantaneous gap
of the interpolating Hamiltonian that governs the annealing, and $b(s) = \left | \left \la 1 \left | \frac{d \hH(s)}{ds} \right | 0 \right \ra \right|$ which, once divided by $\Delta(s)$, gives the instantaneous angular speed of the ground state's eigenvector relatively to the first-excited state's eigenvector.  Then the
condition 
\beq\label{Time_annealing_exact}
\TT > \frac{1}{\epsilon} \frac{b(s)}{\Delta(s)^2} \hspace{1 cm } \forall s \in [0,1]
\eeq
ensures that the probability of not finding the system in 
the ground state of $\hH(1)=\hHf$ at the end of the evolution will be of order at most $\epsilon^2$. We will refer to an evolution time $\TT$ satisfying (\ref{Time_annealing_exact}) as an adiabatic time. In general, $b(s)$ can be thought as half the difference in slopes between the ground state's and first excited state's energies, and has no singular scaling with the system size $N$;
therefore, denoting $\Delta_{\rm min} = \min_{s \in [0,1]} \Delta(s) $ the 
minimum value of the gap during the annealing, (\ref{Time_annealing_exact}) can be replaced by the simpler condition\beq\label{Time_annealing}
\TT \gg {\cal{O}} (N \Delta_{\rm min}^{-2})
\ .
\eeq 
The time of the protocol is governed by the minimum gap and by its
scaling with $N$. 

The condition~(\ref{Time_annealing}) obviously breaks down for any time $\TT$ if the gap of the Hamiltonian vanishes (at finite $N$) for some value of the interpolation parameter, which is not expected to happen for $0<s<1$ for geometrical reasons~\cite{FaGoGuSi00}: the Hamiltonian $\hH(s)$ can be seen as a map from $[0,1]$ to a real space of dimension $2^{2N}$, in which the subspaces of operators with degenerate eigenvalues are hyperplanes of co-dimension 2. Therefore, in the absence of additional symmetries, no strict level crossings are to be expected and $\Delta_{\rm min}$ remains strictly positive.

A more subtle situation is encountered if the ground state of the final Hamiltonian $\hH_f$ is degenerate. In this case, the vanishing of the gap for $s$ getting close to $1$ is obviously not relevant for the adiabatic evolution of the system. The basic idea would be to modify the formulas for $\Delta(s)$ and $b(s)$ to consider only transitions between continuations of the classical ground states and first excited state(s). However, to the best of our knowledge, no precise formulation of the adiabatic theorem exists in this context. For the case in which the ground state of $\hH(s)$ is degenerate with the same degeneracy for \textit{all} values of the interpolation parameter $s$, sufficient and necessary conditions for adiabaticity have recently been proposed in~\cite{RiOr11},  extending the work of Wilczek and Zee~\cite{WiZe84}. Note that classically, for a certain class of NP-complete problem such as $k$-SAT, NP-completeness remains if one conditions on instances with a unique solution~\cite{uniqueSAT}. Hence, for a worst-case analysis, it is meaningful to study the behaviour of the QAA on these instances
with a \textit{Unique Satisfying Assignment} (USA). However, if one is interested in an ensemble of random instances (an average-case study), then
one should be careful that USA instances may not be typical for the problem considered, as will be further discussed in Sec.~\ref{sec:generating_USA}.

Finally, let us note that worst-case bounds building on the adiabatic theorem for diluted spin systems, as the one relevant for the optimization problems considered hereafter, were obtained in~\cite{review_Nishimori}, allowing to prove that, as for thermal annealing, the time for adiabaticity is never larger than an exponential in the system size.

\subsubsection{The finite-time Landau-Zener example}
\label{sec:QAA-LZ}

\begin{figure}
\centering
\includegraphics[width=.3\textwidth]{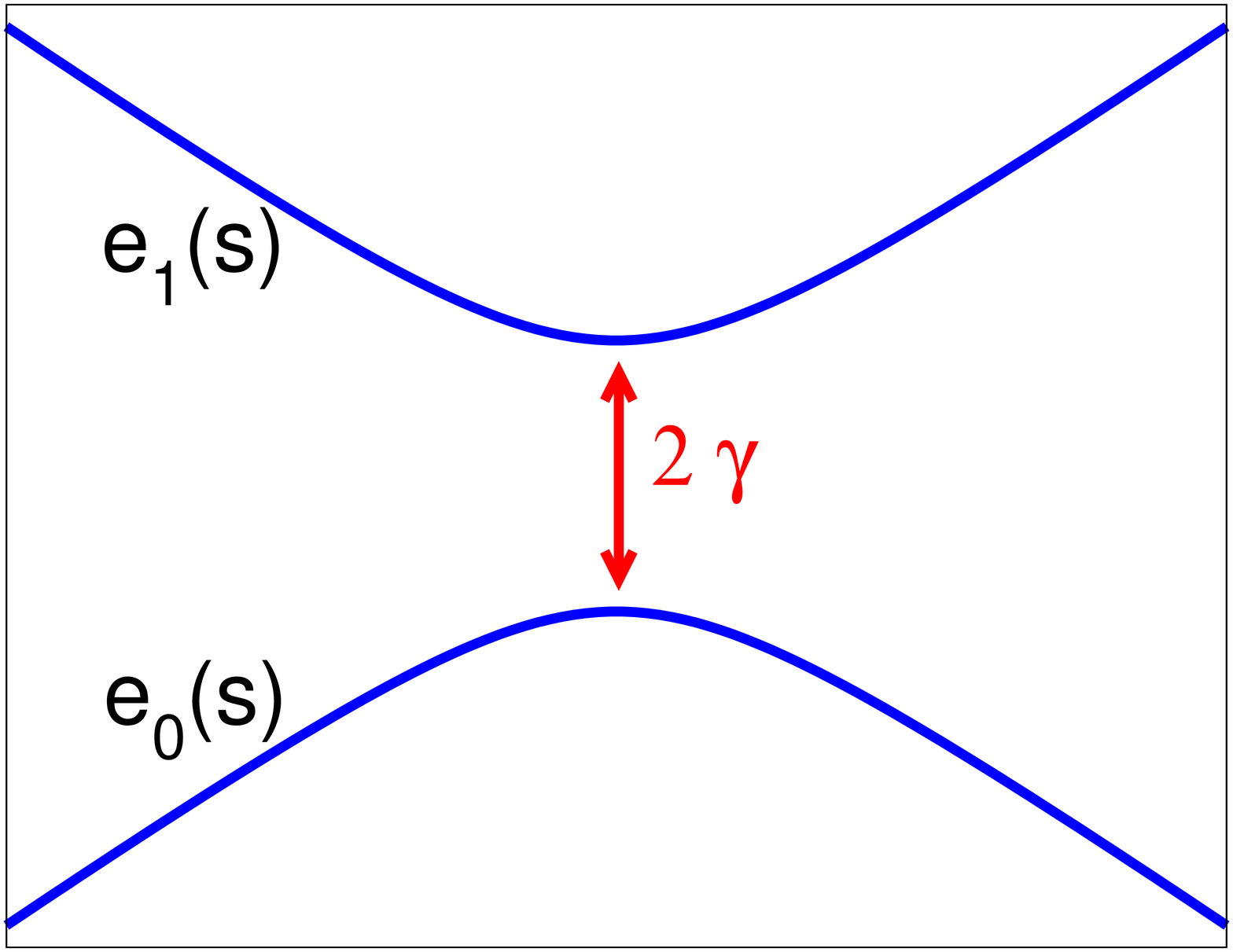}
\caption{Schematic representation of the eigenvalues of (\ref{two_level})
as a function of time $s$. Note the avoided level crossing in correspondence of the
minimum spectral gap. 
}\label{Fig:LZ}
\end{figure}

As explained above, in absence of special symmetries in $\hH(s)$ and when the size
of the system is finite (even if very large), true level 
crossings are not expected; but levels may still get extremely close, defining the appearance of \textit{avoided level crossings}. It is then very useful to consider the following ``reduced" Hamiltonian that describes such an avoided crossing (see Fig.~\ref{Fig:LZ}):
\beq\label{two_level}
\hH_{\rm LZ}(s) =  \left( \begin{array}{cc}
b (s-s^{\ast}) & \gamma \\
\gamma & - b (s-s^{\ast})  \end{array} \right)
\ .
\eeq

The instantaneous gap is 
$\displaystyle \Delta(s) = 2 \sqrt{\gamma^2 + b^2 (s-s^{\ast})^2}$, 
and even when
the two diagonal (``unperturbed") elements are equal in $s=s^{\ast}$ the states do not cross but are split by a gap $\Delta_{\rm min} = 2 \gamma$.
The advantage of this simplified formulation is that it is exactly solvable in the limit of an evolution going from $s= - \infty$ to $s= \infty$, the Landau-Zener formula \cite{landau32,zener32} giving the probability $P$
of a \textit{diabatic} transition to an excited state as $P = e^{ - 2 \TT \pi \gamma^2 /b}$,
which has for consequence the necessary condition for an adiabatic process 
$\TT \gg b \gamma^{-2} \simeq b \Delta_{\rm min}^{-2}$, which is precisely (\ref{Time_annealing_exact}). For evolutions of finite duration, as the ones relevant in our context, this formula has to be corrected~\cite{vitanov96,vitanov99} but the conclusions remain unchanged.

Finally, note that it is possible to extend this formula to consider several level crossings \cite{volkov06}, and to build on these exact results for this simplified model to make predictions for realistic systems involving an extensive or exponential number of levels~\cite{SaMaToCa02,BaSe12}.

\subsubsection{Universality of the quantum adiabatic algorithm}

It could seem at first sight that the quantum adiabatic algorithm has little
to do with the algorithms based on the quantum circuits model described
in Sec.~\ref{sec:quantum_circuits}; the former is based on a continuous
time evolution of the quantum computer, and is aimed at finding the
ground state of the Hamiltonian $\hHf$, while the latter class of algorithms
proceed via a discrete succession of unitary transformations, and encompass
a large variety of computational tasks. An equivalence between the two
paradigms has however been demonstrated, in the following sense. On the
one hand, a continuous time annealing procedure can be approximated, with
an arbitrary precision and with a polynomial overhead, by a series of
discrete transformations~\cite{FaGoGuSi00,DaMoVa01}. In the reverse 
direction, it was shown in~\cite{AhDaKeLaLlRe07} that any quantum circuit
model can be converted into a quantum annealing procedure, by using a
final Hamiltonian $\hHf$ introduced in~\cite{KiShVy02} whose ground state has 
a positive overlap with the final state of the original quantum circuit.
Moreover the minimal gap along this interpolation was proven 
in~\cite{AhDaKeLaLlRe07} to be only polynomially small in the number of
gates of the circuit, hence the requested time for the adiabatic algorithm
is polynomial in the size of the circuit. This result was strengthened
in~\cite{AhGoIrKe09}, which demonstrated that the annealing Hamiltonian
can be written with nearest-neighbor interactions on an unidimensional
lattice (the price to be paid being the internal dimension of the qudits that 
has to be larger than 9).

\subsubsection{Deficiencies and improvements of the quantum adiabatic 
algorithm}
\label{sec:Def-Imp}

The adiabatic theorem stated above provides a very simple and generic condition under which the quantum adiabatic algorithm is guaranteed to find a solution, if any, or at least a minimal energy configuration, to any given optimization problem. However, this does not mean that the algorithm will be efficient in finding this answer; in fact its performance will strongly depend on the possibly fast closing of the gap along the annealing path. 
A detailed discussion of these phenomena is the main focus of this paper. However, it is useful to give here
a short account of the main points to be discussed in the following.
As a first example, specific instances of $k$-SAT on which the QAA is inefficient because of an exponentially small minimum gap were constructed in~\cite{DaMoVa01,DaVa01}. More generally, we shall see in the following that for random optimization problems, two quite general mechanisms may cause gaps closing exponentially fast with the system size, and hamper the performances of the quantum adiabatic algorithm:
\begin{itemize}
\item The low energy states of the adiabatic Hamiltonian for $s=0$ and $s=1$ are very far away one from each other in the Hilbert space. One may in particular expect a \textit{spin glass} phase for $s$ close to $1$, and a \textit{quantum paramagnetic phase} for $s$ close to $0$, separated by a \textit{quantum phase transition}. Such a phase transition generically leads to a vanishing gap in the thermodynamic limit, with a scaling in the system size that depends on the order of the transition: in general, the gap closes polynomially fast if the transition is second-order, and exponentially fast if it is first order. We will come back on such quantum phase transitions in Sec.~\ref{sec:low_energy}, and in particular on their effects on typical constraint satisfaction problems in Sec.~\ref{sec:results}.
\item 
According to classical spin glass theory (Sec.~\ref{sec:classical_mean_field}),
typical difficult problem cost functions are characterized by
the existence of many very ``different'' minima (local or global), leading to a very complicated structure for the low energy phase of the final Hamiltonian.
It may happen that the addition of quantum fluctuations leads to many exponentially small gaps between different states even within the spin glass phase. This phenomenon will be further discussed in Sec.~\ref{sec:Altshuler}, and on a particular example in~\ref{sec:results_coloring}.
\end{itemize}

Before entering the more detailed discussion of the problematics related to 
the efficiency
of QAA for solving real optimization problems, let us note that the QAA setting is fairly general and leaves open a lot of directions for improvement. For instance, we assumed here that $\hH(s)$ interpolates 
linearly in time between $\hHi$ and $\hHf$; but more general interpolation rates can be considered, as will be discussed below. Another important freedom in the setting of the QAA is the choice of the initial Hamiltonian $\hHi$ 
(and of the final one, $\hHf$, if one is only interested in the satisfiability decision problem). 
The most general formulation of a QAA should thus be that of a smooth mapping from some interpolation range, that can be taken without restriction to be $[0,1]$, to the space of Hermitian operators on (some) Hilbert space, with the constraint that $s=1$ is mapped on the classical energy cost function, and $s=0$ to some operator defined \textit{without using information on the ground state of} $\hHf$, and such that its ground state is easy to prepare. Although we do not intend to give a more precise definition of these conditions here, their meaning should be clear on concrete examples. The latter formulation naturally maps the question of finding the best annealing path to a geometrical problem in a Hilbert space~\cite{zanardi10}.

Let us now come back to the unsorted database search introduced in \ref{sec:quantum_circuits}, to show on this simple example how a modification in the interpolation rate can lead to important changes of the adiabatic time of the QAA with fixed initial and final Hamiltonians. We will follow here the works of~\cite{Roland,CaMuFaSa09,DaMoVa01}. We recall that this problem can be seen as the search for the ground state of the classical Hamiltonian $\hHf = \wh I- | \ut \ra \la \ut |$, where $|\ut\ra$ is some fixed vector of the Hilbert space. Let us take for initial Hamiltonian $\hHi = \wh I - | \Psi_0 \ra \la \Psi_0|$, with $|\Psi_0\ra = 2^{-N/2} \sum_{\us} | \us \ra$ 
the uniform superposition of all the states of the Hilbert space defined in \ref{sec:quantum_circuits}. 
It can be seen that any state $|\us\ra - |\us'\ra$ with $\us,\us' \neq \ut$ is an eigenvector of $\hH(s)$ with eigenvalue 1 for all $s$. 
One can construct $2^N-2$ linearly independent such states, that are all orthogonal to both $|\ut \ra$ and $|\Psi_0 \ra$:
thus, only the subspace spanned by $|\ut \ra$ and $|\Psi_0 \ra$ is relevant for the adiabatic evolution. The Hamiltonian restricted to this subspace can easily be diagonalized, leading to a gap $\Delta(s) = \sqrt{1-4\left(1-2^{-N}\right)s(1-s)}$ which is minimal for $s=1/2$, resulting in $\Delta_{\rm min} =2^{-N/2}$ and in a growth of the adiabatic time proportional to $2^N$, which is also the duration of a naive exhaustive search. However, we know from the Grover circuit algorithm~\cite{Grover97} that a quantum computer is able to get a quadratic speed-up and to find the ground state of $\hHf$ in a time growing only as $2^{N/2}$. The reason why the QAA seems, in its naive setting, inefficient is that the condition (\ref{Time_annealing}) is realized only at one particular point of the spectrum ($s=1/2$) but leads to a constraint on the speed of evolution for all values of the interpolation parameter $s$, even when the gap is large and the annealing could be faster without inducing diabatic transitions. Therefore, it is better to make a more precise use of the condition (\ref{Time_annealing_exact}) and to do the evolution with the change of parametrization $\widetilde{H}(s) = \hH(\varphi(s))$, allowing to vary the speed of evolution as a function of the parameter $s$. Then, with the notations of (\ref{Time_annealing_exact}), using $\tilde{b}(s) \equiv  | \langle 1 | d \widetilde{H}/ds  | 0 \rangle | = b(\varphi(s)) d \varphi(s)/ds$ and the simple bound $b(s) \leq 1$, the adiabatic condition (\ref{Time_annealing_exact}) translates into:
\beq \frac{1}{\TT} \left | \frac{d \varphi(s)}{d s} \right| < \epsilon \Delta(\varphi(s))^2 \eeq
Solving this differential equation as a function of $\epsilon$ and $N$ with the boundary conditions $\varphi(0) = 0$ and $\varphi(1)=1$ allows to find the optimal annealing schedule $\varphi(s)$ and fixes the annealing time as $\TT \simeq \frac{\pi 2^{N/2}}{2 \epsilon}$, which is the expected quantum speed-up.
In general it is more difficult to find the best annealing rate for a fixed evolution time $\TT$; such questions are related to quantum optimal control, as presented in~\cite{CaMuFaSa09,CaCaFaSaGiMo11,NeMoEkSmFaCa11}. The intuitive idea is that, at least if the location of the gap is exactly known, it should always be possible to trade the scaling of the adiabatic time with $\Delta_{\rm min}^{-2}$ of~(\ref{Time_annealing}) into a scaling with $\Delta_{\rm min}^{-1}$, in the same fashion as was done for the Grover problem. On the other hand, it is easy to see that one cannot do better; in fact, the regime of $\TT \ll \Delta_{\rm min}^{-1}$ corresponds to the fast passage regime of~\cite{Messiah} in which the system is strongly diabatic. In particular, one cannot hope to change the scaling of the adiabatic time with the system size only by playing on the time-dependence of the evolution Hamiltonian.

The choice of $\hHi$ is more crucial, and for general optimization problems, it is mainly an open problem to understand whether the modification of the annealing path can change the scaling of the time needed for the adiabatic condition to hold. Such a possible change was argued for in~\cite{NiSe12,SeNi12} to avoid gaps at a first order phase transition for fully connected models by introducing a two-parameter annealing path
(see also~\cite{ribeiro06} for another example of a two-parameter annealing path). Alternatively, a randomization of $\hHi$ was proposed in~\cite{FGGGS10} to avoid gaps of the second type in the classification above, that appear within the classical spin glass phase, for a particular problem; but its efficiency for more general optimization problems is still an open question. This proposal will be further discussed in Sec.~\ref{sec:randomization}.  Let us finally emphasize that the existence, for a given problem, of an annealing path allowing for a fast adiabatic evolution is not enough if the time needed to find this particular path grows exponentially fast with the system size~\cite{DicAm11}.

\subsubsection{Quantum annealing without adiabaticity and 
approximation issues}

Finally, an important observation is that most of the works up to now 
focused on the efficiency of the quantum adiabatic algorithm in \textit{solving} exactly the problems, that is finding the ground state of the final Hamiltonian $\hHf$. However, the question of finding approximate solutions to an optimization problem is of great importance, both theoretically~\cite{vazirani2001} and practically. A convenient way to quantify the performance of a given algorithm in finding an approximate solution to an optimization problem is to introduce its \textit{residual energy} on a given time $\TT$, which is the difference between the lowest value of the cost function it can achieve in time $\TT$ and the absolute minimum of the cost function. A zero residual energy means that the algorithm can find a solution in time $\TT$, while the \textit{trivial} residual energy corresponds to the energy of a randomly chosen configuration, that can be achieved with $\TT\simeq0$. Between these two extreme cases, 
\textit{finite} ($N$ independent) and \textit{extensive} (proportional to $N$) residual energies shall also be distinguished.

This leads to the natural question of how the evolution time must grow with the system size for the residual energy to be under a given threshold. Classically, it is known that for certain hard problems such as $k$-SAT or $k$-XORSAT, obtaining a non-trivial residual energy can already require an exponentially long time~\cite{hastad01}. Hence a fast non-adiabatic
evolution has a computational interest if one can find a good compromise 
between the evolution time and the residual energy. In the classical case, hardness of approximation results can more generally be obtained via the PCP theorem \cite{pcp}. In the quantum complexity literature
a quantum analog of the PCP theorem has been conjectured 
in~\cite{qpcp_conjecture}. For recent works on the approximation algorithms in 
the quantum complexity setting we refer the readers 
to~\cite{qpcp_hastings,qapprox_kempe,qapprox_kempe2}.

Still, the performances of QAA in finding approximate solutions
remain widely unexplored.
Already in \cite{SaMaToCa02}, it was shown some evidence that QAA could 
outperform
classical simulated annealing within the same exponential scaling of the 
running time. 
To make more general theoretical predictions on the residual energy obtained by the QAA, it will be necessary to extend the relationship between the spectrum and the behavior of the quantum time evolution beyond the adiabatic criterion that focuses on the gap above the ground state. It has for instance been shown in~\cite{BaSe12} that the metastable continuation of the ground state that emerges after a first order phase transition for fully connected mean field models is particularly relevant for quantum evolution on sub-exponential time scales, and leads to extensive residual energies for evolution times that do not grow as fast as the time for adiabaticity (\ref{Time_annealing}). Such a connection is expected to have a wider range of validity; in particular to hold for random optimization problems as the ones studied hereafter, although no quantitative prediction has been obtained yet for these models.

\section{Classical random optimization problems and their
connection with mean field spin glasses}
\label{sec:classical_mean_field}

\subsection{Optimization in the typical case, and the statistical
  physics of disordered
  systems}
\label{sec:Optimization-Statistical-Mechanics} 

The theory of classical computational 
complexity~\cite{GareyJohnson,Pa83,Papadimitriou94}
that we described in Sec.~\ref{sec:classical_complexity} considers the
difficulty of a problem in the worst-case. For instance the
fact that $q$-coloring belongs for $q\ge 3$ to the NP-complete class
means that at present there is no polynomial-time algorithm able to decide the 
colorability of \emph{every} possible graph. However this does not
mean that all the graphs are equally difficult, and in fact for many
NP-complete problems there exist algorithms that do work efficiently on a 
large set of instances. This raises the question ``where are the
{\it really} hard instances of NP
problems~\cite{CheesemanKanefsky91}?'', and how to construct such hard
instances efficiently.

The idea of using random instances of Constraint Satisfaction
Problems (CSP) as benchmarks for algorithms
emerged in the 80's; however the first ensembles proposed
turned out to contain mostly easy instances for known algorithms, and
it was only at the beginning of the 90's that two seminal papers
by Cheeseman, Kanefsky and Taylor~\cite{CheesemanKanefsky91} and
Mitchell, Selman and Levesque~\cite{MitchellSelman92} introduced the random
ensembles that answered positively the question above. Instances from
these ensembles are actually very simple to describe: in the coloring
case one creates a graph by selecting uniformly at random $M$ edges among
the $\binom{N}{2}$ possible ones, i.e. one constructs 
an Erd\H{o}s-R\'enyi $G(N,M)$ random 
graph~\cite{Janson}. The large-size limit ($N\to \infty$) has to be performed
with $M$ growing like $N$, in other words the thermodynamic limit for
these instances is parametrized by a finite real number $\alpha=M/N$.
For $k$-SAT random instances the construction is generalized to random
hyper-graphs, the $M$ $k$-uplet of indices in Eq.~(\ref{def:SAT})
being chosen uniformly at random among the $\binom{N}{k}$ possible ones,
and the signs $J_a^j$ of the corresponding literals are chosen to be $\pm 1$
with probability $1/2$. Again the large-size limit is taken with
$\alpha=M/N$ fixed. The authors of~\cite{CheesemanKanefsky91,MitchellSelman92}
performed extensive numerical experiments on such randomly generated instances.
Using complete algorithms they determined the probability $P(\alpha,N)$ that
an instance with $N$ variables and $M=\alpha N$ constraints has a ground state
of vanishing energy (i.e. is $q$-colorable, or satisfiable depending on the
case). This probability is obviously a decreasing function of $\alpha$: 
it can only become harder to satisfy all the constraints as their number is increased.
What came as a surprise at that time is the
fact that for larger values of $N$ the probability of satisfiability decreased
in a steeper and steeper way, which suggested the following 
\emph{satisfiability conjecture}:
\beq
\lim_{N \to \infty} P(\alpha,N) = 
\begin{cases} 1 & \text{if} \ \alpha < \alpha_{\rm s} \\
0 & \text{if}\ \alpha > \alpha_{\rm s}
\end{cases} \ ,
\label{eq:SAT-conjecture}
\eeq
where $\alpha_{\rm s}$ is some fixed threshold value, that depends on the
problem considered (coloring or satisfiability), and on the parameters $k,q$.
In more physical terms this threshold phenomenon corresponds to a phase
transition between a SAT (or COL) phase where almost all instances are
satisfiable (colorable) and their ground state energy is zero 
to an UNSAT (UNCOL) phase in which almost none of
them is, and the average ground state energy is positive.
Moreover the hardest instances, in terms of the time required for the
algorithms to decide their satisfiability, are those with 
$\alpha \approx \alpha_{\rm s}$: for $\alpha \ll \alpha_{\rm s}$ the problem
is under-constrained, and it is easy to find configurations satisfying all
the constraints simultaneously, while for $\alpha \gg \alpha_{\rm s}$ there
are so many constraints that it becomes (relatively) easy again to discover
an unavoidable contradiction between them.

Since their introduction these ensembles have been the subject of a very
important research effort in computer science, discrete mathematics, and
statistical physics; they have played (and still do) a prominent role
in understanding the origin of algorithmic hardness. The rigorous
works on this problem were first aimed at the proof of the satisfiability
conjecture~(\ref{eq:SAT-conjecture}) and the determination of the
threshold $\alpha_{\rm s}$. The main outcomes of this line of research have
been a proof of a weaker version of~(\ref{eq:SAT-conjecture}) where
$\alpha_{\rm s}$ is allowed to depend on $N$~\cite{Friedgut}, and rigorous 
upper and lower bounds on 
$\alpha_{\rm s}$~\cite{transition_lb,Achltcs,transition_ub}.
These bounds are asymptotically tight when $k,q$ get 
large~\cite{transition_largek}.
In addition statistical mechanics techniques, starting 
from~\cite{MoZe}, have also been applied to these
problems and have led to quantitative computations of the value of
$\alpha_{\rm s}$~\cite{MeZe,MezardParisi02,MeMeZe,col1}. Moreover these studies
unveiled several new qualitative features besides the
satisfiability transition at the threshold $\alpha_{\rm s}$; it has been shown
in particular that in the SAT phase $\alpha < \alpha_{\rm s}$ there exist
further phase transitions~\cite{MezardParisi02,BiMoWe,KrMoRiSeZd}
that affect the organization of the solutions of 
the random CSP in the configuration space, and that are at least as relevant
as the SAT-UNSAT transition to understand the algorithmic hardness.
Note that even if the statistical mechanics techniques are not rigorous
from a mathematical point of view, many of the insights they offered on the
features of random CSP have later been turned into mathematically
rigorous statements~\cite{FrLe,PaTa,MoraMezard05b,clus_rig_Fede,Coja11}.

Our goal in the remaining of this section is to review the picture of
random CSP that has been obtained by physics methods
(see~\cite{Mo07,MM09,BookCrisMoore} for textbook presentations). 
Let us first
explain in generic terms why statistical mechanics is a natural tool
for their study, besides the superficial analogy between the satisfiability
threshold phenomenon and phase transitions of real materials.
As explained with the dictionary introduced in Sec.~\ref{sec:QAA_def}, 
the cost function $E(\us)$ for one instance of a CSP can be viewed as an 
energy function; turning to random CSP, this energy function becomes itself 
a random object. Physical systems defined via random constructions have been
studied for decades in physics (an early example being the Anderson 
model~\cite{An58} of localization); in that context the randomness in the
energy function of one instance (for instance the choice of the graph in 
random coloring) is usually called quenched disorder of that sample. 
Random CSP can thus be studied from
the perspective of the statistical mechanics of disordered systems. Moreover
they belong to the so-called mean field class of models, because their 
structure is unrelated to a finite-dimensional physical space: in the 
Erd\H{o}s-R\'enyi definition of a random graph all pairs of vertices have the same
probability to become neighbors (i.e. be linked by an edge), there is
no a priori Euclidean distance between them.

In order to make the results on random CSP
accessible to readers not acquainted with the field of statistical mechanics
of disordered systems we shall make a detour and first discuss simpler models,
introducing the necessary ingredients progressively.
In Sec.~\ref{sec:mfsg} we shall introduce the disordered physical systems that 
are most relevant to this discussion, namely spin glasses, and discuss
the various kinds of mean field models. Then in Sec.~\ref{sec:REM} we present
the random energy model~\cite{De81}, the simplest disordered model
that yet displays a phase transition important for the following discussions.
In Sec.~\ref{sec:pspinFC} we move on to a slightly more complicated model,
the so-called fully connected $p$-spin model, and discuss its interpretation
in terms of the physics of glasses. We then come back to the main focus of our
interest, i.e. random CSP; in Sec.~\ref{sec:ch3-XOR} we discuss random
instances of the XORSAT problem, followed in Sec.~\ref{sec:subcubes} by a
presentation of a toy model that exhibits, in a 
controlled way, the transitions of the random $k$-satisfiability and 
$q$-coloring model. The latter are discussed in 
Sec.~\ref{sec:transitions_rCSP}, without entering into technical details
of their derivations, some of which will be given in 
Sec.~\ref{sec:classical_cavity}.
In Sec.~\ref{sec:thermalannealing} we will discuss the consequences
of these transitions for thermal annealing.
Finally in Sec.~\ref{sec:generating_USA} we discuss the generation
process of random CSP, with a particular interest on ensembles of instances with
a Unique Satisfying Assignment (USA); these have a special
interest as benchmarks for the quantum adiabatic algorithm.

\subsection{Mean field spin glasses}
\label{sec:mfsg}

Spin glasses can be prepared as alloys of two elements, with a small fraction
of a magnetic element (Fe for instance) being added to a metallic host with no 
magnetic properties (Au). This mixture is prepared as a liquid phase at high
temperature; when the sample is cooled down and becomes a solid the position
of the magnetic impurities becomes frozen (quenched) to a random location.
The magnetic moments (spins) carried by the impurities interact with one
another, but, depending on the distances between them, their pairwise
interactions can be either ferromagnetic, favoring the alignment of the two
spins, or antiferromagnetic, forcing them to point in opposite
directions. When the temperature is varied there appears in these compounds
a phase transition for the magnetic degrees of freedom. The low
temperature phase is an unusual state, with frozen moments but no periodic
order; hence, the name spin glass, by analogy with amorphous window
glass, slow to respond to changes in external controls, accompanied by
non-ergodicity, behaving differently depending on the order in which
external perturbations, such as magnetic field or temperature, are
applied. Nowadays, the expression ``spin glass'' is however used much
more broadly to refer to systems that exhibit glassiness owing to the
combination of quenched disorder and frustration.

In 1975 Edwards and Anderson (EA) introduced in~\cite{EdwardsAnderson75}
a model for spin glass materials, in which the magnetic moments are modeled 
by Ising variables $\s_i=\pm 1$, lying on a regular finite-dimensional 
lattice and interacting via the energy function
\beq
E_{EA}(\us) = - \sum_{\la i,j \ra} J_{ij} \s_i \s_j \ ,
\eeq
where the sum runs over the pairs of neighboring spins $i$ and $j$.
The couplings $J_{ij}$ are chosen at random from a given distribution (for 
instance a Gaussian one) that allows both positive (ferromagnetic) and
negative (antiferromagnetic) values for $J_{ij}$. Statistical models
defined on finite-dimensional lattices are very difficult to treat 
analytically, even in the pure case without disorder. The situation only
becomes worse with the inclusion of disorder, and no analytic solution
of the Edwards-Anderson model can be hoped for in dimensions larger than 1.

The usual prescription of field theory is to start working out the
mean field version of a model (usually qualitatively correct in large 
dimensions). For spin glasses this was first investigated, 
after~\cite{EdwardsAnderson75}, by Sherrington
and Kirkpatrick (SK) \cite{SK75}. In the SK model
$N$ Ising spins interact with the energy function
\beq 
E_{SK}(\us) = - \sum_{i < j} J_{ij} \s_i \s_j \ , 
\label{eq:E_SK}
\eeq 
where the sum is now over {\it all} couples $i \neq j$ (making the graph of
interaction a fully connected, or complete, one). For the
thermodynamic limit to be well-defined the random couplings
$J_{ij}$ must be individually weak, for instance they can be chosen to be
Gaussian with zero mean and variance of order $1/N$.

This model has played a fundamental role in the theory of spin glasses. 
Despite its mean field character the quenched disorder in its definition
makes the computation of its free energy a very difficult problem, that
was only solved in 1980 by Parisi~\cite{Pa80}, via the development of
the replica method in its Replica Symmetry Breaking (RSB) form. The SK
model exhibits a phase transition from a high temperature, paramagnetic phase,
to a spin glass phase at low temperature, characterized by a proliferation
of metastable states in a very complex free energy landscape. 
The reader is referred to the books~\cite{Beyond,FH91} for details
on the replica method and the original works on the characterization of
the spin glass phase of the SK model. The methods originally employed by
physicists for the resolution of this model were highly non-rigorous.
However the value of the free energy computed by Parisi was rigorously 
proven to be exact, much more recently, by Talagrand~\cite{Ta06}, building
on the interpolation method of Guerra and Toninelli~\cite{GuTo02}.

Let us introduce here some variants of the SK model and set up some terminology
that will often appear
in the rest of the discussion. A first twist on Eq.~(\ref{eq:E_SK}) consists
in promoting the pair-wise interactions to $p$-wise couplings, leading to
\begin{equation}
E(\us)=-\sum_{i_{1}<\ldots <i_{p}} J_{i_{1} \dots i_{p}}
\s_{i_{1}} \dots \s_{i_{p}} \ , 
\label{eq:E_pspinFC}
\end{equation}
where the sum is over all $p$-uplets of spins, and the couplings 
$J_{i_{1} \dots i_{p}}$ are Gaussian random variables of zero mean and variance
of order $1/N^{p-1}$. This model is known as the fully connected $p$-spin
model, and was first introduced and studied in~\cite{De81,GrossMezard84};
the replica theory developed for the SK model is also applicable to this model,
that we shall discuss slightly further in Sec.~\ref{sec:pspinFC}.
The models defined in Eqs.~(\ref{eq:E_SK}), (\ref{eq:E_pspinFC}) have a mean field
nature, because each variable $\s_i$ interacts (weakly) with all the other 
variables, destroying completely any notion of finite-dimensional distance
between the variables; this class of models defined on complete graphs are
usually called fully connected mean field models. 

But as we already mentioned 
there exists another class of mean field models, dubbed sparse, or diluted, or
finitely-connected, in which each degree of freedom interacts strongly (i.e. 
with a coupling of order 1) with a finite number of neighbors, the latter
being chosen in some random way unrelated to a finite-dimensional space.
For instance Viana and Bray~\cite{VB85} considered a model of pairwise 
spin glass interactions along the edges of an Erd\H{o}s-R\'enyi random graphs 
(as defined in Sec.~\ref{sec:Optimization-Statistical-Mechanics}). More
generically a finitely-connected mean field model can be defined
with other types of sparse random graphs (or hypergraphs to 
include interactions between more than two spins), as long as the connectivity
(degree) of each vertex remains finite. One way to define a random graph
probability law is to impose its degree distribution, i.e. to generate
uniformly at random a graph among those that have a prescribed fraction
$q_0$ of isolated vertices, $q_1$ of vertices adjacent to a single edge,
and so on and so forth. A particular case of this construction that will be
met often in the following is the regular one: a random $c$-regular graph
is a graph chosen uniformly at random among all the graphs in which each
vertex has exactly $c$ neighbors (or belong to exactly $c$ $k$-uplets for the
hypergraph generalization). 

All these sparse random graphs models share a
crucial property: they are locally tree-like. In other words if one selects
an arbitrary vertex $i$ in a sparse random graph of size $N$, with a probability
which goes to one in the limit $N\to \infty$ the shortest loop around $i$ will
be larger than any fixed length~\cite{Janson}. Statistical mechanics models defined on trees
are trivial: they can easily be solved by recurrence (somehow like in 
unidimensional models with the transfer matrix method). The richness of the
models defined on random graphs comes from a subtle combination between
their locally tree-like character, and the existence of long loops. The latter
are very important in creating self-consistent boundary conditions, and in
avoiding the pathologic surface to volume ratio of tree models. Sparse random
graphs are sometimes called Bethe lattices, in honour of the Bethe 
approximation that becomes exact on trees; note however that this terminology
can be misleading, some authors using it as a synonym for infinite Cayley 
trees, some restricting it to the case of regular random graphs.

From the introduction to random CSP of 
Sec.~\ref{sec:Optimization-Statistical-Mechanics} it should be clear that
the diluted mean field models will ultimately be more useful in this respect
than the fully connected ones (though other optimization problems, not 
described here, are defined on complete 
graphs~\cite{MezardParisi85,FuAnderson86}). They are 
unfortunately much more difficult from a technical point of view. The
replica method could be adapted to deal with sparse random graphs 
(see~\cite{replica_diluted} and references therein) but yields functional 
equations under a form that is not directly amenable to numerical resolution.
An alternative formulation was developed under the name of cavity 
method~\cite{cavity,cavity_T0} and allowed to bypass this difficulty. This
method, that we shall review in Sec.~\ref{sec:classical_cavity}, yields
formally exact predictions for the thermodynamic limit of the free energy
of models defined on random (hyper)graphs, even if in some cases the 
extraction of actual numbers out of the method can be difficult.

Before getting to the discussion of the picture of random CSP provided by
statistical mechanics studies let us discuss, as announced above, simpler
models of disordered systems.

\subsection{The random energy model}
\label{sec:REM}

By definition the energy function $E(\us)$ of a disordered system, by
contrast with a pure or ordered one, is a random object. For generic local cost functions
(in the sense of Sec.~\ref{sec:QCC}, i.e. that are a sum of terms each
involving a finite number of spins),
the energies of the $2^N$ configurations (for Ising spins) are random variables,
correlated one with the other: for instance in the SK model (\ref{eq:E_SK})
the number of independent couplings $J_{ij}$ is only of order $N^2$. It is
however very instructive to study the simplified (but non-local) Random Energy Model (REM) of
Derrida~\cite{De81}, which keeps the random character of the energy
function but discards the correlations between the energies $E(\us)$ of
various configurations. More precisely,
in the REM one assigns to each of the $2^N$ configurations an energy
$E(\us)$ drawn independently at random with a Gaussian distribution
of density
\beq
P(E)=\frac{e^{-\frac{E^2}N}}{\sqrt{N\pi}} \ .
\eeq
The simplest way to solve this model is to use the micro-canonical
ensemble. Let us denote $n(E) \dd E$ the number of energy levels belonging to 
the
interval $(E,E+\dd E)$; its average over the realizations of the disorder
(the choice of the energies) is easily computed: 
\beq
\overline{n(E)} = 2^N P(E) \sim 
e^{N\left((\log\!2-E^2/N^2\right))} = e^{Ns(E/N)} \ ,
\eeq
where $\sim$ denotes here equality at the leading exponential order
when $N \to \infty$, and the micro-canonical entropy $s(e)$
for the reduced intensive energy $e=E/N$ is
\beq
s(e)=\log\!2-e^2 \ .
\eeq
This function is positive on the interval $[e_0,-e_0]$, with 
$e_0=-\sqrt{\log\!2}$; for energies $E$ corresponding to this interval
$\overline{n(E)}$ is exponentially large and the random variable $n(E)$
is thus typically close to its average (with fluctuations of order 
$\overline{n(E)}^{1/2}$). On the other hand if $E$ is outside the interval
the average number $\overline{n(E)}$ is exponentially small, hence in the
vast majority of samples the number $n(E)$ is equal to zero. The typical
value of the free energy can then be computed by the Legendre transform
of the typical micro-canonical entropy:
\beq
f_{\rm REM}= -\frac{1}{\beta} \lim_{N \to \infty} \log \int_{e_0}^{-e_0}
 e^{N[-\beta e + s(e)] } \dd e = \inf_{e \in [e_0,-e_0]} [e - T s(e) ] \ ,
\label{eq:f_REM}
\eeq
where we evaluated the integral by the Laplace method.
A transition between two regimes thus arises at a critical temperature
$T_{\rm c}$ such that $\frac
1{T_{\rm c}}=\frac{ds(e)}{de}\big|_{e_0}=2\sqrt{\log\!2}$ and the
thermodynamic behavior of the model follows: 
\begin{itemize}
\item {\rm i)} For $T<T_{\rm c}$,
$f_{\rm REM}=-\sqrt{\log\!2}$ and the system is frozen in its lowest
energy states (the integral in (\ref{eq:f_REM}) is dominated by the lower 
edge $e_0$ of the integration domain). One can show that only a finite number of configurations (and
only the ground state at $T=0$) contribute significantly to the partition sum 
(see for instance \cite{MM09,Ta03}). The energy gap
between them is finite.
\item  (\rm ii) For $T>T_{\rm c}$, $f_{\rm REM}=-\frac
1{4T} - T \log\!2$; exponentially many configurations contribute to the
partition sum. 
\end{itemize}
The free energy of the model is thus non-analytic at $T_{\rm c}$. This
phase transition is often called a ``condensation'' transition (or Kauzmann
transition in the context of glasses, see below), because it separates
a high temperature phase in which the Gibbs-Boltzmann distribution is spread
over an exponential number of configurations from a low temperature
phase where this support condenses on a much smaller number of configurations.
This kind of transition appears, with additional subtleties, in many of
the more complicated mean field disordered systems that we shall discuss in the
following.

\subsection{The fully connected $p$-spin model}
\label{sec:pspinFC}

The next model we would like to discuss is the fully connected
$p$-spin model of Eq.~(\ref{eq:E_pspinFC}), studied in particular
in~\cite{GrossMezard84} (see~\cite{CC05} for an extensive pedagogical 
discussion). We assume $p\ge 3$ here, the case $p=2$ of the SK model
being qualitatively different.
At variance with the REM, this is a $p$-local cost function (it is a sum of $p$-spin interactions) and 
the energies of the various configurations are correlated;
flipping one of the spins does not completely change the energy, some 
continuity is preserved in the energy landscape of the configurations. There
is however one common feature with the REM (which is actually the $p\to \infty$
limit of the $p$-spin model~\cite{De81}): it also exhibits a condensation transition at
some temperature $T_{\rm c}$, accompanied by a non-analyticity of the 
free energy. The difference is that in the low temperature phase the 
Gibbs-Boltzmann measure is supported by a small number of ``pure states''.
The latter, that take the place of the low energy single configurations of
the REM, are whole sets of exponentially many correlated configurations. Each
pure state therefore has an extensive ``internal'' free energy with both 
energetic and entropic contributions. 
It is not easy to give a clear-cut
definition of pure states in mean field disordered systems; the reader might
want to think about them as a generalization of the pure states of 
low temperature ferromagnets with positive/negative magnetizations. The 
partition of the configuration space into pure states has both static and
dynamic characterizations: long-distance connected correlation functions
vanish inside one pure state, and the dynamics remains trapped for a long time
in the pure state it started in. For a given realization of the disorder
let us index with $\gamma$ the pure states on which the system is decomposed.
The partition function can be written as a sum over the pure states,
\beq
Z= \sum_\gamma Z_\gamma \ , \quad 
Z_\gamma = \sum_{\us \in \gamma} e^{-\beta E(\us)} \ , \quad
f_\gamma = - \frac{1}{N \beta} \log Z_\gamma \ ,
\label{eq:ps-decomposition}
\eeq
where we denoted $f_\gamma$ the internal free energy density of the pure state
$\gamma$. In mean field disordered systems,
at sufficiently low temperature, there exist exponentially many pure states,
whose internal free energy density can vary in an interval 
$[f_{\rm min},f_{\rm max}]$; one defines a complexity, or configurational entropy,
$\Sigma(f)$, such that $e^{N\Sigma(f)}$ gives, at the leading exponential order,
the number of pure states with internal free energy $f$. Then the computation
of the free energy is a generalization of (\ref{eq:f_REM}),
\beq
f= -\frac{1}{\beta} \lim_{N \to \infty} \log \int_{f_{\rm min}}^{f_{\rm max}}
 e^{N[-\beta f_{\rm int} + \Sigma(f_{\rm int})] } \dd f_{\rm int} = 
\inf_{f_{\rm int} \in [f_{\rm min},f_{\rm max}]} 
[f_{\rm int} - T \Sigma(f_{\rm int}) ] \ .
\eeq
The condensation transition is thus due here to a competition between the
internal free energy of the pure states and their degeneracy (configurational
entropy); at low temperatures the integral becomes dominated by the lower
edge $f_{\rm min}$ of the integration domain, where the configurational entropy 
generically vanishes (note that in general $\Sigma(f)$ also depends on the
external parameters like the temperature), and remains zero at lower
temperatures.

Another difference between the REM and the $p$-spin model is the existence
of another transition at a higher temperature $T_{\rm d}>T_{\rm c}$. This
so-called ``dynamic'' temperature marks the appearance of pure states
inside the Gibbs-Boltzmann measure; for higher temperature the space of 
configurations is essentially connected and ergodic, only for $T<T_{\rm d}$
the decomposition in pure states is relevant and the complexity
$\Sigma(f)$ non-trivial. This transition has thus a direct impact on the
dynamics of the system: for $T<T_{\rm d}$ it is not possible for a physical 
dynamics to equilibrate and the ergodicity is broken.
However the free energy has no singularity at this temperature.

This model has played a very important 
role~\cite{kirkpatrick:87,kirkpatrick:88,KW87,KTW89} in the development of a 
first principle theory of the structural glass transition, known as the
Random First Order Transition (RFOT) theory, see~\cite{LW07,Cavagna09,BB09,BB11} for recent reviews.
In this scenario,
the dynamic transition at $T_{\rm d}$, with no impact on the statics, 
corresponds to the transition of the Mode Coupling Theory 
(MCT)~\cite{gotzebook}, while the condensation transition at $T_{\rm c}$
is an idealization of the thermodynamic glass transition envisioned by
Kauzmann~\cite{Kauzmann48}.

It is worth mentioning that in many models, a third
phenomenon is observed as the temperature is further lowered, called
the Gardner transition~\cite{Gardner85}.  It is a transition towards a
more complicated phase, similar to the one found in the
Sherrington-Kirkpatrick model~\cite{Beyond}.

\subsection{The random XORSAT model}
\label{sec:ch3-XOR}

\begin{figure}
\centerline{\includegraphics[width=8cm]{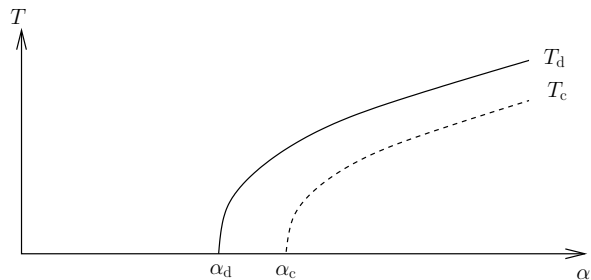}}
\caption{A sketch of the phase diagram of the random XORSAT model in the 
$(\alpha,T)$ plane.}
\label{fig:sketch_XORSAT}
\end{figure}

Let us now come back to our main topic, namely the behavior of random
optimization problems, and consider them in the perspective of the mean field
disordered models we have just discussed. We shall first emphasize
the striking similarity between the energy function (\ref{def:XORSAT}) of the 
XORSAT model and the one of the $p$-spin model (\ref{eq:E_pspinFC}): both
are written as sums of products of Ising spin variables; for historical reasons
the number of spins involved in each interaction is called $k$ or $p$ depending
on the context. Apart from this minor conventional difference, the main
discrepancy between the two cases is the structure of the interactions 
involved: in the random XORSAT problem there are $M=\alpha N$ interactions
with couplings of order 1, defining an Erd\H{o}s-R\'enyi random hypergraphs, while
all the $\binom{N}{p}$ possible couplings are present in (\ref{eq:E_pspinFC}),
with individual strengths vanishing in the thermodynamic limit. Despite this
difference both models are mean field, and share most of their phenomenology.
In the XORSAT case there are two external parameters: the temperature $T$,
and a ``geometrical'' parameter $\alpha$, that controls the number of 
constraints put between the variables. It has been shown, in particular in
\cite{xor_1,xor_2,MoSe}, that the phase diagram in the $(\alpha,T)$ plane is
divided in three regimes, separated by two transition lines 
$\alpha_{\rm d}(T)$ and $\alpha_{\rm c}(T)$, see Fig.~\ref{fig:sketch_XORSAT} for a schematic 
representation. These two lines are the counterpart of the two transition
temperatures $T_{\rm d}$ and $T_{\rm c}$ discussed above in the context of the fully connected
$p$-spin model (that is recovered in the $\alpha \to \infty$ limit). In
the high temperature/low $\alpha$ phase, the configuration space is 
well connected and ergodic; in the intermediate phase it becomes split
in an exponential number of pure states, yet no singularity appears in the
thermodynamic functions on the line $\alpha_{\rm d}(T)$; the thermodynamic
phase transition lies on the condensation line $\alpha_{\rm c}(T)$.

We shall now concentrate on the zero temperature limit of the XORSAT model,
i.e. on the properties of its ground state configurations, that are obviously
the most relevant ones when the model is viewed as an optimization problem. The two 
transition lines have finite limits when $T\to 0$, that we shall denote
$\alpha_{\rm d}$ and $\alpha_{\rm c}$; their expression as a function of $k$
can be found in~\cite{xor_1,xor_2}. It turns out, in this particular case,
that $\alpha_{\rm c}=\alpha_{\rm s}$, where $\alpha_{\rm s}$ is the SAT-UNSAT threshold defined in 
Eq.~(\ref{eq:SAT-conjecture}). The dynamic transition $\alpha_{\rm d}$ is
called clustering transition in this context, and similarly the pure states
introduced above become \emph{clusters} of solutions, i.e. sets of close-by
solutions, well separated from each other.

The XORSAT problem has some specific features that allow for an explicit
definition of clusters which can be explained as follows~\cite{xor_1,xor_2}. Consider an 
arbitrary XORSAT formula, and suppose that one of the variables $\s_i$
appears in a single interaction, call it $a$. A moment of thought reveals
that the formula is satisfiable if and only if the formula, with the
interaction $a$ removed, is satisfiable. One can iterate this process and
reduce further the formula, removing at each step the interactions in which
appears a variable of degree 1. At the end of this ``leaf-removal'' process
one ends up with a reduced formula called the 2-core of the original one.
Two cases can occur: either the 2-core is empty, and then the original formula
is obviously satisfiable. One can assign satisfying values for the variables 
in the last removed interaction, and then reintroduce the interactions in the
reverse order of the removal, using the fact that at each step at least one 
variable (the leaf) can be freely chosen to satisfy the re-introduced
interaction. This case occurs with high probability when 
$\alpha < \alpha_{\rm d}$, and one can 
show~\cite{IKKM2011,achlioptas2011} that all the solutions that
can be constructed from the free choices are in some precise sense close one 
to each other. On the other hand, when $\alpha > \alpha_{\rm d}$ the 2-core
contains typically an extensive number of variables and interactions. This
reduced formula, in which all variables are involved in at least two 
interactions, goes from satisfiable to unsatisfiable at the higher threshold
$\alpha_{\rm s}$~\cite{xor_1,xor_2}. Let us consider the intermediate regime
$\alpha \in [\alpha_{\rm d},\alpha_{\rm s}]$, where the reduced formula on the
2-core is non-trivial but still has some solutions. A very important point
is that two distinct solutions $\us$ and $\us'$ of the 2-core formula are far 
away from each
other, in the Hamming distance sense (the number of different variables between
them). Indeed, because of the form of the constraints of the XORSAT problem,
each interaction must contain an even number of spins $i$ with 
$\s_i\neq \s'_i$. In other words, as the 2-core does not contain leaves, a 
loop of disagreeing spins between $\us$ and $\us'$ has to be closed. As the 
random graphs are locally tree-like such a loop has necessarily a length
diverging with $N$ in the thermodynamic limit. Now, from every solution
of the 2-core reduced formula one can construct different solutions of the
full formula, by reintroducing the interactions in reverse order, as explained
above. All these solutions that emerge from the same seed, i.e. from the
same solution of the 2-core, will be said to belong to the same cluster 
(or pure state); then one realizes that solutions inside one cluster are
close to each other, while solutions belonging to distinct clusters are 
necessarily separated by a large Hamming distance. To finish the connection
with the phenomenology of the $p$-spin fully connected model, let us call
$e^{N\Sigma(\alpha)}$ the number of solutions of the 2-core formula for a random
instance with parameter $\alpha$; $\Sigma(\alpha)$ is defined in the interval
$[\alpha_{\rm d},\alpha_{\rm c}=\alpha_{\rm s}]$, precisely like the intermediate
regime of temperature $[T_{\rm c},T_{\rm d}]$ of Sec.~\ref{sec:pspinFC}. Moreover
the complexity (or configurational entropy) $\Sigma$ that counts the number
of relevant clusters vanishes at the transition $\alpha_{\rm c}$, similarly
to the condensation on a sub-exponential number of pure states for 
$T < T_{\rm c}$.

\begin{figure}
\centerline{\includegraphics[width=.7\textwidth]{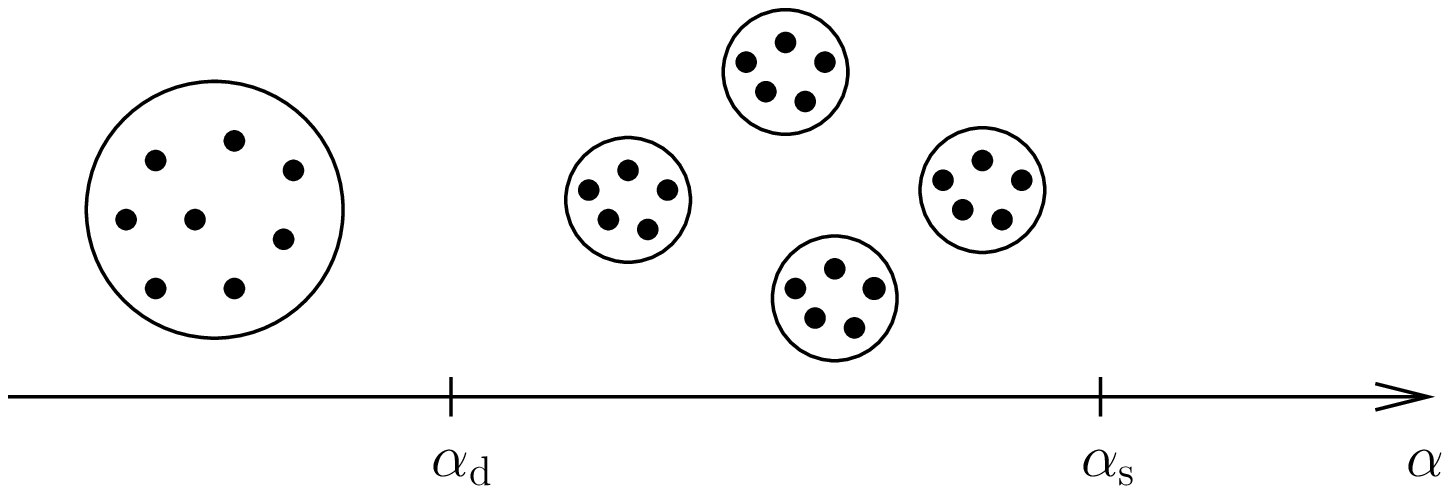}}
\centerline{\includegraphics[width=.7\textwidth]{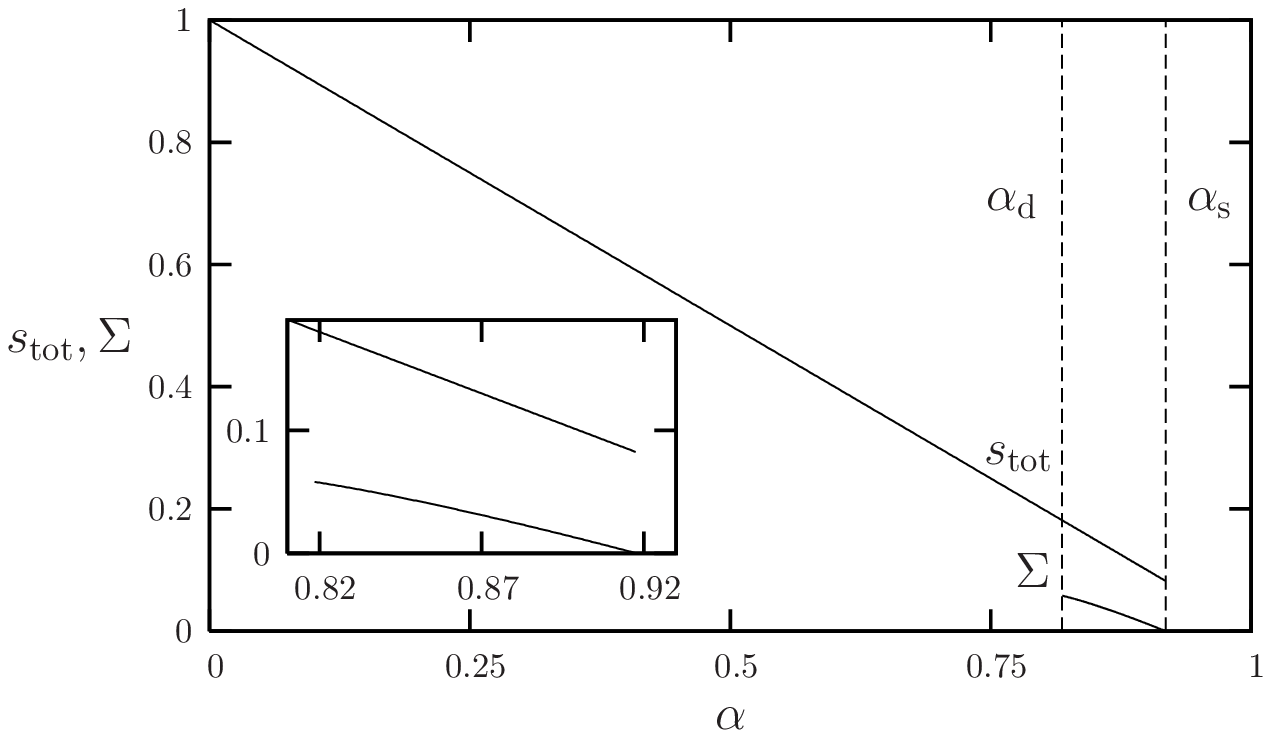}}
\caption{
({\it Top panel})
A sketch of the configuration space of random XORSAT problems.
For low values of $\alpha$ the solutions, represented as black dots, are
evenly spread on the $N$-dimensional hypercube. In the intermediate regime
they are grouped in clusters, symbolized by the circles. For 
$\alpha \ge \alpha_{\rm s}$ there are no more solutions.
\newline
({\it Bottom panel})
The total entropy of solutions $s_{\rm tot}(\a)$ and the complexity
(or entropy of clusters) $\Si(\a)$ for the $3$-XORSAT problem
on an Erd\H{o}s-R\'enyi graph~\cite{xor_1,xor_2}. The inset is a zoom
of the region close to $\a_{\rm d}$ and $\a_{\rm s}$.
}
\label{fig:sketch_XOR_clusters}
\end{figure}

Let us summarize the main messages on the properties of random CSP that 
should be drawn from this particular case 
(see Fig.~\ref{fig:sketch_XOR_clusters} for an illustration). 
For low values of the control
parameter $\alpha$ the exponentially many solutions are spread in the whole
configuration space, and close-by one to the other. Increasing $\alpha$
there appears a clustering transition at $\alpha_{\rm d}$, 
after which there are still 
exponentially many solutions, yet they are grouped in clusters of close-by
configurations, the clusters being separated one from the other; in this 
regime the complexity or configurational entropy $\Sigma$ counts the 
exponential number of clusters. The total entropy density of solutions, 
$s_{\rm tot}$, is the sum
of the complexity $\Si$ and the internal entropy density $s$ of each cluster, which
is here the same for all clusters: $s_{\rm tot}(\a) = \Si(\a) + s(\a)$.
For even larger values of $\alpha$ the
satisfiability transition $\alpha_{\rm s}$ is due to the vanishing of $\Sigma$,
i.e. the disappearance of the clusters of solutions; the last clusters that 
disappear can still contain an exponential number of solutions, i.e. the
internal entropy density $s(\a)$ can be finite right at $\alpha_{\rm s}$. The complexity
$\Sigma(\a)$ and the total entropy $s_{\rm tot}(\a)$ are reported in 
Fig.~\ref{fig:sketch_XOR_clusters} for 3-XORSAT on an Erd\H{o}s-R\'enyi graph.
The plot shows that indeed $s_{\rm tot}(\a_{\rm s}) = s(\a_{\rm s})$ 
is finite at $\alpha_{\rm s}$ for this model.
This is important because it shows that typical instances have an
exponential number of solutions even at the SAT-UNSAT transition, hence
instances with a unique solution (that are particularly important for the analysis
of the quantum adiabatic algorithm)
are everywhere exponentially rare in this model. We will come back to this point
in Sec.~\ref{sec:generating_USA}.

We should also emphasize that XORSAT exhibits some specific features
that are not shared by more complicated random CSP like
$k$-satisfiability or $q$-coloring.  In particular all the clusters of
XORSAT contain exactly the same number of solutions, because of the
linear structure of the set of equations modulo 2 it encodes. In
general there are clusters of different sizes, and because of these
fluctuations the condensation and satisfiability threshold do not
coincide, as will be discussed further below.

\subsection{The random subcubes model}
\label{sec:subcubes}

The more complex phenomenology of random $k$-SAT and $q$-COL
has been first unveiled with rather intricate computations based on the
cavity method, that we shall review in Sec.~\ref{sec:classical_cavity}.
For pedagogical reasons we shall first explain this phenomenology using
a toy model introduced in~\cite{rcm}, the Random Subcubes Model (RSM), which is
a non-local model (in the sense of Sec.~\ref{sec:QCC}) similar to the REM.
The main new ingredient that is introduced in the RSM (to mimic $k$-SAT and $q$-COL)
 is a distribution of
clusters of different sizes. 
While in the XORSAT problem, for a fixed $\a$,
all clusters contain the same number of solutions, in the RSM it is assumed
by construction that each cluster contains a different number of solutions,
given by $e^{N s}$. 
Similarly, the number of clusters of internal
entropy density $s$ is given by $e^{N \Si(s)}$.
Hence the complexity is here a non-trivial function of $s$, like in the
$p$-spin model discussed in Sec.~\ref{sec:pspinFC}, and not just a number
as in the XORSAT problem. 
As discussed in Sec.~\ref{sec:pspinFC}, the fluctuations of internal entropy of clusters have
an important consequence: they induce a new phase transition characterized by a 
{\it condensation} of the Gibbs measure onto a small number of clusters.
Because in the RSM clusters are uncorrelated,
all of its properties, and in particular the condensation transition,
can be extracted with much simpler computations than in random $k$-SAT and $q$-COL.
In addition, the RSM will be very useful in the quantum
setting for understanding the effect of quantum fluctuations
on random optimization problems.
 
For all these reasons, we will discuss the RSM in more detail than we did for the previous models. 
In this section we explain the classical
version of the random subcubes model~\cite{rcm}, its quantum 
extension~\cite{FSZ10} being treated in Sec.~\ref{sec:qsubcubes}.
It will be useful for the discussion of Sec.~\ref{sec:qsubcubes}
to define directly the model in a quantum notation, so we will do this here.

\subsubsection{Definition of the model}

The RSM
distinguishes configurations that belong to a set of low energy clusters
from those that belong to the remaining set of high energy configurations.
It is defined as follows. Consider the
Hilbert space $\HH$ of $N$ spins $1/2$ (qubits), in the basis of the
Pauli matrices $\hs^z_i$, $|\us \rangle = | \s_1, \cdots, \s_N
\rangle$.  A {\it cluster} $A$ is a subset (subcube) of the Hilbert
space \beq A = \{ |\us \rangle \, | \, \forall i \, : \, \s_i \in \pi_i^A
\} \ , \eeq where $\pi^A_i$ are independent random sets defined as
follows: \beq \pi^A_i = \begin{cases} & \left.\begin{aligned}
      -1 & \hspace{1cm}\text{with probability $\frac{p}{2}$ \,} \\
      1 & \hspace{1cm} \text{with probability $\frac{p}{2}$\,}
\end{aligned} 
\hspace{0.65cm}  \right\}  \hspace{0.15cm} \text{ $\s_i$ is ``frozen'' in cluster $A$,}\\
& \{1,-1\} \hspace{0.5cm} \text{with probability $1-p$ \, }  \hspace{0.5cm}  \text{ $\s_i$ is ``free'' in cluster $A$.}
\end{cases}
\eeq
Thus, with probability $p$
the variable $i$ is frozen in $A$ and with probability $1-p$ it is free.
With this definition the number of states, i.e. classical configurations, in a cluster $A$ is a random variable
 equal to $2^{Ns(A)}$, where $N s(A)$ 
is the number of free variables and  we call $s(A)$ the {\it internal entropy} of 
a cluster (for convenience in this section we use $\log_2$ to define entropies).
We next define a set $\SS$ as the union of $2^{N (1-\a)}$ random clusters, and its total
entropy $s_{\rm tot}$:
\beq
\SS = \bigcup_{i=1}^{2^{N (1-\a)}} A_i
\hskip1cm
s_{\rm tot} = \frac1N \log_2 |\SS|
\ .
\eeq
The parameter $\a$ here is analogous to the density of constraints in 
CSP. The probability $p$ that a variable is
frozen instead plays the role of the clause size $k$ in $k$-SAT or the 
number of colors $q$ in the $q$-coloring problem.

For each cluster $A$ we assign a Hamiltonian $\hH_A = N e_0(A)
\sum_{\us \in A} | \us \rangle \langle \us |$ with $e_0(A)\geq0$ and a
``penalty'' Hamiltonian $\hH_V = N V \sum_{\us \notin \SS} | \us
\rangle \langle \us |$ which describes the classical energy of states
not belonging to $\SS$.  The problem Hamiltonian $\hH_P =
\hH_V + \sum_A \hH_A$ is of course diagonal in the basis $|\us \rangle$, and
the associated cost function $E(\us) = \la \us | \hH_P | \us \ra$ is equal
to $N \sum_{A : \us \in A} e_0(A)$ if $\us\in \SS$ and $N V$ if $\us \notin \SS$.
With these definitions we wish to interpret the states in
$\SS$ as ``local minima'' of $\hH_P$ and the others as ``excited
states''.  A sharp distinction between them can be obtained by sending
the positive constant $V$ to infinity, a choice that we adopt in this section.
In Sec.~\ref{sec:qsubcubes} we will consider also a finite $V$, but we will always
assume that $V \gg \max_{A} e_0(A)$.

\subsubsection{Clustering}

\begin{figure}
\centering
\includegraphics{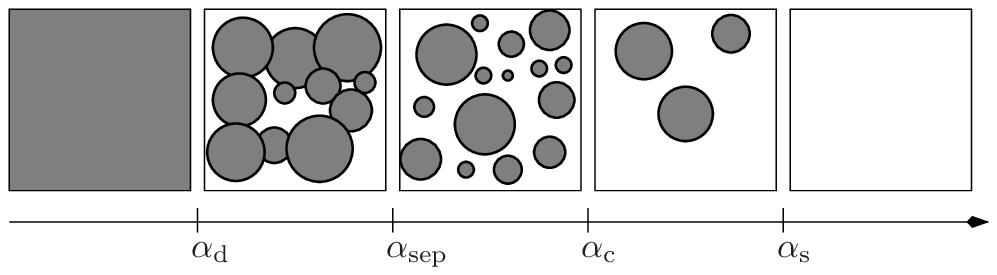}
\caption{Pictorial representation of the different phase transitions
in the set of solutions of the random subcubes model~\cite{rcm}.}
\label{fig:phase_cubes}
\end{figure}

To begin the discussion, we shall briefly 
characterize
the structural changes in the set $\SS$ when $\a$ is
varied, which are shown in
Fig.~\ref{fig:phase_cubes} and have been derived in~\cite{rcm}.
We will assume here for simplicity that all the clusters have $e_0(A)=0$.
The set $\SS$ is then the set of the ``solutions'' of the problem (or of the ground states of zero energy).
We will show
that the transitions $\alpha_{\rm d}$, $\alpha_{\rm c}$ and
$\alpha_{\rm s}$ outlined in
Sec.~\ref{sec:Optimization-Statistical-Mechanics} can be very precisely defined in the RSM.

We will extensively make use of two well known results
of probability theory called the {\it union bound} and the {\it Chebychev inequality}. 
In fact, the properties of $\SS$ can be traced back to 
probability statements concerning $2^{N b}$ 
events $\EE_i$, each having probability ${\cal P}(\EE_i)=2^{- N a}$, 
for some $a$ and $b$. Under these conditions the {\it union bound}
states:
\beq\label{union_bound}
{\cal P}\Big(\bigcup_{i=1}^{2^{N b}} \EE_i\Big) \leq \sum_{i=1}^{2^{N b}} {\cal P}(\EE_i) = 2^{N(b-a)}
\ ,
\eeq
which implies that when $a>b$ the probability ${\cal P}(\cup_i \EE_i)$ is
exponentially suppressed in the size of the system $N$. 
When the events are independent, the number of true events ${\cal N}$ is a random variable with
a binomial distribution with
$\la{\cal N}\ra = 2^{N(b-a)}$ and $\la{\cal N}^2\ra = 2^{N(b-a)}(1-2^{-N a})$. Then
for arbitrary small $\epsilon$ one can apply the {\it Chebychev inequality}:
\beq\label{Chebychev}
{\cal P}\Big(\frac{|{\cal N}-\la{\cal N}\ra |}{\la{\cal N}\ra} > \epsilon\Big) 
\leq \frac{\la{\cal N}^2\ra}{\la{\cal N}\ra^2\epsilon^2} \leq \frac{1}{2^{N (b-a)} \epsilon^2}
\eeq
which ensures that when $a<b$,  ${\cal N}$ is {\it self-averaging} in the large $N$ limit, i.e.
the average is exponentially large and concentration around the average 
${\cal N} \sim \la {\cal N}\ra$ is found.

As a first application, we consider the number of clusters $\NN(s)$ of entropy $s$. 
Because frozen variables are chosen independently, we have
\beq
{\cal P}(s(A)=s) = {N \choose Ns} p^{N(1-s)}(1-p)^{N s} \ .
\eeq
Hence $\NN(s)$ follows a binomial distribution
with parameter ${\cal P}(s)$ and $2^{N(1-\a)}$ terms,
and its average, at the leading exponential order, is
$\la \NN(s) \ra = 2^{N(1-\a)} {\cal P}(s) = 2^{N \Si(s)}$
with the complexity $\Si(s)$ defined by
\beq
\label{sigma_RSC}
\begin{split}
&\Si(s) =  1 - \a - D(s || 1-p) \ , \\
&D(x || y) = x \log_2(x/y) + (1-x) \log_2 [(1-x)/(1-y)] \ .
\end{split}
\eeq
In the region $s\in (s_{\rm min}, s_{\rm max})$ where $\Si(s) > 0$,
we can apply the Chebychev inequality to show that $\NN(s)$ concentrates around the average
when $N\to\io$.
In the region where $\Si(s)<0$, we can apply the union bound to show that with probability
1 there are no clusters of entropy $s$ when $N\to\io$.

Next, we apply similar arguments to identify the following changes of the structure of the space of solutions
when $\a$ is varied (see Fig.~\ref{fig:phase_cubes}). We will only give brief sketches of the proofs; the reader
is referred to~\cite{rcm} for further details.

\begin{itemize}
\item For $\a \leq \a_{\rm d} = \log_2(2-p)$, each state $|\us \rangle$ belongs
to an exponential number of clusters and $\SS = \HH$.
For $\a > \a_{\rm d}$ a random state does not belong to $\SS$ with probability 1 when $N\to \infty$,
thus $\SS \neq \HH$ and $s_{\rm tot} < 1$.

\noindent
{\it Proof:} The probability that a configuration $|\us \rangle$ belongs to a cluster $A$ is
${\cal P}(|\us \rangle \in A)=(1-\frac{p}{2})^N$ and 
\beq\label{prob_us_in_A}
{\cal P}(|\us \rangle \notin \SS)=\Big[1-\Big(1-\frac{p}{2}\Big)^N\Big]^{2^{N(1-\a)}}
\ .
\eeq
Then from the union bound if $\a<\a_{\rm d}= \log_2(2-p)$:
\beq
{\cal P}(\SS \neq \HH)=
{\cal P}(\cup_{|\us \rangle} |\us \rangle \notin \SS) \leq 2^N e^{-2^{N[\log_2(2-p)-\a]}} \to 0
\ ,
\eeq
which implies that all states are in $\SS$ and $s_{\rm tot}=1$.
For $\a > \a_{\rm d}$,
from Eq.~(\ref{prob_us_in_A}) we get
${\cal P}(|\us \rangle \notin \SS) \to 1$. Thus $\SS\neq\HH$ and $s_{\rm tot}<1$.

\item For $\a > \a_{\rm sep} = 1 + \log_2(1-p^2/2)/2$, the clusters are 
well separated, in the sense that with probability 1 for $N\to\io$ the Hamming
distance (minimal number of different spins) between any two clusters is of order $N$.

\noindent
{\it Proof:} We note that ${\cal P}(A \cap A' \neq \emptyset) = (1-\frac{p^2}{2})^{N}$.
Then we can apply the union bound over all possible intersections in the set $\SS$
\beq
{\cal P}(\cup_{ij} (A_i \cap A_j \neq \emptyset)) \leq \frac12 2^{N(1-\a)}( 2^{N(1-\a)} - 1)\left(1-\frac{p^2}{2}\right)^N \to 0
\eeq
for $\a>\a_{\rm sep}$. This means that with probability 1 when $N\to\infty$ 
the clusters are disjoint, i.e. their Hamming 
distance is strictly positive. The probability to find clusters at distance $x$ is finite only
when $x={\cal O}(N)$~\cite{rcm}.

\item For $\a_{\rm d} < \a < \a_{\rm c} = p/(2-p) + \log_2 (2-p)$
 most of the solutions belong to one of the 
exponentially many clusters of size $s^*$, with $\Si(s^*)>0$ and
$s^*\in(s_{\rm min}, s_{\rm max})$. 
On the contrary when  
  $\a > \a_{\rm c}$,
$s^* = s_{\rm max}$ and most of the solutions
belong to the largest clusters whose number is sub-exponential
in $N$ because $\Si(s_{\rm max})=0$.

\noindent
{\it Proof:} 
One can compute the total number of states in $\SS$ by observing that
\beq\label{eq:SS_RSM}
| \SS | = 2^{N s_{\rm tot}} \sim \sum_A 2^{N s(A)} 
\sim \int_{s_{\rm min}}^{s_{\rm max}} {\rm d}s \, 2^{N [ \Si(s) + s]} \ ,
\eeq
therefore
$s_{\rm tot} = \max_{s\in[s_{\rm min},s_{\rm max}]} [ \Si(s) + s ]$.
Studying the function $\Si(s) + s$ it turns out that up to $\a_{\rm c}$
its maximum value, dominating the saddle point in the integral, is taken inside the
allowed interval and thus $\Si(s^{\ast})>0$. 
When $\a>\a_{\rm c}$ instead the maximum is achieved at the boundary of the
interval, implying ${\cal N}(s^{\ast})={\cal O}(1)$.

\item Finally, for $\a > \alpha_s=1$ there are no more solutions.

{\it Proof:} This follows trivially from the definition of the number of clusters, equal to 
$2^{N(1-\a)}$. Then for $\a > 1$ there are no more clusters and the set $\SS$ is empty.
In the language of random CSP, $\alpha_s$ corresponds to the SAT-UNSAT transition. 
\end{itemize}

Note that in this particular model the entropy has a singularity at $\a_{\rm d}$, which
is not present in local random CSP. From the dynamical point of view what characterizes $\a_{\rm d}$
is that for $\a \geq \a_{\rm d}$ there
is ``ergodicity breaking'' in the sense that a local random walk over solutions 
starting in one cluster takes an exponentially long time to reach another 
cluster~\cite{rcm}.

\subsubsection{The partition function at finite temperature}
\label{sec:subcubes_finiteT}

Similar results can be obtained when the clusters have a distribution of energies~\cite{rcm}.
Let us assign to each cluster an i.i.d. random energy $e_0 \in [0,e_{\rm m}]$ in such a way that the total 
number of clusters of energy $e_0$ is $2^{N (1-\a) g(e_0)}$, with $g(e_0)$ an arbitrary increasing
function of $e_0$, because it is reasonable to assume that the number of clusters increases with energy.
Then, the above arguments can be easily generalized for each level of energy $e_0$. 
Following the same reasoning that leads to Eq.~(\ref{sigma_RSC}), 
the number
of clusters of energy $e_0$ and entropy $s$ is $2^{N\Si(e_0,s)}$, with
\beq
\label{sigma_RSC_e}
\begin{split}
&\Si(e_0,s) = (1-\a) g(e_0) - D(s || 1-p) \ ,
\end{split}
\eeq
and is positive in an interval $s\in (s_{\rm min}(e_0), s_{\rm max}(e_0))$.
Of particular interest is the computation of the partition function at finite temperature, that replaces
Eq.~(\ref{eq:SS_RSM}) and reads
\beq\label{eq:ZsubT}
Z = \sum_A 2^{N s(A)} e^{-\b N e_0(A)} 
\sim \int_0^{e_{\rm m}} de_0 \int_{s_{\rm min}(e_0)}^{s_{\rm max}(e_0)} {\rm d}s \, 2^{N [ \Si(e_0,s) + s - \b e_0 \log_2 e]} \ ,
\eeq
and that can be evaluated by a saddle point. When the saddle point values of $e_0$ and $s$ reach the boundary of the integration
interval a condensation transition happens, on a line $\a_{\rm c}(T)$.
Moreover, at each level of energy, the previous analysis of the structure of the union of clusters 
can be repeated, and the same transitions happen at 
energy-dependent values $\a_{\rm d}(e_0)$, $\a_{\rm sep}(e_0)$. Because for each temperature a unique value of $e_0$ dominates
the partition function, these can be converted in lines $\a_{\rm d}(T)$, $\a_{\rm sep}(T)$. It is easy to show that all transition points
increase with $T$, like in the XORSAT case (Fig.~\ref{fig:sketch_XOR_clusters}).

\subsection{The space of solutions of random constraint satisfaction 
problems}
\label{sec:transitions_rCSP}

\begin{figure}
\begin{center}
\includegraphics[width=\linewidth]{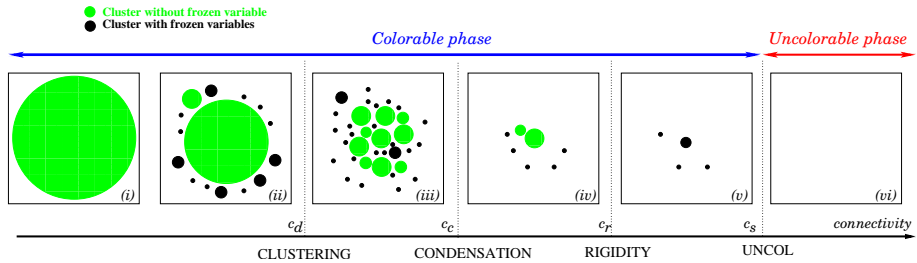}
\end{center}
\caption{\label{fig11} Sketch of the space of solutions ---colored
  points in this representation--- in the $q$-coloring problem on random
  graphs when the connectivity $c$ is increased~\cite{KrMoRiSeZd,col2}.  (i) At low $c$, all
  solutions belong to a single cluster. (ii) For larger $c$, other clusters of
  solutions appear but a giant cluster still contains almost all solutions.  (iii)
  At the clustering transition $c_{\rm d}$, it splits into an exponentially large
  number of clusters.  (iv) At the condensation transition $c_{\rm c}$, most
  colorings are found in the few largest of them. (v) The rigidity transition
  $c_{\rm r}$ ($c_{\rm r}<c_{\rm c}$ and $c_{\rm r}>c_{\rm c}$ are both possible depending on $q$) arises
  when typical solutions belong to clusters with frozen variables (that are
  allowed only one color in the cluster). (vi) No proper coloring exists beyond the
  COL/UNCOL threshold $c_{\rm s}$.}
\end{figure}

By means of non-local toy models such as the REM (Sec.~\ref{sec:REM}) and RSM (Sec.~\ref{sec:subcubes})
we built a lot of intuition about the different
transitions that happen in spin glass models.
The analysis of the XORSAT problem (Sec.~\ref{sec:ch3-XOR}), 
that can be carried out in a relatively straightforward way using rigorous methods, also illustrated
the emergence of clustering in the simplest random CSP.
Having now understood the kind of transitions to be expected, 
one would like to make precise computations in other local random 
CSP such as $k$-SAT or $q$-COL. Unfortunately, this turns out to be a more difficult task
and requires the introduction of sophisticated statistical mechanics methods, in particular the cavity
method~\cite{MM09}. These will be introduced in Sec.~\ref{sec:methods}.

Before turning to quantum versions of spin glass models,
we will give here a summary of all possible transitions that have been found in local random CSP
at zero temperature,
taking as an illustrative example the random $q$-COL problem
with $q\ge 4$ colors (the $q=3$ case being a bit
particular) and a large Erd\H{o}s-R\'enyi random graph whose average
connectivity $c=2\alpha$ increases. 
Different phases are encountered that we
will now describe in order of appearance. The corresponding phase
diagram is depicted in Fig.~\ref{fig11}~\cite{KrMoRiSeZd,col2}.
\begin{itemize}
\item[(i)]\textbf{ A unique cluster exists}: For low enough
  connectivities, all the proper colorings are found in a single
  cluster, where it is easy to ``move'' from one solution to
  another: for any given pair of solutions, one can construct a path of solutions that connects them,
  such that at any step along the path only a sub-extensive number of colors are changed.
  The total entropy of solutions
  can be computed and reads in the large graph
  size $N$ limit:
\beq 
s_{\rm tot} =  \log{q} + \frac {c}{2}
\log{\left(1-\frac{1}{q}\right)}\, .\label{S_RS} 
\eeq

\item[(ii)] \textbf{ Some (irrelevant) clusters appear}: As the connectivity is
   increased, the phase space of solutions decomposes into a large
  (exponential) number of different clusters. It is tempting to identify that
  as the clustering transition, but it happens that all (but one) of these
  clusters contain relatively very few solutions ---as compared to whole set---
  and that almost all proper colorings still belong to one single giant
  cluster.  Clearly, this is not a proper clustering phenomenon and in fact,
  for all practical purposes, there is still only one single cluster.
  Eq.~(\ref{S_RS}) still gives the correct number of colorings at this stage.

\item[(iii)] \textbf{ The clustered phase}: For larger connectivities,
  the large single cluster also decomposes into an exponential number
  of smaller ones: this now defines the genuine clustering threshold
  $c_{\rm d}$. Beyond this threshold, a local algorithm that tries to move
  in the space of solutions will remain prisoner of a cluster of
  solutions. Interestingly, it can be shown that the
  total number of solutions is still given by Eq.~(\ref{S_RS}) in this
  phase.  This is because the
  free energy has no singularity at the clustering transition (which
  is therefore not a true transition in the sense of Ehrenfest, but
  rather a geometrical transition in the space of solutions).

\item[(iv)] \textbf{ The condensed phase}: As the connectivity is further
  increased, a new sharp phase transition arises at the condensation threshold
  $c_{\rm c}$ where most of the solutions are found in a finite number of
  largest clusters. From this point, Eq.~(\ref{S_RS}) is no longer valid, because
  this is a genuine phase transition. The entropy is therefore non-analytic at $c_{\rm c}$ and 
  Eq.~(\ref{S_RS}) becomes just an upper bound.

\item[(v)] \textbf{ The rigid phase}: Two different types of cluster
  exist: in the first type, that we shall call the \textit{unfrozen}
  ones, all spins can take at least two different colors.  In the
  second type however, a finite fraction of spins are allowed only one
  color within the cluster and are  thus ``frozen'' into this
  color. It follows
  that a transition exists, that we call \textit{rigidity}, when
  frozen variables appear inside the dominant clusters (those that
  contain most colorings).  If one takes a proper coloring at random
  above $c_{\rm r}$, it will belong to a cluster where a finite fraction of
  variables is frozen into the same color.  Depending on the value of
  $q$, this transition may arise before or after the condensation
  transition (a list of values can be found in~\cite{rearr_csp,col2}).

\item[(vi)] \textbf{ The UNCOL phase}: Eventually, the connectivity
  $c_{\rm s}$ is reached beyond which no more solutions exist. The ground
  state energy (sketched in Fig.~\ref{fig2}) is zero for $c<c_{\rm s}$ and
  then grows continuously for $c>c_{\rm s}$. The values $c_{\rm s}$ computed
  within the cavity formalism are in perfect agreement with the
  rigorous bounds~\cite{transition_lb,Achltcs,transition_ub,BookCrisMoore} derived using probabilistic methods
  and are widely believed to be exact, although this remains to be
  rigorously proven (see~\cite{FrLe,PaTa} for a proof that they
  are at least rigorous upper bounds).
\end{itemize}

Notice that in specific models some of these transitions coincide.
We have already seen in Sec.~\ref{sec:ch3-XOR} that in XORSAT 
$c_{\rm r} = c_{\rm d}$ and
$c_{\rm c} = c_{\rm s}$,
therefore some of the phases above do not exist: all clusters are frozen, 
and the condensed phase does not exist.
Another example is the $3$-COL problem, which 
is peculiar because $c_{\rm d}=c_{\rm c}$ so that the
clustered phase is always condensed. 
In view of this rich and model-dependent
phase diagram, it is important to get an intuition on the meaning and
the properties of these different phases.

At this point, there are many questions one could ask. First of all:
are these problems hard {\it only} close to the SAT-UNSAT threshold $c_{\rm s}$? 
The answer is no:
for instance in $q$ coloring, when $q$ is large, problems are easy (in
this case, the complexity is linear in the number of nodes) for almost
every algorithm as long as $c < c_{\rm d} \sim q\log{q}$ (to leading order) but suddenly
very hard (so that no algorithm is known that performs provably in
sub-exponential time in the number of nodes) if
$c> c_{\rm d}$. It is known, however, that there exist solutions up to
$c = c_{\rm s} \sim 2q\log{q}$. A similar problem appears in random $k$-SAT between
$2^k\log{k}/k$ and $2^k \log{2}$~\cite{AC08}. 
One could then conclude that the clustering above $c_{\rm d}$ 
is responsible for the hardness of the problem. Yet, for small enough $q$ (e.g. $q=4$), 
many algorithms are able to find solutions
in the clustered phase at $c$ much larger than $c_{\rm d}$~\cite{MezardParisi02,circumspect,PhysRevE.76.021122}.
Why then are some problems hard and some easy?
Does something else explain the sudden onset of hardness?

Unfortunately, the answer to these questions is in large part still open.
Yet, many interesting results on the connection between the above picture and algorithmic hardness have been
obtained. For reasons of space, in the rest of this section
we will focus in particular to simulated (thermal) annealing.

\subsection{Efficiency of the simulated annealing}
\label{sec:thermalannealing}

It turns out that close to the satisfiability threshold $c_{\rm s}$ finding the
solutions to the problem becomes particularly hard and most algorithms
suffer of a dramatical slowing down.  Quite generally this phenomenon
is attributed to the presence of many minima in the energy landscape
and to the organization of the solutions in phase space.  Simulated
(thermal) annealing \cite{KirkpatrickGelatt83} is one of most famous
algorithms designed to tackle complex energy landscapes.  Despite the
fact that it only partially accomplishes this task as it actually
fails when too many clusters dominate the partition function, it
represented a true breakthrough in the domain and it is still
exploited in many applications.  The prescription of simulated
annealing (Sec.~\ref{sec:QAA_def}) is
to initialize the algorithm with a random, high
temperature, configuration. Then, lower the temperature, eventually
down to zero, in discrete steps according to an assigned protocol, and
at each step, perform a given number of local movements in phase space
--Monte Carlo steps-- in order to equilibrate at that temperature and
use the last generated configuration to initialize the search at the
new temperature. Technically this is the implementation of a
\textit{time dependent Markov chain}. An implementation of simulated annealing
in continuous time is also possible.
In this section we want to discuss in more details the relation between the
structural transitions discussed in Sec.~\ref{sec:transitions_rCSP} and the performances
of simulated annealing.

\subsubsection{Effects of the clustering transition on thermal annealing}

\begin{figure}
\includegraphics[width=25pc]{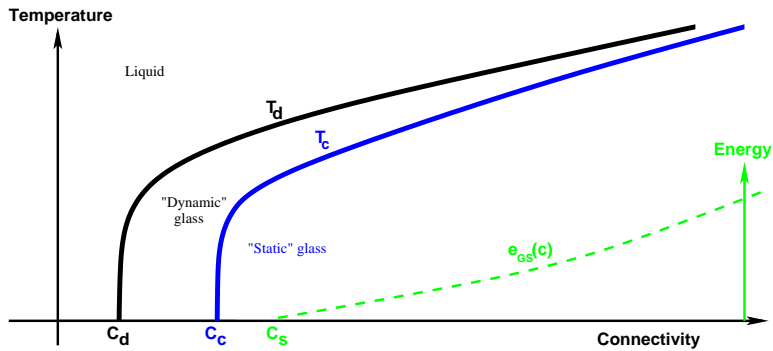}\hspace{1pc}%
\caption{\label{fig2} Typical phase diagram for a spin glass problem
  on a random graph. At $T_{\rm d}$, local Monte Carlo dynamics becomes inefficient
  (``dynamical'' transition \cite{kirkpatrick:88,MoSe,MoSe2}). At $T_{\rm c}$
  the system undergoes an equilibrium glass transition
  \cite{gross:85,KZ08}. $e_{\rm GS}$ represents the ground
  state energy and is positive when the problem is not
  satisfiable. At zero temperature, the phase
  transitions of Sec.~\ref{sec:transitions_rCSP} are recovered.}
\end{figure}

In order to
discuss better the properties of thermal annealing, we need to introduce a temperature into the problem,
as discussed in Sec.~\ref{sec:QAA_def}, and investigate
the  finite temperature phase diagram. The latter is sketched in Fig.~\ref{fig2}~\cite{KZ08}
for $q$-COL with $q\ge4$ on Erd\H{o}s-R\'enyi random graphs as a function of average connectivity $c$. 
At high
temperature the system behaves as a paramagnet in the language of
magnetic systems. The clustering and condensation transitions extend
in lines $T_{\rm d}(c)$ and $T_{\rm c}(c)$. On the contrary, the
rigidity and SAT-UNSAT transitions exist only at zero temperature,
because at finite temperature the notions of ``solution" and ``frozen
variable" cannot be defined: constraints can always be violated with
some finite probability.

 The idea behind a simulated annealing in temperature
is that thermally activated processes allow to overcome the energy
barriers and at the latest stages the zero temperature dynamics converges
towards the solution. However this is only true if the annealing is slow enough. 
Let us first try to apply, as in the quantum case, an
adiabatic strategy. There exist rigorous bounds on the time needed for
a thermal annealing to stay adiabatic, but they yield, in the worst
case, an exponential time \cite{Geman}. Indeed, on random CSP for $c > c_{\rm d}$,
it turns out that equilibration is an exponentially hard task 
when the clustering temperature $T_{\rm d}(c)$ is
reached. Below $T_{\rm d}(c)$ the dynamics falls out-of-equilibrium
\cite{MoSe2} unless one is ready to wait for exponentially long times.
This rigorous result can be intuitively understood by the following three arguments.
First, the probability to overcome the barriers between the pure states
of the Gibbs measure is exponentially small in the size of the system, hence these states 
trap the dynamics for an exponentially large time. 
Secondly,
even if one can go out of a pure state, there are exponentially many of
them, and it thus takes again some time to find the equilibrium ones
\cite{0305-4470-29-14-012,springerlink:10.1007/s00220-008-0565-7}. To
further complicate the problem, a third effect exists: even if one manages to
equilibrate the system at a given temperature $T$, this is not really
useful because the pure states that dominate the partition sum at any
$T'<T$ are completely different ones, so that the hard equilibration
work has to be entirely redone from scratch as soon as the
temperature is slightly changed. This is an effect called temperature
chaos~\cite{BrMo87,FH88,KrMa02,0295-5075-90-6-66002,ZK10} which is the
classical analog of the quantum level crossing that will be discussed in
Sec.~\ref{sec:qsubcubes}. The combination of all these effects is behind 
the exponential hardness of an adiabatic thermal cooling. 

\subsubsection{Non-adiabatic thermal annealing}

If one, however, is not interested in being adiabatic and instead is
just interested in the final configuration reached, then the situation
is different: it may be possible to reach a zero energy state at the
end of the protocol while being out of equilibrium at intermediate stages of the annealing. This is
observed for instance in random coloring~\cite{MourikSaad}. 
There are also many random walk algorithms that are similar
to Monte Carlo Markov chains but do not satisfy the detailed balance condition:
these can find a
solution of random CSP in some part of the clustered phase
\cite{asat,PhysRevE.76.021122}. This seems at first contradictory with
the previous section. However, the problem of finding a
solution is different from sampling solutions uniformly,
which is what is achieved by an adiabatic cooling.
If one just wants to find a solution, it is often possible to succeed up to
much larger connectivities than the clustering one. In fact, it seems
from empirical evidences (see the discussion in~\cite{0295-5075-90-6-66002,ZK10,PhysRevE.76.021122}) 
that the moment the problems become truly hard is when the rigidity transition is
reached, or more precisely when all solutions belong to frozen
clusters.

This can be understood in terms of energy landscape: below the
clustering temperature, one is trapped into a single pure state and
cannot visit the whole space. However, the energy of the pure state
one is trapped in can be lowered when the temperature is reduced, and
maybe even go to zero when $T\to 0$. As shown in
\cite{0295-5075-90-6-66002,ZK10,PhysRevE.76.021122} such
``canyon-like'' states, that reach the zero energy configurations,
become rare after the rigidity transition. In this case, one needs
again to visit many states until a good one is found, and one is back to
the situation discussed in the previous section. In summary, if one
is ready to forget about adiabaticity, annealing and other strategies
can be applied and work well up to connectivities larger than $c_{\rm d}$, but fail
when rigidity is met. According to this empirical analysis, 
the clustered {\it and} frozen region of the
phase diagram contains the hardest possible instances of
random CSP.

\subsubsection{Existence of good paths for classical annealing}

We have discussed so far annealings in temperature, but other
annealing protocols can be used: for instance, an
annealing starting from a large magnetic field and reducing it to zero, 
or any path in a
temperature--field phase diagram that ends at zero field and zero
temperature. Of course, the fact that a thermal annealing is
inefficient does not imply that {\it all} possible paths are,
eventually, inefficient. Indeed, it is easy to see that there are
paths that will make a classical annealing work in a polynomial time,
however hard the problem is. 

Consider for instance the following setting: we take an instance of a CSP defined on 
Ising spin variables,
and one of its solutions $\underline{\tau} = \{\tau_i\}$, where $\tau_i$ is the
value of variable $i$ in the solution. Consider now a protocol in which the following
term is added to the cost function:
\beq
E_{\ut}(\us) = E(\us) - h \sum_i \tau_i \sigma_i \ .
\eeq
The ``local magnetic field'' $h_i = h \tau_i$ points each spin in the direction of the
correct value in the chosen solution. We can now perform a simulated annealing by starting
at large $T$ with finite $h$ and then reducing progressively $T$ and $h$ to zero.
Upon cooling in this field the final configuration will be the solution associated
with the field.  One might (rightfully) argue that using a protocol that knows the 
solution of the problem is equivalent to cheating. Indeed the fact
that such a protocol exists is not very useful, because finding the right field is equivalent to
finding a solution. The problem is then: how difficult is it to find a ``good'' annealing path?

The message here is that the mere fact that some efficient annealing
protocol exists is not conclusive. More generically, if
one studies classical algorithms to solve CSP, there is always a good
one (for instance, if one initialize a random walk algorithm in a
solution) but it is, of course, hard to find it in general. To prove
rigorously this statement is however difficult; this is basically
proving that P is not NP. In short: the question is not to decide whether
there is a classical protocol able to find a solution quickly: for a
given problem, such protocols always exist. The question is rather
how to find them. The only thing one can do is to consider a given
protocol and study it: to the best of our knowledge, the random
instances in the clustered and entirely frozen region are extremely hard
---and can be considered as some of the hardest representatives of
the NP class--- and there has been no practical way to solve them
generically.

One will have to keep these considerations in mind in the analysis of the quantum
annealing. Proving that efficient quantum paths exists is not
enough from an algorithmic perspective: 
one really needs to be able to construct these paths explicitly. 
After all, we are interested here in constructing a specific algorithm
(a specific annealing protocol) and prove its efficiency (or inefficiency).

\subsection{Generating USA instances (locked CSP)}
\label{sec:generating_USA}

Instances with a 
Uniquely Satisfying Assignment (USA) are very practical in
quantum studies, as the absence of degeneracy of their ground state
allows to unambiguously define their gap. 
Generating a problem that has only one single solution
is, however, not so easy a priori. 
Actually, even certifying that an instance is USA requires to count all of
its solutions, which is an exponentially hard task (therefore the decision
problem of whether an instance is USA or not is in general not in NP).

Moreover, the usual protocol, used since~\cite{Fa01}, 
consists in generating many instances of a given
problem (for a fixed size), finding all the solutions of each instance using a complete
solver, and then keeping only the instances with a single
solution. It turns out that the latter is a very difficult task, that is possible
only for small sizes. In fact, in most random CSP, there are
generically exponentially many solutions, so that one needs to generate
exponentially many instances to find the rare ones with a
USA and in the end generating the instances in this way
is at least as hard than finding their solution.
Fortunately, there is a way around this problem, using the so-called
``locked problems'' at the satisfiability threshold, or using a hidden
assignment in their UNSAT phase.

\subsubsection{Locked problems at the SAT-UNSAT threshold}

The concept of ``locked problems'' was introduced in~\cite{ZM08}.
This is a very broad class of CSP
where {\it i)} if one changes one variable (and only one) in a given clause in
a SAT configuration, then the system becomes UNSAT and {\it ii)} each variable
is involved in at least two clauses.
If one wants to
keep the configuration SAT, then one has to flip a variable in a
neighboring clause, and so on, until a loop is found and the chain can
be closed. This means that, on a random graph that is locally
tree-like, in order to go from one solution (or satisfying assignment)
to another one, it is necessary to flip at least a closed loop of
variables in the factor graph representation, which typically involves
$O(\log{N})$ changes. 

There are many such models, such as the XORSAT, 1-in-3 and 1-in-4 SAT problems (Sec.~\ref{sec:examples_optimization})
defined on the ensemble of regular random graphs~\cite{ZM08}. 
Generically, the phase transition in locked models are the same as
in the XORSAT problem on an Erd\H{o}s-R\'enyi graph, 
as illustrated in Fig.~\ref{fig:sketch_XOR_clusters}, but with some important differences:
\begin{itemize}
\item When $\alpha$ is small enough, the set of solution is ``nearly''
  connected: by flipping $O(\log{N})$ variables, one can visit all
  possible solutions starting from a particular one. In fact, if one
  allows also to visit ``quasi solutions'' with $O(1)$ cost, single
  flip spins are enough to visit the whole space of
  solution\footnote{For expert readers: this is the difference between
    reconstruction and small noise reconstruction
    \cite{ZM08,LenkaThesis}.}.  We thus say that the
  space of solutions is made of a single, unique cluster.
\item When $\alpha>\alpha_{\rm d}$, the set of solution undergoes a
  clustering transition and splits into an exponential number of
  components, separated by $O(N)$
  flips.  This is the clustering transition.
  The crucial characteristic of locked model is that {\it each cluster contains a single solution} 
  (the circles in Fig.~\ref{fig:sketch_XOR_clusters} contain only one black dot in this case).
  Therefore, the internal entropy of clusters vanishes, $s(\a)=0$, and the total entropy coincides
  with the complexity, $s_{\rm tot}(\a) = \Si(\a)$.
    \item For $\alpha>\alpha_{\rm s}$, there are no solutions anymore. 
    As usual, the complexity vanishes continuously at the SAT-UNSAT transition $\a_{\rm s}$. Because $s_{\rm tot}(\a) = \Si(\a)$, the total  
  entropy also vanishes at the transition: $s_{\rm tot}(\a_{\rm s})=0$.
\end{itemize}

There are two consequences of these properties that are important for the present
discussion. First, these models are hard to solve not only at the
threshold, but also in the full clustered phase (see
\cite{0305-4470-35-35-301,ZM08}): here, the clustering and rigidity
transitions coincide and the entire clustered phase is always
associated with hard instances. In these models we thus have a clear
and well defined link between the presence of a phase transition and
the computational complexity. The same conclusion is valid in particular for the
XORSAT problem on random regular graphs. 
This problem is in P (while locked problems are generically
NP-complete), as it can be solved by Gaussian elimination. If one,
however, decides to forget about this information, it is a very
difficult problem. In fact, the hardest instances of SAT problems
to this day are obtained by generating a XORSAT formula and by adding to it a
bit of non linearity (such that Gaussian elimination cannot be used,
see \cite{Jarvisalo,Ricci-Tersenghi17122010}).

There is another reason why these models are useful in quantum
annealing studies. When working exactly at the satisfiability
thresholds $\a_{\rm s}$, the entropy is exactly zero as we discussed above: hence, USA instances
appear with a finite probability, making them very easy to generate (see Sec.~\ref{sec:XORSAT_USA} and~\cite{JKSZ10}). 

\subsubsection{Planted locked models}

There exists another way to generate USA instances, not only with
finite probability, but with a probability going to one in the large
size limit: planting a solution in the UNSAT phase of locked
models. This was proposed in \cite{QuietPlanting} and studied with
statistical physics methods and rigorous mathematical proofs.

The idea is to create both an instance and a solution by first
assigning a configuration to all variables, and then choosing only
constraints compatible with this configuration. This creates instances
from the so-called ``planted'' ensemble. As shown in
\cite{QuietPlanting}, for random locked CSP models, instances from the planted ensemble 
have with high probability a single satisfying
assignment (or a pair of them if a global symmetry is present)
beyond the satisfiability threshold. This allows to create USA instances of any
size at zero computational cost. 

The question is whether the instances created in this way are
difficult. This, again, depends on the average density of constraints~\cite{QuietPlanting}: 
an easy-hard-easy pattern for finding a
solution appears in the planted ensemble as the constraint density is
increased. The boundaries of the hard phase are given by the
clustering transition on one side, and by another transition
(called the threshold for the robust reconstruction~\cite{QuietPlanting}) on the other
side. Between these two transitions, hard instances with USA are
generated. Again, XORSAT on random regular graphs is particular in the sense that planted
instances remains hard for arbitrary large connectivity.

We shall therefore discuss in a lot of details the quantum XORSAT model in Sec.~\ref{sec:results}, 
using its double status as a hard benchmark and as a simple model to generate USA instances.

\section{The low energy spectrum of quantum spin glasses}
\label{sec:low_energy}

As described in Sec.~\ref{sec:quantum_computers} the Quantum Adiabatic Algorithm (QAA), or quantum annealing, 
is a procedure designed to find the ground state of a classical Hamiltonian. 
From the point of view of the computational complexity theory, random optimization problems are thus 
natural benchmarks for this algorithm. We described in Sec.~\ref{sec:classical_mean_field} 
the rich phenomenology of these classical disordered models, and explained how the study of 
mean field spin glasses allowed to understand them. We shall now progressively turn towards 
the study of quantum optimization problems, i.e. classical disordered models with some 
non-commuting term (for instance a transverse field) inducing quantum fluctuations. 
As in Sec.~\ref{sec:classical_mean_field} we will start by discussing simpler models for pedagogical reasons.

The quantitative assessment of the performances of the QAA, 
as discussed in Sec.~\ref{sec:QAA-gap}, requires a detailed understanding of the low energy
spectrum of disordered quantum Hamiltonians. Its time complexity is indeed directly related to 
the square of the inverse of the minimum gap $\Delta_{\rm min}$ of the Hamiltonian along the 
annealing path. On the basis of complexity theory we are then particularly interested
in discriminating between Hamiltonians whose minimum gap vanishes  
polynomially and those for which $\Delta_{\rm min}$ is expected to be
exponentially small in~$N$.

It is well known that the gap of the Hamiltonian vanishes upon increasing $N$ in correspondence
with a quantum phase transition~\cite{sachdev2001,JKKM08,JKSZ10,AC09,YKS10}. 
However, the reverse is not true: the gap might vanish even without an underlying 
quantum phase transition. Indeed, some model Hamiltonians display phases such that the gap is 
everywhere exponentially small in $N$, due to a continuum of level crossings~\cite{FSZ10}.
Interestingly, this was associated to a kind of Anderson localization phenomenon 
in phase space~\cite{AKR10}. In these cases, the vanishing of the
gap in the thermodynamic limit is not associated to a singularity 
in the ground state energy~\cite{FSZ10}. 

In this section, we will present a review of quantum phase transitions in several simple 
Hamiltonians,
and discuss the corresponding scaling of the gap. 
We will discuss how level crossings can be induced by disorder, and how their accumulation can
result in a complex spin glass
phase where the gap is everywhere exponentially small.
We will keep the discussion informal, and focus on toy models. 
A more precise discussion on realistic optimization problems will be presented after
the methods to study such complex phenomena will have been introduced in 
Sec.~\ref{sec:methods}. 

For concreteness in this section we shall only consider models of quantum spins $1/2$,
with Hamiltonians $\hH$ made of two terms. The first is 
diagonal in the eigenbasis of the $\hsiz$ operators, it
encodes the problem to be solved and we will refer to it as $\hH_P$. The second term
induces quantum fluctuations of strength $\G$ and we will call it $\Gamma \hH_Q$, in such 
a way that the total Hamiltonian is $\hH = \hH_P + \G \hH_Q$.
For concreteness, we will consider as a quantum term
a transverse field, $\hH_Q = -\sum_{i=1}^N \hsix$. 
Hence we denote here $\Gamma$ the strength of this transverse field, and
when speaking of an annealing it is understood to be from $\Gamma=\infty$ down to $\Gamma=0$.
The connection with the notations of Sec.~\ref{sec:quantum_computers} is easily made: 
the problem Hamiltonian $\hH_P$ corresponds to $\hH_{\rm f}$, the quantum $\hH_Q$ corresponds
to $\hH_{\rm i}$, and the correspondence between the interpolation parameter $s$ and $\Gamma$ is 
$\Gamma=(1-s)/s$.
Quantum fluctuations different from a transverse field are expected to have similar effects,
provided they are simple enough and in particular they do not contain detailed
information on the classical part of the Hamiltonian (for instance, a hopping quantum term 
$-t \sum_{\la ij\ra } (\hsix \widehat{\sigma}_j^x + \widehat{\sigma}_i^y \widehat{\sigma}_j^y)$ was 
considered in~\cite{FSZ11} and led to similar results).

\subsection{Second order transitions}
\label{sec:secondorder}

\subsubsection{Ordered models}

A well known example of a system without disorder 
that exhibits a second order phase transition
associated to a polynomially vanishing gap is the one dimensional Ising ferromagnetic
chain in a transverse field:
\beq\label{eq:1ddef}
\hH = - J \sum_{i=1}^N \hsiz \widehat{\sigma}_{i+1}^z - \G \sum_{i=1}^N \hsix \ .
\eeq
This Hamiltonian is integrable and can be solved completely~\cite{sachdev2001}.
Generically, in finite dimensional ordered models, the gap 
of the system, in the thermodynamic limit, vanishes when one gets close to the quantum phase transition as 
$\epsilon^{z\nu}$, where $\epsilon$ is a measure of the distance to the 
transition, $\nu$ is the exponent for the divergence of the correlation length
and $z$ the dynamic exponent. Right at the transition the gap vanishes only in
the thermodynamic limit, it thus scales with the size $N$ of the system, for
instance as $1/N$ in the one dimensional case.
As our goal is to study mean field models, for the motivations explained in
Sec.~\ref{sec:classical_mean_field}, here we want to mention also the Curie-Weiss 
mean field Hamiltonian~\cite{BJ83,DV04,DV05,DSSC06}:
\beq\label{eq:H_CW}
\hH = - \frac{J}{2 N} \sum_{i,j=1}^N \hsiz \widehat{\sigma}_j^z - \G \sum_{i=1}^N \hsix \ .
\eeq
This Hamiltonian is easily solved by the mean field construction, that amounts to replace
one of the $\widehat{\sigma}^z$ by its average $m$; one then obtains a single site Hamiltonian 
$\hH = - J m \widehat{\sigma}^z - \G \widehat{\sigma}^x$ and the magnetization is computed 
self-consistently, leading to the mean field equation
\beq
m =\frac{J m}{ \sqrt{ J^2 m^2 + \G^2 }} \tanh( \b \sqrt{ J^2 m^2 + \G^2 } ) \ .
\eeq
At zero temperature, this leads to a phase transition between a paramagnetic ($m=0$)
phase at $\G > J$ and a ferromagnetic phase with magnetization
$m = \sqrt{1 - (\G/J)^2}$ for $\G < J$. The order parameter $m$ is continuous at the 
transition and the ground state energy as a function of $\G$ has a singularity in
the second derivative. Hence the transition is of second order.
It is possible to show that the gap of the Hamiltonian vanishes polynomially (more precisely
as $N^{-1/3}$) at the phase
transition point $\G=J$~\cite{BJ83,DV04,DV05,BaSe12}, therefore a quantum annealing can find
its ground state in polynomial time.

\subsubsection{Disordered models}

Let us now consider what happens when disorder is introduced in the Hamiltonians 
described above. 
In the unidimensional case,
Eq.~(\ref{eq:1ddef}) now becomes:
\beq\label{eq:1ddis}
\hH = - \sum_{i=1}^N J_i \hsiz \widehat{\sigma}_{i+1}^z - \sum_{i=1}^N \G_i\hsix \ .
\eeq
Consider for instance the case where the $J_i$ and $\G_i$ 
are i.i.d. random variables, uniformly distributed in $[0,J]$ and $[0,\G]$ respectively
(negative couplings or transverse fields can be eliminated through a simple redefinition of the spins).
Note that in the classical limit $\G=0$ the ground state is $\s_i=1$ or $\s_i=-1$, hence there is no 
frustration in the model.

This random model has been extensively studied by means of the renormalization group
by Fisher~\cite{Fisher}. For our purposes, the main results are the following:
\begin{itemize}
\item A quantum critical point is present at $\G=J$, such that for $\G<J$ the system is ferromagnetic
while for $\G>J$ it is paramagnetic~\cite{Fisher}.
\item At the transition point $\G=J$, the typical gap is exponentially small: $\D = e^{- g \sqrt{N}}$, where $g$ has a finite
probability distribution over disorder (hence the distribution of $\D$ is very broad)~\cite{FiYo98,YoRi96}.
\item Above the transition, $\G > J$, the gap is typically of order one~\cite{FiYo98,YoRi96}. 
\item Below the transition, $\G < J$, the gap is exponentially small, but its logarithm is concentrated around its average: 
it is distributed according to a Gaussian distribution with
 $\overline{\log\D} \propto N$, and $\overline{ (\log\D)^2 } - \overline{\log \D}^2 \propto N$~\cite{FiYo98,YoRi96}. 
 \item The exponentially small gaps for $\G < J$ are due to crossings within different low energy levels, due to the fact
 that these levels are localized (in a sense that will be made more precise below)~\cite{SaMaToCa02,AC09,AKR10}.
\end{itemize}
Hence, a quantum annealing will typically encounter a gap $\D \sim \exp(-g \sqrt{N})$ at the transition $\G=J$,
followed by a series of gaps $\D \sim \exp(-g' N)$ for $\G<J$, and will typically require an exponential time to find
the ground state. However, it has been shown that the residual energy per spin
after a quantum annealing over a finite time $\t$ (as well as after a simulated annealing) goes to zero when
$\t\to\io$~\cite{SaMaToCa02,CaFaSa07}.
This implies that although finding the ground state is exponentially hard in $N$, finding a state whose energy
per spin coincides with the one of the ground state is indeed quite easy 
(it can be done in polynomial time)\footnote{
It was shown in~\cite{SaMaToCa02,CaFaSa07} that the total residual energy goes as 
$\D E \sim N/(\log\t)^\xi$
at large times. This implies that finding the ground state energy 
(i.e. finding a state with energy $\D E$ of order 1)
requires a time $\t \sim e^{N^{1/\xi}}$. However, if one is only interested in finding the ground state
energy {\it per spin}, it is enough to require that $\D E$ grows slower than $N$. For instance one
can choose $\D E = N/\log N$ and in this case a time $\t \sim N^{1/\xi}$ is enough.
}.

The mean field model that corresponds to Eq.~(\ref{eq:1ddis}) is the quantum Sherrington-Kirkpatrick (SK) model, which 
can be obtained either from the quantum Curie-Weiss model of Eq.~(\ref{eq:H_CW}) by including
randomness in the interaction couplings, or from the classical SK model defined in
Eq.~(\ref{eq:E_SK}) by the addition of a transverse field:
\beq
\hH = - \sum_{i<j} J_{ij} \hsiz \widehat{\sigma}_j^z - \G \sum_{i=1}^N \hsix \ .
\eeq
Here, the $J_{ij}$ are quenched i.i.d. Gaussian variables, with zero average
and variance $\overline{J^2_{ij}} = J^2/N$; hence, the couplings can be positive or 
negative, leading to frustration. This model has a quantum critical point at 
$\G=J$,
separating a spin glass phase at $\G<J$ from a paramagnetic one at 
$\G>J$~\cite{BrMo80_qSK}.
The analysis of the gap in this model is however more difficult, and it has not
yet been performed to our knowledge. However, it has been 
shown~\cite{AnMu_future}
that the whole spin glass phase at $\G<J$ is gapless for $N\to\io$. 
Because the spin glass phase is characterized, in the classical limit $\G=0$, by many
almost degenerate low energy minima, it is very natural to expect, like in the finite dimensional
case, the existence of level crossings between the energy levels corresponding to these minima
when quantum fluctuations are switched on~\cite{SaMaToCa02,AC09,AKR10}. These should lead
to an exponentially small gap in the whole spin glass phase, as we will discuss later in a simpler example.

\subsection{First order transitions}
\label{sec:firstorder}

\subsubsection{Ordered models}

We now turn to the discussion of first order phase transitions. Perhaps the simplest
mean field model without disorder 
that shows such a transition is a generalization of the quantum Curie-Weiss
model to 3-spin interactions~\cite{JKKMP10,BaSe12,FiDuVi11}:
\beq\label{eq:H_pCW}
\hH = - \frac{J}{3 N} \sum_{i,j,k=1}^N \hsiz \widehat{\s}_j^z \widehat{\s}_k^z 
- \G \sum_{i=1}^N \hsix \ .
\eeq
Like in the Curie-Weiss model the mean field nature of the model allows for an exact 
computation of its free energy density. A simple way to obtain the self-consistency
equation on the order parameter (that can be formally justified via a path integral 
representation of the model~\cite{JKKMP10,BaSe12}) is to replace
two $\widehat{\s}^z$ by their average $m$ and obtain a single site
Hamiltonian $\hH = -J m^2 \widehat{\s}^z - \G \widehat{\s}^x$, 
from which we get the mean field equation
\beq
m =\frac{J m^2}{ \sqrt{ J^2 m^4 + \G^2 }} \tanh( \b \sqrt{ J^2 m^4 + \G^2 } ) \ .
\eeq
The paramagnetic solution $m=0$ of this equation always corresponds to a local minimum of
the free energy. A ferromagnetic ($m>0$) solution however appears discontinuously on a spinodal
line in the $(\b,\G)$ phase diagram. The free energies of the two locally stable phases
cross on a first order transition line, distinct from the spinodal of the ferromagnetic phase.
The phase transition is here characterized by a jump of the magnetization. Correspondingly,
the first derivative of the ground state energy with respect to $\G$ has a jump, hence
the transition is of first order.

In this case, one can show analytically that the gap is exponentially small at the phase 
transition, $\D_{\rm min} \sim \exp(-\mu N)$, and compute analytically the coefficient 
$\mu$~\cite{JKKMP10,BaSe12}.
The reason behind this is indeed quite simple. The transition is characterized by an (avoided) 
crossing between two different eigenvalues: the state $|P\rangle$ corresponding to the 
paramagnetic solution, which is the
ground state at large $\G$, and the state $|F\rangle$ corresponding to the ferromagnetic 
solution, which is the ground state at small $\G$. One can see that these two states have an 
overlap $\langle F | P \rangle \sim \exp(-\mu N)$. Naturally,
the matrix element of the Hamiltonian between these states is 
$\g = \langle F | \hH | P \rangle \sim \exp(-\mu N)$,
and an analysis of the two level system similar to the one of Sec.~\ref{sec:QAA-LZ} 
leads to the exponential scaling of the gap at the phase transition. Therefore, a quantum 
annealing of this model requires an exponential time to find the ground state (despite the 
latter is again the trivial ferromagnetic one). It is also interesting to remark that performing 
a quantum annealing (as well as a classical annealing) of this model for a finite (with respect 
to $N$) time $\t$ always leads to a final energy that is extensively higher than the ground state
one, even when $\t\to\io$ (after $N\to\io$)~\cite{BaSe12}. This means that even the problem of 
approximating the ground state energy of this model through quantum annealing is very hard.
The effect of spinodals in mean field models with first order phase transitions on quantum
annealing has been recently discussed in~\cite{BaSe12}.

Note that the physics of finite dimensional models undergoing a first order phase transition 
is quite different from the one of mean field models. This is because in finite dimension the 
dynamics around a first order transition is dominated by nucleation events that are absent in 
the mean field treatment~\cite{Cavagna09}; this should affect the scaling of the gap, as well as boundary conditions in some unidimensional models~\cite{LaMoScSo12_2}.
We shall not discuss these issues further because in the following we will be mainly interested 
in mean field models.

\subsubsection{Disordered models}
\label{sec:firstorder_disordered}

We now turn to disordered spin glass models that show a quantum first order phase transition.
The simplest such model is obtained by introducing disordered couplings in Eq.~(\ref{eq:H_pCW}),
which amounts to perform the same step that leads from the Curie-Weiss to the 
SK model, or to add a transverse field to the fully connected $p$-spin
model of Eq.~(\ref{eq:E_pspinFC}). The resulting Hamiltonian is the one of the 
$3$-spin quantum spin glass~\cite{DT90,NR98}:
\beq\label{eq:H_3spin}
\hH = - \sum_{i<j<k} J_{ijk} \hsiz \widehat{\s}_j^z \widehat{\s}_k^z  - \G \sum_{i=1}^N \hsix \ .
\eeq
Here, the $J_{ijk}$ are quenched i.i.d. Gaussian variables, with zero average
and $\overline{J^2_{ijk}} = J^2/(2 N^2)$; hence, the couplings can be positive or 
negative, leading to frustration.
The classical ($\G=0$) thermodynamics of these models is very similar to the one
of random optimization problems, as was discussed in Sec.~\ref{sec:pspinFC},
and is described by the so-called ``Random First Order Transition" (RFOT) theory; for this reason
they will be particularly relevant for the rest of the discussion.

This model and similar ones
were studied through techniques that combine
replica and Suzuki-Trotter methods \cite{DT90,NR98,CGS01,BC01}, 
and it was shown that the model undergoes a first order quantum phase transition at low
temperatures.
In order to avoid introducing the replica method, that is not relevant for the present discussion,
in the following, instead of discussing the Hamiltonian (\ref{eq:H_3spin}), we consider the simplest
representative of the RFOT universality class, namely the Quantum Random Energy 
Model (QREM)~\cite{Go90,JKKM08,FGGGS10}.
This model is just the classical Random Energy Model (REM) introduced in 
Sec.~\ref{sec:REM} to which one adds
a quantum transverse field, and it can be thought as a model similar to 
Eq.~(\ref{eq:H_3spin}), but with interactions involving $p$ spins in the limit 
$p\to\io$~\cite{De81}.
It is described by the Hamiltonian
\beq
\hH = \sum_{\underline{\sigma}} E(\underline{\sigma}) \
|\underline{\sigma}\rangle \langle \underline{\sigma}|  - \Gamma  \sum_i \hat{\sigma}^x_i
\label{eq:H_E_Gamma}
\eeq
where $E(\underline{\sigma})$ are 
i.i.d. random variables, extracted from a Gaussian probability density 
with zero average and variance $N/2$. 

\begin{figure}
\centering
\includegraphics[width=.49\textwidth]{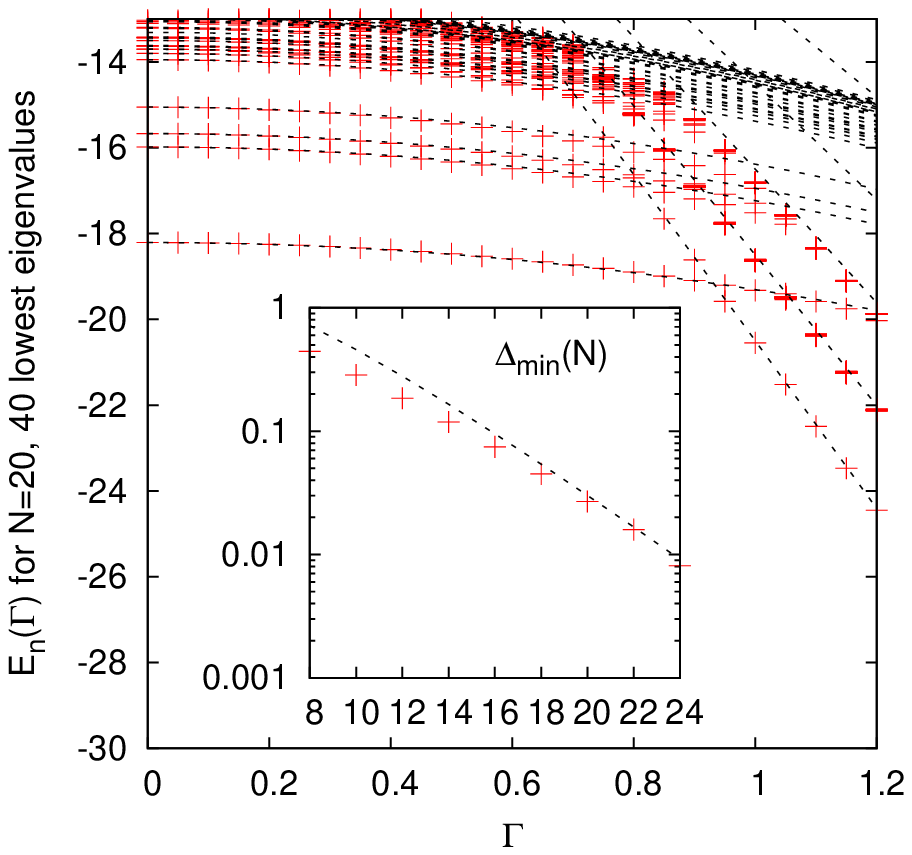}
\includegraphics[width=.49\textwidth]{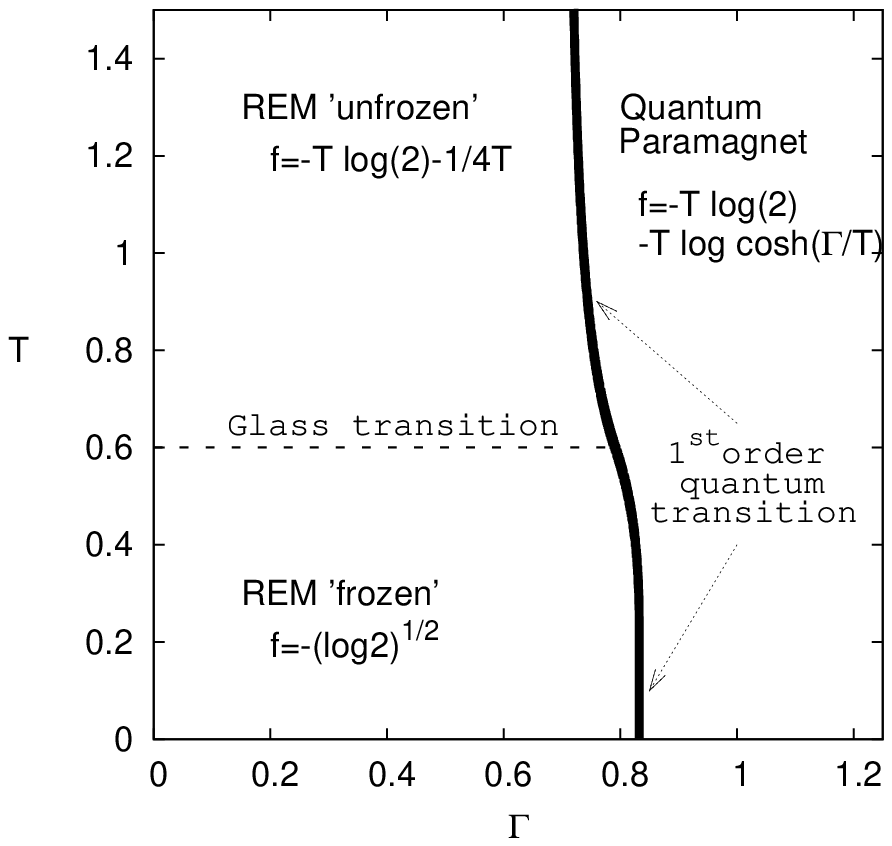}
\caption{\textit{(Left panel)} Spectrum of the QREM as a function of $\Gamma$~\cite{JKKM08}. 
Red dots represent the results 
from exact diagonalization of a system with $N=20$, dotted lines are the analytical values from lowest order perturbation theory. 
The inset shows the
scaling of the minimal gap as a function of the size $N$. 
\textit{(Right panel)}
Phase diagram as a function of temperature and
transverse field~\cite{Go90,JKKM08}.
}\label{Fig:Phase_REM}
\end{figure}

The complete phase diagram of the QREM as a function of $T$ and $\G$ has been obtained in~\cite{Go90}
by means of the replica method, and is reported in Fig.~\ref{Fig:Phase_REM}.
The existence of a first order phase
transition in (a slight variant of) this model has been rigorously confirmed in~\cite{FGGGS10}.
This phase diagram can be obtained via a series of extremely simple
arguments~\cite{JKKM08}:

\begin{itemize}
\item \textit{Extreme cases} 

We start by discussing the two limiting cases (i) $\Gamma$ = 0 and (ii) $\Gamma = \infty$.

(i) When $\Gamma = 0$ the model has a phase transition as a function of the
temperature~\cite{De81}, that we already discussed in Sec.~\ref{sec:REM} and
we briefly recall here. The micro-canonical entropy density is $s(e)=\log 2 - e^2$
on the interval of energy densities $e=E/N$ where it is positive, i.e. $[e_0,-e_0]$ with
$e_0=-\sqrt{\log 2}$. In consequence a condensation (or glass) transition occurs at the critical
temperature $T_{\rm c}=1/(2 \sqrt{\log 2})$: at high temperatures the free energy density is
$f_{\rm REM}(T) = - \frac{1}{4 T} -  T \log 2$ and an exponential number of configurations 
contribute to
the partition function. On the contrary at low temperatures the Gibbs measure is concentrated
on a finite number of configurations of energy density $e_0$, and 
$f_{\rm REM}(T)=e_0 = -\sqrt{\log 2}$;
the entropy density then vanishes.

(ii) When $\Gamma\gg 1$ the system corresponds to $N$ independent spins in a transverse field,
i.e. a simple Quantum Paramagnet (QP). As a consequence the free energy reads 
$f_{\rm QP}(T,\G) = - T \log( 2 \cosh(\Gamma/T))$ and the entropy
density $s(e)$ is the logarithm of a binomial distribution on $e \in [-\Gamma,\Gamma]$.

\item \textit{Perturbation theory}

We now discuss the perturbation theory around these two limiting cases. Consider first
a classical configuration $\us$ with a negative energy density $e(\us)=E(\us)/N<0$. The energy
density $e(\us,\G)$ of the corresponding eigenstate of $\hH$ can be computed order by order in
$\Gamma$, treating the transverse field as a perturbation, and this yields:
\beq\label{eP}
e(\us,\Gamma) = e(\us) + \frac{\Gamma^2}{N e(\us)} + O\Big(\frac{1}{N^2}\Big) \ ,
\eeq
as typical configurations that are reached by a single spin flip from a low energy 
configuration of the REM are the most numerous ones, with vanishing energy density (this will
be further explained in Sec.~\ref{sec:Altshuler}).
 
In the opposite limit $\G \gg 1$, the eigenstate of the pure transverse field are degenerate so we must use
degenerate perturbation theory. Consider the space of eigenstates where $N-k$ spins are aligned in the 
transverse field direction. This space has degeneracy $\binom{N}{k}$. Here we consider the classical part
of the Hamiltonian, $\hH_P = \sum_{\underline{\sigma}} E(\underline{\sigma}) \
|\underline{\sigma}\rangle \langle \underline{\sigma}| $ as a perturbation.
It is easy to show that the restriction of this Hamiltonian to each degenerate subspace is a random matrix,
whose elements are Gaussian variables with zero mean and variance $N/2^N$. Therefore this is a 
random matrix belonging to the Gaussian Orthogonal Ensemble (GOE) and
its spectrum is the usual semi-circle law with eigenvalues $\sim N/2^N$.
We obtain that on the QP side, at first order, 
the energy levels are those of free spins in a transverse field, 
with exponentially small corrections. See~\cite{JKKM10} for the second order computation.

The important outcome of these considerations is the vanishing as $N\to \infty$ of the
perturbative corrections around the two limits $\G=0$ and $\G \gg 1$, hence
the energy, entropy and free energy densities are not modified.
The free energy density of the QREM is thus 
$f_{\rm QREM}=\min[f_{\rm REM}(T),f_{\rm QP}(T,\G)]$, which leads to a first order phase transition 
when the values of the free energies of the two competing phases become equal.
The spectrum and the phase diagram of the model in the plane $(\Gamma,T)$ 
are shown in Fig.~\ref{Fig:Phase_REM}. 

\item \textit{Gap}

A good approximation of the minimum gap is given, as in the ordered case, 
by considering a two level
problem similarly to (\ref{two_level}) where the space is that spanned by the
ground states of the classical part of $\hH$ and of the transverse field, denoted 
respectively $ \ket{E_0}$ (that corresponds to the classical state of minimal energy)
and $\ket{QP} = 2^{-N/2} \sum_{\us} \ket{\us}$. 
The diagonal matrix elements
are the perturbed energies.  The off-diagonal elements
are proportional to the overlap $\langle E_0 \ket{QP} = 2^{-N/2}$. 
From this we obtain that $\Delta_{min} \propto 2^{-N/2}$ and the 
accuracy of this scaling is shown in the inset of Fig.~\ref{Fig:Phase_REM}.

\end{itemize}

The phase diagram of the QREM shares strong analogies 
with the results obtained in other quantum spin glass
models belonging to the RFOT class~\cite{DT90,NR98,CGS01,BC01}, 
consistently with the fact that the 
REM is a good approximation for more complex systems.
In all these problems the classical glass transition occurs as a function
of the temperature. At $T=0$ the system is in the glass phase and
the classical ground state is not extensively degenerate. 
We stress here that this is a crucial difference with respect to 
many other optimization problems, like $k$-SAT, where 
the glass transition also arises at $T=0$ as a function
of the density of constraints. 
In these cases the entropy density is non-vanishing at zero 
temperature and 
the transition is an entropic phenomenon.
We will show next how the role of entropy can be taken into account
in a simple extension of the REM, the random subcubes model we already
introduced in Sec.~\ref{sec:subcubes}.

Note the analogy of the computation of the gap in the QREM with the one of the Grover
problem discussed in Sec.~\ref{sec:Def-Imp}. However, in the QREM
the classical intensive ground state energy has fluctuations of order $1/\sqrt{N}$,
which induce similar fluctuations of the location in $\Gamma$ of the minimal gap.
Hence, in this case the optimal schedule discussed in Sec.~\ref{sec:Def-Imp} for the Grover
problem (that is based on the exact knowledge of the location of the minimal gap) cannot
be applied and a quadratic speedup is not achieved by the QAA in the QREM.
Hence, the behavior of a QAA (as well as that of a classical annealing)
is the same as in the ordered case: finding the ground state takes a
time $\sim 2^N$, exactly as in an exhaustive search. An annealing over
a finite time will lead to an extensive residual energy.

\subsection{Level crossings and localization on the hypercube}

In sections \ref{sec:secondorder} and \ref{sec:firstorder}, we presented an overview of several simple models
that show different phenomenologies: first and second order quantum phase transitions, associated to different scalings
of the gap at the transition. Moreover, we announced that some of these models are characterized by phases where the
gap is everywhere exponentially small due to an accumulation of {\it level crossings}~\cite{SaMaToCa02,AC09,AKR10}.
In this section, we shall give a more detailed description of this level crossing phenomenon.

The approach we shall use here
has its roots in the physics of Anderson localization. 
In~\cite{An58},
Anderson considered a particle hopping on a $d$-dimensional cubic lattice
of $N=L^d$ sites, and 
subject to a random disordered potential, as a good starting point for the 
comprehension of transport properties
of metals and of the metal-insulator transition. The Anderson model
describes an electron hopping in a disordered environment and the Hamiltonian 
reads
\beq\label{Anderson_model}
\hH_{\rm AM} = - t \sum_{\la i,j\ra} (\widehat{c}^{\dag}_i \widehat{c}_j + \widehat{c}^{\dag}_j \widehat{c}_i) + 
\sum_{i=1}^N \epsilon_i \widehat{c}^{\dag}_i \widehat{c}_i
\ ,
\eeq
where the first sum runs over the edges of the lattice, i.e. pairs of sites at distance 1,
the second sum runs over all $N$ sites, $\widehat{c}^\dag_i$ is the creation operator
at vertex $i$ of the lattice, and $\epsilon_i$ are i.i.d. random local energies, 
taken from a given distribution.
Anderson showed that, depending on dimensionality and on the strength of the disorder, 
the eigenstates of the Hamiltonian (\ref{Anderson_model}) can
be extended or localized in real space~\cite{An58}. 
What is particularly interesting for the present discussion is that the spectral properties 
change completely in the extended or localized regions of the spectrum. Indeed, extended
states are typically separated by much larger gaps than localized ones.
In the literature on Anderson localization, this is sometimes referred to as 
\textit{level repulsion}. Level repulsion is suppressed for localized states exactly because
the matrix elements that connect them are much smaller, hence the $\g$ in Eq.~(\ref{two_level})
is much smaller leading to smaller gaps.
Therefore, one might expect that the presence of avoided level crossings leading to exponentially
small gaps could be interpreted as some kind of localization phenomenon~\cite{AKR10}. We discuss
this point of view in detail in the rest of this section.

\subsubsection{A different view on the QREM: the Anderson model on the hypercube}

A simple observation is that the transverse field operator $\hH_Q = -\sum_i \hsix$ has non-zero matrix
elements between two states $|\us \ra$ and $|\us'\ra$ if and only if the Ising spins
configurations $\us$ and $\us'$ differ on exactly one variable $\s_i$. Considering the
$2^N$ configurations $\us \in \{-1,+1\}^N$ as the vertices of the $N$ dimensional hypercube,
and defining the Hamming distance $d(\us,\us')$ as the number of different bits
between the two configurations $\us$ and $\us'$, one can rewrite any Hamiltonian of the form
(\ref{eq:H_E_Gamma}) as
\beq\label{hopping}
\hH =
 \sum_{\underline{\sigma}}
 E(\underline{\sigma}) |\underline{\sigma}\rangle \langle \underline{\sigma}| - \Gamma \sum_{\la\underline{\sigma},\underline{\sigma}'\ra} (
\ket{\underline{\sigma}'}\bra{\underline{\sigma}} + 
\ket{\underline{\sigma}}\bra{\underline{\sigma}'} )
\ ,
\eeq
where the second sum runs over pairs of neighboring configurations on the hypercube, i.e. 
such that $d(\underline{\sigma},\underline{\sigma}')=1$.
In this formulation the QREM discussed in Sec.~\ref{sec:firstorder_disordered}
is exactly the Anderson model on the hypercube, as the energies $E(\us)$ are random i.i.d.
variables, precisely as the $\epsilon_i$ of (\ref{Anderson_model}), the transverse field
playing the role of the particle hopping term of the original Anderson model.

As we discussed in Sec.~\ref{sec:firstorder_disordered}, the energies
$E(\us)$ of the QREM provide a disordered environment
that induces localization on one of the vertices of the hypercube
when the hopping $\G$ is not strong enough. This is shown by the fact
that the low energy eigenstates at small enough $\G$ coincide
with the classical ones, hence with $|\underline{\s}\rangle$, at all
orders in perturbation theory for $N\to\io$, see Eq.~(\ref{eP})~\cite{JKKM08}.
The crucial difference with the Anderson model is that here, the delocalization
transition coincides with the first order phase transition,
and it happens via a level crossing between the localized and extended ground state.
Moreover, in the localized (small $\G$) phase, no level crossings are observed between
different states; this is clearly due to the fact that the energies of the
lowest eigenstates do not depend on $\G$, again due to Eq.~(\ref{eP})~\cite{JKKM08}.

The analogy between the QREM and the Anderson model is strongly appealing
for the physicists community since it brings the field of quantum information
and the one of Anderson localization of interacting systems in close contact~\cite{AKR10}.
On the other hand, the localization in the QREM is ``extreme'' in the sense that in the localized
phase the eigenstates of the Hamiltonian coincide with the classical states. This is due to the fact
that the Hamiltonian is non-local and flipping a spin typically costs an extensive energy.

Therefore, several interesting questions remain open.
In local Hamiltonians (Sec.~\ref{sec:QCC}), one expect that for finite $\G$, states are always delocalized over an 
exponential
number of states of the computational basis (see the discussion in~\cite{BucDelScar11}). What is then the meaning of many-body localization?
Is it possible to observe, in more general models, level crossings in the ``localized'' phase?
Does delocalization always happen through a first order transition?
And finally, what happens when the classical ground state is exponentially degenerate, 
unlike in the QREM?
In the following we try to answer some of these questions.

\subsubsection{A mechanism for level crossings between localized states}
\label{sec:Altshuler}

Let us first consider the case of local Hamiltonians in the sense of Sec.~\ref{sec:QCC}.
A mechanism that induces level crossings between localized states was proposed 
independently by Altshuler et al.~\cite{AKR10} and by Amin and Choi~\cite{AC09}.
It relies on the fact that in optimization problems with local interactions, the diagonal
energies $E(\underline{\s})$ are not uncorrelated as in the QREM. Therefore,
a careful choice of the classical energy function can lead to level crossings.
We now review this construction.

In~\cite{AKR10}, a classical random 
energy function that is a sum of local interactions was considered 
(namely, the Exact Cover problem defined in Sec.~\ref{sec:examples_optimization}).
The analysis starts by choosing an instance of the problem with $M-1$ clauses,
such that there are at least two isolated (i.e. separated by an Hamming distance of order $N$)
solutions $\underline{\sigma}_1$ and $\underline{\sigma}_2$
of all the $M-1$ clauses.  
These configurations represent degenerate
 eigenvectors for $\Gamma=0$. However, as soon as $\Gamma>0$
 the two ground state energies must split. 
 Let us call $\ket{E_1(\Gamma)}$
 and  $\ket{E_2(\Gamma)}$ the two eigenvectors that transform continuously
into $\ket{\underline{\s}_1}$ and $\ket{\underline{\s}_2}$ for $\G\to 0$,
and $E_1(\Gamma)$ and $E_2(\Gamma)$
the corresponding energies.

\begin{figure}
\centering
\includegraphics[width=.49\textwidth]{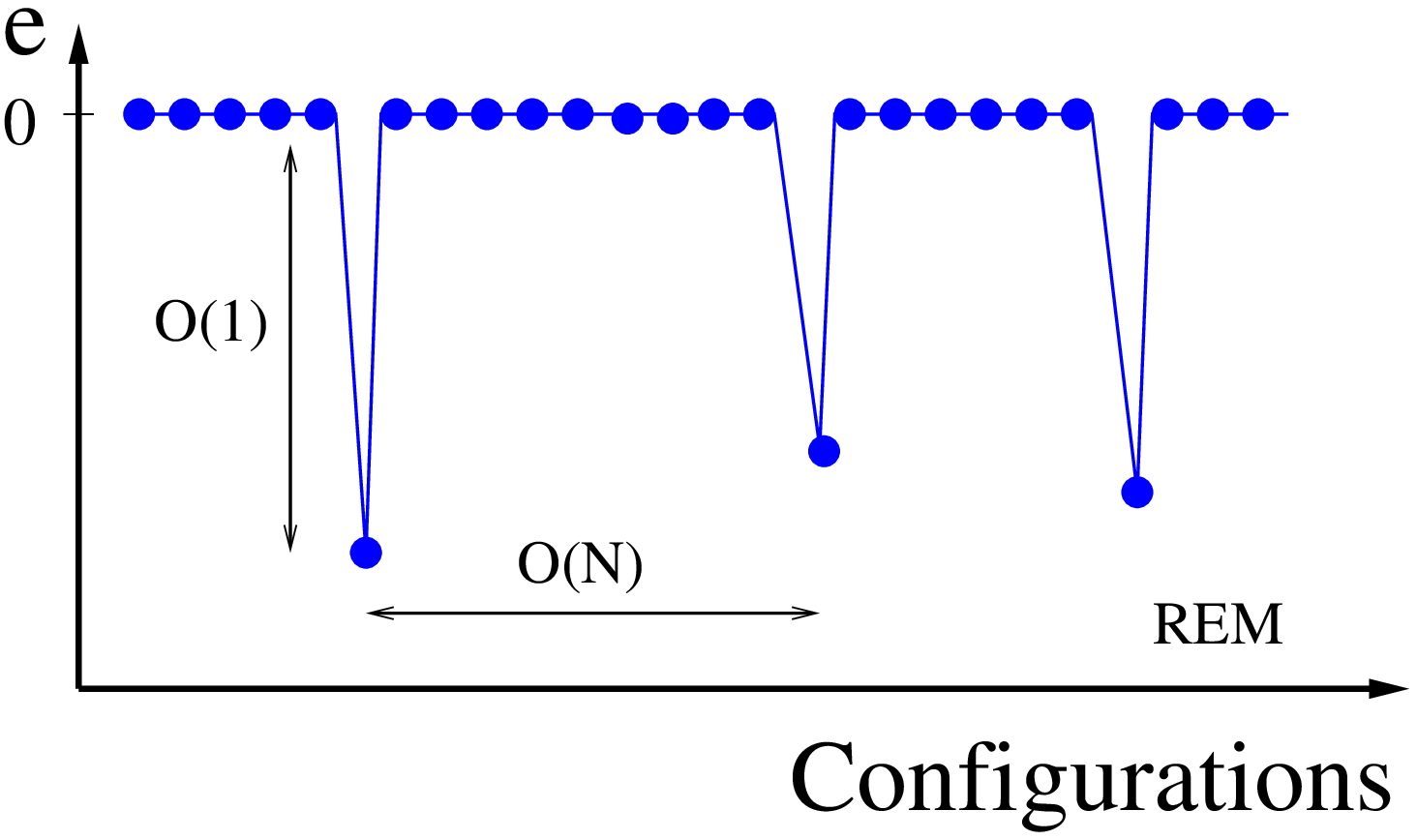}
\includegraphics[width=.49\textwidth]{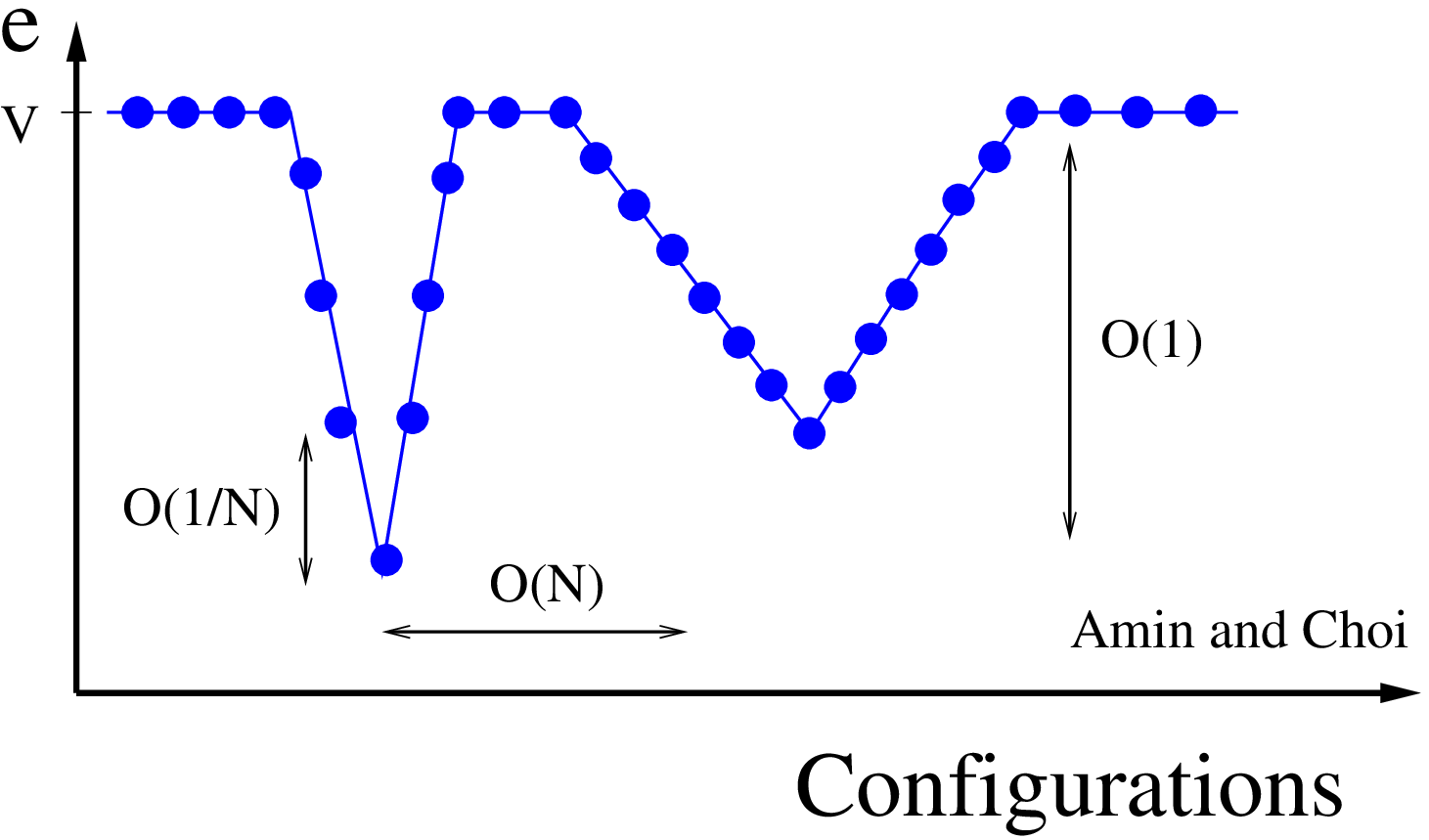}
\caption{
A comparison between the classical energy function of the QREM {\it (left panel)} and
the one of the correlated system studied by Amin and Choi~\cite{AC09} {\it (right panel)}.
The horizontal axis is a sketch for the $2^N$ dimensional space of classical configurations,
while the vertical axis is the corresponding energy density. Each point represents a distinct classical
configuration, and neighboring points have Hamming distance equal to 1.
}
\label{fig:landscape_AminChoi}
\end{figure}

The splitting of the two solutions can be computed using perturbation theory
for small enough $\G$, as we already did for the QREM.
One can write for any given non-degenerate classical eigenstate $\ket{\underline{\s}}$:
\beq
E(\Gamma,\underline{\s}) = E(\underline{\s}) + \sum_{n=1}^\io \G^{2n} F_{n}(\underline{\s}) \ ,
\eeq
with some coefficients $F_n(\underline{\s})$.
The effect of the presence of two degenerate solutions will appear only at order $n \propto N$
if the two solutions $\us_1$ and $\us_2$ have extensive Hamming distance, and therefore
it was neglected in the discussion~\cite{AKR10}.
Let's look to the first order as an example. It has the form
\beq\label{eq:F1}
F_1(\underline{\s}) = 
\sum_{\underline{\s'}  : d(\underline{\sigma},\underline{\sigma}')=1}
\frac{1}{E(\underline{\s}) - E(\underline{\s}')} \ .
\eeq
Fig.~\ref{fig:landscape_AminChoi} highlights the crucial ingredient in the construction of 
Amin and Choi~\cite{AC09} and of Altshuler et al.~\cite{AKR10}, and compares it to the QREM.
In the latter, the classical energies are uncorrelated and for low energy eigenstates, 
the coefficient $F_1$ turns out to have a finite limit for $N\to \io$. This
is because a low energy configuration is connected by a single
spin flip to $N$ configurations (hence there are $N$ terms in the sum), but those typically
have an extensive energy difference above it (hence the denominator is of order $N$). 
Considering intensive energies, the correction is therefore of order $\G^2/N$ as given in
Eq.~(\ref{eP}). 
The crucial difference for correlated energies $E(\underline{\s})$ that are sums of local
terms is that a spin flip always leads
to a {\it finite} energy difference with respect to the
starting point. Therefore, the denominators in the perturbative expansion are of order
1, and $F_1 \sim N$ is of the same order of the classical energy. 
Extending the argument to higher orders one easily sees that all orders in perturbation
theory are proportional to $N$, if the energy is the sum of local interactions, and therefore
contribute to the $\G$-dependence of the energy levels.

Hence, recalling that both $\underline{\sigma}_1$ and $\underline{\sigma}_2$
are assumed to have zero classical energy and calling 
$F_n^{12} = F_{n}(\underline{\s}_1)-F_{n}(\underline{\s}_2)$, we have:
\beq
 E_1(\G)-E_2(\G) = \sum_{n=1}^\io \G^{2n} F_{n}^{12} \ .
\eeq
It was argued in~\cite{AKR10} that for random problems the coefficients $F_{n}(\underline{\s})$
have the same average over the disorder; hence, $F_n^{12}$ has zero mean and naturally one can
assume that $\overline{(F_n^{12})^2} \sim N$. This leads, on average, to
\beq\label{eq:AKRsplit}
 E_1(\G)-E_2(\G) = \sqrt{N} \sum_{n=1}^\io \G^{2n} f_{n}^{12} \ ,
\eeq
where $f_n^{12}$ are finite for large $N$.
Eq.~(\ref{eq:AKRsplit}) shows that, if the first non-zero coefficient $f^{12}_{n^\ast}$ is negative, 
one can find a small enough
$\Gamma^{\ast}$ such that 
 $E_2(\Gamma^{\ast})-E_1(\Gamma^{\ast}) > \D E$ for any finite $\D E$. In fact,
\beq
\G^\ast \sim \left|\frac{\D E}{f^{12}_{n^\ast} \sqrt{N}}\right|^{\frac1{2 n^\ast}} \ .
\eeq

Now, we can add an $M$-th clause to the problem and fix $\D E$ in such a way that the new clause
introduces  at most a penalty $\Delta E$ on the classical energy (in the Exact Cover problem, $\D E=4$). 
Then, it still holds that $E_2(\Gamma^{\ast}) > E_1(\Gamma^{\ast})$.
At the same time, there is a finite probability
 that the additional clause will be satisfied by $\underline{\sigma}_2$ but not by $\underline{\sigma}_1$,
 so that $E_1(0) > E_2(0)$.
 Because in the Hamiltonian there are no particular symmetries, the matrix element between $\ket{E_1(\Gamma)}$
 and  $\ket{E_2(\Gamma)}$ will be non-zero and
the introduction of the additional clause induces an avoided level crossing, as in Eq.~(\ref{two_level}). 

Once again, the avoided level crossing is associated with an exponentially small gap because by assumption
the two solutions have an Hamming distance of order $N$, hence they can only be connected at order $N$
in perturbation theory, leading to a matrix element of order $\G^N$, i.e. exponentially small.
It is important to stress that in this construction the crossing happens for $\G < \G^\ast \sim N^{-1/(4 n^\ast)}$,
hence at very small $\G$ in the thermodynamic limit. For this reason, these crossings have been
called {\it perturbative crossings} in the literature~\cite{AKR10,FGGGS10,Dic2011}.
This scaling was not clearly found in the numerical experiments, but this was attributed to
the small exponent, visible only for large $N$~\cite{AKR10}.

A similar phenomenon was discussed by Amin and Choi~\cite{AC09}. Their construction is based on
a system whose classical energy function has a deep but narrow energy minimum, and a secondary
local minimum which is higher in energy but wider (see Fig.~\ref{fig:landscape_AminChoi}, right panel). 
This means that around the secondary minimum,
flipping a spin costs less energy. The denominators in the perturbation theory, Eq.~(\ref{eq:F1}),
are smaller around the secondary minimum than around the global one. In turn, the secondary minimum
lowers its energy more quickly under the action of the transverse field and eventually crosses
the classical minimum.

The analysis of~\cite{AKR10,AC09} points out an important 
mechanism
that can induce level crossings representing a serious bottleneck for quantum algorithms, that is
a missing ingredient of the QREM, which is the simplest Anderson-like model of localization on the hypercube:
namely,
the fact that in systems with local interactions the diagonal energies $E(\underline{\s})$ are
correlated, and most importantly a single spin flip always lead to a finite change in the classical
energy, which is not the case in the QREM. Thanks to this, the energy densities of the classical eigenstates have a non-trivial perturbative expansion in $\G$. Therefore, one can find particular realizations of the disorder, such
that the energy of a classically excited state decreases faster, as a function of $\G$, 
than the ground state energy, leading to a crossing at small $\G$. 

This mechanism is for the moment only understood in perturbation theory (whose validity
for these systems has been criticized~\cite{KnySme10}), and therefore it holds whenever perturbation 
theory holds, that is, if at small enough $\G$ the full eigenstates of the quantum problem remain
close enough to the classical eigenstates. This is what has been called
a ``many-body localization'' phenomenon in~\cite{AKR10}.
The other important ingredient is a very particular 
construction of the instances of the problem, that admit only two 
solutions.
But, as we already discussed in Sec.~\ref{sec:generating_USA},
typical instances of generic random optimization problems,
even close to the satisfiability threshold, have an exponentially large number of solutions, 
and so we expect that non-degenerate perturbation theory should not hold, 
and the spectrum should be much more complex. 
Therefore we would like to understand what happens generically 
in problems that exhibit multiple and not necessarily
isolated solutions. In particular, do the avoided crossings remain finite and isolated in $\G$
(hence leading to singularities in the ground state energy for $N\to\io$)
or do they proliferate and accumulate, leading to a continuum of level crossings and a gapless
phase?
The latter question is particularly important, because it has been argued that a finite number
of level crossings can be eliminated by
suitable redefinitions of the quantum 
Hamiltonian~\cite{Ch10,Ch10b,Ch11,DicAm10,DicAm11,Dic2011,FGGGS10}.
We will address it in the next section.

\subsection{Level crossings and the role of entropy: the random subcubes model}
\label{sec:qsubcubes}

\begin{figure}
\centering
\includegraphics[width=.49\textwidth]{landscape_REM.eps}
\includegraphics[width=.49\textwidth]{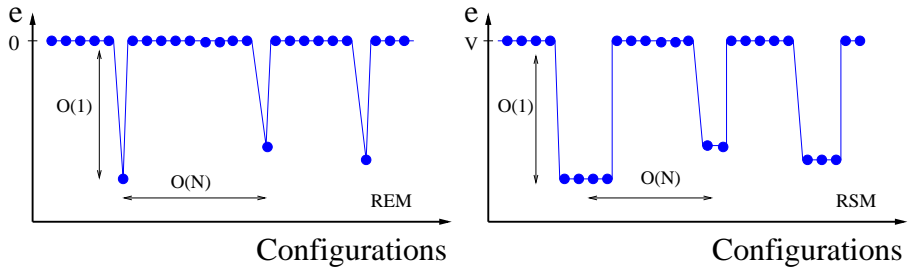}
\caption{
A comparison of the energy density landscape of the REM {\it (left panel)} and
RSM {\it (right panel)}, using the same conventions as in Fig.~\ref{fig:landscape_AminChoi}.
The main difference is that in the REM, low energy configurations are isolated and typically surrounded
by configurations that have extensively larger energy. Conversely, in the RSM low energy configurations
are arranged in clusters, each containing $2^{N s}$ degenerate neighboring configurations.
}
\label{fig:landscape_entropy}
\end{figure}

Motivated by the previous discussion, we now analyze a simple extension of the REM 
that takes into account the role of the massive ground state degeneracy: the Random
Subcubes Model (RSM) introduced in its classical version in~\cite{rcm}. 
Given that this model can be fully solved and reproduces most of the phenomenology
we are interested in this section, we will discuss its properties in some detail. 
Some of these results have been published in~\cite{FSZ10}.

The cost function (or problem Hamiltonian $\hH_P$) of the classical model
has been defined in Sec.~\ref{sec:subcubes}.
We recall here that there are $2^{N(1-\a)}$ random clusters (that have the topology of sub-hypercubes of the total Hilbert space, hence the name 
subcubes); when these clusters are 
disjoint (the regime of interest here) configurations belonging
to a cluster $A$ have a random classical energy $e_0(A)$. Configurations that do not belong to any
cluster have classical energy $V$ and we assume that $V \gg \max_A e_0(A)$.
The classical properties of the model have been discussed in Sec.~\ref{sec:subcubes},
where it was shown that at large enough $\alpha>\a_{\rm sep}$ the space of low energy configurations
is decomposed into a set of disconnected clusters separated by large energy
barriers. This is illustrated in Fig.~\ref{fig:landscape_entropy} that should make the difference
between the REM and RSM evident.
All our analysis of the Quantum RSM (QRSM) will be restricted to the region $\alpha>\a_{\rm sep}$
when the clusters are well disjoint (see Sec.~\ref{sec:subcubes}).

The main result of this section will be that quantum fluctuations, combined with the cluster structure, 
give rise to a series of level crossings induced by a combined energetic-entropic effect.
Before going into the details, it is useful to give an overview and
illustrate the way in which the model will be analyzed.
In Sec.~\ref{spectrum_clusters_A3} we will discuss the 
spectrum of the clusters at finite $N$ in the limit $V\to\infty$. 
In this limit the Hamiltonian is block diagonal, each
block corresponding to one cluster. The spectrum is characterized 
by true level crossings between states belonging to different clusters. 
The level crossings are due to the interplay of the classical energy and the classical entropy 
of the clusters. Quantum fluctuations, indeed, favor more entropic clusters. 
In Sec.~\ref{quantum_PM_A3} we will consider the case of finite $V$ (still at finite $N$). 
A finite $V$ reintroduces a lot of additional states that have to be taken
into account. They allow to connect the clusters by single spin flips,
therefore the Hamiltonian is no longer block diagonal. We will treat this
situation by perturbation theory and variational arguments to show
that a finite (large) $V$ induces only minor modifications with respect to the
infinite $V$ case. 
In Sec.~\ref{exact_diagonalization_A3}  we will investigate
the low energy spectrum obtained by exact diagonalization
for finite system sizes, and show that it is characterized by several avoided level crossings.
In Sec.~\ref{Thermodynamic_limit_A3} and \ref{finite_temperature_A3} 
we will consider the thermodynamic limit $N\to\io$.
In Sec.~\ref{Thermodynamic_limit_A3} we will focus on the ground state, at $T=0$. 
We will show that there is a first order phase transition that separates a Quantum
Paramagnetic (QP) phase, at large $\Gamma$, from a Spin Glass (SG) phase, 
at smaller $\Gamma$, like in the QREM. 
In the spin glass phase the ground state continuously changes
from one cluster to the other as a function of $\Gamma$, because of the level crossings between
different clusters; the latter accumulate for $N\to\io$ giving rise to a unique SG phase,
and are therefore distinct phenomena with respect to the first order transition.
Finally in Sec.~\ref{finite_temperature_A3} we will consider the case $T>0$:
we will show in particular that quantum fluctuations promote the 
glass transition. 
In Sec.~\ref{discussion_A3} we summarize and we comment the results.

\subsubsection{Spectrum of the cluster Hamiltonian}\label{spectrum_clusters_A3}

\begin{figure}
\centering
\includegraphics[width=.4\textwidth]{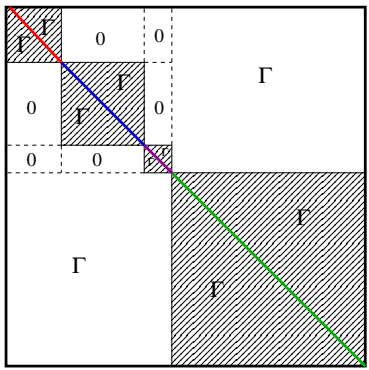}
\caption{
Schematic Hamiltonian matrix representing
a finite size realization of the QRSM with 3 clusters 
(red, blue and purple components). 
The biggest green sector represents states that
do not belong to $\SS$. 
The Hamiltonian has zero matrix elements between 
states belonging to different clusters because 
 for $\a>\a_{\rm sep}$ their Hamming distance is bigger than one.
 The size $n_A$ of each cluster block is fixed by its entropy $n_A=2^{N s(A)}$
 while the size of green component is much larger $n_V \sim 2^N$.
 We indicated with $\Gamma$ the sectors of the Hamiltonian where
 there are non-zero off-diagonal matrix elements; still $\Gamma$
 connects only classical configurations at Hamming distance 1, therefore the matrix
is very sparse in these blocks.
}
\label{fig:Hamiltonian_subcubes}
\end{figure}

We will now study the spectrum of the quantum 
Hamiltonian $\hH = \hH_P + \G \hH_Q$ as
a function of $\G$, and from now on 
we focus on the region $\a > \a_{\rm sep}$ where clusters are well separated
(see Sec.~\ref{sec:subcubes}),
which is the most interesting for our purposes.
The computation of the spectrum for $\a< \a_{\rm sep}$ is more complicated,
because in this region the clusters have overlaps and the arguments below do not apply
straightforwardly
(although they might be generalized for $\a > \a_{\rm d}$ where the overlaps
are exponentially small~\cite{rcm}).
A schematic example of the Hamiltonian describing a 
finite system with three clusters in the regime
where the clusters are well-separated 
is shown in Fig.~\ref{fig:Hamiltonian_subcubes}.

\begin{figure}
\centering
\includegraphics[width=.48\textwidth]{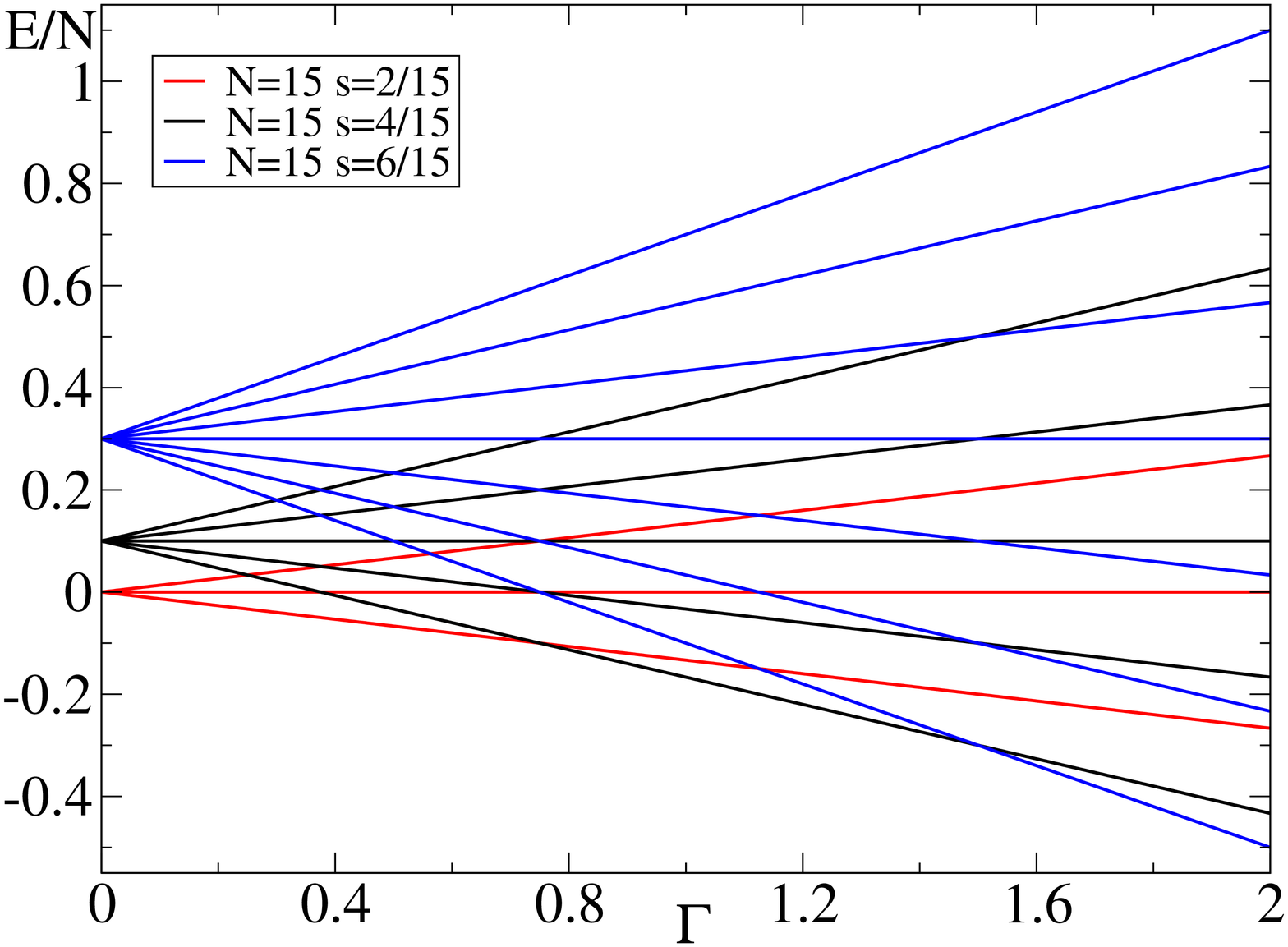}
\includegraphics[width=.48\textwidth]{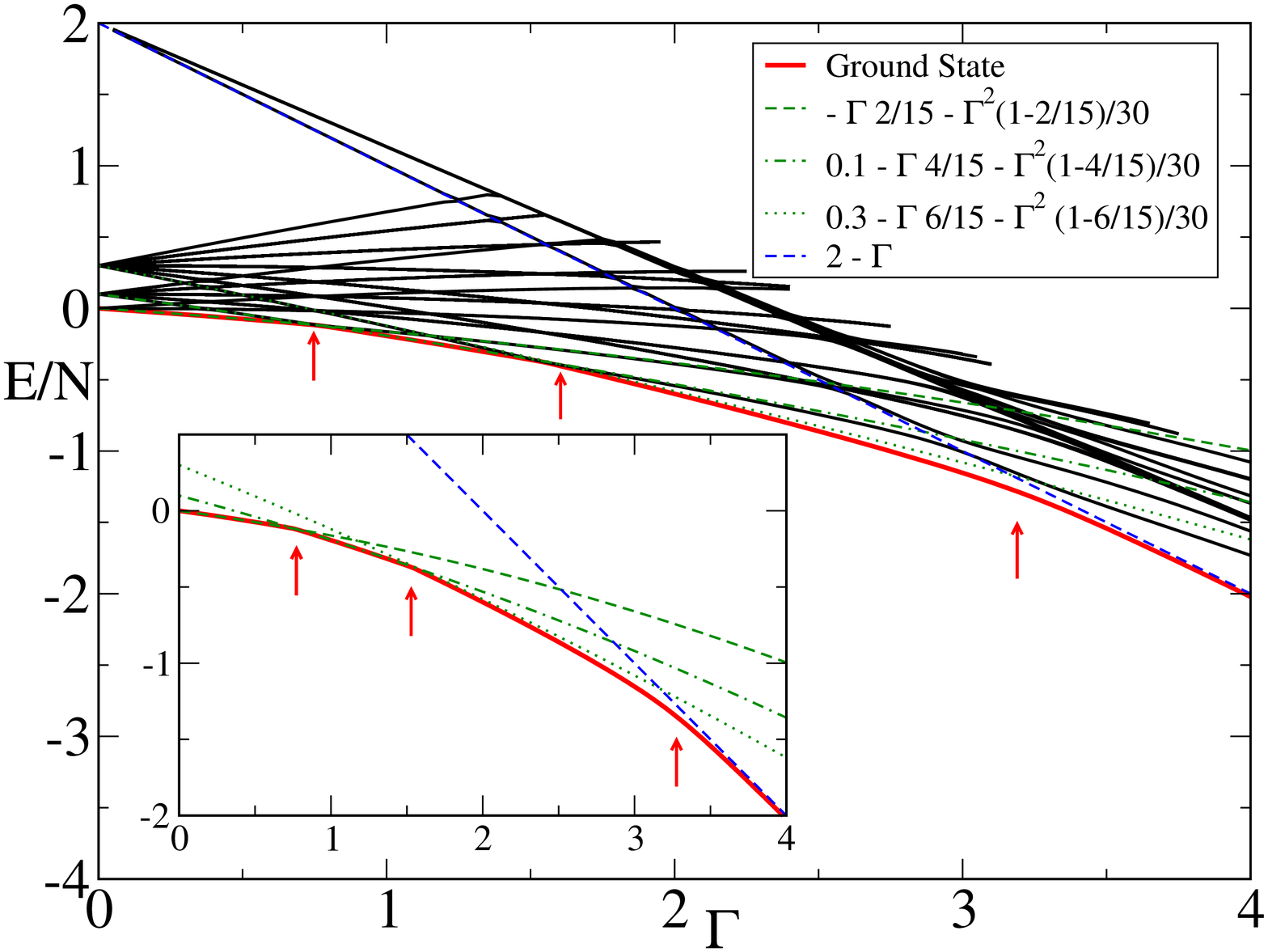}
\caption{Low energy spectrum for a system with $N=15$ and 3 clusters 
at Hamming distance larger than 1
such that $\{(s(A_i),e(A_i))_{i=1,2,3}\}=\{(2/15,0);(4/15,0.1);(6/15,0.3)\}$. 
\newline
{\it (Left panel)} Spectrum in the $V=\infty$ case. To each cluster $A_i$
corresponds the spectrum of $N s(A_i)$ free spins in a magnetic field.
\newline
{\it (Right panel)} Partial spectrum for finite $V=2$ obtained by exact diagonalization,
with a zoom on the ground state in the inset.
Green lines are the results of second order perturbation theory, the blue line is $e_{\rm QP} =V-\G$. 
At small $\G$ the low energy spectrum is
in good agreement with that at $V=\infty$. 
For larger values of $\G$, avoided level crossings appear (marked by red arrows).
The crossings at smallest $\G$ involve different clusters. The largest crossing, instead,
involves the ground state of the spectrum connected to the classical low 
energy spectrum (set $\SS$) and the ground state of the
$V$-band; this crossing becomes a true first order phase transition
in the $N\to\infty$ limit.
}
\label{fig:spectrum_subcubes}
\end{figure}

Remember that we call $\SS$ the set of all classical 
configurations that belong to at least one cluster ($\hH_A$ being the Hamiltonian of a cluster), 
and have therefore classical energy extensively
smaller than $N V$;
configuration that do not belong to $\SS$ have energy $N V$ (and are described by the Hamiltonian $\hH_V$).
The total Hamiltonian is $\hH = \sum_A \hH_A + \hH_V + \G \hH_Q$.
We consider first the (``hard'')
$V\to \infty$ limit where $\hH_P$ is infinite for the states that do not 
belong to
$\SS$: then we can project out these states from the Hilbert space and look to the
restriction of $\hH = \sum_A \hH_A + \G \hH_Q$ on $\SS$, which contains $2^{N s_{\rm tot}}$ 
states.
Because the matrix $\hH_Q$ only connects configurations at unit Hamming distance, and
different clusters have distance of order $N$, the
Hamiltonian $\hH$ has no matrix elements connecting different clusters.
Therefore we can diagonalize $\hH$ separately in each cluster. 
The restriction of $\hH$ to a given cluster $A$
with $N s(A)$ free spins is equal to $\hH_A$ plus the Hamiltonian of $N s(A)$ 
uncoupled spins in a transverse
field, its spectrum is hence made of levels 
\beq\label{Ek_spectrum_Vinfty}
E_k(A) = N e_0(A) + (2 k - N s(A) ) \G \ , \ \ \ k = 0,\cdots,Ns(A) \ ,
\eeq
each $\binom{N s(A)}{k}$
times degenerate. In particular the lowest level has energy per spin 
$e_{GS}(A) = e_0(A) - \G s(A)$, therefore
the energy of clusters with larger entropy 
decreases faster with $\G$. 
In this regime then one expects level crossings 
between states belonging to different clusters.
In the situation where bigger clusters at $\G=0$
have larger classical energy, which is the case for 
most random optimization problems, the level crossings concern the 
ground state and at $T=0$ each crossing corresponds to a
global rearrangement of the system.  
A simple example of a spectrum in the $V=\infty$ limit regime
for a finite system containing three clusters is shown 
in the left panel of Fig.~\ref{fig:spectrum_subcubes}. Note that
as long as the clusters are
well separated, due to the $V\to\infty$ limit,
there are no corrections in the size of the system. The crossings
are not avoided and the degeneracy of the states is not removed, 
due to the complete independence of the 
Hamiltonian sectors describing each cluster.

\subsubsection{Quantum paramagnetic state}
\label{quantum_PM_A3}

Next, we consider a ``soft'' version of the model in which $V$ is finite
(still with $V \gg \max_{A} e_0(A)$).
Therefore now $\hH$ is defined on the full Hilbert space $\HH$.
In this case, in addition to the $2^{N s_{\rm tot}}$ energy levels discussed above
(that we shall refer to as the $\SS$-band), 
there exists another set
of $2^{N} - 2^{N s_{\rm tot}} \sim 2^{N}$ levels (the $V$-band), whose energy is
expected to be of order $V$ at small $\G$.

For the states in the $\SS$-band we use perturbation theory in $\G$.
As soon as the transverse field is switched on 
a first order correction in $\G$ to the states in 
the $\SS$-band is present. This correction 
comes from 
the partial lifting of the degeneracy within the cluster
and it is given by the spectrum $E_k(A)$ in Eq.~(\ref{Ek_spectrum_Vinfty}).
A second order correction is induced by the 
presence of the $V$-band at finite $V$. 
In order to compute it one can apply perturbation theory
assuming as unperturbed basis the one that diagonalizes
the perturbation $\hH_Q$ inside each cluster.
In particular we are interested in the correction to the lowest
energy level $e_{GS}(A)$ in the clusters
whose state $| GS(A) \rangle$ is given by all free spin 
polarized in the direction of the field.
Then the correction is 
\beq
\Delta E_A^{\G^2} = \sum_{|\psi\rangle \notin A} \frac{| \langle \psi | \hH_Q | GS(A) \rangle |^2}{E_{\psi}-E_{GS(A)}}
= \frac{\G^2 (1-s(A)) N}{NV - N e_{GS}(A)} 
\eeq
and  at any finite order $n$ the correction to the energy per spin is
${\cal O}((\G^{2}/(NV))^n)$, so it vanishes in the thermodynamic limit. 
This mechanism is similar to the QREM and is due to the fact that states 
at the boundary of the clusters have extensive larger energy.

To study the lowest energy level in the $V$-band $e_{GS(V)}$ 
it is convenient to rewrite the Hamiltonian in the following way:
\beq
\hH = \underbrace{N V \widehat{I} - \G \sum_i \hs_i^x }_{\hH_{QP}}  -  
\underbrace{N \sum_{A} ( V - e_0(A) )  | A \rangle \langle A |}_{\hH_{\SS}} = \hH_{QP} -\hH_{\SS}
\ ,
\eeq
where $\widehat{I}$ is the identity and 
$|A \rangle \langle A |=\sum_{\us\in A}|\us \rangle \langle\us |$ indicates the projector over 
the cluster $A$. $\hH_{QP}$ acts on the entire  Hilbert space while 
$\hH_{\SS}$ acts only on the subspace spanned by the
clusters. 
This form aims to interpret $\hH_{\SS}$ as a ``perturbation''
over $\hH_{QP}$ which describes a system of $N$ free spins in a transverse field
(with a shift $NV$ in the energy). 
However the ``perturbation'' is not in the strength of the energy, which
may be large, but in the number of states that are involved.
Note, in fact, that Rank($\hH_{\SS}$)$\ll$Rank($\hH_{QP}$), 
being  Rank($\hH_{\SS}$)$={\cal R}=2^{N s_{\rm tot}}$ and 
Rank($\hH_{QP}$)$=2^{N}$. This, together with the fact that the perturbation matrix is
positive defined (it shifts some states all in the same direction) allows to apply
the results of small rank perturbation analysis~\cite{smallrank}
in order to study $e_{GS(V)}$.
From these results we can safely say that 
\beq\label{small_rank_spectrum}
 E_{QP}^{k-{\cal R}} \leq E_H^k \leq E_{QP}^k \hspace{0.5cm} \text{ for $k=1,\dots,2^N$}
 \ , 
\eeq
where $E_{QP}^k$ and $E_{H}^k$ are respectively the $k$-th eigenvalues of $\hH_{QP}$ and $\hH$, and
we assume $E_{QP}^k=-\infty$ when $k\leq 0$. 
In particular when $\G$ is small, $e_{GS(V)}$  is
 larger than all the energies in the $\SS$-band. 
This implies that 
 \beq\label{lower_bound}
V-\G \leq e_{GS(V)}
\ .
\eeq
The results from small rank perturbation (\ref{small_rank_spectrum})
also shows that the spectrum of the $V$-band is close to 
the one of $N$ free spins 
in transverse field with classical energy~$N V$:
$$
E^k_V = N V + (2 k - N ) \G \ , \ \ \ k = 0,\cdots,N \ ,
$$
with degeneracy close but not equal to $\binom{N}{k}$.
We expect that the unperturbed ground state of $\hH_{QP}$, 
$|QP\rangle = 2^{-N/2} \sum_{\us}\ket{\us}$
 describes well the lowest energy level of the $V$-band $e_{GS(V)}$ and
 remains unaffected by the presence of the 
 states in $\SS$ for all $\G$ except from the region
 where it crosses the spectrum of $\SS$.
The reason for this comes from the intuition that in absence of $\hH_{\SS}$
the spectrum of $\hH_{QP}$ is highly degenerate, especially in the middle of the band.
Then, also comforted by the results of exact diagonalization, we expect that
the states that recombine the most in order to create the $S$-band when $\hH_{\SS}$ is applied,
are those belonging to the more degenerate part of the spectrum. 
On the contrary, $|QP\rangle$ is made of all spins aligned along $\G$
 without degeneracy and thus it is weakly perturbed by $\hH_{\SS}$.
A rigorous study of this energy level is not possible, but we can use
a variational argument to understand its behavior.
The state $|QP\rangle$ has exponentially small overlap
with any state in the $\SS$-band  $\langle \psi(A) | QP \rangle \sim {\cal O}(2^{-Ns(A)/2})$  
and thus it gives an expectation
value of $\hH$ equal to $\langle QP | \hH | QP \rangle = N(V-\G)+{\cal O}(2^{- \gamma N})$
for some $\gamma$.
If we interpret this as a variational upper bound on the 
true ground state of the $V$-band we get:
\beq\label{upper_bound}
e_{GS(V)} \leq V - \G
\ .
\eeq
Combining (\ref{lower_bound}) and (\ref{upper_bound}) we obtain 
\beq\label{energy_QP}
e_{GS(V)} = V - \G + {\cal O}(2^{- \gamma N})
\ ,
\eeq
and the corresponding eigenvector remains up to
exponentially small corrections the same $|QP\rangle$,
which is uniformly extended in the basis $| \us \rangle$.

\subsubsection{Exact diagonalization results}\label{exact_diagonalization_A3}

We checked these predictions for the spectrum by means
of exact diagonalizations for a system made of $N=15$ spins.
The results are shown in Fig.~\ref{fig:spectrum_subcubes},
in the right panel. There we have plotted the spectrum 
of a system made by three clusters characterized by
classical energy and entropy 
$\{(s(A_i),e(A_i))_{i=1,2,3}\}=\{(2/15,0);(4/15,0.1);(6/15,0.3)\}$
and $V=2$.
The plot shows that for small $\G$ the states in the $V$-band
do not affect those in the set $\SS$, whose spectrum is
in good agreement with that at $V=\infty$, in the left panel. At larger $\G$, 
avoided level crossings first appear between the ground states of
different clusters.
Finally, an avoided crossing happens with the
ground state of the $V$-band, whose slope in $\G$ is much larger
due to the big entropy which characterizes this sector.
We have also plotted in green the analytical result that we obtain
up to second order in perturbation theory for
the lowest energy level of each cluster and in blue
the energy of the quantum paramagnetic state.
We see that the true ground state, crossing after crossing, well interpolates
between all these curves.
Since the clusters have Hamming distance proportional to $N$, we expect 
all these crossings to be avoided at finite $N$  producing 
exponentially small gaps~\cite{AC09, AKR10, FGGGS10}.

\subsubsection{Level crossings in the thermodynamic limit}
\label{Thermodynamic_limit_A3}

\begin{figure}
\centering
\includegraphics[width=.47\textwidth]{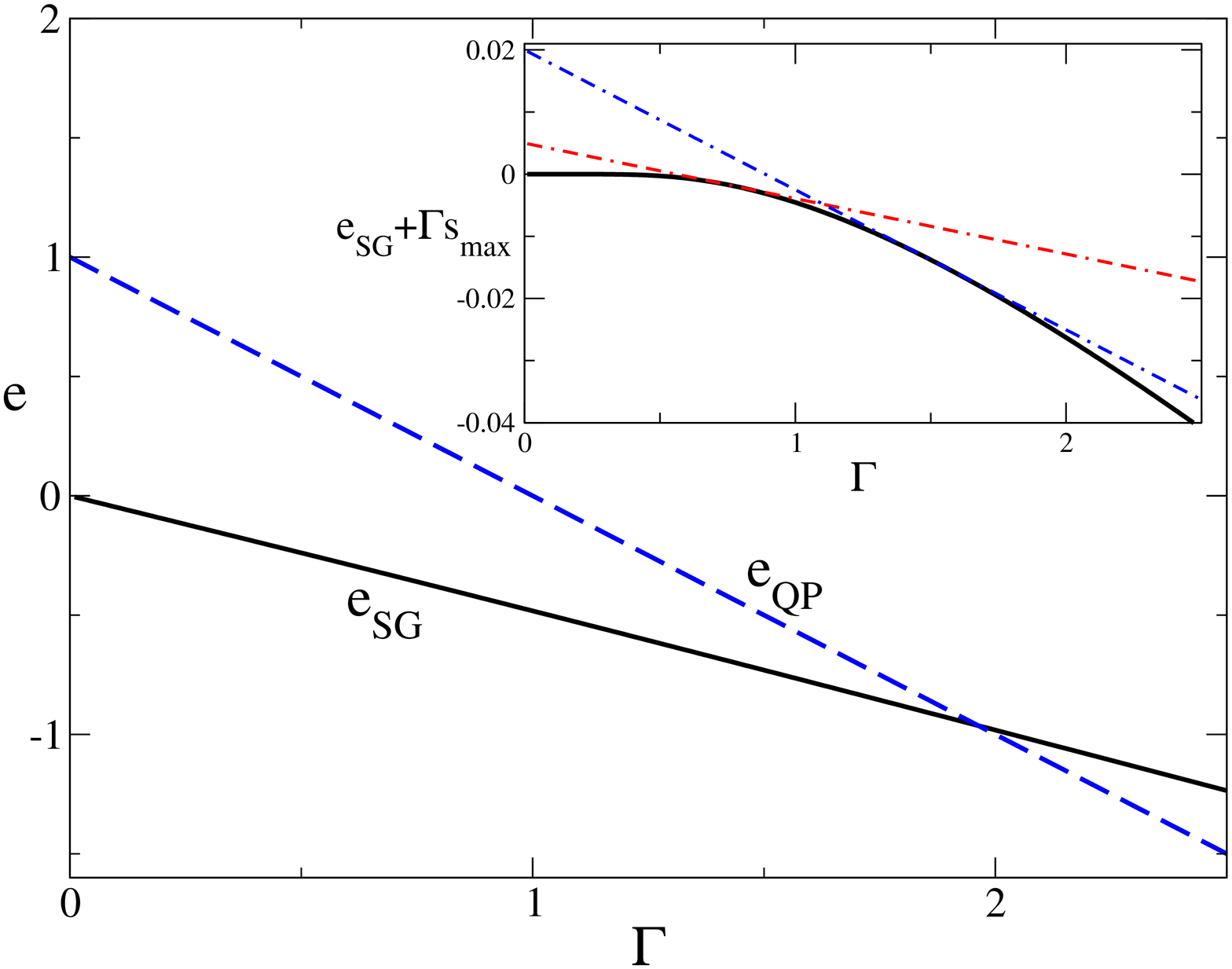}
\includegraphics[width=.47\textwidth]{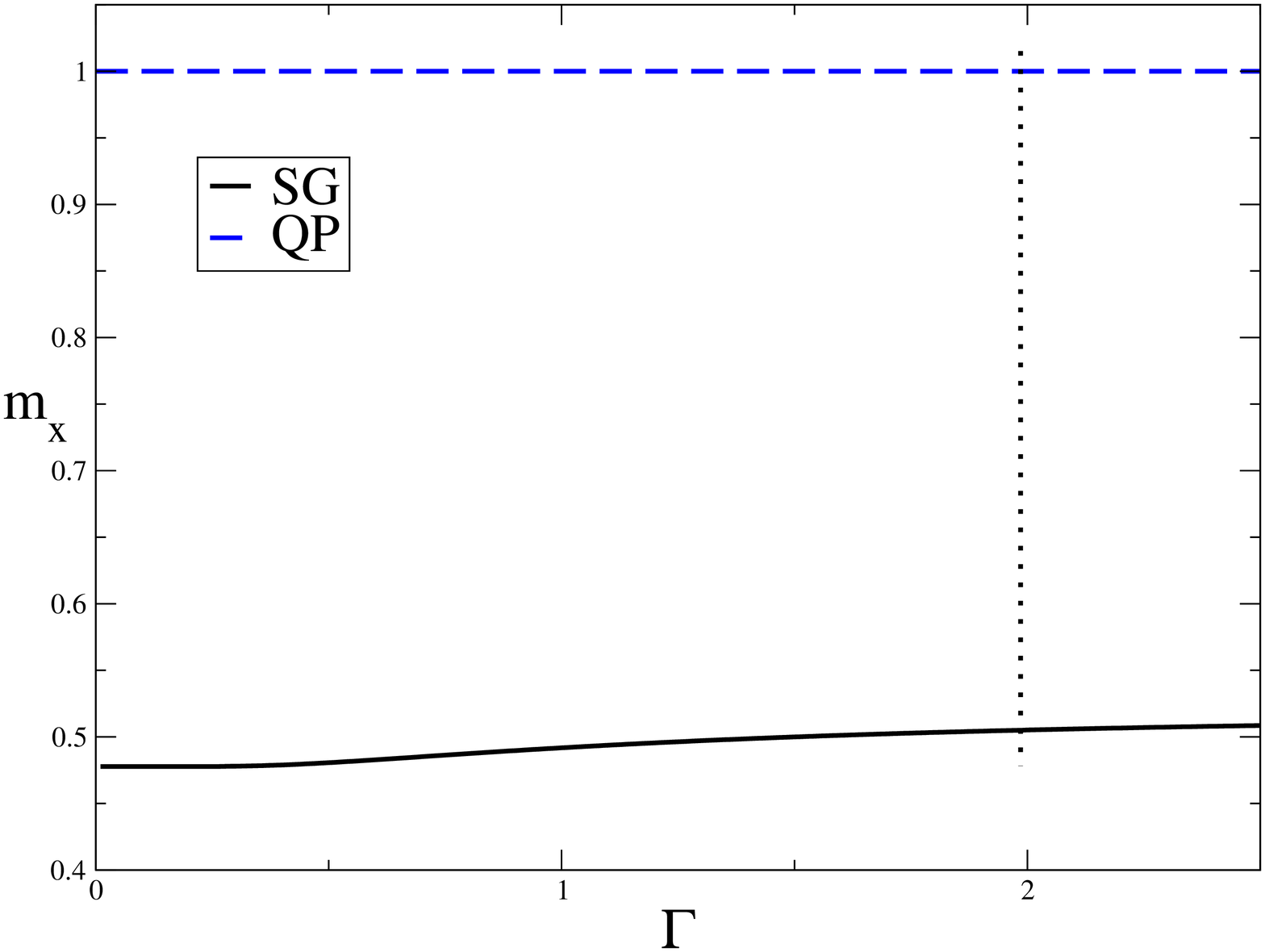}
\caption{ Results for the QRSM in the region $\a_{\rm sep}<\a<\a_{\rm s}$.
As an example we choose (following~\cite{rcm}) $p=0.7$, $\a=0.85$, and
$g(e_0) = [2 + e_0/e_{\rm m} - (e_0/e_{\rm m}) \log(e_0/e_{\rm m})]/3$
for $e_0 \in [0,e_{\rm m}]$ with $e_{\rm m}=0.1$. 
\newline
{\it (Left panel)} Energy of the SG ground state
[Eq.~(\ref{SGe}), full line]
and of the QP state $e_{\rm QP} = V - \G$ 
for $V=1$ (dashed line).
A first order transition between the two states happens at $\G \sim 2$.
{\it (Inset)} Level crossings in the SG state.
For better readability we plot $e_{\rm SG}+\G s_{\rm max}(0)$ [Eq.~(\ref{SGe}), full line] 
and show the energy
$e_0 - \G [s_{\rm max}(e_0) - s_{\rm max}(0) ]$ of two different clusters
with $e_0 = 0.05, 0.2$ (dot-dashed lines).
\newline
{\it (Right panel)} Transverse magnetization $m_x$ as a function of $\Gamma$ 
for the same parameters. The first order phase transition between the SG and
the QP is manifested by a jump
in $m_x$, shown by a vertical dotted line. 
The value of $m_x$ for the SG is the solid black line, while that of the QP is the dashed blue line. 
Note that the latter is bigger because it corresponds
to a more entropic phase.
}
\label{fig_energy_cube}
\end{figure}

We discuss now the zero temperature phase diagram of the model
for $\a > \a_{\rm sep}$ and $N \to \io$. 
Following Sec.~\ref{sec:subcubes},
to get a meaningful
thermodynamic limit,
the number of clusters of energy $e_0$ is
set to $2^{N (1-\a) g(e_0)}$, where $g(e_0)$ is an arbitrary increasing function 
of $e_0 \in [0,e_{\rm m}]$
(as in most random optimization problems).
We assume that $g(e_{\rm m}) = 1$ so the total number of 
clusters in $\SS$ is still $2^{N(1-\a)}$.
As discussed in Sec.~\ref{sec:subcubes},
the complexity of clusters of energy 
$e_0$ and entropy $s$ is
$\Si(e_0,s) = (1-\a) g(e_0) - D(s || 1-p)$, and
it vanishes at $s_{\rm max}(e_0)$ which is also an increasing function of $e_0$.
The $\SS$-band, or spin glass (SG),
ground state energy is
\beq\begin{split}\label{SGe}
e_{\rm SG} &= \min_{e_0 \in [0,e_{\rm m}]} 
\left[ \min_{s\in [s_{\rm min}(e_0),s_{\rm max}(e_0)]} 
(e_0 - \G s) \right]
\\ & =\min_{e_0\in [0,e_{\rm m}]}\big[ e_0 - \G s_{\rm max}(e_0) \big] \ .
\end{split}\eeq
The minimum is in $e_0 =0$ as long as $\G < \G_{\rm lc} =1/(s_{\rm max}'(0))$. Above this value,
the minimum is in a different $e_0$ for each value of $\G$: in this region the ground state
changes abruptly from one cluster to another upon changing $\G$ by an infinitesimal amount,
similarly to what is called temperature chaos in spin glasses~\cite{BrMo87,KrMa02}.
Note that in some relevant cases the slope of $g(e_0)$ in $e_0=0$ is infinite, therefore
$\G_{\rm lc} =0$ and level crossings happen at all $\G$.

The energy $e_{\rm QP}$ crosses
the SG ground state given by Eq.~(\ref{SGe}), giving rise to a first order phase
transition between the SG and the QP~\cite{JKKM08,Go90,NR98,BC01,CGS01,JKSZ10} at
a critical $\G \propto V$. As a consequence, the transverse magnetization $m_x = de/d\G$
has a jump at the transition~\cite{JKSZ10} (see the right panel of Fig.~\ref{fig_energy_cube}). 
Note that $m_x = s$, thus the
transverse magnetization is determined by the entropy of the ground state, and
the entropy of the $V$-component is much larger than 
those of the clusters.

\subsubsection{Finite temperature: the condensation transition}\label{finite_temperature_A3}

\begin{figure}
\centering
\includegraphics[width=.47\textwidth]{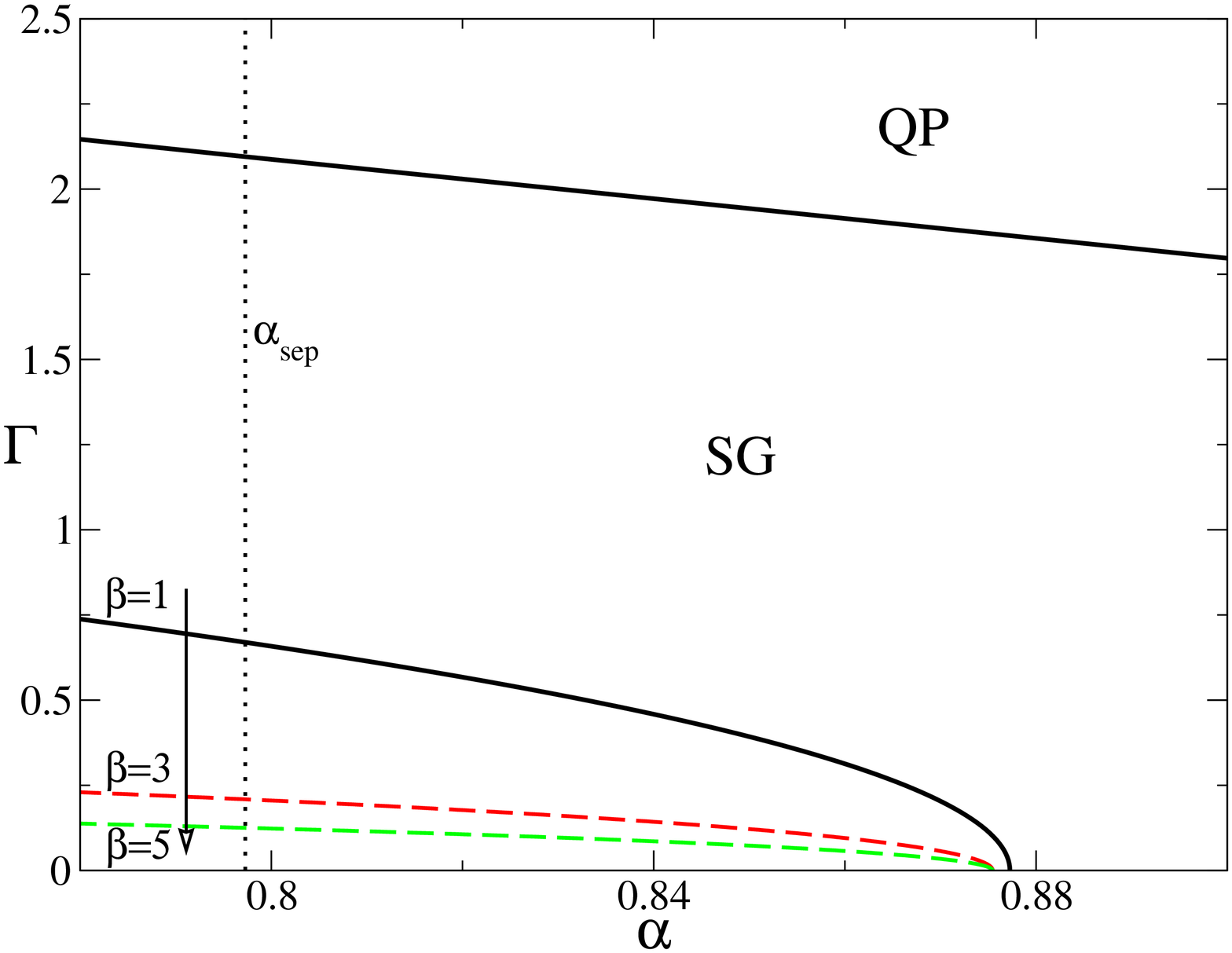}
\caption{Phase diagram of the model for $p=0.7$,
$g(e_0)$ as in Fig.~\ref{fig_energy_cube}, and $\b=1$ (full lines).
The vertical line corresponds to $\a_{\rm sep}=0.797$ for this value of $p$.
The higher $\G$ line is the first order transition between SG and QP.
Above the lower $\G$ line $\a_{\rm c}(\G,\b=1)$ the system is in the condensed
phase.
The condensation transition lines $\a_{\rm c}(\G,\b)$ are also reported (dashed lines) for 
different values of $\b$, showing that the non-condensed phase disappears
for $\b \to \io$.
The complexity of the zero-energy clusters is $(1-\a) g(e_0=0)=2(1-\a)/3$, hence
one has $\a_{\rm c}(\G=0,\b=\io)=\frac{2p-1}{2-p} + \frac32 \log_2(2-p) = 0.875$.
}
\label{fig_diagram_cube}
\end{figure}

The previous analysis shows that in the region 
$\a_{\rm sep} < \a < \a_{\rm c}$ the perturbation
$\G \hH_Q$ has a dramatic effect. At $\G =0$, most of 
the states in $\SS$ belong to one of exponentially
many small clusters, while at any $\G > 0$ the few largest clusters of entropy $s_{\rm max}$ have
the smallest energy. This is related to the fact that the presence of 
a transverse field introduces a correction to the energy 
that favors the more entropic clusters. 
A more complete picture is obtained by studying the model at finite
temperature (recall that the classical model at finite temperature was studied in Sec.~\ref{sec:subcubes_finiteT}).
It is convenient to separate the contribution of the two parts of the spectrum
(the $\SS$-band corresponding to the SG phase and the $V$-band corresponding to the QP phase) 
to the partition
function, $Z = \Tr \, e^{-\b \hH} = Z_{\rm SG} + Z_{\rm QP}$,
with $c = 2 \cosh(\b \G)$:
\beq\nonumber
\begin{split}
Z_{\rm QP} & \sim \sum_k e^{-\b E_k^V} = e^{-\b N V} c^{N} \ , \\
Z_{\rm SG} & \sim
\sum_{A,k} e^{-\b E_k(A)} = 
\int {\rm d}e_0 {\rm d}s \, 2^{N \Si(e_0,s)} e^{-\b N e_0}
c^{N s} \ .
\end{split}\eeq
Of course, $Z_{\rm SG}$ reduces to the classical partition function in Eq.~(\ref{eq:ZsubT})
for $\G=0$.
The free energy is $f_{\rm QRSM} = -(T/N) \log Z = \min\{f_{\rm SG}, f_{\rm QP}\}$, 
analogously to what was found in \cite{JKKM08} for the 
QREM 
(see the discussion in Sec.~\ref{sec:firstorder_disordered}),
with $f_{\rm QP} = V - T \log c$ and
\beq\label{fSG}
f_{\rm SG} = - T \hskip-15pt \underset{
\substack{e_0 \in [0,e_{\rm m}] \\ s\in[s_{\rm min}(e_0),s_{\rm max}(e_0)]}
} 
{\max}
\hskip-10pt
[ \Si(e_0,s) \log 2 - \b e_0 + s \log c] \ .
\eeq
The first order transition happens when the free energies $f_{\rm SG}$ and $f_{\rm QP}$ 
cross, while the condensation transition $\a_{\rm c}(T,\G)$ happens when the maximum 
in Eq.~(\ref{fSG}) is attained in $s_{\rm max}$ 
for the first time. In Fig.~\ref{fig_diagram_cube} we plot the
lines $\a_c(T,\G)$ versus $\G$ for several 
temperatures. We observe that
in the limit $\b \to \io$, the lines $\a_c(T,\G)$ shrink to the 
horizontal axis and the system is in the condensed phase 
for any~$\G~>~0$. The first order transition to the QP phase happens
for larger values of $\G$ at fixed temperature, 
and it is reported in the plot for $\beta=1$.

\subsubsection{Summary}\label{discussion_A3}

Before presenting a more general perspective on random optimization problems,
let us summarize the results of this section.
We introduced a simple toy model of a quantum optimization problem,
the QRSM based on the RSM of~\cite{rcm}.
In the classical case $\G=0$, the model captures the essential
structure of the space of solution of random optimization problems, and
displays several phase transitions
that are present also in more realistic problems such as $k$-SAT,
at least at large $k$. 
We explored the consequences of this complex
structure on the spectrum of the quantum Hamiltonian at $\G>0$, and we showed that:
(i) Quantum fluctuations lower the energy of a cluster proportionally to
its size.
(ii) Because the energy and the entropy vary from cluster to cluster,
level crossing between different clusters
are induced as a function of $\G$ in the SG phase, due
to a competition between energetic and entropic effects.
These crossings accumulate for $N\to\io$ 
in a continuous range of $\G$, giving rise to a complex SG
phase characterized by a continuously changing
ground state and an everywhere exponentially small gap.
(iii) At large $\G \sim V$ the SG phase undergoes 
a first order transition towards a
QP phase,
corresponding to the complete delocalization of the
ground state in the computational basis $| \us \rangle$.
(iv) At finite temperature, there is a line of condensation 
transitions $\a_{\rm c}(\Gamma)$ that shrinks to $\Gamma=0$ at low temperatures:
indeed, at zero temperature 
the condensation transition becomes abrupt. 
While at $\G=0$ the space of solutions is dominated by an exponential number of 
clusters of intermediate size, for any $\G > 0$ the biggest
clusters contain the ground states.

\subsection{Phase transitions in quantum optimization problems: an attempt towards a general perspective}

Overall, the discussion of the previous sections
shows that the low energy spectrum
of quantum optimization problems can be very complex, and 
characterized by different level crossings:
internal level crossings in the SG phase, or the crossing 
between the SG and the QP giving rise to a first order phase transition. Moreover, 
both entropic and energetic effects are important. 

Yet the previous discussion was based on a series of toy models
(such as the QREM or the QRSM) or on the analysis of extremely simplified
instances of random optimization problems. These problems were basically
constructed {\it ad hoc} to exhibit the desired phenomenology.
The next task is therefore to demonstrate that these phenomena indeed happen
in {\it typical} instances of realistic random optimization problems,
such as those defined in Sec.~\ref{sec:examples_optimization}. 
This will be the subject
of Sec.~\ref{sec:results}, but it requires the introduction of sophisticated
quantum statistical mechanics tools that we will discuss in Sec.~\ref{sec:methods}.
Before proceeding, in this section we want to complete the picture 
by presenting coherently 
what are the expected properties of the spectrum of generic random optimization problems.

\begin{figure}
\centering
\begin{tabular}{cc}
\includegraphics[width=.35\textwidth]{landscape_REM.eps}&
\includegraphics[width=.35\textwidth]{landscape_RSM.eps}\\
\includegraphics[width=.35\textwidth]{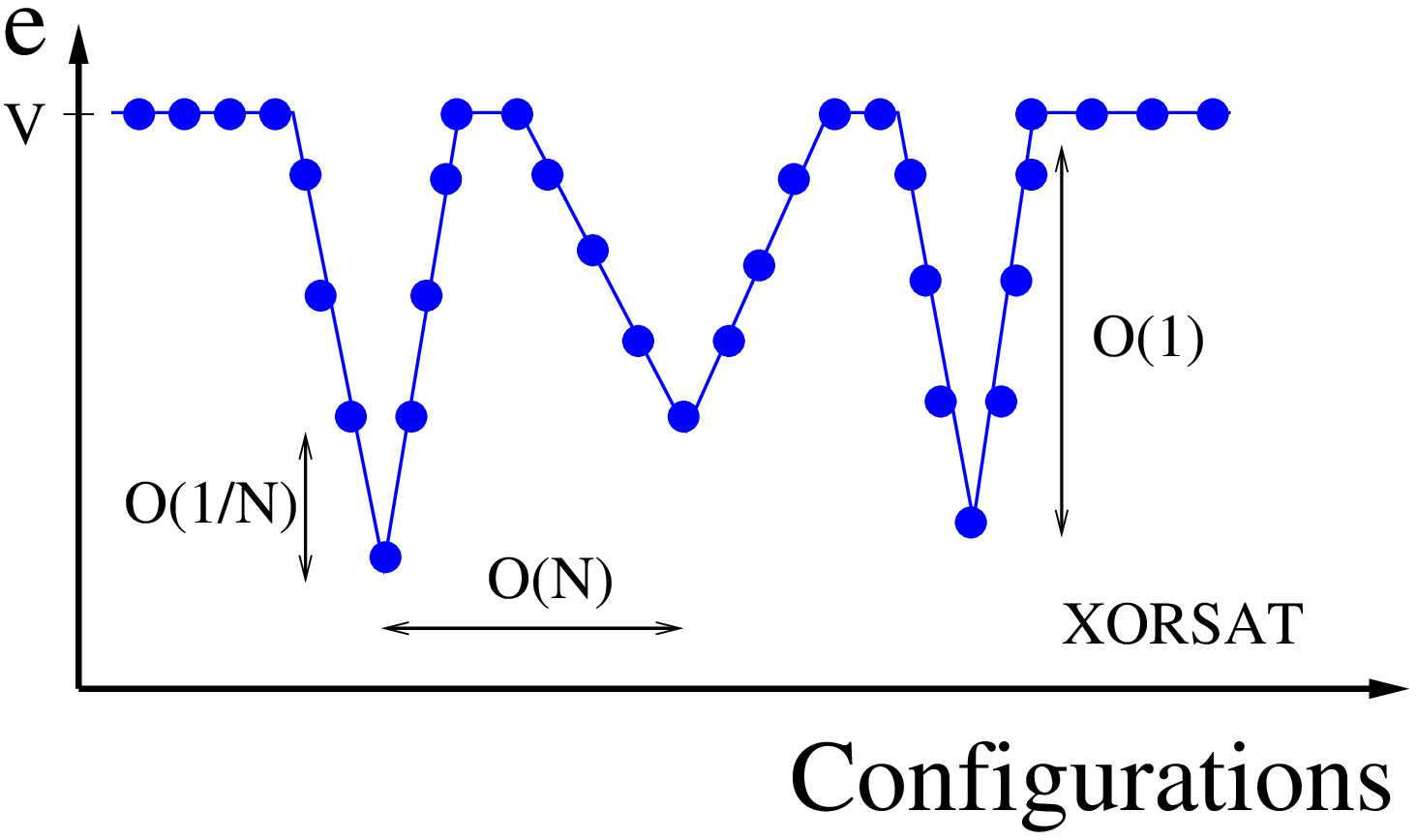}&
\includegraphics[width=.35\textwidth]{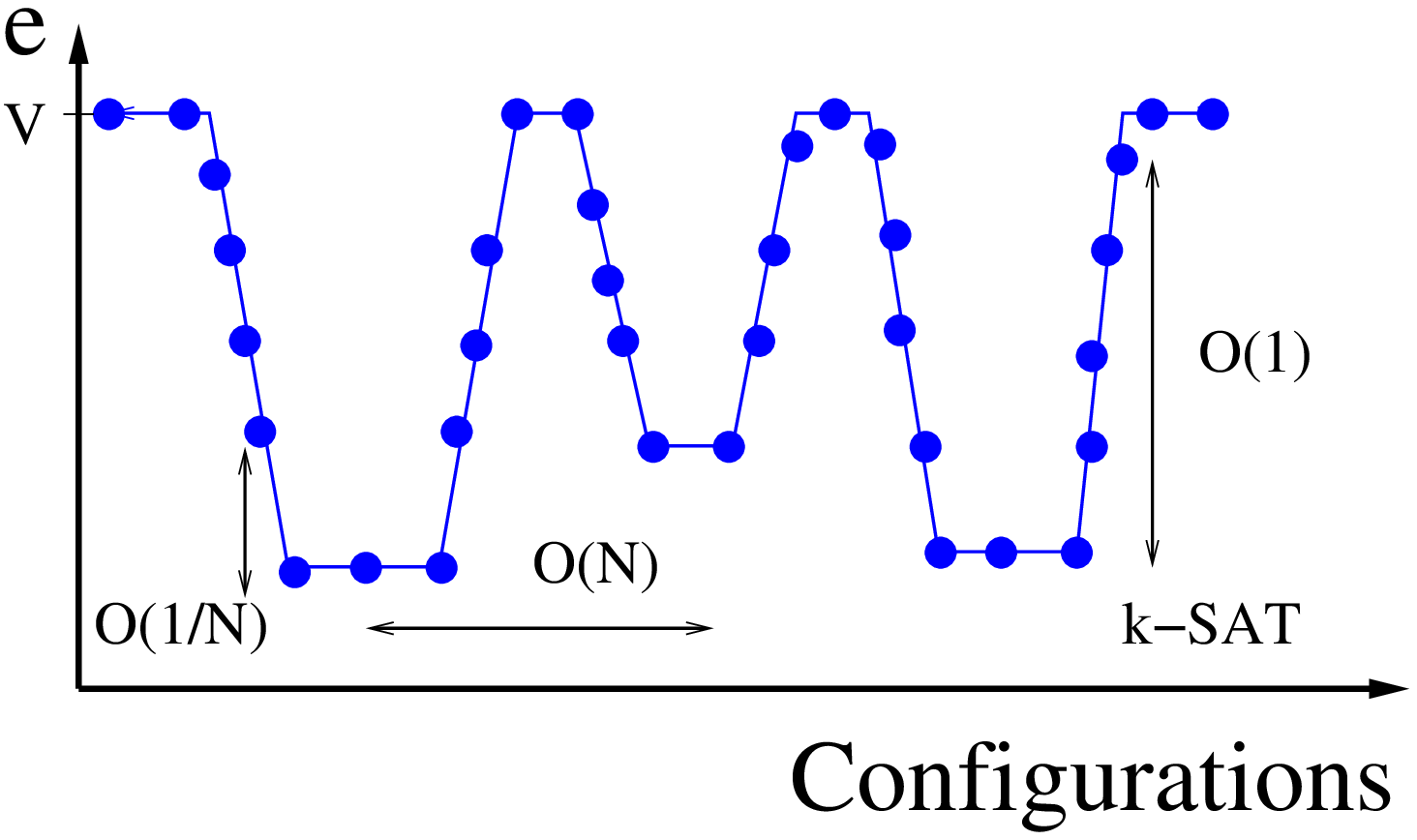}
\end{tabular}
\caption{
Pictorial energy landscape of the REM {\it (upper left panel)},
 RSM {\it (upper right panel)}, XORSAT on a satisfiable random regular graph 
 {\it (lower left panel)} and $k$-SAT {\it (lower right panel)}, using the same conventions
as in Fig.~\ref{fig:landscape_AminChoi}.
}
\label{fig:landscape_subcubes}
\end{figure}

As was explained in Sec.~\ref{sec:classical_mean_field},
the classical energy landscapes of several
random optimization problems have been recently characterized in much detail,
thanks to important developments in the analysis of 
classical spin glasses~\cite{Beyond,BiMoWe,cavity,MeZe,Mo07,MM09}. 
These studies show that the classical energy is characterized by
many ``valleys'' at the bottom of which local minima are found;
we are now able to obtain a quite
detailed quantitative characterization of the shape of these valleys~\cite{ZK10}.
We can use as running examples the random regular XORSAT problem, in the satisfiable
phase where solutions exist,
which is a representative of the class of locked models
discussed in Sec.~\ref{sec:generating_USA}, 
and the random $k$-SAT problem, which instead displays (for $k>3$) all the transitions
discussed in Sec.~\ref{sec:transitions_rCSP}.
In Fig.~\ref{fig:landscape_subcubes} we sketch pictorially the energy
per spin as a function of the configuration for those models, and compare with 
the previously investigated toy models, the REM and RSM.
The picture shows some very general aspects of mean field spin glasses,
that are shared by all the models considered here:
\begin{itemize}
\item[a.] The energy landscape contains many local minima.
\item[b.] The distance between low energy configurations belonging to the basin of attraction of
different minima is ${\cal O}(N)$.
\item[c.] The height of the energy barriers separating two different minima is ${\cal O}(N)$.
\end{itemize}
At the same time, there are some crucial features that are model dependent also within mean field
models:
\begin{itemize}
\item[d.] The number of configurations around a given local minimum 
might be exponentially large (the entropy is positive) or not (the entropy is zero). 
\item[e.] The (intensive) energy change associated to a spin flip starting from a low energy configurations
can be either ${\cal O}(1)$ or ${\cal O}(1/N)$. The latter case is the rule for Hamiltonians that 
are the sum of local terms, and in this case the ``steepness'' of the energy around a local minimum
can depend on the minimum itself.
\end{itemize}

Based on the previous analysis of toy models,
we expect that each low energy cluster of the classical energy function $E(\us)$ gives rise, 
under the action of a quantum term like a transverse field, 
to a set of states (whose size roughly corresponds to the classical entropy of the cluster), 
with an energy density of the form
\beq\label{perturbation_SAT}
\displaystyle e(\Gamma) =   e + \underbrace{\G {\cal O}(1)}_{\text{Entropic effects}} 
+   \underbrace{\G^2 {\cal O}(1)}_{\text{Energetic effects}} \dots .
\eeq
The coefficients of both terms depend on the shape of the classical energy around the local minimum. 
It was shown explicitly for the QRSM that the coefficient
of the linear term is directly proportional to the intensive entropy of the cluster. Therefore, 
if local minima are exponentially degenerate, this coefficient
is ${\cal O}(1)$ with respect to $N$,
and its fluctuations from cluster to cluster are also ${\cal O}(1)$.
Moreover, it was
shown in Sec.~\ref{sec:Altshuler} that the coefficient of the quadratic term depends on
the neighborhood of the local minimum. Fluctuations of the latter 
coefficient among different minima are ${\cal O}(1/\sqrt{N})$
in the example of~\cite{AKR10} but might be ${\cal O}(1)$ in other models.

The cluster to cluster fluctuations of the quantum corrections will generically lead to avoided
level crossings. Because of the huge number of different clusters, it is reasonable to expect
that these crossings will accumulate for $N\to\io$ leading to an everywhere gapless spin glass
phase, as it was shown explicitly for the QRSM. This phase will cease to exist at large enough $\G$,
when the completely delocalized state, corresponding to the quantum paramagnetic phase, will
cross the cluster ground state. Generically we expect this crossing to become a first order 
transition in the thermodynamic limit (as in the QREM and in the QRSM), but the transition
might also be of higher order depending on the model. We will discuss concrete examples in 
Sec.~\ref{sec:results}. In summary, we expect generic problems to display a complex spin glass
phase for small $\G$, separated by a quantum critical point from a simple quantum paramagnetic phase.

\section{Methods}
\label{sec:methods}

The aim of this section is to review various methods that can be used to 
investigate quantum spin glass models. 
We will give a particular emphasis to the methods that are most efficient 
to treat the quantum versions of the family of random 
optimization problems introduced 
in Sec.~\ref{sec:examples_optimization} and Sec.~\ref{sec:Optimization-Statistical-Mechanics}. 
We will use these methods in Sec.~\ref{sec:results} to obtain results on 
specific
models such as the XORSAT and coloring problems. This section is organized
as follows: we will start in Sec.~\ref{sec:classical_cavity} by introducing
the classical cavity method, a framework that has been developed to study
random ensembles of optimization problems and which has led to the understanding
of their complex phenomenology, that we explained in Sec.~\ref{sec:transitions_rCSP}.
The next three sections (\ref{sec:PIQC}, \ref{sec:OQC} and \ref{sec:VQC}) 
will be devoted to different approaches to the generalization of the 
classical cavity method to quantum models. Finally in 
sections~\ref{sec:ed} and~\ref{sec:qmc} we will give some details on some 
more standard numerical methods, such as exact diagonalization and quantum 
Monte Carlo, that have also been used to obtain important informations
on these problems. Before entering in the core of the discussion let us
give a more detailed overview of the rest of this section.

As already explained in Sec.~\ref{sec:mfsg}, 
disordered mean field models~\cite{Beyond} can be roughly classified in two
categories: fully connected ones, where each degree of freedom interacts
weakly with all others, and finitely connected ones, with a finite number 
of strong interactions for each degree of freedom. The replica method
has been originally devised for the former family, most notably for
the Sherrington-Kirkpatrick model~\cite{SK75}. Its extension to the 
finite connectivity case~\cite{replica_diluted,BiMoWe} has been more
conveniently reformulated in terms of the cavity method~\cite{cavity}, 
and applied
in particular to random ensemble of Constraint Satisfaction 
Problems (CSP)~\cite{MezardParisi02,KrMoRiSeZd}. 
The classical cavity
method is by now a well established technique, with many presentations
in original research papers~~\cite{cavity,MRS08} and in a textbook~\cite{MM09},
and rigorous proofs of validity in some cases~\cite{FrLe,PaTa,DM10}. 
For the sake
of completeness, in Sec.~\ref{sec:classical_cavity} 
we provide a quick survey of the classical cavity method,
before turning to the specificities of its quantum version.
In this section we also discuss the general definition of 
random graph models and the key concept of replica symmetry breaking.

Several quantum extensions of the cavity
method have been recently proposed in a series of 
papers,
able to treat for instance spin 1/2 
models in presence of a transverse field. 
Roughly speaking, these methods can be divided in three groups.
Path Integral Quantum Cavity (PIQC) methods
exploit a path integral representation in order to map the quantum problem into
a classical one and then make use of the classical cavity method~\cite{LSS08,KRSZ08,leifer2008,JKSZ10}.
Operator Quantum Cavity (OQC) methods work directly with quantum 
operators~\cite{poulin2008,Bil-Poul2010,Poulin_Hasting2011,IM10,dimitrova2011}.
Finally, Variational Quantum Cavity (VQC) methods propose a variational ansatz for the ground
state wavefunction that can be represented in terms of a set of local parameters,
and then use the classical cavity method to optimize the energy of the variational state~\cite{ramezanpour2012}.

PIQC is at the moment the only analytical method that was used to obtain 
concrete results on one of the random CSP defined in
Sec.~\ref{sec:examples_optimization}, namely XORSAT in presence of a transverse field~\cite{JKSZ10}.
The goal of Sec.~\ref{sec:PIQC} is to explain
the technical details of the PIQC at the level of one step 
of replica symmetry breaking, that is needed for the solution of these problems~\cite{JKSZ10}, and to generalize it
to arbitrary discrete quantum degrees of freedom\footnote{
Note that the PIQC has also been used in a condensed matter context, namely to investigate
quantum glassy phases of disordered interacting bosons on random lattices;
a detailed explanation of the method 
in this case can be found in~\cite{STZ09,FSZ11}.
}.
We will rely on this method in Sec.~\ref{sec:results} to present detailed results on the XORSAT
model~\cite{JKSZ10} and original results on the coloring problem.

The main drawbacks of PIQC are that {\it i)} as any path integral sampling method,
it is restricted to Hamiltonians that are not plagued by the ``sign problem'' (or in other words
that admit a path integral representation where the trajectories have positive weights), {\it ii)} it does not allow to work exactly at $T=0$, but only
to perform an extrapolation to $T\to 0$ from finite temperatures and {\it iii)} 
the resulting functional equations have to be solved using a statistical representation, that is
affected by fluctuations and/or finite size (of the representation) effects.
These drawbacks could in principle be overcome by working directly with operators.
In Sec.~\ref{sec:OQC} we will describe several 
attempts to construct OQC methods~\cite{poulin2008,Bil-Poul2010,Poulin_Hasting2011,IM10,dimitrova2011}.
In Sec.~\ref{sec:VQC} we will describe VQC methods that are specifically designed to obtain 
direct information at $T=0$~\cite{ramezanpour2012}.
Although these attempts are extremely promising and already allowed to obtain very interesting results
for simple ordered and disordered models, it turns out that for the moment 
they are too computationally demanding to
be used to solve the problems of interest here (e.g. the XORSAT problem). 
This will be discussed in Sec.~\ref{sec:results}.

\subsection{The classical cavity method}
\label{sec:classical_cavity}

\subsubsection{Factor graph models} 

\begin{figure}
\centerline{\includegraphics[width=.48\columnwidth]{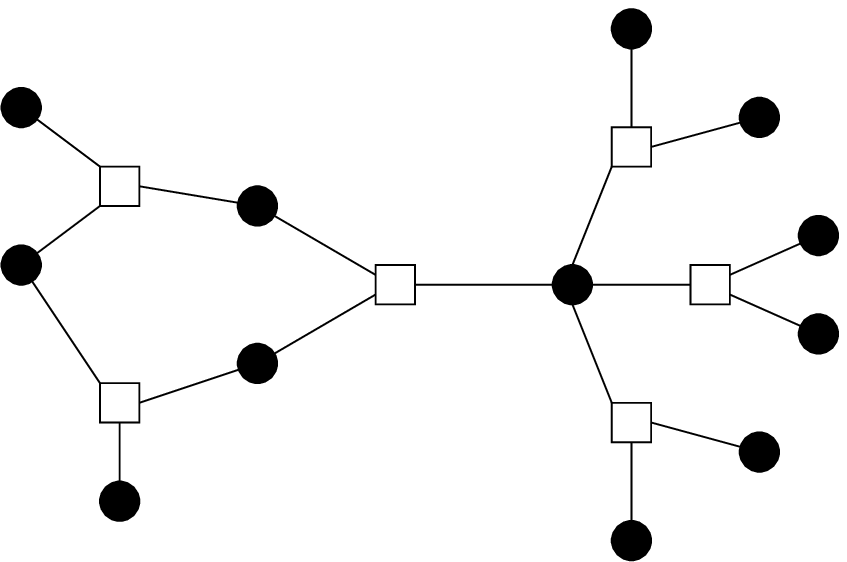}
\hspace{8mm}
\includegraphics[width=.35\columnwidth]{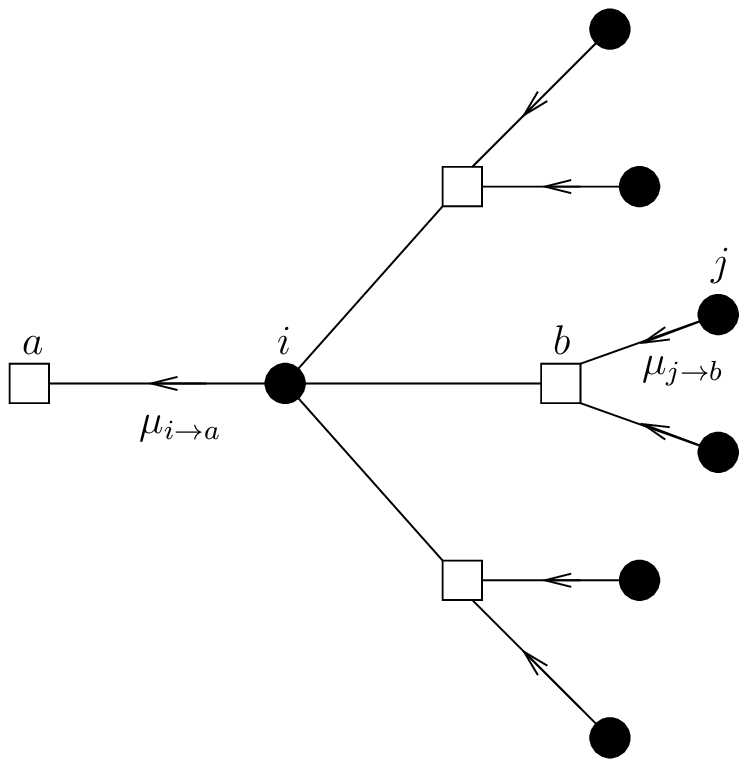}}
\caption{
{\it (Left panel)} An example of a factor graph. 
{\it (Right panel)} Graphical representation
of Eq.~(\ref{eq_BP}).}
\label{fig_factor_graph_and_BP}
\end{figure}

Let us recall some definitions of Sec.~\ref{sec:examples_optimization} 
and put them in a more general context.
We consider a model with $N$ degrees of freedom $\s_i$ taking values
in a finite alphabet ${\cal X}$, for instance ${\cal X}=\{-1,+1 \}$ for 
Ising spins. We denote $\us=(\s_1,\dots,\s_N)$ the global
configuration, and for a subset $S \subseteq \{1,\dots,N\}$ we write
$\us_S = \{\s_i | i \in S \} $ the configuration of those variables.
The model is further defined by its energy function $E(\us)$, that contains
$M$ interactions labeled by $a=1,\dots,M$:
\beq
E(\us)= \sum_{a=1}^M \e_a(\us_\da) \ .
\label{eq_E_class}
\eeq
For each of the interactions $\da$ denotes the set of variables that interact
through $a$ with the energy term $\e_a$; a convenient representation of 
such an energy function is
provided by factor graphs~\cite{factorgraph}, 
see left panel of Fig.~\ref{fig_factor_graph_and_BP}.
Each variable $i$ is associated to a circle vertex (variable node), while
interactions $a$ are represented by squares (function nodes), an edge being 
present between interaction $a$ and variable $i$ if and only if the value
of $\e_a$ depends on $\s_i$, i.e. if and only if $i \in \da$. We shall also
use the notation $\di$ for the set of interaction nodes linked to the variable
$i$, and call graph distance between two variables $i$ and $j$ the minimal 
number of interactions in a path along the factor graph between $i$ and $j$.

The Gibbs-Boltzmann measure at inverse temperature $\beta$ of the model 
can be written as
\beq
\mu(\us)= \frac{1}{Z} \prod_{a=1}^M w_a(\us_\da) \prod_{i=1}^N v_i(\s_i) \ ,
\qquad 
Z = \sum_{\us \in {\cal X}^N} \prod_{a=1}^M w_a(\us_\da) \prod_{i=1}^N v_i(\s_i)
\ ,
\label{eq_mu}
\eeq
where the partition function $Z$ ensures the normalization of the probability 
law, and where the interaction weights are given by $w_a = e^{-\beta \e_a}$.
For future convenience we shall treat a slightly generalized case with
weights $v_i$ on the variables. 
For Ising spins, the latter can be thought as originating from local magnetic fields.

Let us assume momentarily that the factor graph representing the model
under study is a tree. Then the problem of characterizing the 
measure~(\ref{eq_mu}) and computing the associated partition function $Z$
can be solved exactly in a simple, recursive way. One introduces on each 
directed edge $i \to a$ from
one variable $i$ to an adjacent function node $a$ a ``message'' $\mu_{i \to a}$,
which is a probability measure on the alphabet ${\cal X}$, that would be
the marginal probability of $\s_i$ if the interaction $a$ were removed from
the graph. These messages are easily seen to obey the recursive
(so-called Belief Propagation) equations depicted on the right panel
of Fig.~\ref{fig_factor_graph_and_BP},
\beq
\mu_{i \to a}(\s_i) = 
\frac{1}{z_{i \to a}} v_i(\s_i) \prod_{b\in \dima} 
\sum_{\us_\dbmi} w_b(\us_\db)
\prod_{j \in \dbmi} \mu_{j \to b}(\s_j) \ ,
\label{eq_BP}
\eeq
with $z_{i \to a}$ ensuring the normalization of the law $\mu_{i \to a}$.

On a tree factor graph there exists a single solution of these equations, which
is easily determined starting from the leaves of the graph (for which the empty
product above is conventionally equal to 1) and sweeping towards the inside of
the graph. Once the messages $\mu_{i \to a}$ have been determined all local
averages with respect to $\mu$ can be computed, for instance the marginal
law for a single variable and for the set of variables around one interaction
read respectively
\bea
\mu_i(\s_i) &=& \frac{1}{z_i} v_i(\s_i) \prod_{a\in \di} 
\sum_{\us_\dami} w_a(\us_\da)
\prod_{j \in \dami} \mu_{j \to a}(\s_j) \ , \\ \nonumber
\mu_a(\us_\da) &=& \frac{1}{z_a} w_a(\us_\da) \prod_{i \in \da} 
\mu_{i \to a}(\s_i) \ ,
\eea
with $z_i$ and $z_a$ defined by normalization. Moreover the partition
function can be computed as
\bea\nonumber
\log Z &=& \sum_{i=1}^N \log z_i - \sum_{a=1}^M (|\da|-1) \log z_a \\
&=&
 \sum_{i=1}^N \log \left[ \sum_{\s_i}  v_i(\s_i) \prod_{a\in \di} 
\sum_{\us_\dami} w_a(\us_\da) \prod_{j \in \dami} \mu_{j \to a}(\s_j)
\right] \\ \nonumber
 &-& 
\sum_{a=1}^M (|\da|-1) \log \left[ \sum_{\us_\da} w_a(\us_\da) \prod_{i \in \da} 
\mu_{i \to a}(\s_i) \right] \ .
\eea

\subsubsection{Random ensembles} 

In the definition of the energy function (\ref{eq_E_class}) and of the
associated Gibbs-Boltzmann measure (\ref{eq_mu}) we considered a single 
realization of the model. We turn now to random ensembles of such models. Their
definition involves a probability law on the integers, that we shall denote
$q_d$; for simplicity of notation we assume that all interactions $a$ involve
the same number $k$ of variables. The factor graphs are then supposed to be
chosen uniformly at random among those with $N$ variables and such that a 
fraction $q_d$ of variables is involved in $d$ interactions. The number of interactions
is then $M = \alpha N$, with $\alpha$ related to the degree distribution 
$q_d$ by the relation $\alpha k =\sum_d d q_q$. 
This definition is an hypergraph generalization
of the random graph models with prescribed degree distribution, studied for
instance in~\cite{MolloyReed95}. The Erd\H{o}s-R\'enyi construction described
in Sec.~\ref{sec:classical_mean_field}, where the $M$ interactions are chosen
uniformly at random among the $\binom{N}{k}$ possible ones, is essentially
equivalent to the one just described, with $q_d$ the Poisson distribution of
mean $\alpha k$.
In the definition of the random energy function we also assume
that the interaction and variable weights 
$w_a$ and $v_i$ are chosen independently at random from two
probability laws. We will denote $\E[\cdot]$ the average with respect to the
whole construction; the main objective in the field of disordered systems is the
computation of the average free energy density $f$ in the thermodynamic 
(large size) $N\to \infty$ limit,
\beq
-\beta f = \lim_{N \to \infty} \frac{1}{N} \E [ \log Z ] \ ,
\eeq
as the free energy is a self-averaging quantity, which yields all relevant 
thermodynamic quantities by suitable derivatives with respect to its 
parameters.

The cavity method allows to compute the thermodynamic limit of the free energy
density for models defined on random graphs. It exploits some properties
of these random graphs, in particular their tree-like character: in the
thermodynamic limit these random graphs converge locally to random trees.
This means that the neighborhood within a fixed graph distance $t$ of an
arbitrary variable node $i$ is, with a probability going to one as $N\to \infty$
with $t$ fixed, a tree. The latter can be described by $q_d$ and an associated
distribution $\tq_d = (d+1) q_{d+1} / \alpha k$. Indeed the reference node
$i$ will have $d$ interaction nodes around it with probability $q_d$. Then
each of the $(k-1) d$ variables at distance 1 from $i$ will have a number
of descendants drawn independently from the law $\tq_d$, and so on and so
forth until $t$ generations of vertices have been drawn (all degrees are drawn
from $\tq_d$ except the first one with $q_d$). One can notice
that $q_d$ is the degree distribution of a variable node chosen uniformly at
random, while $\tq_d$ corresponds to the selection procedure where one edge of
the factor graph is chosen at random, say between $i$ and $a$; then $i$ belongs
to $d$ interaction besides $a$ with probability $\tq_d$.

We explained above how statistical mechanics models defined on trees could be
solved recursively. On the other hand we have just recalled that random graphs
are locally tree-like; the cavity method is a set of prescriptions on how
to exploit the local properties of the random graphs to make global predictions
on the free energy density. Depending on the models, in particular on the
amount of frustration between the variables they induce, different level of
sophistications of the cavity method are necessary. 

\subsubsection{Replica symmetric cavity method} 

The simpler one, exact for instance for ferromagnetic, unfrustrated, models, 
goes under the name of Replica Symmetric (RS). In that case there exists a 
single pure state (or a small number of them simply related by explicit 
symmetries of the model) in the Gibbs measure, which enjoys in consequence 
spatial correlation decay properties. The effect of the long loops which are 
present in the random graphs is then negligible in the thermodynamic limit, 
only providing a self-consistent boundary condition. The recursive equations
(\ref{eq_BP}) valid for a single tree factor graph are given a probabilistic
meaning to describe the thermodynamic limit of random graphs. More precisely,
the order parameter of the RS cavity method is $\cP(\eta)$, the probability (over the disorder)
that the messages $\mu_{i\to a}$ in Eq.~(\ref{eq_BP}) (which are themselves probability 
distributions on ${\cal X}$) are equal to $\eta$. $\cP$ obeys a self-consistent functional equation,
which is more simply written in a distributional form as
\beq
\eta \eqd g(\eta_{1,1},\dots,\eta_{1,k-1},\dots,\eta_{d,1},\dots,\eta_{d,k-1},v,w_1,\dots,w_d)
\ .
\label{eq_eta_RSclass}
\eeq
In this equation all the $\eta$'s are drawn independently from $\cP$, and
$\eqd$ denotes the equality in distribution between random variables. Moreover
$d$ is drawn according to the law $\tq_d$, the $v$ and $w_a$'s are independent
copies of the variable and interaction random weights, and the function $g$ 
in the r.h.s. is defined by
\bea
\eta(\s) &=& \frac{1}{z(\{\eta_{a,i}\},v,\{w_a\})} v(\s) \times 
\nonumber \\
&&\sum_{\{ \s_{a,i} \}_{a\in[1,d]}^{i\in[1,k-1]}} 
\left( \prod_{a,i} \eta_{a,i}(\s_{a,i}) \right) \ 
\prod_{a=1}^d w_a(\s,\s_{a,1},\dots,\s_{a,k-1}) \ ,
\label{eq_g_RSclass}
\eea
$z(\{\eta_{a,i}\},v,\{w_a\})$ ensuring the normalization of $\eta$.
The RS prediction for the free energy in the thermodynamic limit is then
\bea
-\beta f = \lim_{N \to \infty} \frac{1}{N} \E[\log Z] 
&=& \E\left[\log\left(   
z_{\rm v}(\{\eta_{a,i} \}_{a\in[1,d]}^{i\in[1,k-1]},v,\{w_a \}_{a\in[1,d]})
\right)  \right] \nonumber \\
&-& \alpha (k-1) \E\left[\log \left( 
z_{\rm f}(\{\eta_i \}_{i\in[1,k]},w) \right) \right]  \ , 
\label{eq_f_RSclass}
\eea
where the expectation is with respect the independent choices of $d$ (with the
degree distribution $q_d$), of the $\eta$'s (according to the law $\cP$)
and of the random weights $v$ and $w_a$'s, and where the contribution of
the variable and function nodes are defined by
\bea
z_{\rm v}(\{\eta_{a,i} \},v,\{w_a \})&=&
\sum_{\s, \{ \s_{a,i} \}_{a\in[1,d]}^{i\in[1,k-1]}} v(\s) \times \nonumber \\
&&\left( \prod_{a,i} \eta_{a,i}(\s_{a,i}) \right) 
\prod_a w_a(\s,\s_{a,1},\dots,\s_{a,k-1}) \ ,
\label{eq_zv_RSclass}\\ \nonumber
z_{\rm f}(\{\eta_i \},w)&=&
\sum_{\s_1,\dots,\s_k}  
\left( \prod_i \eta_i(\s_i) \right) w(\s_1,\dots,\s_k) \ .
\label{eq_zf_RSclass}
\eea

 One can also obtain the
disorder average of local observables, for instance the average of
the marginal probability for a single variable $\s_i$ reads
\beq
\E [\mu_i(\s)] = \E \left[
\frac{ 
\underset{ \{ \s_{a,i} \}_{a\in[1,d]}^{i\in[1,k-1]}}{\sum}  v(\s)
\left( \underset{a,i}{\prod} \eta_{a,i}(\s_{a,i}) \right)
\underset{a}{\prod} w_a(\s,\s_{a,1},\dots,\s_{a,k-1})}
{
\underset{\s', \{ \s_{a,i} \}_{a\in[1,d]}^{i\in[1,k-1]}}{\sum} v(\s')
\left( \underset{a,i}{\prod} \eta_{a,i}(\s_{a,i}) \right) 
\underset{a}{\prod} w_a(\s',\s_{a,1},\dots,\s_{a,k-1})
}\right] \ .
\nonumber
\eeq
Other thermodynamic observables can be obtained in a similar fashion; one
way to derive their expressions is to observe that the expression 
(\ref{eq_f_RSclass}) for the free energy is variational, in the sense that
the stationary conditions with respect to $\cP$ coincide with the 
self-consistent condition of Eq.~(\ref{eq_eta_RSclass}). In consequence
one can use the partition function as a generating function of the observables
to be computed, and take only explicit derivatives with respect to their
conjugated fields in (\ref{eq_f_RSclass}).

\subsubsection{Replica symmetry breaking} 
\label{sec:class_RSB}

The assumption of correlation decay that underlies the RS cavity method
can fail in presence of frustration, for instance in the case of random
CSP with $\alpha > \alpha_{\rm d}$, the clustering transition.
Indeed the configuration space of these
models gets fractured in a large number of pure states, called clusters in
the context of random CSP, as explained in Sec.~\ref{sec:transitions_rCSP}
and sketched in Fig.~\ref{fig11}. 
The correlation decay hypothesis only holds for
the Gibbs measure restricted to one pure state, not for the complete Gibbs 
measure.
This complication can be handled by the cavity method with ``replica symmetry
breaking'' (RSB). It amounts to make further self-consistent hypotheses on
the organization of these pure states, and on the correlated boundary 
conditions they induce on the tree-like portions of the factor graph.
Inside each pure state the RS computation holds true, the RSB computation
is then a study of the statistics of the pure states. Let us explain
how this is done in practice at the first level of RSB (1RSB cavity method).
The partition function is written as a sum over the pure states $\gamma$,
$Z=\sum_\g Z_\g$, where $Z_\g$ is the partition function restricted to
the pure state $\g$ (recall the decomposition of 
Eq.~(\ref{eq:ps-decomposition})).
It can be written in the thermodynamic limit as
$Z_\g = e^{-N\beta f_\g}$, with $f_\g$ the associated (internal) 
free energy density. One further assumes that the number of pure states with 
a given value $f$ of the internal free energy density is, at the leading 
exponential order, $e^{N\Si(f)}$, where $\Si$ is called the complexity, or
configurational entropy. The latter is assumed to be a concave function of $f$,
positive on the interval $[f_{\rm min},f_{\rm max}]$. In order to compute
$\Si$ one introduces a parameter $m$ (called Parisi breaking parameter)
conjugated to the internal free energy, and the generating function of the
$Z_\g$ as $Z(m) = \sum_\g Z_\g^m$. In the thermodynamic limit its dominant 
behavior is captured in the 1RSB potential $\Phi(m)$,
\beq
\Phi(m)=-\frac{1}{m\b} \lim_{N \to \infty} \frac{1}{N}\log Z(m) = 
\inf_f \left[f - \frac{1}{m\b}\Si(f)\right] \ ,
\label{eq_Phi_1RSBclass}
\eeq
where the last expression is obtained by a saddle-point evaluation of the
sum over $\g$. The complexity function is then accessible via the inverse
Legendre transform of $\Phi(m)$~\cite{Mo95}, or in a parametric form
\beq
\Si(f(m)) = m^2 \b \Phi'(m) \ , \qquad f(m) = \Phi(m) + m \Phi'(m) \ ,
\eeq
where $f(m)$ denotes the point where the infimum is reached in 
Eq.~(\ref{eq_Phi_1RSBclass}). 

The actual computation of $\Phi(m)$
is done as follows. One introduces on each
edge of the factor graph a distribution $P_{i \to a}(\eta)$ of messages,
which is the probability over the different pure states that $\mu_{i \to a} = \eta$,
where $\mu_{i\to a}$ are the messages that appear in Eq.~(\ref{eq_BP}).
Because $P_{i \to a}(\eta)$ fluctuates from instance to instance,
the order parameter now becomes the distribution of $P_{i \to a}(\eta)$ with respect to
the disorder, which we call $\cP^{(1)}(P)$ and is solution of a self-consistent
functional equation written as
\beq
P \eqd G(P_{1,1},\dots,P_{1,k-1},\dots,P_{d,1},\dots,P_{d,k-1},v,w_1,\dots,w_d)
\ .
\label{eq_P_1RSBclass}
\eeq
Similarly to the RS case the $P$'s are independent copies drawn from 
$\cP^{(1)}$, and the random weights $v$ and $w_a$ are also independently
generated. The r.h.s. of this equation stands for:
\bea
P(\eta) &=& \frac{1}{Z(\{P_{a,i} \},v,\{w_a\},m)} \times \nonumber\\
&&\int \prod_{\underset{i\in[1,k-1]}{a\in[1,d]}} \dd P_{a,i}(\eta_{a,i})\,
\delta(\eta-g(\{\eta_{a,i}\} ) ) \, z(\{\eta_{a,i}\},v,\{w_a\})^m \ ,
\label{eq_G_1RSBclass}
\eea
with $g$ and $z$ defined above in Eq.~(\ref{eq_g_RSclass}), and
$m$ is the Parisi parameter. From the solution of this equation one
computes the 1RSB potential $\Phi(m)$ via an expression similar to 
(\ref{eq_f_RSclass}), namely
\bea
- m \beta \Phi(m) &=& \E\left[\log\left(  
\int \prod_{\underset{i\in[1,k-1]}{a\in[1,d]}} \dd P_{a,i}(\eta_{a,i})\,
z_{\rm v}(\{\eta_{a,i} \},v,\{w_a \})^m \right)\right] \nonumber \\
&-& \alpha (k-1) \E\left[\log \left( \int \prod_{i\in[1,k]} \dd P_i(\eta_i)\,
z_{\rm f}(\{\eta_i \},w)^m \right) \right] \ .
\eea
with the functions $z_{\rm v}$ and $z_{\rm f}$ defined in 
Eqs.~(\ref{eq_zv_RSclass}), (\ref{eq_zf_RSclass}). As in the RS case this
expression is variational, the implicit dependence of $\cP^{(1)}$ on the various
parameters ($m$ in particular) can be discarded when computing derivatives.
Various physical situations
translate in different behaviors of the 1RSB equations. 
\begin{itemize}
\item It can happen that only trivial solutions of (\ref{eq_P_1RSBclass}) 
exist, i.e. the $P$'s are supported
on a single value of $\eta$. It is then easy to check that this random $\eta$
obeys precisely the RS equation (\ref{eq_eta_RSclass}), and that the 1RSB 
potential $\Phi(m)$ is equal to the RS prediction for the free energy of
Eq.~(\ref{eq_f_RSclass}), for all values of $m$. This is the translation of 
the existence of a single pure state, in which case the whole 1RSB machinery 
reduces to the RS case. This case is realized at high temperatures/low
connectivity parameter $\alpha$, i.e. on the left of the line
$T_{\rm d}(\alpha)$ of Fig.~\ref{fig2}. In more physical terms it corresponds
to a ``liquid'' phase, for instance the high temperature phase 
$T >T_{\rm d}$ of the fully connected $p$-spin model described in 
Sec.~\ref{sec:pspinFC}.

\item If on the contrary non-trivial solutions of the
1RSB equations appear, one has to investigate them more carefully in order
to obtain the 1RSB prediction for the free energy density. From the
definition of $Z(m)$ one would naturally take $\Phi(1)$ for it. This is
indeed the case, provided the corresponding complexity $\Si(f(m=1))$ is
positive, in other words if $f(m=1)\in[f_{\rm min},f_{\rm max}]$. The physical
interpretation is that an exponential number $e^{N\Si(f(1))}$ of pure states
contribute to the Gibbs-Boltzmann measure, each with an internal free energy 
density $f(m=1)$. It turns out that in this case the prediction $\Phi(m=1)$
coincides with the RS one; this form of replica symmetry breaking is not
seen in the thermodynamic properties of the model, yet it has drastic
consequences on its dynamics~\cite{CuKu93,MoSe2}. Such a phase is usually called
for this reason a dynamic 1RSB (d1RSB) phase, and is realized in the
fully connected $p$-spin model in the intermediate temperature
regime $T_{\rm c} < T < T_{\rm d}$, or more generically for diluted mean field
spin glasses in the part of their phase diagram enclosed by the lines
$T_{\rm d}(\alpha)$ and $T_{\rm c}(\alpha)$ (see Fig.~\ref{fig2}).

\item If there are non-trivial solutions of the 1RSB equations, but
with $\Si(f(1))<0$, one has to find the value $m_{\rm s}\in[0,1)$ which solves 
the equation $\Si(f(m_{\rm s}))=0$, i.e. $f(m_{\rm s})=f_{\rm min}$, and the 
1RSB cavity method predicts that the free energy density is precisely this value 
$f_{\rm min}$. In more physical terms this is called a condensed (or true 1RSB)
phase, many pure states exist in the system yet only the sub-exponentially
numerous ones with $f=f_{\rm min}$ do contribute to the thermodynamic
behavior of the system, as for instance in the $T<T_{\rm c}$ phase of
the fully connected $p$-spin model.

\end{itemize}

As far as the free energy density is concerned, one can give a general formula
that encompasses and summarizes the three cases above: it is
given by the maximization of the potential $\Phi(m)$ with respect to $m$,
\beq
 f = -\frac{1}{\beta} \lim_{N \to \infty} \frac{1}{N} \E [ \log Z ] =
\underset{m \in [0,1]}{\max} \Phi(m) \ .
\eeq

\subsubsection{Population dynamics}
\label{sec:popu_dynamics}

In general there is no hope to find an analytical solution of
the cavity equations, neither at the RS level (\ref{eq_eta_RSclass})
nor at the 1RSB level (\ref{eq_P_1RSBclass}). These equations are
however amenable to a numerical resolution in a relatively simple way.

Let us first consider the RS case. The variables $\eta$ are probability
distributions over the discrete space ${\cal X}$, each of them can thus
be represented by $|{\cal X}|-1$ real numbers (thanks to their normalization).
In particular for Ising spins a single real is enough, that can be interpreted
as an effective magnetic field acting on a spin. In the RS equation 
(\ref{eq_eta_RSclass}) the unknown is $\cP(\eta)$, a probability
distribution over such effective fields. A convenient way to represent
it numerically~\cite{abou1973,cavity} is to use a sample, also called 
population, of representative fields, i.e. to write
\beq
\cP(\eta) = \frac{1}{\Next} \sum_{i=1}^\Next \delta (\eta - \eta_i) \ .
\label{eq:popu_representation}
\eeq
This representation thus uses a number $\Next$ of representants $\eta_i$,
each of them encoded as a single real for Ising spins, or $|{\cal X}|-1$ 
real numbers in the generic case, and is obviously more and
more accurate as $\Next$ gets larger. As the equation (\ref{eq_eta_RSclass})
has the form of a fixed point condition, it can be solved by iteration: one
starts from an arbitrary initial assignments of the sample $\{\eta_i\}$,
plugs the representation (\ref{eq:popu_representation}) in the r.h.s. of
(\ref{eq_eta_RSclass}), and constructs a new set $\{\eta'_i\}$ that represents
the l.h.s.. This is simply done by repeating $\Next$ times the following 
steps:
\begin{itemize}
\item draw an integer $d$ from the law $\widetilde{q}_d$.
\item draw $d(k-1)$ integers $j_{1,1},\dots,j_{d,k-1}$ uniformly at random in
$[1,\Next]$.
\item draw a random vertex weight $v$ and $d$ random interaction weights 
$w_1,\dots,w_{d}$.
\item compute $\eta'$ from Eq.~(\ref{eq_g_RSclass}), where $\eta_{a,i}$ is
taken to be the $j_{a,i}$'th element of the current population representation
of $\cP$.
\end{itemize}
The $\Next$ elements $\eta'$ thus generated form a sampled representation
of the l.h.s. of (\ref{eq_eta_RSclass}); this process can be iterated, i.e.
this new representation can be plugged in the r.h.s., and so on and so forth
until convergence towards the fixed point of (\ref{eq_eta_RSclass}) is
achieved. Then the expectation values over $\cP$, as for instance in 
Eq.~(\ref{eq_f_RSclass}), can be simply computed by taking an empirical average
over the sample of the $\Next$ representants of $\cP$, which gives access
to all the physical observables of the model, at the replica symmetric level.

Let us now discuss the generalization of this method to the 1RSB level.
The equation (\ref{eq_P_1RSBclass}) has exactly the same structure as
(\ref{eq_eta_RSclass}), and one can thus follow the same strategy as above,
with a population of $\Next$ representants $P_i$ of the distribution 
$\cP^{(1)}$.  The last step in the algorithm explained above becomes
\begin{itemize}
\item compute $P'$ from Eq.~(\ref{eq_G_1RSBclass}), where $P_{a,i}$ is
taken to be the $j_{a,i}$'th element of the current population representation
of $\cP^{(1)}$.
\end{itemize}
This step itself deserves some explanation. 
The additional difficulty is that $P_i$ is itself a distribution
over fields; but this can be handled in a similar way, by representing each
$P_i$ by a sample of $\Nint$ fields $\eta_{i,i'}$, with $i\in [1,\Next]$,
\hbox{$i' \in [1,\Nint]$}. Now to generate a representation of $P'$
from Eq.~(\ref{eq_G_1RSBclass}) one has to construct $\Nint$ fields
$\eta$, by extracting the $\eta_{a,i}$ from their respective distributions
$P_{a,i}$ and computing $\eta=g(\{ \eta_{a,i}\})$. However these $\Nint$ fields
should not be given an equal importance in the representation $P'$: they
have to be weighted according to the replica symmetry breaking weighting
factor $z(\{\eta_{a,i}\},v,\{w_a\})^m$. Several reweighting schemes can be
used to perform this task~\cite{cavity}, and this is a well-studied problem
in the field of statistics where this kind of representation is called
``particle approximation''~\cite{particle_filters}. A simple idea
to perform this reweighting will be given for a similar equation in 
Sec.~\ref{sec:PIQC_QC}, we refer the reader to the original 
literature for more details on this issue.

To summarize, the generic RS cavity equation is handled numerically by a
population (of $\Next$ elements) of fields (each corresponding to $|\chi|-1$
reals, i.e. a single one for Ising spins), while the resolution of the 1RSB 
equation involves a population of $\Next$ populations of $\Nint$ fields.
Higher levels of replica symmetry breaking~\cite{Beyond} can be formally treated
by increasing the number of generation in this hierarchical construction,
but  in general the numerical cost of their resolution increases too fast to allow one to go beyond
the first step.

There are however some cases, to be encountered in the following, where one 
level of complexity disappears. For models defined on regular random 
(hyper)graphs, the distribution $q_d$ is concentrated on a single integer;
then, if the vertex and interaction weights $v$ and $w$ are not random or
are sufficiently symmetric, the RS equation (\ref{eq_eta_RSclass}) 
(resp. the 1RSB equation (\ref{eq_P_1RSBclass})) admits a solution where
$\cP$ (resp. $\cP^{(1)}$) is concentrated on a single field $\eta$ (resp. on
a single distribution $P$). In other words the external size $\Next$ can be
reduced to 1 in these cases, which greatly reduces the numerical cost of the
resolution of the equations (in particular in the quantum case, as will be 
discussed below); this situation is often termed \emph{factorized}
in the mean field spin glass literature.

\subsubsection{Analyzing a mean field spin glass model}
\label{sec:flowchart}

The phase diagram of mean field spin glass models, as a function of the
temperature, magnetic field, connectivity, or other control parameters,
can vary qualitatively from
model to model. We will not attempt here to provide a general classification,
but only propose a flowchart that one should follow when confronted with a
new model, for each point of its phase diagram.
As a first step one should solve the RS equation of the model,
and compute the physical observables associated to it. Then the validity of the
RS hypothesis should be tested; in some cases its violation is apparent from
the inconsistency of the RS results (a negative entropy for instance), but
not always. In any case the 1RSB equation has to be solved, first with the
breaking parameter $m$ set to 1. Two cases can then appear: 
\begin{itemize}
\item if there are no
non-trivial solutions of the $m=1$ 1RSB equations, the RS results should be
conjectured to be valid, the system is in a liquid phase.

\item if there is a non-trivial solution of the $m=1$ 1RSB equations, then
one has to check the sign of the associated complexity.
\begin{itemize}
\item if $\Sigma(f(m=1)) > 0$, the system is in a dynamic 1RSB phase. The RS
prediction for the free energy is correct, yet the Gibbs-measure is split
on an exponentially large number of pure states (clusters), and the dynamics
is non-ergodic.

\item if $\Sigma(f(m=1)) < 0$, the system is in a true 1RSB glass phase, with a 
sub-exponential number of pure states dominating the equilibrium measure.
It is then necessary to make a study as a function of $m\in [0,1]$, and
to find the value $m_{\rm s}$ where the complexity vanishes. The 1RSB
prediction for the free energy is then $f(m_{\rm s})$.
\end{itemize}
\end{itemize}

As a matter of fact this analysis can be further complicated by the 
coexistence of multiple solutions to the RS or 1RSB equations (that should
be discriminated by comparing the free energies they yield), and by the
instability of the 1RSB solutions towards higher levels of replica symmetry
breaking. This second point is particularly difficult to handle in the
case of diluted models~\cite{MoPaRi}; however, in fully connected models where any level
of RSB can be solved, it is found that the 1RSB results are quantitatively 
very good approximations to exact ones.

\subsection{The path integral quantum cavity method}
\label{sec:PIQC}

\subsubsection{Path integral representation of discrete quantum models} 

We shall now define the quantum version of the models studied in the
following. To the space ${\cal X}^N$ of classical configurations and 
energy function $E(\us)$ we associate the Hilbert space spanned by the vectors
$\{|\us \ra | \us \in {\cal X}^N\}$, and an operator $\hH_P$ (we keep here the notations of Sec.~\ref{sec:low_energy}), diagonal in this 
basis, according to
$\hH_P = \sum_{\us} E(\us) |\us \ra \la \us | $. 
We then define the quantum Hamiltonian
\beq
\hH = \hH_P - \sum_{i=1}^N \G_i \, \hT_i \ , 
\eeq
where in the second term the operator $\hT_i$ acts on the $i$-th variable only,
according to
\beq
\la \us | \hT_i | \us' \ra = T_{\s_i,\s'_i} \prod_{j\neq i} \delta_{\s_j,\s'_j}
\ .
\eeq
Without loss of generality we assume that the matrix $T$ of order $|{\cal X}|$ 
has vanishing diagonal elements (these can be incorporated in the 
classical part $\hH_P$). We make the further hypotheses that all its
off-diagonal elements are real non-negative (this is crucial to avoid the
``sign problem''), and that $T$ is symmetric (this ensures the Hermitian 
character of $\hH$). 
The simplest example is provided by
the spin 1/2 case, with ${\cal X}=\{-1,+1\}$; the above basis is taken to be the
eigenvectors of the Pauli matrices $\hs_i^z$, and one can take 
$\hT_i = \hs_i^x$, the parameters $\G_i$ are then the transverse fields of the
model.
The quantum statistical mechanics study of the model amounts to the 
computation of the partition function $Z = \Tr e^{-\beta \hH}$. The additional
technical difficulty, with respect to the classical models, arises from
the non-commutativity of the two terms in $\hH$. The standard way to handle
this difficulty is to introduce a path integral representation of the
partition function, of the form:
\beq
Z = \underset{\ubs(0)=\ubs(\beta)}{\int} \left[ \prod_{i=1}^N D \bs_i \, v_i(\bs_i) \right]
\, \, e^{-\int_0^\beta E(\us(t)) \dd t } \ .
\label{eq_Z_quantum_PI}
\eeq
Let us precise the notations we introduced. Here and in the following
bold symbols will be used for functions of an ``imaginary'' time 
$t\in[0,\beta]$; in particular here $\bs_i$ is a piecewise constant function 
$\s_i(t) : [0,\beta] \to {\cal X}$. The integration measure $D\bs_i$ is
decomposed as a sum over the number $n$ of discontinuities of the function
$\s_i(t)$, the times $t_1 \le \dots \le t_n$ at which they occur, and the
values $\s_i^0,\dots,\s_i^n$ the function takes on the $n+1$ intervals
$[0,t_1]$, $[t_1,t_2]$, \dots, $[t_n,\beta]$:
\beq
\int D\bs_i \equiv \sum_{n=0}^\infty \sum_{\s_i^0,\dots,\s_i^n}
\int_{0}^\beta \dd t_1 \int_{t_1}^\beta \dd t_2
\dots \int_{t_{n-1}}^\beta \dd t_n \ .
\label{eq_def_Dbsi}
\eeq
For such a function the weight $v_i(\bs_i)$ reads
\beq
v_i(\bs_i) = (\G_i)^n \prod_{j=1}^n T_{\s_i(t_j^-),\s_i(t_j^+)}
= (\G_i)^n \prod_{j=1}^n T_{\s_i^{j-1},\s_i^j} \ ;
\label{eq_v_quantum}
\eeq
as the diagonal elements of $T$ are supposed to vanish the only contributing
paths are those with $\s_i^{j-1} \neq \s_i^j$. Such a path integral 
representation can be devised for any matrix element of $e^{-\b \hH}$, by 
simply fixing the initial and final configurations $\us(0)$ and $\us(\beta)$.
Here we let them free, under the condition $\us(0)=\us(\beta)$, to compute
the trace of $e^{-\b \hH}$.

There are several ways to obtain such a path integral representation.
The most pedestrian one is to use the Suzuki-Trotter formula, decomposing
the imaginary time interval $[0,\beta]$ in $\Ns$ slices,
\beq
e^{\hX_1 + \hX_2} = \lim_{\Ns \to \infty} \left( 
e^{\frac{1}{\Ns} \hX_1} e^{\frac{1}{\Ns}\hX_2} \right)^\Ns \ ,
\eeq
with the two non-commuting operators $\hX_1=-\beta \hH_P$ and 
$\hX_2 = \beta \sum_i \G_i \hT_i$. One can then introduce $\Ns$
representations of the identity between the terms of the product,
as sums over the configurations $\us(t=\a \b/\Ns )$ with $\a$ a discrete 
time index. The continuous time limit $\Ns \to \infty$ then yields the
path integral (\ref{eq_Z_quantum_PI}).
A maybe more elegant way to obtain this result is to use the following
operator identity,
\beq
e^{\hX_1 + \hX_2} = \sum_{p=0}^\infty \int_0^1 dt_1 \int_{t_1}^1 \dd t_2 \dots 
\int_{t_{p-1}}^1 \dd t_p \,  e^{t_1 \hX_1} \,
\hX_2 \, e^{(t_2-t_1) \hX_1} \,  \hX_2 \, \dots \, \hX_2 \, e^{(1-t_p) \hX_1} 
\ ,
\nonumber
\eeq
with the same values of $\hX_{1,2}$ as above; one then inserts $p$ 
representations of the identity in the $p$-th term of the sum, rescales the 
time integrals and reorders the summation according to the number of
times each flipping operator $\hT_i$ is picked in the expansion of $\hX_2$.
Finally these path integral representations can also be handled
in a mathematically rigorous way, see for 
instance~\cite{Gi68,GMR68,FG92,AiNa94,ChCrIoLe08,Ioffe09,MaWo12}, by
starting from Poisson point processes for the candidate times of 
discontinuities of the variable trajectories, a distribution which is then
properly biased by the classical energy terms.

\subsubsection{Representation of the cavity messages}
\label{sec:PIQC_QC}

The path integral representation of the partition function given in 
Eq.~(\ref{eq_Z_quantum_PI}) is valid for any classical energy function 
$E(\us)$. Suppose now that $E$ is decomposed as a sum of local interaction
terms, according to Eq.~(\ref{eq_E_class}). The quantum partition function
(\ref{eq_Z_quantum_PI}) can then be rewritten as
\beq
Z = \sum_{\ubs \in \widehat{\cal X}^N} \delta_{\ubs(0),\ubs(\beta)}
 \prod_{a=1}^M w_a(\ubs_\da) 
\prod_{i=1}^N v_i(\bs_i) \ , \qquad w_a(\ubs_\da)=e^{-\int_0^\b \dd t \,
\e_a(\us_\da(t))} \ ,
\eeq
where we introduced $\widehat{\cal X}$ the space of periodic piecewise constant 
functions from $[0,\beta]$ to ${\cal X}$, and used the notation
$\sum_{\bs_i \in \widehat{\cal X}}$ as a synonym of the integration 
$\int D \bs_i$ defined in Eq.~(\ref{eq_def_Dbsi}). This notation was
chosen to emphasize the similarity with the classical partition function
(\ref{eq_mu}): the quantum computation is reduced to a classical one,
the cost to be paid being the replacement from a discrete variable $\s_i$ in
${\cal X}$ to a function (trajectory) $\bs_i=\{ \s_i(t) | t \in[0,\b] \}$ in 
$\widehat{\cal X}$ as basic degrees of freedom. The partition function can
also be interpreted as the normalization constant of a probability
measure over $\widehat{\cal X}^N$, namely
\beq
\mu_{\rm Q}(\ubs) = \frac{1}{Z} \delta_{\ubs(0),\ubs(\beta)} 
\prod_{a=1}^M w_a(\ubs_\da) \prod_{i=1}^N v_i(\bs_i) \ .
\label{eq_mu_quantum}
\eeq
Note that apart from the change of the nature of the degrees of freedom,
the ``spatial'' structure of the interactions $w_a$ encoded in the factor graph 
is the same in the classical and in the quantum case. 
In particular as soon as the
classical energy part of the quantum Hamiltonian falls into the category of
models that can be solved by the cavity method (i.e. sparse random graphs), 
then this is also true for the quantum problem. This observation was first
exploited in~\cite{LSS08} to study the spin 1/2 spin glass on the Bethe lattice,
within a finite number of Suzuki-Trotter slices. The formulation of the
quantum cavity method in continuous imaginary time was then presented 
in~\cite{KRSZ08} at the RS level, for a ferromagnetic model, and~\cite{STZ09} 
for the Bose-Hubbard model of bosonic particles. Results at the 1RSB level
were given in~\cite{JKSZ10}, and a complete exposition can be found in~\cite{FSZ11}
for interacting particle models.

The general structure of the quantum cavity method is thus exactly the same
as the classical one exposed in Sec.~\ref{sec:classical_cavity}. 
In the rest of this section we explain
the additional technical points that arise when going from $\cal X$ to
$\widehat{\cal X}$ as the base space for degrees of freedom. First of all
one has to find an efficient way to represent the probability distributions
$\eta(\bs)$ over $\widehat{\cal X}$, which are the basic objects of the
method. As should be clear from the discussion of 
Sec.~\ref{sec:popu_dynamics},
the simplest way to do that is to approximate them by a weighted 
sample representation of a large number $\Ntraj$ of elements of 
$\widehat{\cal X}$ , namely
\beq
\eta(\bs) = \sum_{j=1}^\Ntraj a^{(j)} \, \delta(\bs - \bs^{(j)}) \ , \qquad
\text{with} \ \ \ \sum_{j=1}^\Ntraj a^{(j)} = 1 \ .
\label{eq_eta_quantum}
\eeq
Each representative trajectory $\bs^{(j)}$ is itself encoded in a compact way
by the number of discontinuities it contains, their times of occurrence, and
the constant values of the function between the discontinuities.
Sampling an element from $\eta$ means drawing a number $j\in[1,\Ntraj]$ with 
probability $a^{(j)}$, and returning the trajectory $\bs^{(j)}$.

Secondly one must devise a way to implement the quantum equivalent of 
Eq.~(\ref{eq_g_RSclass}),
\bea
&&\eta(\bs) = \frac{1}{z(\{\eta_{a,i}\},\G,\{\e_a\})} v(\bs) \times
\nonumber \\
&&\sum_{\{ \bs_{a,i} \}_{a\in[1,d]}^{i\in[1,k-1]}} 
\left( \prod_{a,i} \eta_{a,i}(\bs_{a,i}) \right) \ 
\prod_{a=1}^d 
e^{-\int_0^\b \dd t \, \e_a(\s(t),\s_{a,1}(t),\dots,\s_{a,k-1}(t))} \ , \ \
\label{eq_eta_quantum_1}
\eea
with the quantum variable weight defined in Eq.~(\ref{eq_v_quantum}),
\beq
v(\bs) = \G^n \prod_{j=1}^n T_{\s(t_j^-),\s(t_j^+)} \ ,
\eeq
$n$ denoting the number of discontinuities of $\bs$, at times 
$t_1\le\dots \le t_n$. To rewrite this equation under a more
convenient form we shall introduce time-varying ``longitudinal fields''
$\bvh$, which are functions $h_\s(t)$ of an index $\s \in {\cal X}$ and
of the imaginary time $t$. Indeed the dependency on $\bs$ of the
energy terms in Eq.~(\ref{eq_eta_quantum_1}) can be conveniently expressed
in terms of such a field defined by
\beq
\bvh(\{ \bs_{a,i} \},\{\e_a\}) \ : \ 
h_\s(t) = -\sum_{a=1}^d \e_a(\s,\s_{a,1}(t),\dots,\s_{a,k-1}(t)) \ .
\label{eq_bvh}
\eeq
We can then rewrite (\ref{eq_eta_quantum_1}) as
\bea
\eta(\bs) &=& \sum_{\{ \bs_{a,i} \}_{a\in[1,d]}^{i\in[1,k-1]}} 
\left( \prod_{a,i} \eta_{a,i}(\bs_{a,i}) \right) \times \nonumber \\
&& p(\bs |\G, \bvh(\{ \bs_{a,i} \},\{\e_a\}) ) 
\frac{{\cal Z}(\G,\bvh(\{ \bs_{a,i} \},\{\e_a\}))}
{z(\{\eta_{a,i}\},\G,\{\e_a\})} \ ,
\label{eq_eta_quantum_2}
\eea
where we defined
\bea
p(\bs|\G,\bvh) &=& \frac{1}{{\cal Z}(\G,\bvh)} v(\bs) e^{\int_0^\b \dd t \,
h_{\s(t)}(t)} \ , \label{eq_p_bs} \\
{\cal Z}(\G,\bvh) &=& \sum_{\bs \in \widehat{\cal X}} 
v(\bs) e^{\int_0^\b \dd t \, h_{\s(t)}(t)} \ .
\nonumber
\eea
In this way $p(\bs|\G,\bvh)$ is a well-normalized probability distribution
over $\widehat{\cal X}$. Suppose that one is able to sample from
this distribution, and to compute the associated normalization 
${\cal Z}(\G,\bvh)$ (we shall see in a short while that this is indeed 
possible). Then the resolution of Eq.~(\ref{eq_eta_quantum_2}) is at hand.
This amounts in fact to the generation of a sampled representation of its
l.h.s., assuming the knowledge of the r.h.s., in particular the ability to
draw the trajectories from the probability laws $\eta_{a,i}$ on 
$\widehat{\cal X}$. To construct the representation of the l.h.s., one has 
to repeat $\Ntraj$ times independently, for $j\in[1,\Ntraj]$, the following
steps:
\begin{itemize}
\item draw the $\{ \bs_{a,i} \}_{a\in[1,d]}^{i\in[1,k-1]}$ from their respective
distributions $\eta_{a,i}$.
\item compute the field $\bvh$ according to Eq.~(\ref{eq_bvh}).
\item extract a trajectory $\bs^{(j)}$ from the law $p(\cdot|\G,\bvh)$ and 
set $a^{(j)}={\cal Z}(\G,\bvh)$.
\end{itemize}
Once these steps have been performed $\Ntraj$ times, normalize the new weights,
\beq
a^{(j)} \leftarrow \frac{a^{(j)}}{a^{(1)}+\dots+a^{({\cal \Ntraj})}}\ .
\eeq
A moment of thought reveals that this is indeed the correct algorithmic 
translation of Eq.~(\ref{eq_eta_quantum_2}).

\subsubsection{Path generation}
\label{sec:PIQC_pathgeneration}

We are thus left with the problem of generating a path according to the law
$p(\bs|\G,\bvh)$ defined in Eq.~(\ref{eq_p_bs}) and of computing the 
normalizing factor ${\cal Z}(\G,\bvh)$. As all trajectories $\bs_{a,i}$ are
piecewise-constant, this will also be the case of the relevant realizations of
the field $\bvh$. Let us call $p$ the number of discontinuities on $[0,\b]$
of $\bvh$, that occur at times $0\le t^{(1)}\le \dots \le t^{(p)} \le \b$, 
and denote 
$\vh^{(0)},\vh^{(1)},\dots,\vh^{(p)}$ the values of $\vh(t)$ on the time
intervals $[0,t^{(1)}]$, $[t^{(1)},t^{(2)}]$, \dots, $[t^{(p)},\b]$. We also denote 
$\l^{(0)}=t^{(1)}$, $\l^{(1)}=t^{(2)}-t^{(1)}$, \dots, $\l^{(p)}=\b-t^{(p)}$ the
duration of these intervals. Let us introduce some further notations;
we consider a $|{\cal X}|$ dimensional Hilbert space spanned by 
$\{| \s\ra  | \s \in {\cal X}\}$, on which we define an operator
$\tH(\G,\vh)$ by its matrix elements,
\beq
\la \s | \tH(\G,\vh) | \s' \ra = - h_\s\, \delta_{\s,\s'} -\G \, T_{\s,\s'} \ .
\eeq
We shall write $\tW(\G,\vh,\l) = e^{-\l \tH(\G,\vh)}$ its associated propagator
on an interval of imaginary time of length $\l$, and
$W(\G,\vh,\l)_{\s,\s'} = \la \s | \tW(\G,\vh,\l) | \s' \ra$ the matrix elements
of the propagator. It is then possible to prove that the sought-for
normalizing factor ${\cal Z}(\G,\bvh)$ reads
\beq
{\cal Z}(\G,\bvh) = \Tr \left[\prod_{i=0}^p \tW(\G,\vh^{(i)},\l^{(i)} ) \right]
\ .
\eeq
This is a computationally affordable expression: it requires diagonalizing
$p$ matrices of (small) dimension $|{\cal X}|$, exponentiating them and 
multiplying them together. Finally the process of generation of $\bs$ with
the law $p(\cdot|\G,\bvh)$ can be implemented as follows:
\begin{itemize}
\item draw the values $\s^{(0)},\dots,\s^{(p)}$ that $\bs$ assumes at times
$0,t^{(1)},\dots, t^{(p)}$.
\item on each of the $p+1$ intervals $[t^{(i)},t^{(i+1)}]$, draw a trajectory
representative of the evolution $\tW(\G,\vh^{(i)},\l^{(i)})$ in a constant
field $\vh^{(i)}$, with boundary conditions $\s(t^{(i)})=\s^{(i)}$, 
$\s(t^{(i+1)})=\s^{(i+1)}$ (we set $t^{(0)}=0$ and \hbox{$\s^{(p+1)}=\s^{(0)}$}).
\end{itemize}
More precisely, the first step consists in extracting these $p+1$ values
from the joint law
\bea
p(\s^{(0)},\dots,\s^{(p)} ) &=& \frac{1}{{\cal Z}(\G,\vh)} 
W(\G,\vh^{(0)},\l^{(0)})_{\s^{(0)},\s^{(1)}}
W(\G,\vh^{(1)},\l^{(1)})_{\s^{(1)},\s^{(2)}}
\dots \nonumber \\ 
&&\quad \dots W(\G,\vh^{(p)},\l^{(p)})_{\s^{(p)},\s^{(0)}} \ .
\eea
This can be easily done by first drawing $\s^{(0)}$ from its marginal
probability, then $\s^{(1)}$ conditioned on the value of $\s^{(0)}$, and
so on until $\s^{(p)}$ has been extracted.

The procedure to follow for the second step is more apparent once an integral
equation on $W$ is written:
\beq
W(\G,\vh,\l)_{\s,\s'} = e^{\l h_\s} \delta_{\s,\s'} + 
\G \int_0^\l \dd t \, e^{t h_\s} \sum_{\s''} T_{\s,\s''} 
W(\G,\vh,\l-t)_{\s'',\s'} \ .
\eeq
In terms of the path integral representation of $\tW$, the two terms in this
equation represent respectively the contribution of a constant path 
(possible only if the boundary conditions are the same at time $t=0$ and 
$t=\l$) and of a path whose first discontinuity occurs at time $t$, where
$\s(t)$ jumps from $\s$ to $\s''$.
In consequence, the procedure to draw a path from $\s(t=0)=\s$ to
$\s(t=\l)=\s'$ in presence of a constant field $\vh$ reads
\begin{itemize}
\item if $\s=\s'$, with probability $e^{\l h_\s}/W(\G,\vh,\l)_{\s,\s}$,
exit with the constant path $\s(t)=\s$ $\forall t \in [0,\l]$.
\item otherwise
\begin{itemize}
\item draw a random time $u \in [0,\l]$ with the cumulative distribution 
\beq
G(u) = \frac{
\int_0^u \dd t \, e^{t h_\s} \sum_{\s''} T_{\s,\s''} 
W(\G,\vh,\l-t)_{\s'',\s'} }
{\int_0^\l \dd t \, e^{t h_\s} \sum_{\s''} T_{\s,\s''} 
W(\G,\vh,\l-t)_{\s'',\s'} }
\label{eq_G_of_u}
\eeq
\item draw an element $\s''$ with probability
\beq
\frac{ T_{\s,\s''} W(\G,\vh,\l-u)_{\s'',\s'}}{
\sum_{\s'''} T_{\s,\s'''} W(\G,\vh,\l-u)_{\s''',\s'}}
\eeq 
\item set $\s(t)=\s$ for $t\in[0,u]$, and call recursively
the same procedure to generate the path on $[u,\l]$, with boundary 
conditions $\s(u)=\s''$, $\s(\l)=\s'$.
\end{itemize}
\end{itemize}

This procedure for the generation of imaginary time paths in presence
of a constant transverse operator $T$ and piecewise constant
longitudinal fields was presented in the case of spins 1/2 in~\cite{KRSZ08};
its recursive nature has the advantage of making it a rejection-free method, 
the inconvenience being the necessity to draw random variables from rather
complicated distributions (\ref{eq_G_of_u}) for the interval of times between
spin flips. An alternative method was proposed in~\cite{FGGGS10}: the
time intervals between spin flips are drawn from exponential distributions with
well chosen averages, which is much easier, but the price to be paid is a 
rejection if the parity of the number of spin flips on the interval $[0,\beta]$
does not satisfy the boundary conditions on $\s(0)$ and $\s(\beta)$. The
rejection rate can in particular become quite high if $\s(0)\neq \s(\beta)$
for low transverse fields. The two methods can actually be combined to gain
both their advantages, by drawing for anti-periodic trajectories
the time of first flip from the recursive method, then continue with the
rejection one.

\subsubsection{Discussion}

The Path Integral Quantum Cavity (PIQC) method described above is an exact way of
dealing with quantum spin models on sparse random graphs. The analysis
of such models proceeds along the same lines as explained in the
classical case in Sec.~\ref{sec:flowchart}, i.e. via an interpretation
of the solutions of the RS and 1RSB equations. In Sec.~\ref{sec:XORSAT} and
\ref{sec:results_coloring} we shall present explicit results on two models
of random constraint satisfaction problems obtained in this way.

It is however important to mention the limitations of the method.
As already explained at the beginning of this section, it
can only handle quantum systems that do not suffer from the sign problem,
and it is a finite temperature method; the ground state properties are
necessarily obtained by an extrapolation to zero temperature. Moreover
the numerical resolution of the quantum cavity equations is a numerically
costly task. The generic quantum 1RSB case involves indeed a representation by
$\Next\times\Nint\times\Ntraj$ imaginary time trajectories\footnote{An alternative approach,
that will not be further discussed here, consists in a systematic perturbative
expansion in the transverse field $\Gamma$; any finite order of the 
expansion can be expressed in terms of the classical cavity computation,
thus strongly reducing the numerical cost with respect to the fully quantum
approach. This however does not give access to non-perturbative effects like
phase transitions.} (recall the 
discussion of Sec.~\ref{sec:popu_dynamics} in the classical case and the
additional population level due to the quantum nature of the model
explained in Eq.~(\ref{eq_eta_quantum})). Fortunately for factorized
models (with regular degrees) this is reduced with $\Next=1$, see the end
of Sec.~\ref{sec:popu_dynamics}. In any case the memory available on present
days computers limit the numbers $\Nint$ and $\Ntraj$ to relatively small
values (examples will be given on concrete cases in Sec.~\ref{sec:XORSAT} and
\ref{sec:results_coloring}). This induces both systematic deviations of the
empirical mean from the exact value and noise in its estimation; 
extrapolations to $\Nint,\Ntraj \to \infty$ via finite size analysis can 
however be performed to reduce these effects. A specific difficulty comes
from the weighted representations of probability distributions used
in Eq.~(\ref{eq_eta_quantum}) for instance; one must take care by resampling
methods of the tendency that these weights have to flow towards very inhomogeneous 
repartitions, which leads to situations where the number
of effective representants of the distribution becomes much smaller than
$\Ntraj$~\cite{particle_filters}.

In the following sections we shall describe alternative approaches
to quantum models on sparse random graphs that, even if approximate,
allow to bypass some of these limitations.

\subsection{Operator quantum cavity methods}
\label{sec:OQC}
\label{sec:IM}

In this section we shall describe
alternative formulations of the quantum cavity method
that do not make use of the path integral formulation but work directly 
with quantum operators~\cite{hastings2007,leifer2008,poulin2008,Bil-Poul2010,Poulin_Hasting2011,IM10,FIM10,dimitrova2011}.
These approaches have been sometimes called ``quantum belief propagation'' but we will refer
to them here as Operator Quantum Cavity (OQC) methods.
They all share common features and ideas whose connections are still only partially understood. 
They represent approximated methods, and their level of accuracy
is not completely controlled yet. However, compared to path integral methods they have the important
advantage that the $T=0$ limit can be taken explicitly. Moreover, the cavity messages are represented
as finite matrices, therefore there are no sampling errors, unlike in the PIQC where the messages are represented
through finite samples of probability distributions over an infinite-dimensional space.

For the sake of simplicity,
we will present these methods in the simpler case of an Hamiltonian that is the sum of two-body interactions:
\beq
\hH = \sum_{\la i,j \ra} \hH_{i,j} \ .
\eeq
The sum over $\la i,j \ra$ runs on the links of a regular lattice of
degree $c$. We will mainly focus on the case $c=2$ of a one dimensional
chain, and $c=3$ with the underlying lattice being a 3-regular random graph.
Moreover we will consider the case of an Ising model in a transverse field for 
which
$\hH_{i,j} = -J_{ij} \hs_i^z \hs_j^z - ( \G_i \hs_i^x + \G_j \hs_j^x)/c$. 
The generalization to more complex Hamiltonians is straightforward.

\subsubsection{Operator cavity messages}

We start our presentation by following the derivation of~\cite{Bil-Poul2010} and considering for simplicity a
finite one-dimensional chain with open boundaries.
The quantum partition function is
\beq
Z = \Tr e^{-\b\hH} = \Tr \left( e^{-\b \hH_{1,2}} \odot e^{-\b \hH_{2,3}} \odot \cdots \odot e^{-\b \hH_{N-1,N}} \right) \ ,
\eeq
where $e^A \odot e^B = e^{A+B}$.
As in the classical case, we can define an operatorial message that acts on the Hilbert space of spin $i$ only:
\beq
\h_{i \to i+1} = \frac{1}{z_{i \to i+1}} \Tr_{1,\cdots,i-1} \, e^{-\b \sum_{k=1}^{i-1} \hH_{k,k+1}} \ ,
\eeq
where the normalization is determined by $\Tr_i \, \h_{i \to i+1} = 1$.

We can derive an approximate recurrence equation for these messages by following the same steps as in the 
classical case:
\beq\label{eq:OQC_1}
\begin{split}
\h_{i \to i+1} & \propto \Tr_{1,\cdots,i-1} \left( e^{-\b \sum_{k=1}^{i-2} \hH_{k,k+1}} \odot e^{-\b \hH_{i-1,i}} \right) \\
& = \Tr_{i-1} \left\{ 
 \Tr_{1,\cdots,i-2} \left[e^{-\b \sum_{k=1}^{i-2} \hH_{k,k+1}} \odot e^{-\b \hH_{i-1,i}}    \right]
\right\} \\
& \sim \Tr_{i-1} \left\{ 
\left[ \Tr_{1,\cdots,i-2} e^{-\b \sum_{k=1}^{i-2} \hH_{k,k+1}} \right] \odot e^{-\b \hH_{i-1,i}}  
\right\} \\
& \propto \Tr_{i-1} \left( \h_{i-1 \to i} \odot e^{-\b \hH_{i-1,i}} \right) \ ,
\end{split}\eeq
and the proportionality constant is determined by normalization as in the classical case.
The crucial point is that, unlike in the classical case, here we made an approximation when we changed the position of 
the square brackets moving from the second to the third line of the above equation.

Indeed, consider a system made of three parts $a, b, c$ and operators $\hH_{a,b}$, $\hH_{b,c}$, 
acting only on $a \otimes b$ and $b \otimes c$ respectively.
Due to quantum entanglement
\beq
 \Tr_a \left[ e^{-\b \hH_{a,b}} \odot e^{-\b \hH_{b,c}} \right] 
\neq 
\left[ \Tr_a e^{-\b \hH_{a,b}} \right] \odot e^{-\b \hH_{b,c}}  \ .
\eeq
However, the above equation is an equality if the ``conditional mutual information''
$I(a : c | b) = S(a,c) + S(b,c) - S(b) - S(a,b,c)$ vanishes (here $S$ is the von Neumann entropy),
indicating that all correlations between $a$ and $c$ are mediated through $b$ (as in the classical case).
It has been argued that this condition holds when the region $b$ is 
sufficiently ``thick''~\cite{Bil-Poul2010}. The problem is that in Eq.~(\ref{eq:OQC_1}) the region $b$
coincides with a single spin, $b = \{i-1 \}$.

This observation motivates the introduction of new messages, that are operators on the space
of spins $\{i-\ell+1,\cdots,i\}$. Repeating the above derivations:
\beq\begin{split}
\h^{(\ell)}_{i \to i+1} & = \frac{1}{z_{i \to i+1}} \Tr_{1,\cdots,i-\ell} \, e^{-\b \sum_{k=1}^{i-1} \hH_{k,k+1}} \\
& \propto  \Tr_{i-\ell} \left\{ 
 \Tr_{1,\cdots,i-\ell-1} \left[e^{-\b \sum_{k=1}^{i-2} \hH_{k,k+1}} \odot e^{-\b \hH_{i-1,i}}    \right]
\right\} \\
& \sim \Tr_{i-\ell} \left\{ 
\left[ \Tr_{1,\cdots,i-\ell-1} e^{-\b \sum_{k=1}^{i-2} \hH_{k,k+1}} \right] \odot e^{-\b \hH_{i-1,i}}  
\right\} \\
& \propto \Tr_{i-\ell} \left( \h^{(\ell)}_{i-1 \to i} \odot e^{-\b \hH_{i-1,i}} \right)
\end{split}\eeq
The crucial difference is that now the region $b = \{ i - \ell,\dots, i-1 \}$ has thickness $\ell$ and
one can hope that the error is much smaller. An argument in favor of this has been discussed
in~\cite{Bil-Poul2010}. The drawback is of course that now the messages are operators acting
on $\ell$ spins, and therefore they have to be represented by matrices of size $2^\ell$.

The generalization of this procedure to a tree is straightforward. 
Let us call $\TT_{i \to j}$ the partial tree rooted at $i$ obtained by cutting the link $\la i,j \ra$,
and $d(i,j)$ the distance on the tree between $i$ and $j$.
The message from $i$ to $j$ is defined as 
\beq 
\h^{(\ell)}_{i \to j}  = \frac{1}{z_{i \to j}} \Tr_{\{ k \in \TT_{i\to j}, d(i,k) \geq \ell \}  } e^{-\b \sum_{\la k,l \ra \in \TT_{i\to j} } \hH_{k,l}} \ , 
\eeq
and we get as in the classical case:
\beq\label{eq:OQC_tree}
\h^{(\ell)}_{i \to j}  \propto \Tr_{  \{ k \in \TT_{i\to j}, d(i,k) = \ell \}   } \left\{ 
\underset{k \in \partial i \setminus j}{\odot} \left( \h^{(\ell)}_{k\to i} \odot e^{-\b \hH_{k,i}} \right)
\right\} \ .
\eeq
Here, the messages are operators acting on $1+(c-1) + (c-1)^2 + \cdots + (c-1)^{\ell-1}$ spins,
so they must be represented by matrices whose size $2^{\sum_{k=0}^{\ell-1} (c-1)^k}$
grows much faster than in the one dimensional case.

With similar reasonings one can obtain the approximate expression for the free energy, 
which is exactly the same as in the classical case
(here specialized to a system with two-body interactions only), with
sums replaced by traces and the normal product replaced by the $\odot$ product:
\beq\label{eq:fRS_OQC}
\begin{split}
- \b F &= \sum_i \log z_i - \sum_{\la i,j \ra} \log z_{ij} \ , \\
z_i &= \Tr_{ i , \cup_{j \in \partial i} \{ k \in \TT_{j\to i}, d(j,k) < \ell \} } \left[
\underset{ j \in \partial i }{\odot} \left( \h^{(\ell)}_{j\to i} \odot e^{-\b \hH_{j,i}} \right)
\right]
\ , \\
z_{ij} & = \Tr_{ \{ k \in \TT_{i\to j}, d(i,k) < \ell \} \cup \{ k \in \TT_{j \to i}, d(j,k) < \ell \} } \left( \h^{(\ell)}_{i \to j} \odot \h^{(\ell)}_{j \to i} \odot e^{-\b \hH_{i,j}} \right) \ .
\end{split}
\eeq

\subsubsection{Explicit equations for single-spin messages}
\label{sec:OQC_single}

Let us now consider more explicitly the above OQC formulation on a tree with $\ell=1$. 
In this case the messages are operators on a single spin, i.e. $2 \times 2$ Hermitian matrices
normalized to have trace 1.
We can parametrize them
by two local fields:
\beq\label{eq:OQC_1_param}
\h_{i \to j} = \frac1{z_{i\to j}} e^{\b ( b_{i \to j} \hs^x_i + h_{i \to j} \hs^z_i ) } \ ,
\eeq
omitting a term proportional to $\hs^y_i$ that vanishes by symmetry.
Equivalently we can describe the message $\h_{i \to j}$ in terms of the magnetizations
\beq\label{eq:IM1}
\begin{split}
&m^x_{i \to j} = \Tr_i ( \hs_i^x \h_{i \to j} ) = \frac{b_{i\to j}}{\sqrt{h_{i\to j}^2 + b_{i\to j}^2}} 
\tanh\left[ \b  \sqrt{h_{i\to j}^2 + b_{i\to j}^2} \right] \\
&m^z_{i \to j} = \Tr_i ( \hs_i^z \h_{i \to j} ) = \frac{h_{i\to j}}{\sqrt{h_{i\to j}^2 + b_{i\to j}^2}} 
\tanh\left[ \b  \sqrt{h_{i\to j}^2 + b_{i\to j}^2} \right]
\end{split}\eeq
Plugging this in Eq.~(\ref{eq:OQC_tree}) with $\ell=1$ we obtain
\beq
 e^{\b ( b_{i \to j} \hs^x_i + h_{i \to j} \hs^z_i ) } \propto 
\Tr_{k \in \partial i \setminus j} 
e^{\b 
\sum_{k \in \partial i \setminus j} 
[ b_{k \to i} \hs^x_k + h_{k \to i} \hs^z_k
- \hH_{k,i} ] 
}
\eeq
which can be recast in the following form:
\beq\label{eq:IM2}
m^x_{i \to j} =\frac{ \Tr_{i,k \in \partial i \setminus j}  ( \hs_i^x e^{-\b \hH_{\rm eff}} )}{ \Tr_{i,k \in \partial i \setminus j}  ( e^{-\b \hH_{\rm eff}} )}
\ ,
\eeq
and similarly for $m^z_{i \to j}$, where
\beq\label{eq:IM3}
\begin{split}
\hH_{\rm eff} & = \sum_{k \in \partial i \setminus j} 
[\hH_{k,i} - b_{k \to i} \hs^x_k - h_{k \to i} \hs^z_k]  \\
& =- \sum_{k \in \partial i \setminus j} 
\big[ J_{ik} \hs_i^z \hs_k^z + ( \G_i \hs_i^x + \G_k \hs_k^x)/c
 + b_{k \to i} \hs^x_k + h_{k \to i} \hs^z_k \big]
\end{split}\eeq
is an effective Hamiltonian acting on spin $i$ and its neighbors (except $j$).
Iteration of these equations then requires at each step the diagonalization of a Hamiltonian
acting on $c$ spins. Note that taking the $T=0$ limit is straightforward and simplifies the computation, 
because in this case we only need to find the ground state of $\hH_{\rm eff}$.

One can actually take a different approach and substitute Eq.~(\ref{eq:OQC_1_param}) in the free energy
Eq.~(\ref{eq:fRS_OQC}), obtaining then a function of the set of all fields $b_{i\to j}$ and $h_{i\to j}$. One can
then derive equations for these fields by imposing stationarity of the free energy with respect to variations
of any field, as in the classical case. However, because the OQC is only approximate, the stationarity equations
{\it do not coincide} with the equations obtained from cavity iteration, Eqs.~(\ref{eq:IM1}), (\ref{eq:IM2}), (\ref{eq:IM3}).
It can be shown on specific examples (e.g. the ferromagnetic case $J_{ij} = J$ and $\G_i =\G$) that
imposing stationarity of the free energy is slightly more accurate than the iteration scheme.

Let us also mention a further approximation that has been proposed in~\cite{IM10,FIM10,dimitrova2011}, which amounts to replace
the operators $\hs_k^x, \hs_k^z$ in Eq.~(\ref{eq:IM3}) by their averages $m_{k\to i}^x, m_{k\to i}^z$ in Eq.~(\ref{eq:IM1}). One thus obtains the following equations:
\beq\begin{split}
& b_{i \to j} = \frac{c-1}c \G_i  \ , \\
& h_{i \to j} = \sum_{k \in \partial i \setminus j} J_{ki} m^z_{k \to i} 
= \sum_{k \in \partial i \setminus j}
\frac{ J_{ki}  h_{k\to i}}{\sqrt{h_{k\to i}^2 + b_{k\to i}^2}} 
\tanh\left[ \b  \sqrt{h_{k\to i}^2 + b_{k\to i}^2} \right]
\ . \\
\end{split} \nonumber
\eeq
These are closed and relatively simple equations for the fields $h_{i\to j}$ and have been exploited in~\cite{IM10,FIM10} to obtain detailed information
on a disordered system that would have been extremely hard to obtain from the numerical solution of the OQC or PIQC equations. 
Additionally, it is clear from these equations that one can take the $\b\to\io$ limit without problems just by dropping the hyperbolic tangent term.
A drawback of this approach is that these equations are approximate, even
in the classical case $\G_i=0$. 
It has been argued in~\cite{IM10,FIM10} that they become exact for $c\to\io$, 
see~\cite{dimitrova2011,Mu11} for a detailed discussion of this delicate point.

\subsubsection{Relation with the PIQC}

The OQC has been introduced  in~\cite{IM10,FIM10}, independently from~\cite{Bil-Poul2010}, to study the
metal-insulator transition in disordered superconductor
and later used in~\cite{dimitrova2011} to discuss the properties of disordered ferromagnets.
The derivation of~\cite{IM10,FIM10,dimitrova2011} starts from the
PIQC formulation and makes a simple ansatz on the functional form of
the distribution of imaginary time trajectories.
In turn, this can be reinterpreted as an ansatz over the Hamiltonian 
governing a reduced part of the system, consisting of neighboring spins, 
and gives back the OQC.

The PIQC leads to the following equation
(which is the specialization of the treatment of Sec.~\ref{sec:PIQC} to Ising spins,
see also~\cite{KRSZ08}):
\beq\label{cav_eq_discr}
\eta_{i\to j} (\bs_i) = \frac{\G_i^{|\bs_i|}}{z_{i\to j}} \prod_{k \in \partial i \setminus j}
\int D\bs_k \, \eta_{k\to i}(\bs_k)\, e^{J_{ik} \int_0^\b \s_i(t) \s_k(t) \dd t} \ ,
\eeq
where $|\bs_i|$ is the number of spin flips in the imaginary time trajectory 
$\bs_i$.
In order to simplify the solution of these self-consistent equations, in~\cite{IM10,FIM10,dimitrova2011}
it was suggested to consider the following ansatz:
\beq\label{ansatz_eta}
\eta_{i\to j} (\bs_i) \propto (b_{i\to j})^{|\bs_i|}
 e^{\int_0^\b h_{i\to j} \s_i(t) \dd t} \ .
\eeq
Once inserted in the right hand side of Eq.~(\ref{cav_eq_discr}) this ansatz 
doesn't give back in the left hand side a message of the same form.
However one can take its ``projection'' over the distributions of trajectories
described by (\ref{ansatz_eta}), by fixing the new fields $h_{i\to j}$ and $b_{i \to j}$
in such a way that the expectation values 
of $\hs^x_i$ and $\hs_i^z$ on the two sides of Eq.~(\ref{cav_eq_discr}) are the same.
Not surprisingly, it is easy to show that this procedure gives back\footnote{ 
Actually, there is a slight difference due to the fact that in the OQC formulation above we chose to
symmetrize the local Hamiltonian $\hH_{i,j}$. The PIQC leads naturally to a non-symmetric formulation
where $\hH_{i,j} = -J_{i,j} \hs^z_i \hs^z_j - \G_j \hs^z_j / (c-1)$. This difference should not be crucial, especially
for large $c$ where approximation~(\ref{ansatz_eta}) is better justified. 
}
the same equations
as the OQC for $\ell=1$, Eqs.~(\ref{eq:IM1}), (\ref{eq:IM2}), (\ref{eq:IM3}).
It was shown in~\cite{dimitrova2011} that this approximation gives 
quite good results when compared with the exact PIQC solution, and the quality of the approximation
increases with increasing connectivity $c$.

For $\ell >1$, the connection between OQC and PIQC is less obvious. We will not discuss it in detail, but roughly speaking
the idea is the following. 
The Markovian ansatz in Eq.~(\ref{ansatz_eta}) neglects all imaginary time correlations in the path integral description of spin $i$.
Therefore, a more refined ansatz would include, for instance, a Gaussian term 
$\int_0^\b \dd t \dd t' G_{i\to j}(t - t') \s_i(t) \s_i(t')$ in the exponent~\cite{LSS08}.
In presence of such a term, the PIQC equations cannot be cast in an operator formulation using only local operators.
The reason is that these imaginary time correlations are obtained by tracing out the neighboring spins. In the PIQC representation, 
this could be represented by considering a Markovian ansatz acting not only on $i$ but also on a neighboring shell of size $\ell$,
and then integrating out the neighbors to obtain an imaginary time correlated message on spin $i$. In the OQC language, this 
should correspond indeed to an operator message acting on spin $i$ and a set of neighbors. We conclude that messages with $\ell > 1$
in the OQC should roughly correspond to adding some imaginary time correlations in the PIQC. This is very reminiscent of what is done in 
dynamical mean field theory where imaginary time correlations are often represented by an Hamiltonian thermal bath of phonons~\cite{DMFT}.

\subsubsection{Discussion}

OQC~\cite{hastings2007,leifer2008,poulin2008,Bil-Poul2010,Poulin_Hasting2011,IM10,FIM10,dimitrova2011} 
(or Quantum Belief Propagation) 
is a very promising approach to the solution of spin glass models on locally tree-like graphs.
First of all, this method is not affected by the ``sign problem'' and therefore can be applied to Hamiltonians that do not admit
a path integral representation with positive weights 
(e.g. the QSAT problem~\cite{LaMoScSo10}).
Another important advantage is that for a given $\ell$ the cavity messages are finite matrices that can be parametrized by a finite set
of real numbers. The accuracy of this representation is only limited by machine precision, unlike in the case
of PIQC where sampling introduces systematic numerical errors and noise. For a given $\ell$, the limit $T=0$
can be taken easily by replacing everywhere the traces at finite temperature by a ground state average.

Its main drawback is that it is an approximate method:
its accuracy is expected to increase with the size of the block $\ell$. 
If one requires a given accuracy, then the block size must be increased when decreasing $T$ and 
$\ell\to\io$ for $T\to 0$ \cite{Bil-Poul2010} (however, there is some hope to combine OQC with local renormalization group methods
to avoid this problem~\cite{Bil-Poul2010}). At the same time, for a fixed block size, the limit $T\to 0$ exists,
is simpler to handle than the finite $T$ computation, and should provide qualitatively correct results~\cite{IM10,FIM10,dimitrova2011}.

The other important problem is that it requires the numerical diagonalization of matrices of large size (especially on a 
tree with large $c$, or if the local variables are not Ising spins, or if there are many-body interactions). This is for the moment
a strong limitation to its applicability
to interesting problems such as XORSAT or the coloring problem as we will see in Sec.~\ref{sec:results}.

To conclude this discussion, it is important to stress that OQC can lead to rigorous bounds on the true free energy of a given problem.
Indeed, in~\cite{Poulin_Hasting2011} it was shown,
based on the strong subadditivity property of
the von Neumann entropy, that slightly modified OQC
equations can lead to a lower bound of the free energy.
Similar results were first derived in the context of 
classical systems~\cite{globerson}.
For the moment, this variational technique has been only applied to low dimensional systems~\cite{Poulin_Hasting2011}. Its generalization
to random graphs with tree-like geometries seems a challenging problem.

\subsection{Variational quantum cavity methods}
\label{sec:VQC}

A promising approach to overcome the sign problem and to investigate directly the $T=0$ limit consists in
using the cavity method to optimize variational wavefunctions.
We will refer to this approach as Variational Cavity Method (VQC).
These methods are based on the well known fact that, given a
system described by a Hamiltonian $\hH$, the expectation value of the energy 
over an arbitrarily chosen wavefunction $|\psi_v \rangle$
provides an upper bound for the true ground state energy.
We present in this section two different attempts in this direction.
For simplicity, as in Sec.~\ref{sec:OQC} 
we will focus on the transverse field Ising Hamiltonian, 
$\hH = \sum_{\la i,j \ra} \hH_{i,j}$ with
$\hH_{i,j} = -J_{ij} \hs_i^z \hs_j^z - ( \G_i \hs_i^x + \G_j \hs_j^x)/c$
on unidimensional chains ($c=2$) and $3$-regular graphs ($c=3$).

\subsubsection{Optimization of Jastrow wavefunctions}

In \cite{ramezanpour2012} a very simple trial wavefunction was considered:
\beq
\la \us | \psi_v \ra = \frac{1}{\sqrt{Z}} e^{ \frac12 \sum_i b_i \s_i + \frac12 \sum_{\la i,j \ra} K_{ij} \s_i \s_j }  \ ,
\eeq
where the constant $Z$ is determined by normalization and $b_i$, $K_{ij}$ are real numbers
(it is shown in~\cite{ramezanpour2012} that adding an imaginary part only increases the energy).
Then, the square of the wavefunction is the Gibbs probability measure of a classical ``auxiliary'' system,
that turns out to be a classical Ising model with couplings $K_{ij}$ and random fields $b_i$ at unit
temperature. 
Such a variational wavefunction (often called Jastrow wavefunction) has been widely used
in the study of quantum systems and for many problems it works as a very 
good approximation~\cite{Jastrow55}.

Once that the wavefunction is chosen one needs to 
express the expectation value of $\hH$
as a function of the variational parameters. For general graphs this step can not be carried on
in an exact way, but on locally tree-like graphs it can be done with the use of the cavity method 
on the auxiliary system.
For a given instance of the problem, this leads to the following expression of the variational energy:
\beq\label{eq:E_jastrow}
\begin{split}
E_v = \la \psi_v | \hH | \psi_v \ra & = 
- \sum_{\la i, j \ra} J_{ij} \frac{\underset{\s_i, \s_j}{\sum} \s_i \s_j e^{K_{ij} \s_i \s_j} \h_{i\to j}(\s_i) \h_{j\to i}(\s_j)}
{\underset{\s_i, \s_j}{\sum}  e^{K_{ij} \s_i \s_j} \h_{i\to j}(\s_i) \h_{j\to i}(\s_j)} \\
& - \sum_i \G_i 
\frac2
{ \underset{\s_i, \{\s_j\}_{j \in \partial i}}{\sum} e^{b_i \s_i} \underset{j \in \partial i}{\prod} e^{K_{ij} \s_i \s_j} \h_{j\to i}(\s_j) }
\end{split}\ ,
\eeq
where the cavity messages satisfy the usual equations:
\beq\label{eq:VQC_ram}
\h_{i\to j}(\s_i) = \frac1{z_{i\to j}}\, e^{b_i \s_i}
\prod_{k \in \partial i \setminus j} \sum_{\s_k} e^{K_{ik} \s_i \s_k} \h_{k\to i}(\s_k) \ .
\eeq
Unfortunately, $E_v$ is a complex non-local function of the variational parameters $K_{ij}$ and $h_i$,
because the cavity messages depend implicitly on far away couplings through the cavity equations.

One way to minimize $E_v$ is to introduce the following
partition function:
\beq\label{eq:Z_jastrow}
\ZZ(\tb) = \int D\{K_{ij}\} D\{b_i\} D\{\eta_{i \to j}\} \, e^{-\tb E_v(\{K_{ij}\}, \{ b_i \}, \{\h_{i\to j}\})} \, 
I_{\rm cavity \ eq} \ ,
\eeq
where $I_{\rm cavity \ eq}$ is a set of delta functions that impose the cavity equations (\ref{eq:VQC_ram})
and $\tb$ is a fictitious inverse temperature.
In this way, considering the cavity messages as independent variables, the computation of $\ZZ$
(more precisely, of the average of $N^{-1} \log\ZZ$ over the disorder, for $N\to\io$)
can be done through a message-passing algorithm whose equations resemble
those of the 1RSB classical computation. From $\ZZ(\tb)$ it is easy to extract the minimum of the 
variational energy by sending the fictitious inverse temperature
$\tb\to\io$. In~\cite{ramezanpour2012} this procedure has been carried out explicitly
and reasonable results for the ground state energy of the Ising model in transverse field have been obtained; this method was then applied to a model of interacting fermions on the Bethe lattice in~\cite{ramezanpour2012fermions}, bypassing the usual sign problem.

\subsubsection{Matrix product states}
\label{sec:MPS}

Within the variational approach,
matrix product states (MPS) represent a very promising way to study the 
properties of quantum systems defined on tree-like structures.
A MPS is a representation of a state of a quantum lattice model 
that is based on a set of tensors defined on the sites of the original model.
As we will discuss below,
this representation is exact on tree-like structures as long as the size
of the tensors is large enough. A truncation of the tensors size usually gives
a very good approximation of the state provided the entanglement is not too large.

For one-dimensional systems, MPS are at the basis of 
many numerical algorithms, most notably the Density
Matrix Renormalization Group (DMRG)~\cite{white1992} or the 
Time Evolving Block Decimation (TEBD)~\cite{vidal2003}.
These methods are widely applied in the study of one-dimensional systems
where they are known to work efficiently while the understanding of their
generalization to higher dimensions is still the subject of intense research. 

These methods have been generalized to tree geometries in different 
works~\cite{NFGSS08,Shi2006,Nagy12,Li12}. 
The method proposed by~\cite{NFGSS08} is a generalization 
of the algorithm developed by Vidal in~\cite{Vidal07}
in order to study translational invariant systems
in the thermodynamic limit.
For finite trees, related algorithms have been proposed in
\cite{Shi2006,Nagy12}. A DMRG algorithm was used in~\cite{Lepetit}
to derive the ground state properties of the Hubbard model on the Bethe lattice. 

All these algorithms exploit the MPS representation of the ground state, combined
with parallel updates that descend from such representation and the local
properties of the Hamiltonian under study.
Once the expression of the state in terms of MPS is given one can apply unitary transformations
involving local operators with update rules that are local~\cite{Shi2006,Vidal07,NFGSS08}.
This is a crucial property on which the efficiency of the algorithm relies. 
An important point is that during such updates an exact calculation generally brings to
increasing tensor sizes. However the size of the tensors can be kept fixed exploiting
a {\it block decimation} technique that aims to project on a restricted subspace that 
carry most of the information.

Beyond unitary operations the same techniques are used to perform the imaginary time evolution,
which is exploited in the search for the ground state. This operation introduces new errors
that originate from the normalization of the tensors that is spoiled by the non-unitarity of the evolution.
Different methods have been proposed~\cite{NFGSS08,Shi2006,Vidal07} to account for this effect.

We refer the reader to the references mentioned above for a more detailed discussions of these algorithms.
At the end of this section, we will take a slightly different perspective by showing how MPS can in principle
be used within a variational cavity approach.

\paragraph{MPS for chains}

For a chain of $N$ spins with open boundary conditions (a finite tree with connectivity 2) 
a MPS is a state of the form:
\beq\label{MPS}
|\psi\rangle = \sum_{\sigma_1,\dots,\sigma_N} c_{\sigma_1 \dots \sigma_N} |\sigma_1\rangle\dots |\sigma_N\rangle \ ,
\eeq
with
\bea
c_{\sigma_1\dots\sigma_N}&=&\sum_{\alpha_1=1}^{\chi_1} \dots
\sum_{\alpha_N=1}^{\chi_N}
\gamma^{[1]\sigma_1}_{\alpha_1}\lambda_{\alpha_1}^{[1]}\gamma^{[2]\sigma_2}_{\alpha_1\alpha_2}\lambda_{\alpha_2}^{[2]}\gamma^{[3]\sigma_3}_{\alpha_2\alpha_3}\dots\gamma^{[N]\sigma_N}_{\alpha_{N-1}} \nonumber \\ &=& \prod_{i=1}^{N} \Big( \sum_{\alpha_i}^{\chi_i}\gamma^{[i]\sigma_i}_{\alpha_{i-1}\alpha_i} \lambda_{\alpha_i}^{[i]} \Big) \ ,
\nonumber
\eea
where $\gamma^{[i]}$ are $N$ matrices defined on the sites of the chain,
$\lambda^{[i]}$ are $N-1$ vectors defined on the links of the chain,
and in the last equality we set
$\lambda_{\alpha_N}^{[N]}=1$, 
$\gamma^{[1]\sigma_1}_{\alpha_{0}\alpha_1} = \delta_{\alpha_{0},\alpha_1}\gamma^{[1]\sigma_1}_{\alpha_{1}} $
and
$\gamma^{[N]\sigma_N}_{\alpha_{N-1}\alpha_N} = \delta_{\alpha_{N-1},\alpha_N}\gamma^{[N]\sigma_N}_{\alpha_{N-1}} $.
The vectors $\lambda^{[i]}$ are the Schmidt coefficients that appear in the Schmidt decomposition
of the system when it is divided into two disjoint parts $1,\dots,i$ and $i+1,\dots,N$
(by ``cutting'' the link between $i$ and $i+1$).

Given a bipartition of the system into two disjoint subparts $A(i) = \{1,\dots,i\}$ and $B(i+1)=\{i+1,\dots,N\}$,
 the Schmidt theorem states that for every vector $|v\rangle$ it is possible to find an orthonormal
basis for the Hilbert space defined over $A$ (resp. over $B$), such that:
\beq
|v\rangle = \sum_{\alpha=1}^{\chi_{i}} \lambda_{\alpha}^{[i]} |v_{\alpha}\rangle_{A(i)} | v_{\alpha}\rangle_{B(i+1)} \ ,
\eeq
with $\chi_i = 2^{\min[ i, N-i]}$ and $\lambda_\a^{[i]} \geq 0$ are positive real numbers such that
$\sum_{\a=1}^{\chi_i} (\lambda_\a^{[i]})^2=1$.

Performing the same decomposition between the sites $A(i-1) = \{1,\dots,i-1\}$ and $B(i)=\{i,\dots,N\}$ one obtains
\beq
|v\rangle = \sum_{\beta=1}^{\chi_{i-1}} \lambda_{\beta}^{[i-1]} |v_{\beta}\rangle_{A(i-1)} | v_{\beta}\rangle_{B(i)} \ .
\eeq
Using the basis defined for the subspace $B(i+1)$ one can write:
\beq
 | v_{\alpha}\rangle_{B(i)} = \sum_{\sigma_i} \sum_{\beta=1}^{\chi_i} 
 \gamma^{[i]\sigma_i}_{\alpha,\beta} \lambda_{\beta}^{[i]} |\sigma_i\rangle | v_{\beta}\rangle_{B(i+1)} \ ,
\eeq
which defines the matrix $\gamma^{[i]\sigma_i}_{\alpha,\beta}$ used above.
In the same way one obtains
\beq
 | v_{\alpha}\rangle_{A(i)} = \sum_{\sigma_i} \sum_{\beta=1}^{\chi_{i-1}} 
 \lambda_{\alpha}^{[i-1]}  \gamma^{[i]\sigma_i}_{\alpha,\beta} |\sigma_i\rangle | v_{\beta}\rangle_{A(i-1)} \ .
\eeq
The orthonormality of the basis $|v_{\alpha}\rangle_{A(i)}$ and $|v_{\alpha}\rangle_{B(i)}$ imposes
normalization conditions on the $\lambda$'s and $\gamma$'s.
They must indeed satisfy, $\forall i= 1,\dots,N$:
\beq\label{eq:MPS_norm}
\begin{split}
\sum_{\a=1}^{\chi_i}(\lambda_\a^{[i]})^2 & =1 \ ,\\
{}_{B(i)}\langle v_{\alpha'} | v_{\alpha} \rangle_{B(i)} & = 
\sum_{\sigma_i} \sum_{\beta=1}^{\chi_i} 
 \gamma^{[i]\sigma_i}_{\alpha,\beta} \lambda_{\beta}^{[i]} 
 ( \gamma^{[i]\sigma_i}_{\alpha',\beta} )^{\ast} \lambda_{\beta}^{[i]} 
=\delta_{\alpha,\alpha'} \ ,\\ 
{}_{A(i)}\langle v_{\alpha'} | v_{\alpha} \rangle_{A(i)} & = 
\sum_{\sigma_i} \sum_{\beta=1}^{\chi_{i-1}} 
 \gamma^{[i]\sigma_i}_{\beta,\alpha} \lambda_{\beta}^{[i-1]} 
 ( \gamma^{[i]\sigma_i}_{\beta,\alpha'} )^{\ast} \lambda_{\beta}^{[i-1]} 
=\delta_{\alpha,\alpha'}  \ .
\end{split}
\eeq
The average values of local observables can be easily computed.
Let us consider a one-body operator
\beq
O^{[i]} = \sum_{\sigma_i,\sigma'_i} O_{\sigma'_i,\sigma_i}^{[i]} |\sigma'_i\rangle\langle \sigma_i| \ .
\eeq
Then
\beq
\begin{array}{c}
\langle\psi|O^{[i]}|\psi\rangle 
 = \sum_{\sigma_i,\sigma'_i} O_{\sigma'_i,\sigma_i}^{[i]} \sum_{\beta=1}^{\chi_i} 
\sum_{\alpha=1}^{\chi_{i-1}} \lambda_{\beta}^{[i-1]} ( \gamma^{[i]\sigma'_i}_{\beta,\alpha} )^{\ast}
 \lambda_{\alpha}^{[i]} \lambda_{\beta}^{[i-1]} ( \gamma^{[i]\sigma_i}_{\beta,\alpha} )
 \lambda_{\alpha}^{[i]}
 \end{array} \ ,
\nonumber
\eeq
while for a product of two one-body operators
\beq
O^{[i]}O^{[j]} = \sum_{\sigma_i,\sigma'_i,\sigma_j,\sigma'_j} O_{\sigma'_i,\sigma_i}^{[i]}
O_{\sigma'_j,\sigma_j}^{[j]} |\sigma'_i \sigma'_j\rangle\langle \sigma_j\sigma_i|
\eeq
the expectation value is:
\beq
\begin{split}
\langle\psi|O^{[i]}O^{[j]}|\psi\rangle
 &=  \sum_{\sigma_i,\sigma'_i,\sigma_j,\sigma'_j} O_{\sigma'_i,\sigma_i}^{[i]}
O_{\sigma'_j,\sigma_j}^{[j]}  \sum_{\alpha=1}^{\chi_{i-1}} 
 \sum_{\beta_i,\beta_i'=1}^{\chi_{i}}\dots   \sum_{\beta_{j-1},\beta_{j-1}'=1}^{\chi_{j-1}}\sum_{\delta=1}^{\chi_j} (\lambda_{\alpha}^{[i-1]})^2 \times \\ &\times
 ( \gamma^{[i]\sigma'_i}_{\alpha,\beta_i'} )^{\ast} \gamma^{[i]\sigma_i}_{\alpha,\beta_i} \lambda_{\beta_i}^{[i]} \lambda_{\beta_i'}^{[i]}\dots
 ( \gamma^{[j]\sigma'_j}_{\beta_{j-1}',\delta} )^{\ast} \gamma^{[j]\sigma_j}_{\beta_{j-1},\delta} (\lambda_{\delta}^{[j]})^2 \ .
   \end{split} 
\eeq

The expression~(\ref{MPS}) is exact if the dimensions
$\chi_i$ of the $\lambda$'s and $\Gamma$'s are
large enough, however
the same definition can be used in a variational way, for fixed (small) sizes, and still provides 
a good representation of the ground state. The entanglement is usually used as a measure
of the accuracy of such representation. The Schmidt coefficients in fact are directly related to
the so-called entanglement entropy through the formula:
\beq
S_A = - \Tr[\hat{\rho}_A \log \hat{\rho}_A] 
= - \sum_{\a=1}^{\chi_{i}} (\lambda_\a^{[i]})^2 \log [(\lambda_\a^{[i]})^2 ]
= S_B \ , \hspace{0.5cm} \hat{\rho}_A \propto \text{Tr}_B \, e^{- \beta \hat{H}}
\ .
\eeq
In the limit in which $\chi_i =1$, i.e. if the two parts of system are separable 
$|v\rangle  = |v\rangle_{A(i)} |v\rangle_{B(i+1)} $, 
then $S_A=0$.

\paragraph{MPS for trees}
Trees have no loops, thus, removing an edge divides the system into disjoint 
parts. The Schmidt decomposition can be applied and it allows to naturally define
MPS also in this context. We refer the reader to~\cite{NFGSS08,Shi2006,Nagy12}
for more details and state in the following the expressions derived in these
works.

In the case of trees the expression~(\ref{MPS}) is generalized 
using a vector $\lambda^{\la ij \ra}_{\alpha}$ for each edge and tensors $\gamma^{[i] \s_i}_{\alpha_1,\dots,\alpha_c}$
with $c$ lower indices (where $c$ is the connectivity of the graph) plus one spin index as in the one-dimensional case.
The normalization conditions generalize for all the $c$ indices of the tensors.
In order to derive the vector $\lambda^{\la ij \ra}_{\alpha}$ one has to perform the Schmidt decomposition
on the corresponding edge $\langle i j\rangle$. This divides the system into two 
disjoint subtrees $\TT_{i \to j}$ and $\TT_{j \to i}$, where each subset
 contains the connected component made of sites connected to  $i$ and $j$ respectively:
 \beq
|v\rangle = \sum_{\alpha=1}^{\chi_{\langle i j \rangle}} \lambda_{\alpha}^{\langle i j\rangle} 
|v_{\alpha}\rangle_{\TT_{i\to j}} | v_{\alpha}\rangle_{\TT_{j \to i}} \ .
 \eeq
The tensors $\gamma^{[i] \s_i}_{\alpha_1,\dots,\alpha_c}$ are obtained
performing the Schmidt decomposition for the $c$ bonds surrounding the site $i$
and then expressing one of the orthonormal basis that derive in terms of basis obtained with the other $c-1$
decomposition and the spin $|\sigma_i\rangle$, similarly to the one-dimensional case:
\beq
|v_{\alpha_{\langle i j \rangle}}\rangle_{\TT_{i\to j}} = \sum_{\sigma_i} \sum_{ \{ \alpha_{\langle i k \rangle} \} : k \in \partial i\backslash j} 
 \gamma^{[i] \sigma_i}_{\{ \alpha_{\langle i k \rangle} \}_{k \in \partial i}} \prod_{k \in \partial i\backslash j}\Big[ \lambda_{\alpha_{\langle  i k \rangle}}^{\langle  i k \rangle}
 | v_{\alpha_{\langle  i k \rangle}}\rangle_{\TT_{k \to i}} \Big]  |\sigma_i\rangle  \ ,
\eeq
where each $\alpha_{\langle i k \rangle}$ is summed from $1$ to $\chi_{\langle i k \rangle}$,
with normalization conditions completely analogous to Eq.~(\ref{eq:MPS_norm}).
In this way one arrives at the following form for a general vector, in terms of matrix product states:
\beq
\begin{array}{c}
\displaystyle |\psi\rangle =  \sum_{\{\sigma_i\}} 
 \sum_{ \{ \alpha_{\langle  i k \rangle} \} } 
 \Big(  \prod_{i=1}^N 
  \gamma^{[i] \sigma_i}_{\{\alpha_{\langle  i k \rangle}\}_{k\in\partial i}}  \Big)
   \Big( 
   \prod_{\langle i k\rangle} \lambda_{\alpha_{\langle i k \rangle}}^{\langle i k \rangle}
   \Big) |\sigma_1\rangle\dots |\sigma_N\rangle \ .
      \end{array}
\eeq
Expectation values of one-body operators are given by
\beq
\langle\psi|O^{[i]}|\psi\rangle 
 = \sum_{\sigma_i,\sigma'_i} O_{\sigma'_i,\sigma_i}^{[i]} 
   \sum_{ \{ \alpha_{\langle i k \rangle} \} : k \in \partial i} 
 \gamma^{[i] \sigma_i}_{\{ \alpha_{\langle i k \rangle} \}_{k \in \partial i}}
 [ \gamma^{[i] \sigma'_i}_{\{ \alpha_{\langle i k \rangle} \}_{k \in \partial i}}]^{\ast}
  \prod_{k \in \partial i}[ \lambda_{\alpha_{\langle  i k \rangle}}^{\langle  i k \rangle}]^2
  \ .
\nonumber
\eeq
In order to compute the expectation value of a two-body operator 
$O^{[i]} O^{[j]}$ acting on two sites $i$ and $j$
we denote with ${\cal S}$ the set of sites on the unique path joining
$i$ and $j$, with ${\cal P}$ the edges that are adjacent to at least
one vertex in ${\cal S}\cup \{i,j\}$, and with ${\cal R}$ the set of the edges 
that are adjacent to exactly one vertex in ${\cal S}\cup \{i,j\}$.
More explicitly, ${\cal P} \setminus {\cal R} $ are the edges of the path
between $i$ and $j$. Then one has 
\beq
\begin{split}
&\langle\psi|O^{[i]} O^{[j]}|\psi\rangle 
 = \sum_{\sigma_i,\sigma'_i,\sigma_j,\sigma'_j} 
\sum_{\{\s_l \}_{l \in {\cal S}} }
O_{\sigma'_i,\sigma_i}^{[i]} O_{\sigma'_j,\sigma_j}^{[j]} \\
&\times
   \sum_{ \{ \alpha_{\langle l k \rangle} \}, \{ \alpha'_{\langle l k \rangle} \} : \langle l k \rangle \in {\cal P}} 
 \gamma^{[i] \sigma_i}_{\{ \alpha_{\langle i k \rangle} \}_{k \in \partial i }}
 [ \gamma^{[i] \sigma'_i}_{ \{ \alpha'_{\langle i k \rangle} \}_{k \in \partial i}}]^{\ast}  
 \gamma^{[j] \sigma_j}_{\{ \alpha_{\langle j k \rangle} \}_{k \in \partial j}}
 [ \gamma^{[j] \sigma'_j}_{\{ \alpha'_{\langle j k \rangle} \}_{k \in \partial j}}]^{\ast} \\ & \times
   \prod_{l \in {\cal S}}
   \gamma^{[l] \sigma_l}_{\{ \alpha_{\langle l k \rangle} \}_{k \in \partial l}}
 [ \gamma^{[l] \sigma_l}_{\{ \alpha'_{\langle l k \rangle} \}_{k \in \partial l}}]^{\ast}
  \prod_{ \langle k l \rangle \in {\cal P}} \lambda_{\alpha_{\langle  k l \rangle}}^{\langle  l k \rangle}
  \lambda_{\alpha'_{\langle  k l \rangle}}^{\langle  l k \rangle}   
   \prod_{  \langle l k \rangle  \in {\cal R}} \delta_{\alpha_{\langle l k \rangle},\alpha'_{\langle l k \rangle}} \ .
\end{split}
\nonumber
\eeq

\paragraph{The disordered Ising model in transverse field}
To show why MPS are particularly useful, we can consider again the simplest example of an Ising model 
on a regular graph of connectivity $c$, with Hamiltonian
$\hH = -\sum_{\la ij \ra} J_{ij} \hs_i^z \hs_j^z - \sum_i \G_i \hs^x_i$, and a variational MPS $|\psi_v\rangle$
with tensors of fixed size $\chi$.
Then the variational energy is
\beq\label{eq:MPS_Ising}
\begin{split}
E_v & = \langle\psi_v | \hH | \psi_v\rangle = 
-\sum_i \G_i  \sum_{\sigma, \{ \alpha_k \}_{k \in \partial i}} 
 \gamma^{[i] \sigma}_{\{\a_k\}}
 [ \gamma^{[i] -\sigma}_{\{\a_k\}}]^{\ast}
  \prod_{k\in\partial i} (\lambda^{\la ik \ra}_{\alpha_k})^2 \\
& -\sum_{\la ij \ra} J_{ij} \sum_{\s,\s'} \s\s' 
\sum_{\{ \alpha_k \}_{k \in \partial i\setminus j},\d,\d', \{ \b_l \}_{l \in \partial j\setminus i}   }
\left( \prod_{k\in\partial i \setminus j} (\lambda^{\la ik \ra}_{\alpha_k})^2 \right)
\g^{[i] \s}_{\{ \alpha_k \}\d} (\g^{[i] \s}_{\{ \alpha_k \}\d'})^* \\
& \times \l^{\la ij \ra}_\d \l^{\la ij \ra}_{\d'}
\g^{[j]\s'}_{\d\{ \b_k \}} (\g^{[j]\s'}_{\d\{ \b_k \}})^*
\left( \prod_{l\in\partial j \setminus i} (\lambda^{\la jl \ra}_{\b_k})^2 \right) \ ,
\end{split}
\eeq
with normalization conditions (on each directed link):
\beq\label{eq:MPS_norm_Ising}
\begin{split}
& \sum_\a (\l^{\la ij \ra}_\a)^2 = 1 \ , \\
& \sum_\s \sum_{\{ \alpha_k \}_{k \in \partial i\setminus j}} 
\left( \prod_{k\in\partial i \setminus j} (\lambda^{\la ik \ra}_{\alpha_k})^2 \right)
\g^{[i] \s}_{\{ \alpha_k \}\b} (\g^{[i] \s}_{\{ \alpha_k \}\b'})^*
 = \d_{\b \b'} \ .
\end{split}\eeq

If we consider first a model without disorder, i.e. with $J_{ij}=J$
on all edges of an infinite tree, and $\G_i=\G$ on all vertices, then
all sites are equivalent and we can assume that the tensors 
and vectors do not depend on the site and link indices.
The variational energy is then a function of a finite set of variational parameters
and one can devise several strategies to minimize it, either based on numerical 
minimization routines, or on simulated annealing. Alternatively, one can use
the strategy of~\cite{NFGSS08,Shi2006,Nagy12} by applying the imaginary time evolution
to the variational state. This procedure gives the variational energy in the thermodynamic limit.

\paragraph{Cavity optimization of MPS}

In the case of a disordered model with randomness in the couplings and/or
local transverse fields one should keep all vectors $\l$ and tensors $\g$
as variational parameters in the expression~(\ref{eq:MPS_Ising}).
As a direct comparison of Eq.~(\ref{eq:MPS_Ising}) and Eq.~(\ref{eq:E_jastrow})
shows very explicitly, there is a crucial advantage of
MPS with respect to e.g. Jastrow wavefunctions. For MPS, the variational energy
can be written as an explicit function of the variational parameters, 
which is a sum of local terms involving the tensors 
on a site $i$ and its neighbors and the vectors on the links among these sites.
On the contrary, for the
Jastrow wavefunction a cavity computation is needed to write $E_v$, which is therefore
a very implicit expression of the variational parameters.

For a generic MPS describing a disordered system,
the variational energy can then be interpreted as a ``classical Hamiltonian''
for a system whose classical variables are the tensors and vectors of the MPS.
The partition function of such a model would read schematically as
\beq\label{eq:Z_MPS}
\ZZ(\tb) = \int d\{\g^{[i]}\}d\{\l^{\la ij \ra}\}
e^{-\tb E_v[\{\g^{[i]}\},\{\l^{\la ij \ra}\}]} \, I_{\rm normalization}[\{\g^{[i]}\},\{\l^{\la ij \ra}\}]
\eeq
with $E_v$ given in Eq.~(\ref{eq:MPS_Ising}), $I_{\rm normalization}$ a set of delta functions
enforcing the normalizations in Eq.~(\ref{eq:MPS_norm_Ising}), and
$\tb$ a fictitious inverse temperature to be sent to $\infty$ at the end.

The latter is much simpler than Eq.~(\ref{eq:Z_jastrow}) where a functional delta over the cavity
messages appears; this is not required here because the variational energy for MPS is an explicit function
of the tensors. The partition function (\ref{eq:Z_MPS}) can then 
in principle be computed via the standard classical cavity method\footnote{
The presence of interactions involving a variable and all of its neighbors requires 
using a ``trick'' consisting in creating a copy of each variable on its neighboring sites:
see~\cite{FSZ11} for a more detailed discussion of this point.
}. The price to pay is of course that the basic classical variables are tensors of size $\chi$, 
and the cavity messages are therefore distributions over the space of such matrices.
The limit $\tb\to \infty$ can be performed explicitly in this case following~\cite{cavity_T0},
leading to the optimized variational energy.
Although this strategy was not turned into a concrete calculation for the moment, we believe
that it is a very promising way to compute the zero temperature properties of quantum random
optimization problems, and therefore complementary to the finite temperature PIQC.

\subsection{Exact diagonalization and numerical integration of the Schr\"odinger equation}
\label{sec:ed}

We now present briefly exact numerical techniques to study the statics and the dynamics of finite quantum systems.
If the size of the Hilbert space is small enough,
thermodynamic and spectral properties of $\hH$ can be obtained from exact diagonalization techniques. 
Such techniques are a whole field of research in themselves so we just give a brief overview 
here. If one is interested in finding all the eigenvectors and eigenvalues of a given matrix, 
the most commonly used techniques are the Jacobi and the Gauss-Seidel methods. 
If one is interested only in the low energy part of the spectrum, it is possible to use Lanczos 
type methods to obtain more quickly the lowest lying eigenvectors. Moreover, these methods 
require only to be able to compute the multiplication of a vector of the Hilbert space by $\hH$. 
In the case of sparse matrices as the ones relevant for optimization problems, 
this can be done without keeping the whole matrix $\hH$ in memory; 
therefore the limitation of this method comes from the size $d^N$ of a vector of the Hilbert space 
(where $d$ is the dimension of a single qudit), 
rather than from the size $d^{2N}$ of the Hamiltonian operator. 
For quantum $1/2$ spins where $d=2$, this allows for computations up to $N=25$ on standard computers. 
The results reported in Sec.~\ref{sec:results} were obtained by using the Arpack package~\cite{arpack}.
Note that in some cases, the presence of exact symmetries 
(operators that commute with $\hH$) allows to reduce the size of the Hilbert space
and thus increase the sizes of the systems that can be exactly diagonalized.

For the analysis of the quantum adiabatic algorithm it is particularly interesting 
to simulate exactly the real time evolution of a quantum system following
the time-dependent Schr\"odinger equation defined in Eq.~(\ref{Hquantum}):
\beq
\label{eq_schrodinger_ch5}
\frac{i}{\TT} \frac{\dd}{\dd s} |\psi(s) \ra = \hH(s) |\psi(s) \ra \ .
\eeq
This is a linear differential equation, which can thus be solved using standard numerical integration techniques, such as Runge-Kutta or Adams-Bashforth methods. However, it is better in practice to make use of the Hermitian nature of the generator of the dynamics~\cite{hatano05}. In fact, (\ref{eq_schrodinger_ch5}) can be rewritten as: 
\beq
|\psi(s) \ra = \mathbf{T} \left( e^{-i \TT \int_{0}^s \hH(s') ds'} \right)  |\psi(0) \ra \equiv \mathcal{U}(0,s) | \psi(0) \ra  
\eeq
where $\mathbf{T}$ denotes the time-ordering operator. Because $\hH(s)$ is Hermitian, $\mathcal{U}(0,s)$ is unitary; one is then interested in finding unitary approximations to $\mathcal{U}(0,1)$. The simplest way to do it is to write:
\beq 
\label{eq_apB_break1} 
\mathcal{U}(0,1) =  \mathbf{T} \left(e^{-i \TT \int_0^1 \hH(s) \dd s}\right) 
= \prod_{i=1}^n \mathbf{T} 
\left(e^{-i \TT \int_{s_i}^{s_{i+1}} \hH(s) \dd s}\right) 
= \prod_{i=1}^n \mathcal{U}(s_i,s_{i+1})  \ .
\eeq
We are interested in the particular case of a linear dependency of $\hH(s)$ on
$s$: $\hH(s) = (1-s)\hH_{\rm i} + s \hH_{\rm f}$. The approximation
\beq 
\label{eq_apB_break2} 
\begin{split} \mathcal{U}(s,s+\Delta s) &= 
\mathbf{T} \left(e^{-i \TT \int_{s}^{s+\Delta s} \hH(s') \dd s'}\right) \\
&\sim  \left(e^{-i \TT \int_{s}^{s+\Delta s} s' \hH_{\rm f} \dd s'}\right)  
\left(e^{-i \TT \int_{s}^{s+\Delta s} (1-s') \hH_{\rm i} \dd s'}\right) \\ &=
\left(e^{-i \TT \frac{2 s \Delta s + \Delta s^2}{2} \hH_{\rm f}}\right)  
\left(e^{-i \TT \frac{2 (1-s) \Delta s - \Delta s^2}{2} \hH_{\rm i}}\right) \\ & 
\equiv \widetilde{U}_{\Delta s}(s) \end{split} 
\eeq
gives rise to an error in operator norm 
$\| A \| \equiv \sup_{X, \| X = 1\|} \| AX \|$ bounded by~\cite{huyg90, poulin11}: 
\begin{equation} 
\| \mathcal{U}(s,s+\Delta s) - \widetilde{\mathcal{U}}_{\Delta s}(s) \| \leq 
\| [\hH_{\rm i},\hH_{\rm f}] \| \frac{\TT (\Delta s)^2}{2} + O(\Delta s^3) 
= O(N\TT\Delta s^2) \ . 
\end{equation}
Indeed in all the cases of interest here the commutator of the initial and
final Hamiltonian has a norm of order $N$.
We define the approximate evolution operator 
$\widetilde{U}(0,s_i) \equiv \prod_{j=0}^{i-1} 
\widetilde{U}_{\Delta s}(s_j)$. The triangle inequality
\begin{equation}\begin{split} 
&\| \mathcal{U}(0,s_{i+1}) - \widetilde{U}(0,s_{i+1}) \| = \\
 &= 
\| \mathcal{U}(0,s_i) (\mathcal{U}(s_i,s_{i+1}) - \widetilde{U}_{\Delta s}(s_i)) 
+(\mathcal{U}(0,s_i)-\widetilde{U}(0,s_i) ) \widetilde{U}_{\Delta s}(s_i)
\| \\ &\leq 
\|\mathcal{U}(s_i,s_i+\Delta s) - \widetilde{U}_{\Delta s}(s_i) \| 
+\|\mathcal{U}(0,s_i)-\widetilde{U}(0,s_i)  \| 
\end{split}
\end{equation}
leads by recurrence to
\begin{equation} 
\| \mathcal{U}(0,1) - \widetilde{\mathcal{U}}(0,1) \| \leq O(nN\TT\Delta s^2)=
O(N\TT/n) \ .
\end{equation}
One can thus replace the exact evolution operator $\mathcal{U}(0,1)$ by
its approximation $\widetilde{U}(0,1)$ with a precision of order $\epsilon$
in the evaluation of intensive observables if the number of discretization 
steps $n$ is of order $N \TT/\epsilon$. Note that this bound does not involve the spectral gap of the system, which is thus not directly the bottleneck for the simulation of the quantum evolution.

Let us evaluate the total complexity of the procedure. Doing the computation in $\hH_{\rm f}$ eigenbasis, the multiplication by an operator $e^{\alpha \hH_{\rm f}}$ is trivial and can be realized in a time proportional to $d^N$, where $d$ is the dimension of the Hilbert space of a single qudit. On the other hand, taking for $\hH_{\rm i}$ a sum of identical operators acting on single qudits, $\hH_{\rm i} = \sum_{j=1}^N \hat{h}_j$, we can write $ e^{\alpha \hH_{\rm i}} = \prod_{j=1}^N e^{\alpha \hat{h}_j}$; the exponential one has to compute is the same for any site and its action on a vector of the Hilbert space can be computed with less than $N d^2$ operations. Therefore we finally see that the action of $\widetilde{U}_{\Delta s}(s)$ on a vector can be computed within $O(d^{N})$ operations; leading finally to a number of operations bounded by $N \TT d^{N} / \epsilon$  to obtain a precision $\epsilon$ on the final result of the evolution. Practically, the resources limitations of this method come both from the size $d^N$ of the vector one has to keep in memory, and from the large time $\TT$ one is interested in for quantum adiabatic computations.

As a side-remark let us mention that the matrix product states approximate
parametrization of quantum vectors, discussed in Sec.~\ref{sec:MPS}, 
can also be used to study the
real-time (Schr\"odinger) dynamics of quantum systems, 
see~\cite{tdDMRG1,tdDMRG2,tdDMRG3} for details.

\subsection{Quantum Monte Carlo}
\label{sec:qmc}

In this section we discuss how Quantum Monte Carlo (QMC) simulation algorithms can 
be used to extract relevant information on random optimization problems,
in particular their energy gap.

Path Integral Quantum Monte Carlo (PIMC) 
simulations have a very long history and were initially
performed to study continuum systems with particular focus on Helium 4~\cite{Ce95}.
Rapidly, several implementations were developed to study lattice systems, made
of spin or bosonic degrees of freedom.
Early implementations were based on a Suzuki-Trotter
path integral in discrete imaginary time~\cite{KTC91,RY94,BS96},
but rapidly it was realized that the continuum imaginary time limit could be
taken explicitly~\cite{BW96,PST98,RK99,KRSZ08,FGGM11}.
A similar approach is based on an exact sampling of the perturbative expansion
and goes under the name of Stochastic Series Expansion (SSE)~\cite{Sa99,HY11}.

PIMC is a Monte Carlo method that produces configurations of imaginary time
trajectories sampled from the measure $\mu_{\rm Q}$ we introduced in 
Eq.~(\ref{eq_mu_quantum}). In the case of spins 1/2 in a transverse field,
i.e. for $\hH=\sum_{\us} E(\us) | \us \ra \la \us| - \G \sum_i \hsix$,
it reads more explicitly:
\beq
\mu_{\rm Q}(\ubs) = \frac{1}{Z} \delta_{\ubs(0),\ubs(\beta)} 
\prod_{i=1}^N \G^{|\bs_i|} \, \, e^{-\int_0^\beta E(\us(t)) \dd t }\ ,
\eeq
with
\beq
Z = \Tr e^{-\b \hH} = \underset{\ubs(0)=\ubs(\beta)}{\int}
\prod_{i=1}^N D \bs_i \, \G^{|\bs_i|} 
\, \, e^{-\int_0^\beta E(\us(t)) \dd t } \ ,
\eeq
where $|\bs_i|$ is the number of flips in the trajectory $\s_i(t)$ of the
$i$'th spin. Empirical averages over the trajectory configurations
allows to compute the thermodynamic (both thermal and quantum) averages
$\la \bullet \ra = \Tr[\bullet \, e^{-\beta \hH}] / Z$.
The various versions of PIMC differ in the allowed moves
between configurations $\ubs$, that are usually required to fulfill
the detailed balance condition with respect to the measure $\mu_{\rm Q}$,
to ensure the stationarity of the latter. The PIMC results reported in 
Sec.~\ref{sec:results} have been obtained by using the heat-bath algorithm 
introduced in~\cite{KRSZ08}, in which the PIMC updates consist in drawing a
new trajectory for a randomly chosen spin, according to its conditional 
probability induced by its neighbors. This is possible thanks to the path
generation procedure explained in Sec.~\ref{sec:PIQC_pathgeneration};
a rigorous proof of fast convergence for this algorithm can be found 
in~\cite{MaWo12} for the Ising ferromagnetic model in transverse field on an 
infinite tree.

As explained in Sec.~\ref{sec:QAA-gap} the efficiency of the quantum adiabatic
algorithm is controlled by the gap between the two lowest lying eigenstates
of the interpolating Hamiltonian; in the following we discuss two different 
strategies to extract such a gap from QMC simulations.

\subsubsection{Extracting the gap from correlation functions}

The first strategy~\cite{YKS08,YKS10,HY11,Hen2012} is based on the computation of imaginary time correlations.
Consider for example the spin-spin correlation
\beq
\la \hs_i^z(\t) \hs_i^z(0)\ra = \frac1Z \Tr\left[ e^{-(\b-\t) \hH} \hs^z_i e^{-\t \hH} \hs^z_i \right] \ ,
\eeq
that as any other observable can be easily computed via PIMC.
Let us denote by $E_0 < E_1 < \dots$ the distinct eigenvalues of $\hH$,
with associated eigenvectors $|n\rangle$.
In the limit $\b\to\io$ with $\tau$ fixed, 
inserting the representation of the identity $I = \sum_n | n \rangle \langle n |$,
one obtains the spectral representation of this function as
\beq
\la \hs_i^z(\t) \hs_i^z(0)\ra = \sum_{n} | \langle 0 | \hs^z_i | n \rangle |^2 e^{- \t (E_n - E_0) } \ .
\eeq
Then, if the limit $\t\to\io$ is taken ({\it after} $\b\to\io$), we have
\beq
\la \hs_i^z(\t) \hs_i^z(0)\ra - \langle \hs^z_i \rangle^2 \sim e^{-\t \D } \ ,
\eeq
where we denoted $\Delta=E_1-E_0$ the energy gap between the two lowest 
levels (this formula is easily generalized if the ground state and first excited level are degenerate).
Even though PIMC simulations can only be performed at finite $\b$, 
in the regime $1/\D \ll \t \ll \b$ the plot of the logarithm of 
$\la \hs_i^z(\t) \hs_i^z(0)\ra_{\rm c}$ versus $\t$ is a straight line that can
be fitted to extract $\D$. We refer the reader to~\cite{YKS08,YKS10,HY11,Hen2012} for 
details and concrete examples of such computations, in particular to~\cite{Hen2012} where an optimal choice of observables in the computed correlation function is discussed.

\subsubsection{Extracting the gap from the specific heat}

Another possible strategy to compute the energy gap $\Delta$
is based on the evaluation of the specific heat:
\beq
C = \frac{\partial}{\partial T} \la \hH \ra = 
\beta^2 (\la \hH^2 \ra - \la \hH \ra^2 ) \ .
\eeq
If again $E_0 < E_1 < \dots$ are the distinct eigenvalues of $\hH$,
with associated degeneracies $g_0,g_1,\dots$, one sees that in the 
low temperature limit the specific heat behaves as
\beq
C \sim (\beta \Delta)^2 \frac{g_1}{g_0} e^{-\beta \Delta} \ .
\eeq
The value of $\Delta$ can thus be obtained from the behavior of
the specific heat at low temperatures. Moreover $C$ can be computed from
a QMC simulation. Let us introduce a fictitious parameter
$x$ and define $Z(x) = \Tr e^{-\b x \hH}$, in such a way that
\beq
\la \hH \ra = - \frac{1}{\beta} \left. \frac{Z'(x)}{Z(x)} \right|_{x=1} \ ,
\qquad 
\la \hH^2 \ra = \frac{1}{\beta^2} \left. \frac{Z''(x)}{Z(x)} \right|_{x=1} \ .
\eeq
As we explained above, with the path integral representation we have
\beq
Z(x) = \underset{\ubs(0)=\ubs(\beta)}{\int} 
\prod_{i=1}^N D \bs_i \, (x \G)^{|\bs_i|} 
\, \, e^{-x\int_0^\beta E(\us(t)) \dd t } \ .
\eeq
On this form it is very easy to take
the derivatives with respect to $x$, which leads to
\bea
\la \hH \ra &=& \int D\ubs \, \mu_{\rm Q}(\ubs) 
\left\{ \frac{1}{\beta} \int_0^\beta
E(\us(t)) \dd t - \G \sum_{i=1}^N \frac{|\bs_i|}{\beta \G} \right\} \ ,\\
\la \hH^2 \ra &=& \int D\ubs \, \mu_{\rm Q}(\ubs) 
\left\{\left[ \frac{1}{\beta} \int_0^\beta
E(\us(t)) \dd t - \G \sum_{i=1}^N \frac{|\bs_i|}{\beta \G} \right]^2 
- \frac{1}{\beta^2} \sum_{i=1}^N |\bs_i|
\right\} \ .
\nonumber
\eea
These two quantities, and in consequence the specific heat, can thus be
directly determined from configurations of paths generated in a QMC simulation.
The advantage of this procedure with respect to the previous one is that
in principle the specific heat should be easier to compute than the 
time-dependent
correlation functions, and that there is no need to identify the correct regime
$ 1/\D \ll \t \ll \b$ in $\t$. The disadvantage is however that one 
has to perform simulations at several temperatures $\b \gtrsim 1/\D$ to 
extract the slope of $\log(C)$ plotted
as a function of $\b$.

\subsubsection{Imaginary time annealing}
\label{sec:qmc_ann}

Let us finally mention two numerical approaches that, although very different,
can be both termed ``imaginary time annealings''.

A first ``imaginary time annealing'' consists in solving the
Schr\"odinger equation in imaginary time (which can be done either by exact diagonalization or by QMC), i.e. to study the evolution of
a vector of the Hilbert space $|\psi(s) \ra$ according to
\beq
-\frac{1}{\TT} \frac{\dd}{\dd s} |\psi(s) \ra = \hH(s) |\psi(s) \ra \ , \qquad
\hH(s) = (1-s) \hHi + s \hHf \ ,
\eeq
to be compared with Eq.~(\ref{Hquantum}). This evolution is not unitary,
hence would not be realizable with a quantum computer,
but can be implemented numerically.
The reader will find in~\cite{qa_review_santoro} and references
therein a detailed comparison of the real and imaginary time version
of the quantum annealing.

A distinct procedure, that we shall use in Sec.~\ref{sec:results}, is more precisely
an annealing of the Path Integral Monte Carlo procedure.
In PIMC,
configurations of imaginary time paths are generated with the measure 
$\mu_{\rm Q}$ of Eq.~(\ref{eq_mu_quantum}), by performing many Monte Carlo
updates on the configuration to approach this stationary measure. 
In other words, as in any Monte Carlo procedure,
one starts with a given initial configuration, and repeatedly apply
to it a given operation to produce new configurations. 
One therefore introduces a fictitious time $t_{\rm MC}$ that describes the number
of such operations that were done since the beginning of the procedure.
In a PIMC annealing,
the parameters of the measure (inverse temperature $\beta$ and transverse
field $\G$) slowly evolve during the PIMC simulation, and become 
``Monte Carlo time''-dependent parameters $\beta(t_{\rm MC}),\G(t_{\rm MC})$.
Note that
classical simulated annealing is a particular case of this procedure
where $\G=0$ at all Monte Carlo times; on the other hand one can set
$\beta$ to a very large fixed value (very small temperature) and let
$\G$ evolve. In the limit where $\beta$ is infinite, and the rate of 
variation of $\Gamma$ vanishes, this coincides with an adiabatic Schr\"odinger
evolution: at all Monte Carlo times the configuration of paths is drawn from
the measure $\mu_{\rm Q}$ which encodes the instantaneous ground state of the
original quantum Hamiltonian. 
Therefore, PIMC annealing allows for an interesting interpolation between classical simulated annealing
and zero temperature quantum annealing.
Note however that the condition of adiabaticity
for the PIMC annealing has a priori nothing to do with the one
of the original Schr\"odinger evolution (see however~\cite{MMBMRR10} and
references therein for a discussion of this point). 
The relevant gap is in the latter
case the one of the quantum Hamiltonian $\hH$, while in the former case
it is the one of the Fokker-Planck generator of the PIMC dynamics on the space
of path configurations; we refer
to~\cite{review_Nishimori,MaWo12} for further analysis of this PIMC
annealing. 
As a final important remark, note that the clustering transition (or ``dynamic transition''), 
which is signaled
by the appearance of a non-trivial solution of the 1RSB cavity equations at $m=1$
(both in the classical and quantum cavity method), is directly related
to a glassy lack of equilibration of the PIMC 
(if the thermodynamic limit is taken before the limit of infinitely slow
annealing)~\cite{MoSe2}. The decorrelation time of PIMC dynamics
in $t_{\rm MC}$ becomes infinite at this transition. This provides a very useful way
to detect this transition using PIMC, that we will illustrate in Sec.~\ref{sec:results}.

\section{Results on specific random optimization problems}
\label{sec:results}

The aim of this section is to apply the methods outlined in Sec.~\ref{sec:methods}
to the study of random instances of real optimization problems, such as those 
defined in Sec.~\ref{sec:examples_optimization}.
At variance with the ``toy'' models investigated in Sec.~\ref{sec:low_energy}, the
problems that we will discuss in this section are standard problems in computer science.
Still we expect that their phenomenology is close to the one of the toy models.
An important lesson that we learned from Sec.~\ref{sec:classical_mean_field} and Sec.~\ref{sec:low_energy} is that
there is a wide variety of behaviors in the different classical
optimization problems, that only become more complicated when quantum
fluctuations are added. In consequence we shall not try here to propose
a complete classification, but present results on a few specific models (or ``case studies''),
that illustrate the main phenomena that were discussed in Sec.~\ref{sec:low_energy} and lead
to exponentially small gaps, namely first order transitions and level crossings.
Part of these results were already published, part are 
original. The analysis is based on the methods that have been described in Sec.~\ref{sec:methods},
we concentrate here on the physical results.

\subsection{Early results}

To put the discussion in a historical perspective, 
it is worth to mention that most of the current interest
in the Quantum Adiabatic Algorithm (QAA) was triggered by a series of early numerical works that found evidence for a polynomial
scaling of the minimum gap in the 1-in-3 SAT (or Exact Cover) problem (see Sec.~\ref{sec:examples_optimization})
with a transverse field, suggesting an exponential speedup with respect to classical algorithms~\cite{Fa01,YKS08}.
These studies considered a particular ensemble of random Exact Cover instances, constructed to have a 
Unique Satisfying Assignment (USA); see~\cite{GY11} for a detailed description of the procedure.
We call this ensemble EC-USA in the following. This choice was made because in this case the minimal gap
between the ground state and the first excited state can be unambiguously defined.

The original work by Farhi et al. \cite{Fa01} was based on exact diagonalization of
very small ($N \leq 20$) EC-USA instances. A polynomial scaling of the gap for those instances was
detected. This was initially confirmed by a Quantum Monte Carlo (QMC) study of EC-USA instances with $N\leq 128$~\cite{YKS08}.
However, it was shown later by the same authors~\cite{YKS10}, by using a more refined
analysis and larger ($N\leq 256$) sizes, that
an increasing (with $N$) number of instances display a first order transition 
and an exponentially small gap. 

These numerical studies were extremely difficult not only because of the numerical cost of the
exact diagonalization step (Sec.~\ref{sec:ed}), but also because
EC-USA instances have an exponentially small probability in the fully random 
ensemble of Exact Cover~\cite{GY11}. 
Therefore, already constructing EC-USA instances is an exponentially hard task (Sec.~\ref{sec:generating_USA}) and constitutes 
one of the main limitations to access large sizes~\cite{Fa01,YKS08,YKS10}. 
Furthermore, strong finite size effects were detected on EC-USA instances~\cite{YKS10}.

Another important problem, that has been discussed in more details in 
Sec.~\ref{sec:generating_USA}, is the following.
Suppose that we take a fully random ensemble of instances of a given problem, that are
exponentially hard for known classical algorithms with probability one when $N\to\io$. 
If one selects only the USA instances of this ensemble, and if the latter have exponentially small probability, then
one is conditioning on extremely rare instances and it is not obvious anymore that these instances
remain exponentially hard for classical algorithms~\cite{QuietPlanting}.
Therefore, despite some numerical 
evidence for an exponential scaling of the running time of classical algorithms for 
EC-USA of $N\leq 256$~\cite{YKS08,GY11} has been reported,
the classical computational complexity of EC-USA at much larger $N$ remains an open problem,
even if rather academic as the instances of this problem essentially do not exist in the
thermodynamic limit.

These early studies highlighted the importance of being able to construct USA instances of
much larger sizes (Sec.~\ref{sec:generating_USA}), and of being able to investigate such
instances in the thermodynamic limit via the quantum cavity methods (Sec.~\ref{sec:methods}). 
Both these goals can be
achieved by using the so-called ``locked models''~\cite{ZM08}. We consider one of such models in
the next section.

\subsection{Locked models: XORSAT on a regular graph}
\label{sec:XORSAT}

In this section, we examine the XORSAT problem on a random regular graph,
a typical representative of the class of locked models~\cite{ZM08} that were discussed in Sec.~\ref{sec:generating_USA}.
The main property of these models is that clusters of ground state configurations
do not have internal entropy: they are isolated 
points. Therefore we
do not expect level crossings induced by 
the energy-entropy competition discussed in 
Sec.~\ref{sec:low_energy}, 
which simplifies a lot the analysis of the models.
Moreover, these are the simplest models
to study with the cavity method, allowing
us to illustrate the usefulness of the method
in the simplest non-trivial setting.
A crucial properties of these models is that the parameters can be tuned
in such a way that USA instances have a finite probability for $N\to\io$ in the random ensemble,
see Sec.~\ref{sec:generating_USA}.

Preliminary results we obtained on the quantum XORSAT model were 
reported in~\cite{JKSZ10}, where we showed the existence of a first
order transition associated to an exponentially small gap in $N$.
We present these results in much more detail
in the rest of this section, together with some previously 
unpublished results. 
Other locked models have been studied in~\cite{HY11,FGHSSYZ12} and
showed a similar behavior.

\subsubsection{Definition of the model and its classical properties} 
 
We focus on the $k$-XORSAT problem, defined on a 
random $c$-regular graph, which has been studied
in the classical case in~\cite{FLRZ01,FMRWZ01}, to which 
we add a quantum transverse field.
In quantum spin language, 
the model is defined by the following Hamiltonian
(where we omit a factor of 2 with respect to the definition of Sec.~\ref{sec:examples_optimization}):
\beq
\label{H}
 \hH = \hH_P + \G \hH_Q = \sum_{a=1}^M (1- J_a
\hs_{i^a_1}^z \dots \hs_{i^a_k}^z) - \G \sum_{i=1}^N \hsix \ .
\eeq 
Here, $J_a = \pm 1$ with equal probability. The $k$ spins $i^a_1,
i^a_2, \cdots, i^a_k$ involved in clauses $a = 1, \cdots, M = N c/k$ are
chosen uniformly at random among all possible choices such that each
spin enters {\it exactly} in $c$ clauses. This defines a regular random
graph structure where variables have connectivity $c$ and interactions
have connectivity $k$.

As usual, in the classical limit
$\G=0$, a given instance of the problem (defined by the choice of the random
graph and of the couplings $J_a$) is called {\it satisfiable}
(SAT) if there is a ground state of zero energy, UNSAT otherwise.
It is easy to see that the annealed entropy (i.e. the logarithm of the average 
number of solutions) density is $\log(2) \left(1-c/k\right)$ when $\G=0$.
It has been shown in~\cite{FLRZ01,FMRWZ01} that when the annealed entropy is
positive ($c < k$) the model is SAT with a probability going to 1 
as $N \to \infty$ and the typical number of solutions is exponential in $N$, concentrated
around its mean predicted by the annealed entropy. On the contrary if
the annealed entropy is negative (for $c>k$) satisfiable instances are exponentially rare,
and typically the model is UNSAT.
In the marginal case $c=k$ the model is 
SAT with finite probability, and when it is SAT the number of solutions is 
typically finite. 

We are particularly interested in USA instances of $\hH_P$.
Based on the above discussion, it is clear that these instances 
are exponentially rare
if $c \neq k$, and it is natural to expect that they have a finite probability for $c=k$.
We have indeed found numerically that for $c=k=3$, in the limit
$N\!\to\!\io$, the fraction of SAT and USA 
instances are $f_{\rm SAT} = 0.609 \pm 0.003$
and $f_{\rm USA} = 0.2850 \pm 0.0022$, as determined by 
using either Gaussian elimination, or 
a Davis-Putnam-Logemann-Loveland--like algorithm~\cite{DPLL}, to 
count the number of solutions of 40000 instances of different 
sizes and extrapolating the result to $N\to\io$~\cite{JKSZ10}.
Similar values are obtained for $c=k=5$ and $c=k=7$
(for even values of $k$ solutions always come in pairs due to the spin flip symmetry).
It is worth to mention that in the limit $k = c \to \io$ (taken after $N\to\io$), 
USA instances of the model should approach a particular
Quantum Random Energy Model (QREM) where the distribution of the classical
energies is a binomial. This model was analyzed in~\cite{FGGGS10}, 
and the existence of a first order transition was shown rigorously, supporting
the results obtained with the cavity method at finite $k$ and $c$.
See~\cite{GY11} for numerical studies of $f_{\rm USA}$ in other locked models.

For this model,
Path Integral Quantum Cavity (PIQC) computations have been performed following the method
described in Sec.~\ref{sec:PIQC}. The model is factorized (Sec.~\ref{sec:popu_dynamics}) thanks to
the regular structure of the hypergraph, hence the replica symmetric (RS) computation
requires a single population of $\Ntraj$ imaginary time trajectories (we used $\Ntraj=10^5$).
For 1-step replica symmetry breaking (1RSB) computations, we used
$\Nint=$4000 populations of $\Ntraj=$4000 trajectories. 
In both cases, each data point has been obtained as an average over 100 cavity 
iterations. We checked that finite population size effects are negligible with
this choice of parameters.
In addition, we will report the behavior of the spectral gap, determined 
by means of Exact Diagonalization (ED), and some
Path Integral Quantum Monte Carlo (PIMC) results.

\begin{figure}
\centering
 \includegraphics[width=.32\textwidth]{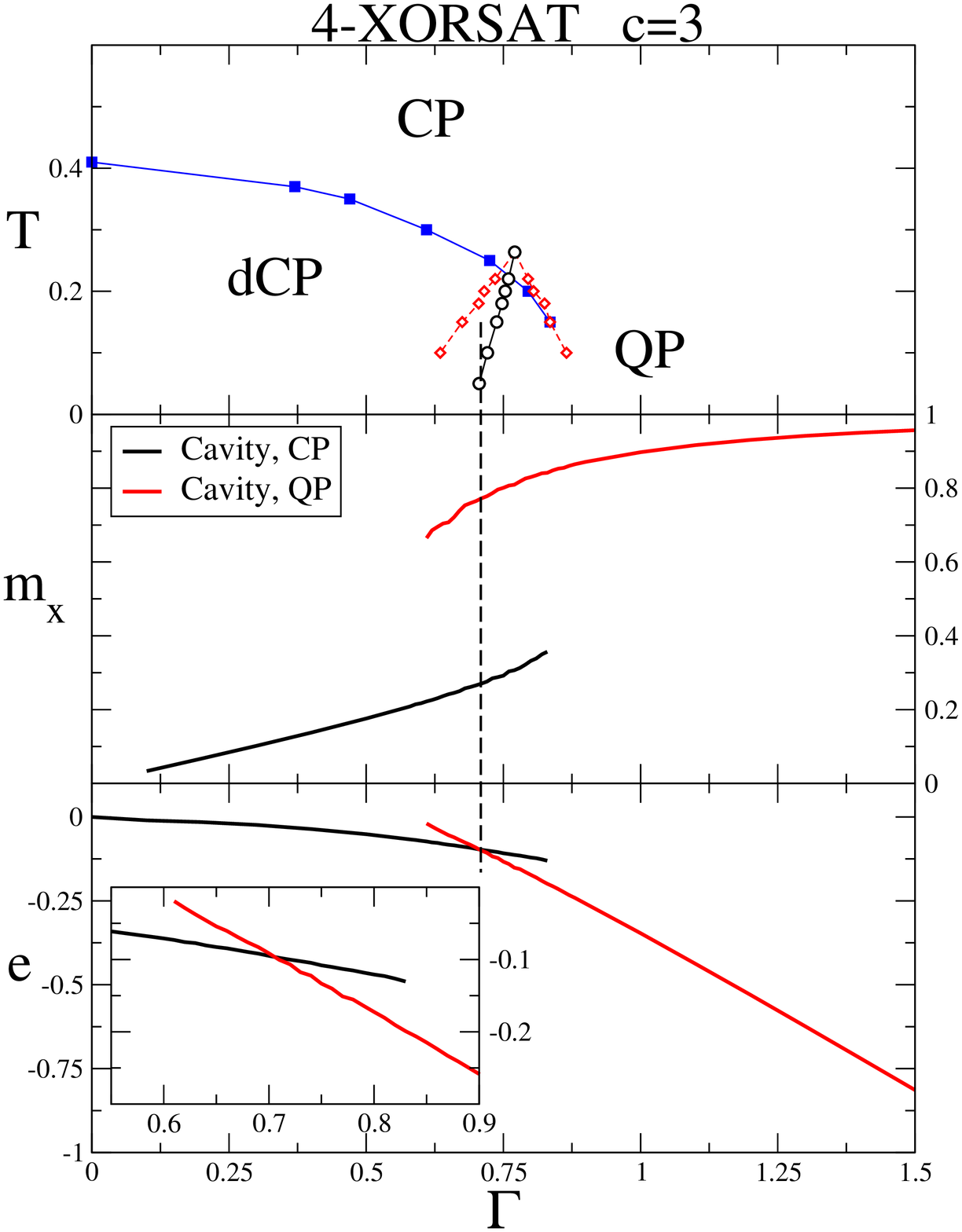}
 \includegraphics[width=.32\textwidth]{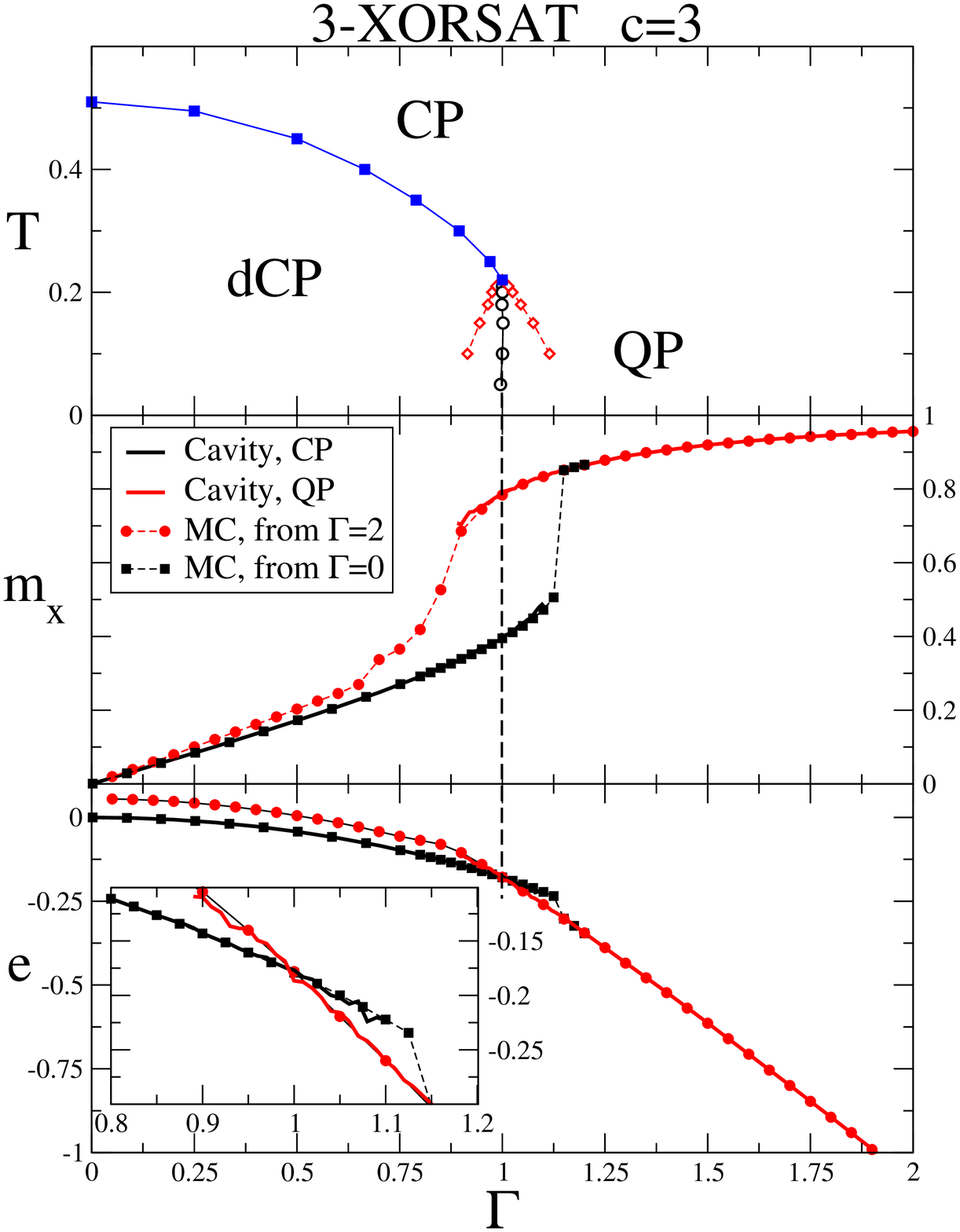}
  \includegraphics[width=.32\textwidth]{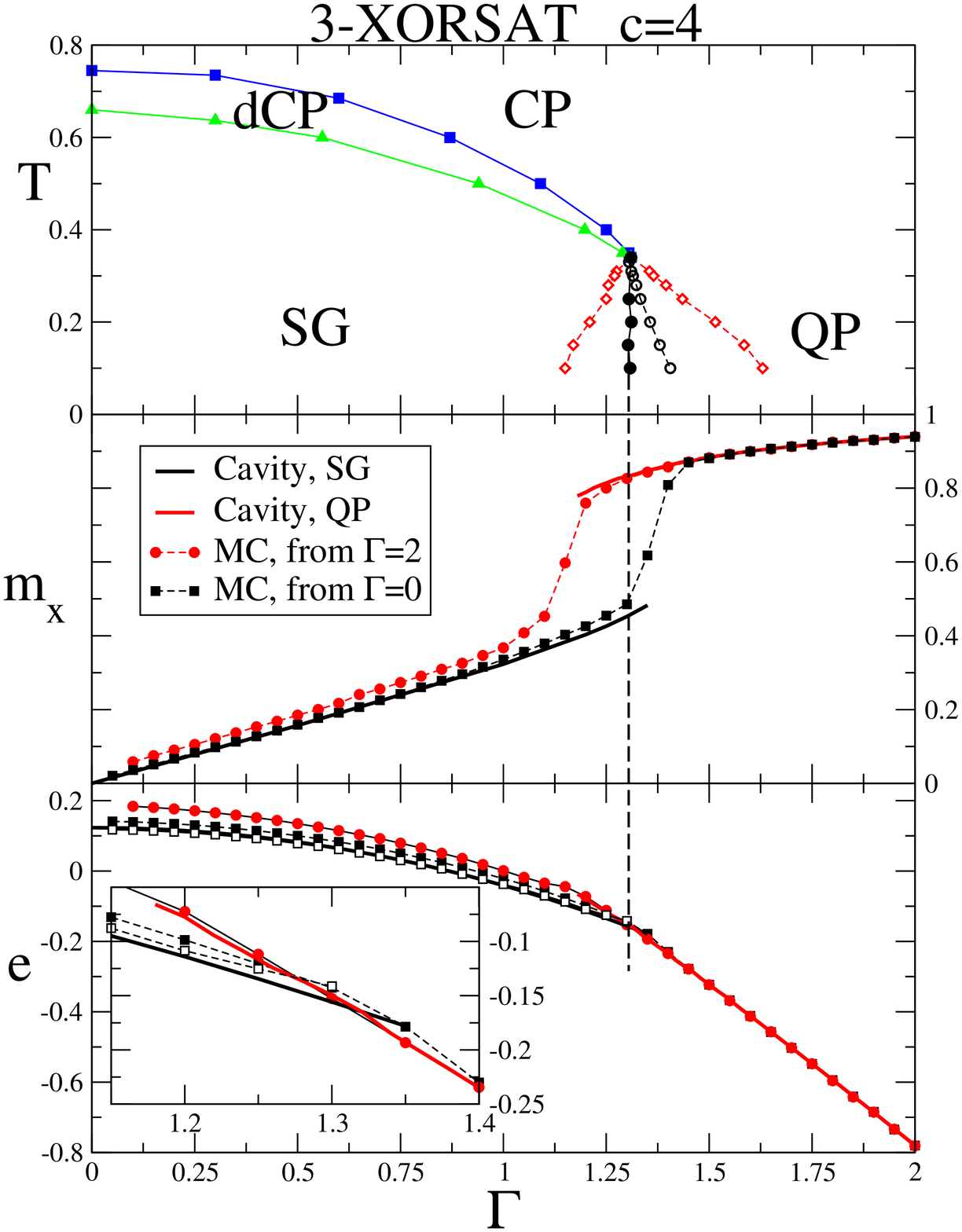} 
  \caption{
    Phase diagram of the $k$-XORSAT model Eq.~(\ref{H}) on a $c$-regular random graph.
    The top panel represents the $(T,\G)$ phase diagram, displaying four possible phases: CP, dCP, QP, SG (see text).
    Open symbols are RS results: first order transition
    line $T_{\rm fo}(\G)$ (open circles) separating the CP and QP, 
    with the corresponding
    spinodals (open diamonds). Full symbols are 1RSB results: 
    clustering transition $T_{\rm d}(\G)$ (full squares) separating the CP and dCP, 
    condensation transition $T_{\rm c}(\G)$ (full triangles) separating dCP and SG, $T_{\rm fo}(\G)$ (full circles) separating the SG and QP.
    The middle panel reports the transverse magnetization $m_x$ as a function of $\G$, 
    and the bottom panel reports the free energy density from PIQC or
    the energy density from PIMC as a function of $\G$, all at
    fixed temperature $T=0.05$. 
    In these panels, full lines are PIQC results 
    (RS or 1RSB) while symbols are PIMC results. \newline
    {\it (Left panel)} $k=4$ and $c=3$. There is no SG phase. The CP phase becomes
    a dCP at low enough temperature, while a first order transition separates the CP (or dCP) and
    QP phases. $m_x$ jumps at the first order transition. \newline
    {\it (Center panel)} $k=c=3$. Also here there is no SG phase.
    PIMC data are reported, for a sample with $N\!=\!2049$: red diamonds are obtained
    starting from the QP ($\G=2$) and decreasing $\G$, while black squares are obtained
    starting from a classical ground state (found using Gaussian elimination) 
    and increasing $\G$. \newline
    {\it (Right panel)} $k=3$ and $c=4$. Here a SG phase is present, and $T_{\rm fo}(\G)$ separates the SG and QP phases.
    PIMC data are reported for $N=120$ and averaged over 20
    samples (full symbols) and extrapolated in $1/N$ to the $N\!\to\!\io$ limit (open symbols).  
    Black curve, starting from the classical ground state
    found using an exact MAXSAT solver \cite{max1}. Red curve,
    starting from the QP.  
}
 \label{fig:dia3_3}
  \label{fig:dia3_4}
  \label{fig:dia4_3}
\end{figure}

\subsubsection{Exponentially degenerate ground state: $c<k$}
\label{sec:res_xor_expdegen}

\begin{figure}
\centering
  \includegraphics[width=.5\columnwidth]{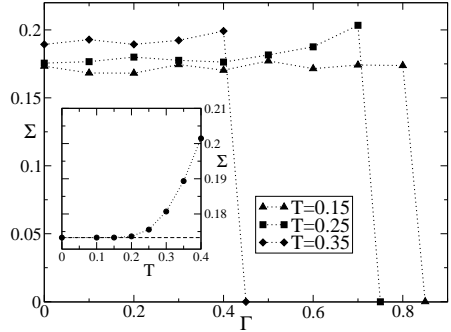}
  \caption{
    Complexity of the $4$-XORSAT model at $c=3$.
    ({\it Main panel}) Equilibrium complexity 
    as a function of $\G$ at fixed $T$. The complexity remains finite
    up to the dynamic transition line $\G_{\rm d}(T)$, where it jumps abruptly to zero.
    ({\it Inset}) Equilibrium complexity of the classical model ($\G=0$) as 
    a function of temperature. The dashed line marks the $T=0$ value, $\log(2)/4$.
  }
  \label{fig:sigma43}
\end{figure}

As a representative of the case $c<k$, we take here $k=4 > c=3$.
The classical ground state is exponentially degenerate with entropy
$\log(2) \left(1 - c/k\right) = \log(2)/4$. It can be shown via the cavity
and replica methods~\cite{FLRZ01,FMRWZ01,ZM08} 
or using rigorous methods~\cite{xor_1,xor_2}
that the ground states are arranged in {\it isolated clusters}. Therefore,
the internal entropy of each cluster is zero, the complexity of clusters
is $\Sigma = \log(2)/4$, and typically the clusters (solutions) have Hamming
distance of order $N$, therefore they are very far away in configuration space.
The classical equilibrium complexity as a function of temperature is plotted
in the inset of Fig.~\ref{fig:sigma43}. The model is SAT with probability
one and the typical number of ground states is $\exp(N \Sigma) = 2^{N/4}$.

The phase diagram of the model,
as obtained from PIQC,
is reported in Fig.~\ref{fig:dia4_3}.
The RS computation predicts, at low
enough temperature $T \lesssim 0.3$, a first order transition between
two different paramagnetic ($m_z = \la \s^z_i \ra =0$) phases: the
{\it Classical Paramagnet} (CP) characterized by a small value of
transverse magnetization $m_x = \la \s_i^x\ra$, and the {\it Quantum
  Paramagnet} (QP) that has a larger value of $m_x$. The first order
transition is signaled by a jump of $m_x$ and a crossing of the free
energies of the two phases, that can be clearly seen in 
Fig.~\ref{fig:dia4_3}. As in any first order transition for a mean
field model, the two phases can be continued in the region where they
are metastable until a well defined spinodal point.
The transition line
and the corresponding spinodals are shown in the $(\G,T)$ phase
diagram in Fig.~\ref{fig:dia4_3}; the transition is found
at a slightly temperature-dependent 
$\G_{\rm fo}(T) \approx 3/4 = c/k$. 
We also report in Fig.~\ref{fig:dia4_3} the
cavity method predictions for $m_x$ and the
free energy density $f=-T \log(Z)/N$ at very low temperature $T=0.05$. 

Next, we discuss the outcome of the 1RSB computation. 
As discussed in Sec.~\ref{sec:methods},
the key quantity that is computed in this approach is 
the equilibrium complexity $\Si_{\rm eq}(\G,T)$ (at $m=1$),
which is reported in Fig.~\ref{fig:sigma43} 
as a function of $T$ and $\Gamma$.
Below the classical dynamical transition 
$T_{\rm d}(\G=0) \sim 0.41$, the complexity is positive in the classical case
(see the inset of Fig.~\ref{fig:sigma43}). Increasing $\G$, we found that
$\Si_{\rm eq}(\G,T)$ remains independent of $\G$, until the dynamic transition line
$\G_{\rm d}(T)$ is met, then the 1RSB solution disappears discontinuously,
as revealed by the abrupt jump of the complexity curves in Fig.~\ref{fig:sigma43}.
The fact that $\Si_{\rm eq}$ is independent of $\G$ is not surprising, based on
the discussion of the random subcubes model (Sec.~\ref{sec:subcubes}).
If the clusters do not have any internal entropy, and if their relative Hamming
distance is of order $N$, different solutions are not mixed at any finite order
of perturbation theory in $\G$.
Therefore, each classical ground state is continuously
transformed in a quantum eigenstate. 
Moreover, the local environment around each
ground state is the same for $N\to\io$, hence at any finite order of perturbation theory
the quantum energy is the same for all ground states, and the degeneracy is not
lifted. This can be shown as follows: suppose that $\underline{\t} = \{ \t_i \}$ is an assignment of the classical
spins corresponding to a ground state. Then $J_a \t_{i_a^1} \cdots \t_{i_a^k}= 1$, $\forall a$.
We can apply to the Hamiltonian (\ref{H}) the unitary transformation
$U_{\ut} \equiv \{\hsiz \to \t_i \hsiz\}$ and we obtain
\beq
\hH = \sum_{a=1}^M (1-
\hs_{i^a_1}^z \dots \hs_{i^a_k}^z) - \G \sum_{i=1}^N \hsix \ .
\eeq 
Therefore, for each classical ground state $\underline{\t}$ there is a symmetry that allows to map the Hamiltonian
into a ferromagnetic one and map $\underline{\t}$ to the ferromagnetic state. This shows that perturbation
theory around each classical ground state gives identical results.
The number of ground states remains constant and equal to its classical
value, $2^{N/4}$, so that the complexity is constant as a function of $\G$. 
Moreover, the fact that the complexity
remains approximately constant at all temperatures suggests 
that there are no level crossings
between states of different intensive energy, as in the 
QREM~\cite{JKKM08}.

The main result is then that the equilibrium complexity at $m=1$ is
positive or zero everywhere. The implications of this result are 
twofold: first of all, it confirms that the RS
computation of the thermodynamic observables is in this case correct
in the whole phase diagram $(\G,T)$ (remember that as discussed in 
Sec.~\ref{sec:methods}, a true 1RSB phase is signaled 
by a negative complexity at $m=1$).
Therefore, the only thermodynamic singularity
is on the first order RS transition line. 
Secondly, the complexity is strictly positive for low
enough values of $T$ and $\G$, implying that the CP phase is 
actually a {\it dynamical CP} (dCP, the meaning and motivation for this 
name have been discussed in Sec.~\ref{sec:methods}) 
where an exponential number of states
coexist. The point where the equilibrium complexity becomes positive
is the {\it clustering temperature}
$T_{\rm d}(\G)$, and is reported in Fig.~\ref{fig:dia4_3}.
We recall that equilibrium
thermodynamic properties are unaffected as one crosses the transition
between CP and dCP (Sec.~\ref{sec:class_RSB}).

\begin{figure}
\centering
  \includegraphics[width=.45\columnwidth]{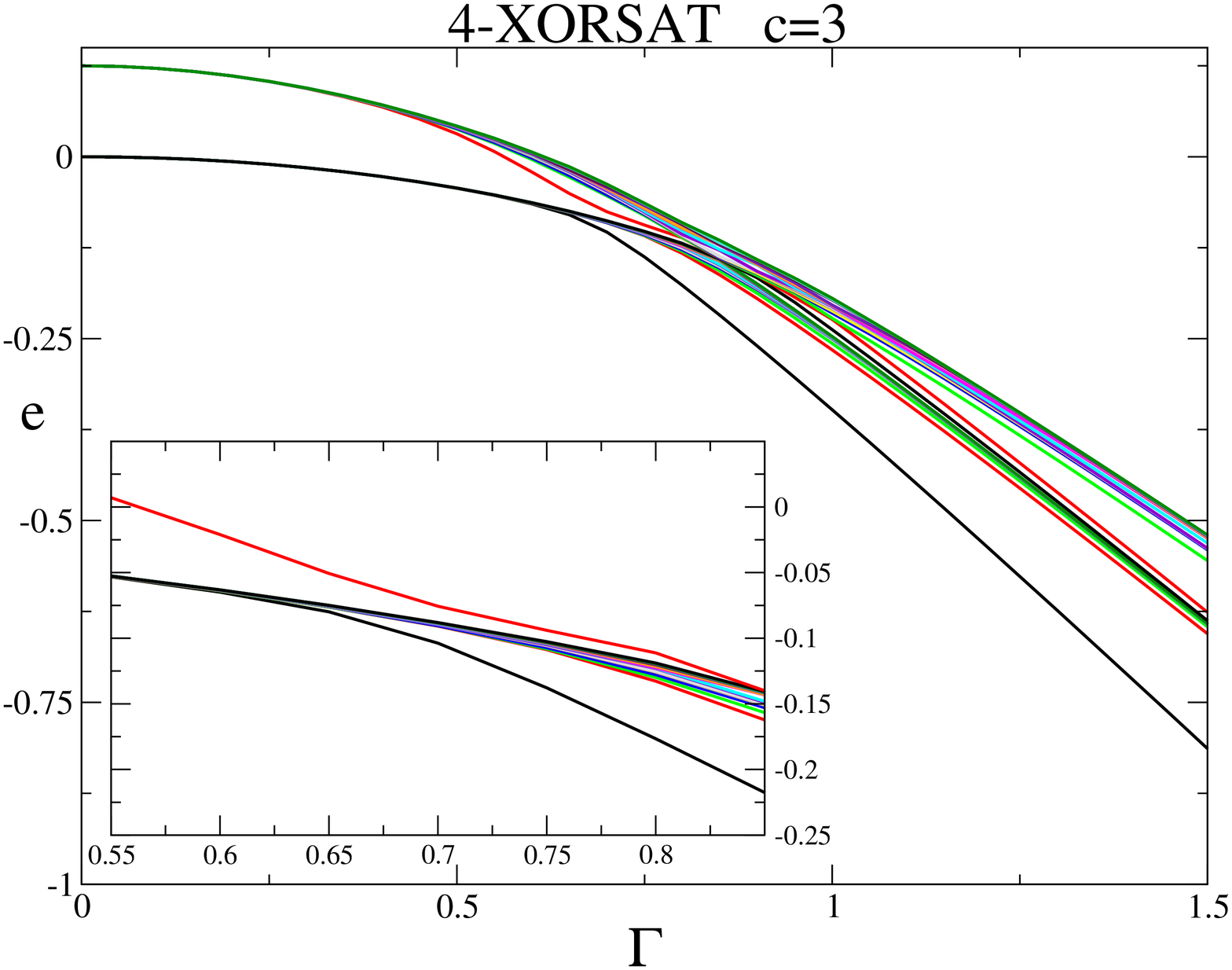}
  \includegraphics[width=.45\columnwidth]{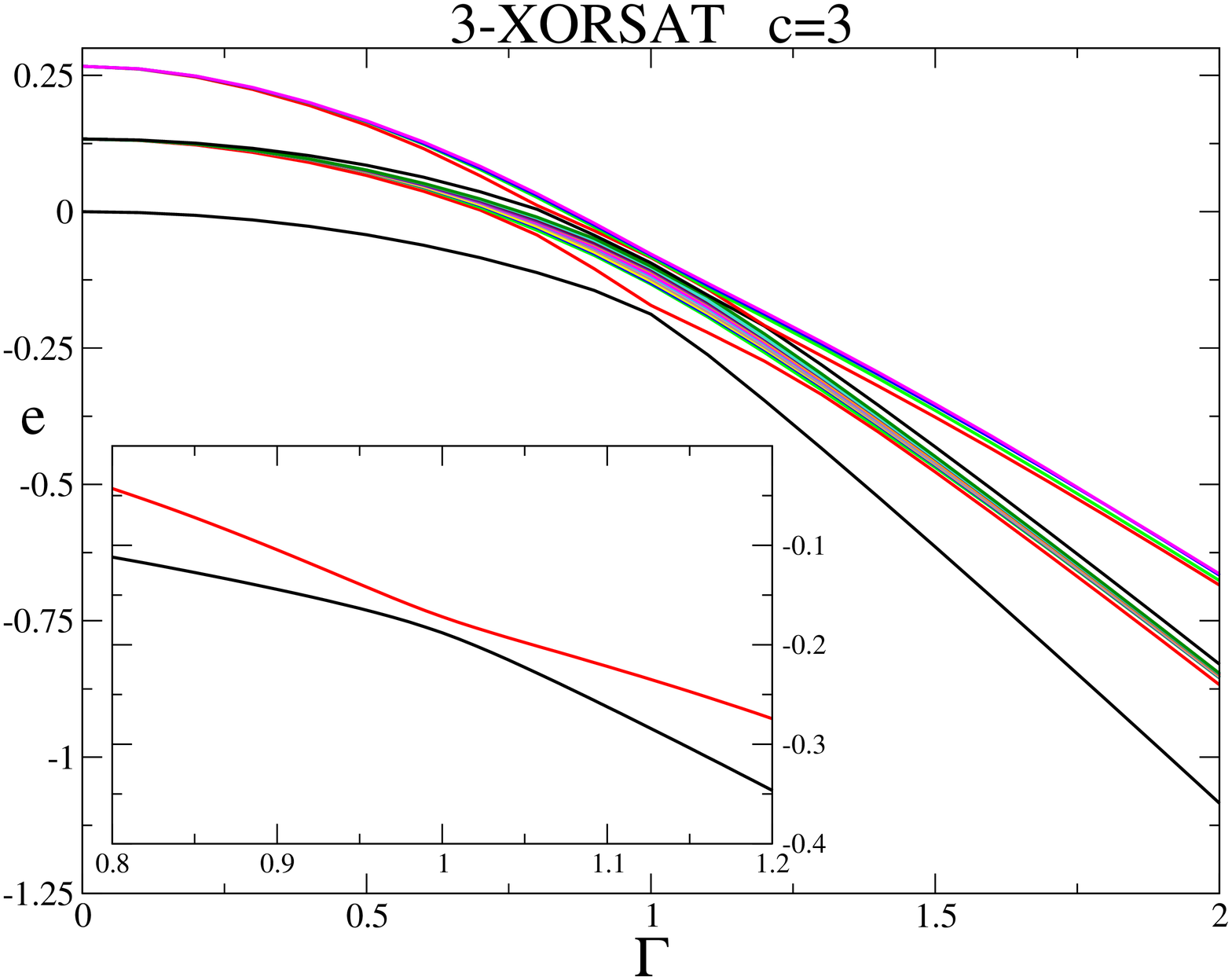}
  \caption{ 
    Lowest energy levels of $k$-XORSAT on a $c$-regular random graph, 
    from exact diagonalization. In the inset the region close to
    the phase transition is magnified. 
    ({\it Left panel}) Typical instance of the $4$-XORSAT model at $c=3$,
    with $N=16$. The classical ground state degeneracy is $2^{4} = 16$.
    ({\it Right panel}) USA
    instance of $3$-XORSAT at $c=3$, with $N=15$. 
  }
  \label{fig:3}
  \label{fig:ED43}
\end{figure}

In Fig.~\ref{fig:ED43} we show ED results 
for this case. The lowest part of the spectrum of a typical instance
with $N=16$ is plotted as a function of $\G$. The instance we show has
a ground state degeneracy $\NN = 2^{N/4} = 16$ at $\G=0$, which is the
most probable value.
Increasing $\G$, we
see that the lowest 16 levels remain extremely close in energy 
(the difference is expected to be exponentially small), up to a value
of $\G \approx 0.75$, the location of the first order transition in
the thermodynamic limit. At this value of $\G$ we observe that the 
17th state (the first classical excited state) goes down in energy
and approaches the bunch of ground states. The figure suggests the
presence of an avoided crossing between
this state and the set of ground states. These data confirm the cavity
prediction: the ground state remains exponentially degenerate at any
finite $\G < \G_{\rm fo}$, while at $\G_{\rm fo}$ a first order transition
happens, caused by a level crossing between the degenerate pure states of the dCP
and the QP. 

The determination of the gap is complicated by the fact that for
a given instance the ground state has degeneracy $\NN$, the 
average of $N^{-1} \log\NN$ over instances being equal to the
zero temperature complexity $\log(2)/4$.
Therefore, the interesting gap to determine the performances
of QAA is the minimal gap (over $\G$) 
between the lowest
energy state and the $(\NN+1)$-th excited state, that we call $\D_{\rm min}$: 
transitions to
any lower energy state are not dangerous because these states will
continuously transform into one of the classically 
degenerate ground
states. A more detailed discussion of the QAA
dynamics in presence of almost degenerate levels
might be in order here but, as discussed in Sec.~\ref{sec:QAA-gap}, 
no precise results are available at present.
Because $\NN$ increases exponentially fast in $N$ (it concentrates quickly 
around the value $2^{N/4}$), one has to compute
an exponentially large number of levels, which slows down considerably
the exact diagonalization code\footnote{This problem could be removed by
making use of the symmetries $U_{\ut}$ discussed above. It is reasonable to assume
that both the ground state and the $(\NN+1)$-th excited state belong to the subspace of the
Hilbert space that is completely symmetric under all the $\NN$ such symmetries.
Then one could restrict the Hamiltonian to this subspace and compute the gap between the two lowest
levels. The resulting reduced matrix will not be sparse, so this strategy is not efficient 
for exact diagonalization and it was
not used here. 
However, a similar strategy turns out to be very useful if one wants to compute the relevant gap via 
QMC, see \cite{HY11} for a detailed discussion
in the case where the ground state is doubly degenerate.} and in practice limits us to $N\leq 20$.
In Fig.~\ref{fig:4} we report data for the minimum gap as defined above.
Despite the strong size limitations, the scaling of the gap appears to
be exponential in $N$, as expected at a first order transition.
We observed that fluctuations in $\NN$ induce large fluctuations of the
gap: indeed, restricting the average to instances having exactly $2^{N/4}$
classical ground states reduces a lot the fluctuations, and the curve is
much closer to an exponential, at the same time the difference at large $N$
being extremely small (we don't show the corresponding data).
We will discuss further the behavior of the gap at the transition
in the simpler case $c=k$, which we analyze next.

\subsubsection{Finitely degenerate ground state: $c=k$}
\label{sec:XORSAT_USA}

We now turn to the case $c=k$, where the complexity at $T=0$ vanishes
and the number of ground states is finite with finite probability.
We choose as the simplest example $c=k=3$.
The phase diagram, reported in Fig.~\ref{fig:dia3_3}, 
is qualitatively identical to the one we obtained
for $c<k$, the only difference being that the equilibrium complexity
now vanishes for $T=0$ and any $\G>0$, so the number of ground states
is finite for any $\G$.
Another quantitative difference is that the first order transition line
looks exactly vertical and equal to $\G_{\rm fo} = c/k =1$, suggesting
the existence of a hidden duality relating the model at large and small $\G$, 
which was indeed proven in~\cite{Gosset,FGHSSYZ12}. 
Note that the spinodal lines merge where the first order transition disappears;
this point is different from the point where
the dynamical transition line crosses the first order transition line,
even if this is not visible in Fig.~\ref{fig:dia3_3}.

\begin{figure}
\centering
  \includegraphics[width=.45\columnwidth]{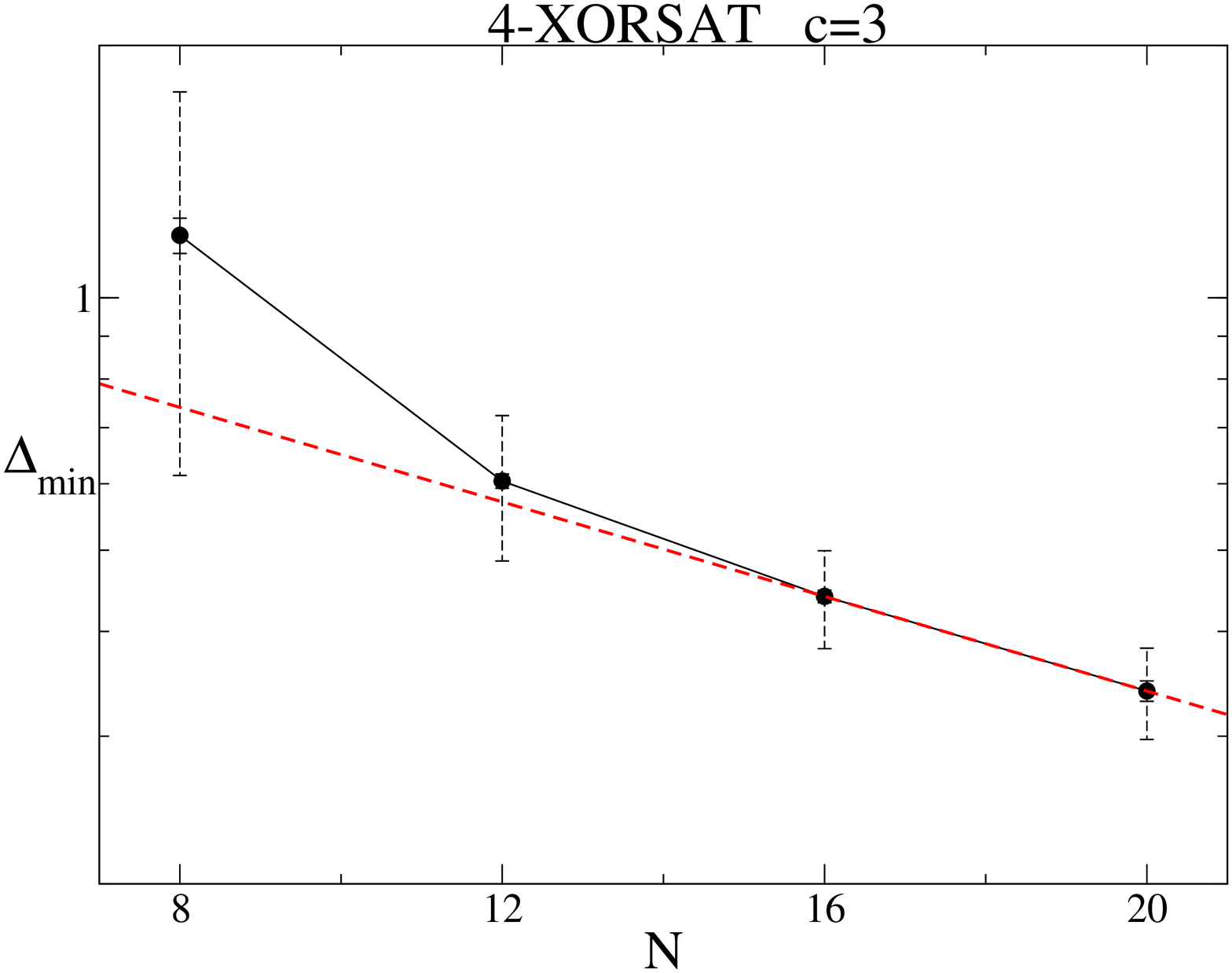}
  \includegraphics[width=.45\columnwidth]{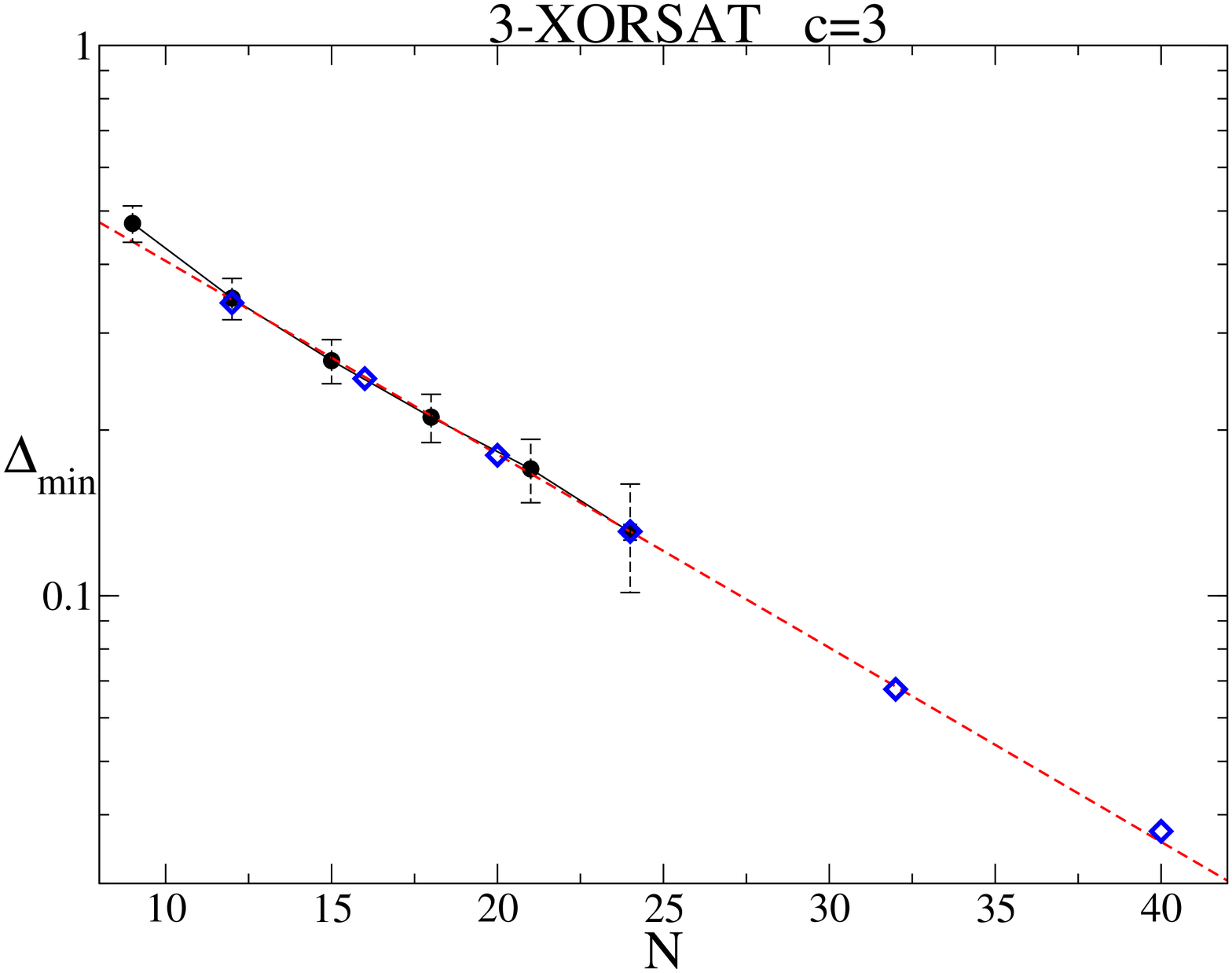}
  \caption{
    Exact diagonalization data for the minimum gap for
    $k$-XORSAT on $c$-regular random graphs.
    \newline
    {\it (Left panel)} $k=4, c=3$, random instances.
    The classical ground state has an instance-dependent 
    degeneracy $\NN$, so the relevant gap is the one between the ground state and the 
    $(\NN+1)$-th state, see Fig.~\ref{fig:ED43}. 
    Full circles represent the average over instances (100 for $N=8,12$ and 60 for $N=16,20$) 
    of $\D_{\rm min}$.
    Full bars represent statistical errors on the average, while dashed bars represent
    the standard deviation over the instances of a single realization of the 
    random variable $\D_{\rm min}$.
    Fluctuations are extremely large at small $N$, the main contribution
    being due to the fluctuations of $\NN$. 
    Dashed line is $\D_{\rm min}(N) = 1.244 \exp(-0.065 N)$ that
    describes well the large $N$ behavior.
    \newline
    {\it (Right panel)} $c=k=3$, USA
    instances. Full black points represent
    the average of $\D_{\rm min}$. 
    Here ED can be performed up to $N=24$ and fluctuations are reduced: 
    error bars are of the order of the symbol size 
    except when explicitly shown ($N=24$). Dashed bars represent
    the standard deviation of a single realization of the random variable
    $\D_{\rm min}$. 
    Open blue diamonds are QMC data for the {\it median} minimal gap from~\cite{FGHSSYZ12}.
    Here again error bars are of the order of symbol size.
     Dashed line is a fit to $\D_{\rm min}(N) = 0.911
    \exp(-0.081 N)$.
      }
  \label{fig:4}
\end{figure}

We now discuss the behavior of the gap at the first order transition point $\G_{\rm fo}$.
There is a definite advantage in this case, namely that
a finite fraction $f_{\rm USA} \sim 0.28$ of instances
have a single ground state, as discussed above (while in previous
studies~\cite{Fa01,YKS08,YKS10} USA instances were exponentially rare).
We report ED results only on USA instances, in order to unambiguously 
define the gap $\D(\G)$ between the ground
state of $\hH$ and its first excited state at all values of $\G$.
The spectrum of a
typical USA instance of $N=15$ spins is reported in
Fig.~\ref{fig:3}. We observe, as expected, that the gap $\D(\G)$
has a minimum $\D_{\rm min}$ close to the phase transition at $\G_{\rm fo}$
(recall that $\G_{\rm fo} = 1$ for $c=k$ at $N\to\io$).  
In Fig.~\ref{fig:4} we show the behavior of the average $\D_{\rm min}$
as a function of $N$. Our data are
clearly consistent with an exponential scaling of the gap, as
expected based on the discussion of Sec.~\ref{sec:low_energy}.
The probability distribution over instances
of $\D_{\rm min}$ has a unique peak close
to the average, and its variance is also reported in Fig.~\ref{fig:4} (dashed bars).
This shows that all instances undergo a first order transition of the same kind
in the thermodynamic limit. These results have been confirmed in~\cite{FGHSSYZ12}
by computing the minimal gap via QMC (as described in Sec.~\ref{sec:qmc}) at larger sizes.
The QMC results confirm the exponential trend. Note that in the QMC study the median minimal gap
was considered, instead of the average. The coincidence of the results confirms that the distribution
of $\D_{\rm min}$ is unimodal and strongly peaked.

Next, we can compare the cavity results with PIMC.
As we already stressed several times, in the case $c=k=3$, 
instances have a finite probability of being SAT, and otherwise
have a minimal energy per spin of order $1/N$ (see~\cite{FLRZ01,FMRWZ01}). 
Moreover, a ground state of SAT instances can be found in
polynomial time using the Gauss elimination algorithm.
This crucial observation allows us to find a classical ground state
of SAT instances for very large sizes ($N=2049$).
We can therefore run a PIMC starting from
the classical ground state at $\G=0$ and slowly increasing $\G$, 
thereby following the evolution of the classical ground states upon
introduction of quantum fluctuations.
We find that the PIMC data follow closely the cavity result up to 
$\G_{\rm fo}$, see Fig.~\ref{fig:dia3_3}.
Then, as expected for a first order
transition, we find hysteresis around $\G_{\rm fo}$ before the
system finally jumps to the QP phase. Next, we consider a second 
PIMC run that is performed starting from large $\G=2$ in the QP phase, and
slowly decreasing $\G$. In this case, PIMC data follow the cavity
ones down to the transition $\G_{\rm fo}$, but then the energy
remains {\it extensively higher} than the ground state energy for any
$\G < \G_{\rm fo}$. This is obviously due to the difficulty in following
adiabatically (in the fictitious PIMC dynamics)
the ground state across an exponentially small gap,
as discussed in Sec.~\ref{sec:qmc_ann}.
This result is an important indication of the
difficulty of finding the ground state, even
in presence of quantum fluctuations, and it
is also an important proof of the usefulness of the cavity
method: in fact, if it were not for the Gaussian elimination that allowed
us to find the classical ground state and run the PIMC starting from it,
we would never be able to compute the quantum ground state using PIMC.
In some models (of which we give an example just below), 
finding the classical ground
states is extremely difficult for any classical algorithm. The cavity method
allows to compute the ground state energy even when Monte Carlo methods
fails because of equilibration problems.

\begin{figure}
\centering
  \includegraphics[width=.45\textwidth]{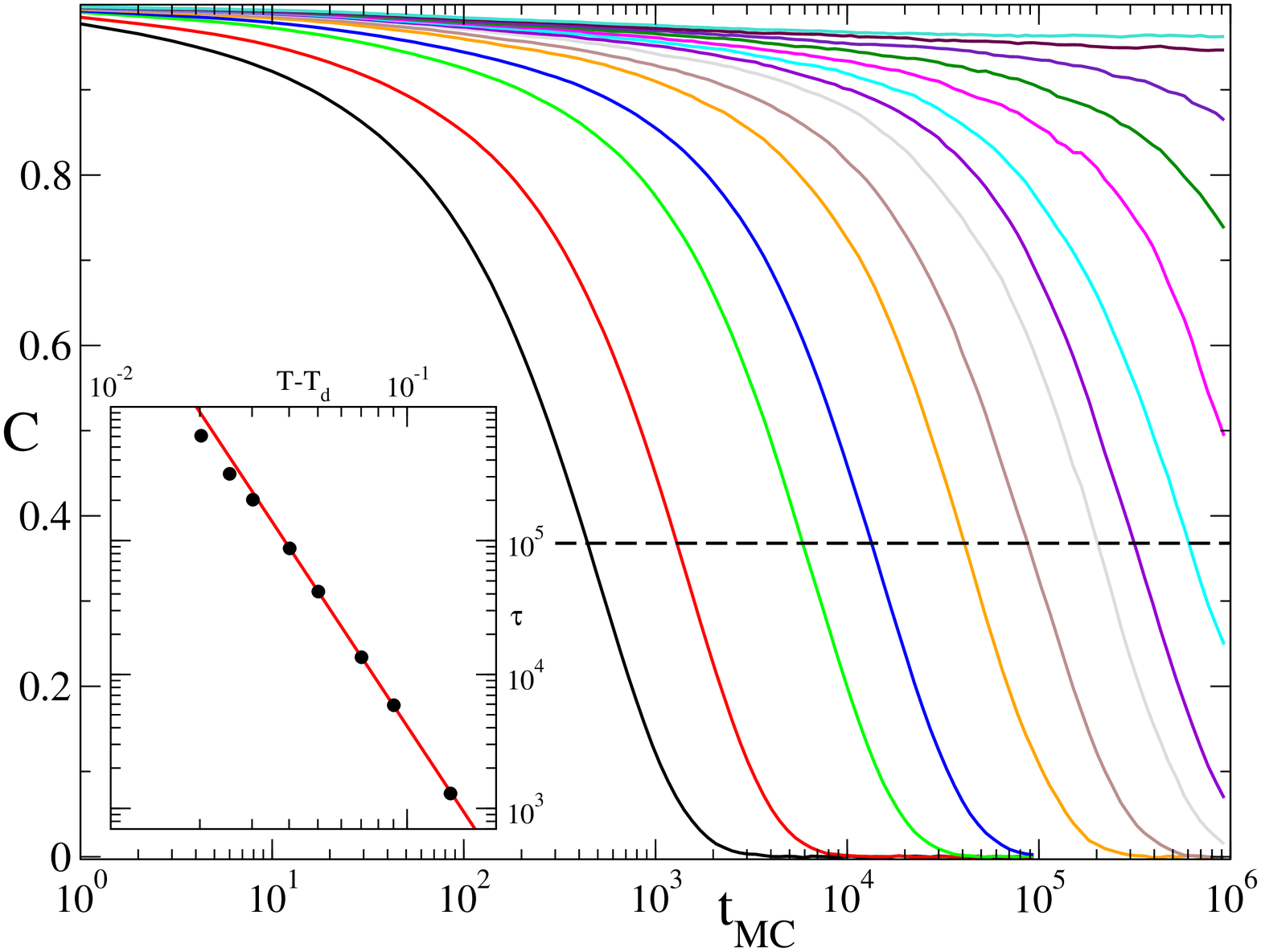}
  \includegraphics[width=.45\textwidth]{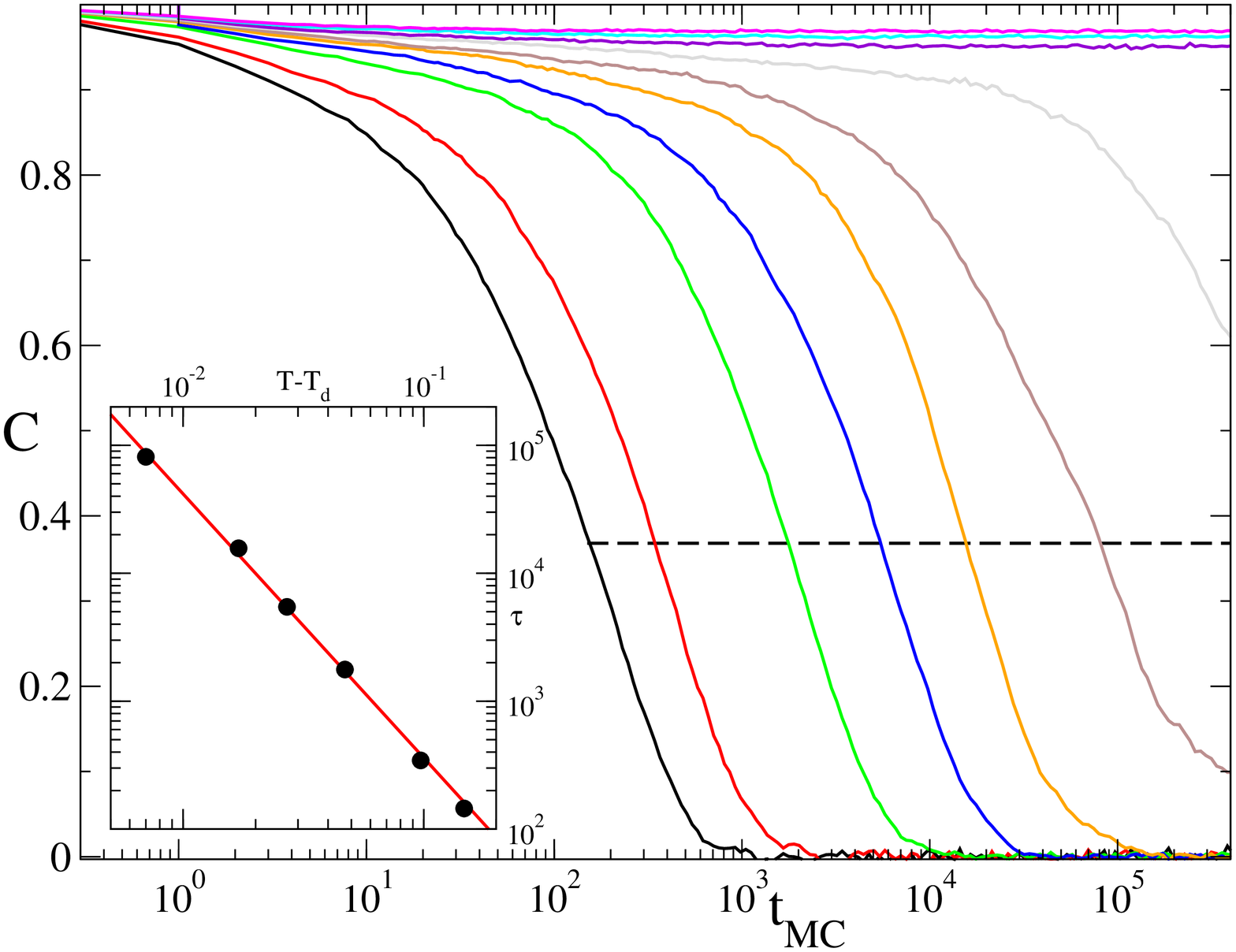}
  \caption{
 Equilibrium normalized spin-spin correlation functions $C(t_{\rm MC}, T)/C(0,T)$ for the Monte Carlo dynamics as functions of 
$t_{\rm MC}$ for several temperatures across the clustering transition $T_{\rm d}$, and
 averaged over several realizations of the random graph and the initial planted configuration.
 In the inset of each figure, the decorrelation time $\t(T)$ such that $C(\t,T) = 1/e$ (dashed line)
is plotted versus $T-T_{\rm d}$ to show the power-law divergence $\t(T)\sim A (T - T_{\rm d})^{-\g}$. 
\newline
 {\it (Left panel)} Classical case $\G=0$, with standard Metropolis dynamics and $N=60000$ spins.
From left to right, $T$=0.7, 0.65, 0.6, 0.58, 0.56, 0.55, 0.54, 0.535, 0.53, 0.525, 0.52, 
0.515, 
0.51, 
0.505.
  Here $T_{\rm d} \sim 0.5098$, $A \sim 1.82$, $\g \sim 3.35$. 
\newline
 {\it (Right panel)} Quantum case for $\G=0.9$, with the PIMC dynamics of~\cite{KRSZ08} and $N=10000$. 
From left to right, $T$=0.45, 0.4, 0.35, 0.33, 0.32, 0.31, 0.30, 0.29, 0.28, 0.27.
Here $T_{\rm d}\sim 0.303$, $A \sim 3.11$, $\g \sim 2.06$. 
The values of $T_{\rm d}$ are consistent with Fig.~\ref{fig:dia3_3} in both cases.
 }
  \label{fig:dyn_XOR}
\end{figure}

Another demonstration of the difficulty of PIMC to equilibrate (and therefore to find the ground state energy)
in this problem is the following.
Let us consider first the classical limit $\G=0$.
As we discussed in Sec.~\ref{sec:thermalannealing}, a classical Monte Carlo simulation 
will never be able to equilibrate in polynomial time below the clustering transition 
$T_{\rm d}(\G=0)$~\cite{MoSe2}. This can be shown by considering a classical Monte Carlo simulation
that starts in an equilibrium configuration at temperature $T$ (which can be generated by the planting technique~\cite{MoSe,QuietPlanting})
and computing the spin-spin correlation in Monte Carlo time:
\beq
C(t_{\rm MC}, T) = \frac1N \sum_{i=1}^N \la \s_i(t_{MC}) \s_i(0) \ra \ ,
\eeq
where the average is done over many realizations of the process. This correlation function is
reported in Fig.~\ref{fig:dyn_XOR}. One can see that on approaching the dynamical transition the 
time over which this correlation function goes to zero increases as a power law, and it diverges 
at $T_{\rm d}$ as $(T - T_{\rm d})^{-\gamma}$. Below $T_{\rm d}$, the correlation does not decay
to zero anymore, indicating that the dynamics is trapped into a state.
The same analysis can be repeated at finite $\G$ by considering the PIMC dynamics. Again, we start
by an equilibrium configuration of the paths\footnote{
In the quantum case one can still use the planting technique to prepare equilibrium configurations:
the trick consists in doing an initial PIMC run in which both the paths and the coupling are changed, i.e. the
couplings are treated as dynamical variables. This amounts to an ``annealed'' computation and is equivalent
to planting~\cite{QuietPlanting}.
}
at temperature $T$, and we perform a PIMC evolution. We call $t_{\rm MC}$ the PIMC time and
\beq
\overline{\s}_i = \frac1\beta \int_0^\beta dt \, \s_i(t) 
\eeq
the imaginary time average of a spin at a given $t_{\rm MC}$. We define as in the classical case
\beq
C(t_{\rm MC}, T) = \frac1N \sum_{i=1}^N \la \overline{\s}_i(t_{MC}) \overline{\s}_i(0) \ra \ ,
\eeq
but in this case $C(t_{\rm MC}=0, T) \neq 1$ so it is convenient to normalize it by the value in $t_{\rm MC}=0$.
This normalized correlation function is reported in Fig.~\ref{fig:dyn_XOR} and it shows exactly the same behavior as the classical
one, with a decorrelation time that diverges as $(T - T_{\rm d}(\G))^{-\gamma}$ when the quantum clustering transition
is approached.
This results shows that PIMC is trapped in a metastable state below $T_{\rm d}(\G)$ and does not equilibrate. 
It also shows that $T_{\rm d}$ can be detected via a PIMC simulation: this is important because it might allow to estimate
$T_{\rm d}$ for generic quantum systems for which the cavity method cannot be used,
like in the classical case~\cite{Cavagna09,BB11}.

\subsubsection{UNSAT case: $c>k$}

Here we discuss the UNSAT case $c>k$, taking as an example $c=4$
and $k=3$.
The results for the phase diagram are displayed in
Fig.~\ref{fig:dia3_4}. 
In this case the model has a richer phenomenology,
very similar to the one of fully connected mean field
models~\cite{Go90,NR98,BC01,CGS01}.
At the RS level, the phenomenology is unchanged and a first order is found
between the CP and QP phases.
However, in this case the equilibrium complexity becomes negative below a temperature
$T_{\rm c}(\G)$
in the dCP phase, signaling that the RS solution becomes
incorrect (Sec.~\ref{sec:flowchart}). 
The dCP phase 
then undergoes a thermodynamically
second order phase transition at $T_{\rm c}(\G)$ to a true {\it Spin Glass} (SG) phase,
where replica symmetry is broken and the Gibbs measure is dominated by
a finite number of pure states. 
Therefore, at low
enough temperature the first order thermodynamic transition happens directly
between the SG and QP phases. 
For this reason, the RS computation
gives a wrong result for the first order transition line, see top
panel of Fig.~\ref{fig:dia3_4}. The correct result is obtained by finding the crossing
between the QP free energy and the SG free energy, and the latter has to be computed
by optimizing over $m$ the 1RSB free energy and is in general higher than the CP free
energy as obtained from the RS calculation (Sec.~\ref{sec:flowchart}).
Still, also in this case we conclude on the existence
of a first order quantum phase transition at $\G = \G_{\rm c}$
and zero temperature, separating the SG from the QP. 
The transition extends in a line $\G_{\rm c}(T) \approx c/k = 4/3$ 
at low enough temperature, and is almost independent of $T$ (at variance with
the RS result).

We tried to repeat the PIMC simulation using the same protocol
as in the $c=k$ case. However, in this case the problem is
typically UNSAT~\cite{FLRZ01,FMRWZ01}: the classical ground states have a finite energy
per spin, finding them is extremely hard (actually, harder than any NP-complete problem, see Sec.~\ref{sec:classical_complexity_classes}) and the quiet planting technique cannot be used.
Therefore, in this case
we are severely limited in the search of the classical ground state, and we can only find it for
quite small sizes ($N \leq 120$) using an exact MAXSAT solver~\cite{max1}. Still, we could repeat the PIMC procedure of increasing $\G$
starting from the classical ground state, and in this way compute the ground state energy at finite
$\G$. A good extrapolation in $1/N$ to the thermodynamic limit is possible, and the result agrees
well with PIQC result, see Fig.~\ref{fig:dia3_4} and~\cite{FGHSSYZ12}.
As in the previous case, we find
that a PIMC run starting at large $\G$ and reducing $\G$ fails to find the ground state at small
$\G$.

Finally, it is worth to note that the case $k=2$ (and any $c>k=2$) belongs to this class but displays 
a very different phenomenology. This model has two-body interactions and is very close to the Sherrington-Kirkpatrick (SK) model 
(see~\cite{cavity, cavity_T0} and Sec.~\ref{sec:low_energy}), and the
transition to the spin glass phase happens via a standard second order phase transition instead
of a random first order transition. In this case, as in the SK model, a second order phase transition line in the $(T,\G)$ plane
separates the CP and SG phases~\cite{LSS08}. The transition line 
extends to a quantum critical point at $T=0$. In this case there is evidence for a polynomially small
gap at the critical point~\cite{FGHSSYZ12}. However, as in the SK model~\cite{AnMu_future}, the spin glass phase seem
to be everywhere gapless with an exponentially small gap~\cite{FGHSSYZ12}. This should be related
to energy-induced level crossings inside the spin glass phase, as discussed in Sec.~\ref{sec:Altshuler}. 
We refer the reader to~\cite{FGHSSYZ12} for a more detailed discussion.

\subsubsection{Randomization of the transverse fields}
\label{sec:randomization}

It has been proposed in~\cite{FGGGS10} that small gaps
induced by level crossings at low values of the transverse field $\G$~\cite{AC09,AKR10}
might be avoided by introducing local random fluctuations of the field, i.e. use a different
random $\G_i$ on each spin.
Motivated by this proposal, we analyzed whether the first
order transition can be washed out by a random transverse
field. We therefore considered a generalization of Eq.~(\ref{H}),
where
\beq
\hH_Q = - \sum_i \ee_i \hsix \ ,
\eeq
$\ee_i$ being a random variable. We choose 
$\ee_i \in \{ 1/2,3/2 \}$ with equal probability, and
independently for each spin, in such a way that the average
of $\G_i = \G \ee_i$ is equal to $\G$. We solved the model for $k=c=3$ 
using the cavity method at the RS level, which we expect to give
the exact solution in this case,
even in presence of random transverse field. The presence of the 
latter forced us to introduce an additional level of population
(the model is not factorized anymore, see Sec.~\ref{sec:popu_dynamics}). 
We used $\Next=1000$ populations, each containing 
$\Ntraj=1000$ trajectories. Again we
did not see observable finite population size effects.
The average free energy
as a function of $\G$ is qualitatively similar to the one with a constant transverse field,
showing that the first
order phase transition is present also with random transverse field.
Moreover, because the free energy is self-averaging
for large $N\to\io$, a deviation of order $1$ in the
intensive free energy is exponentially rare: this implies
that finding a rare sample that does not show the transition should
be exponentially improbable for large $N$.
We conclude therefore that the first order quantum phase transition
observed in this model is robust against randomization of the transverse
field.

\subsubsection{Other approaches}

\begin{figure}
\centering
  \includegraphics[width=.75\columnwidth]{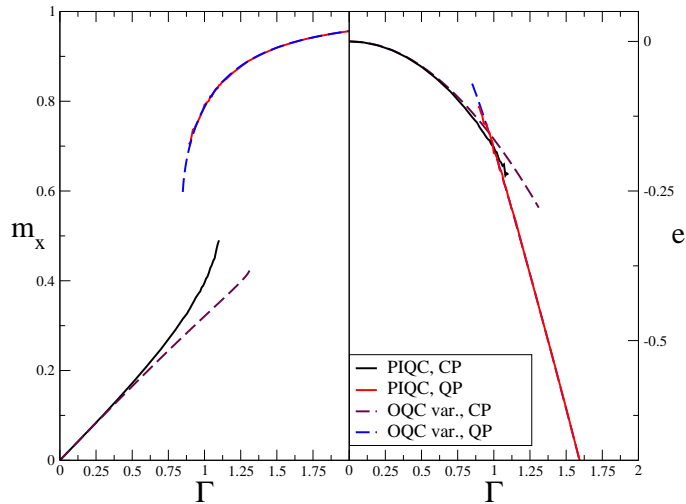}
  \caption{
    Comparison between the free energy {\it (right panel)}
    and the transverse magnetization {\it (left panel)} for XORSAT at $k=c=3$,
    obtained via the path integral quantum cavity method (PIQC) and the operator cavity method (OQC).
The approximation is almost perfect for
    the QP phase, while it is not very accurate for the CP at large $\G$. 
  }
  \label{fig:MIapp_XORSAT}
\end{figure}

To conclude this section, we discuss the applicability of the other approaches discussed
in Sec.~\ref{sec:methods} to the XORSAT problem. We did not attempt to use 
variational cavity approaches for this problem, because these methods have yet
to be tested in simpler cases.
We discuss here the Operator Quantum Cavity (OQC) methods presented in Sec.~\ref{sec:IM}.
Already at the simplest level $\ell=1$ (see Sec.~\ref{sec:OQC_single}),
the applicability of these methods is severely limited by the need of diagonalizing the effective
Hamiltonian $\hH_{\rm eff}$ of Eq.~(\ref{eq:IM3}). This is because each spin in this model
interacts with $c (k-1)$ other spins, therefore $\hH_{\rm eff}$ is a matrix of size $2^{c (k-1) +1}$ which
is quite large already for $c=k=3$. 
For this reason, we could only perform a RS variational calculation, which consists in taking operator
cavity messages at $\ell=1$ of the form (\ref{eq:OQC_1_param}) with site-independent longitudinal
and transverse fields, substituting them into the RS free energy, and optimizing the latter with respect
to the fields.
The results are reported in Fig.~\ref{fig:MIapp_XORSAT} for $c=k=3$.
We found that this approximation gives an extremely good description
of the QP. On the other hand, the description of the CP at large $\G$ is not excellent: the value of the energy is still
an upper bound of the true energy, but there is an observable difference
at large $\G$, while at the same time the transverse magnetization is underestimated
by the approximation. Moreover, the value of $\G$ at
spinodal point of the CP
is largely overestimated.
Despite these quantitative discrepancies, we conclude 
that the variational OQC at $\ell=1$ gives a very good
qualitative description of the RS phase diagram.
Unfortunately, a 1RSB computation within this approach is out of reach of present computational capabilities.

\subsubsection{Discussion}

In this section, we discussed the phase diagram of a typical locked
problem: the $k$-XORSAT model on a $c$-regular random graph.
We have obtained the full phase diagram of the quantum model 
as a function of $T$ and $\G$. The main results are: 
\begin{enumerate}
\item
For $k>2$, 
there is a first order quantum phase transition at $T=0$
between the low temperature classical phase (which can be either
a CP or a SG phase)
and a QP phase, at
a critical value of $\G=\G_{\rm fo}$~\cite{JKSZ10}.
\item
The transition is due to a crossing between the low-$\G$ 
classical-like ground state(s), and
the high-$\G$ quantum paramagnetic state. It is of very different
nature from the level crossing at small $\G$ between different
spin glass ground states discussed in~\cite{AC09,AKR10,FGGGS10}.
\item
The first order transition is observed for almost all instances, even for
very small $N$. In general, finite size effects are extremely small
in this model, and they are mainly due to the fluctuations in the number
of classical ground states.
\item
The first order transition is generically associated to an
exponentially vanishing gap of $\hH$~\cite{JKKM08,JKSZ10}, 
hence, in this model, the QAA requires a run time scaling exponentially with the system size
to find the ground state.
\item
The case $k=2$ is special: here the model has a second order transition with
a polynomially small gap, but level crossings at small $\G$ are present
and induce an exponentially small gap, hence also in this case the QAA is not
efficient~\cite{FGHSSYZ12}.
\end{enumerate}
The main missing ingredient in locked models with respect to the 
general picture outlined in Sec.~\ref{sec:subcubes} is the internal
entropy of the clusters. 
In these models clusters are isolated
configurations, and they are very far away from each other.
If the local environment around each solution is the same, as in the XORSAT problem,
the degeneracy is not lifted at any order in perturbation theory.
Then, the degeneracy
is only lifted by (non-perturbative) quantum tunneling 
leading to an exponentially small splitting. 
Level crossings at small $\G$
are therefore very difficult to observe in these models, and in the thermodynamic limit 
the ground states remain closely degenerate
on increasing $\G$ up to the first order phase transition.

\subsection{The coloring problem}
\label{sec:results_coloring}

In order to discuss the effect of the existence of exponentially many clusters
(pure states) with a non-trivial distribution of internal energy and entropy
densities, we consider here 
the quantum version of the simplest classical model that shows this effect: the coloring problem
on random regular graphs.
The phenomenology of this problem is rather different from the
one of XORSAT: contrary to the latter, the coloring problem is not a locked model.
Hence, in the classical limit the number of solutions is exponentially large in $N$ in the SAT phase.
As it has been discussed in Sec.~\ref{sec:transitions_rCSP}, the structure of the solution space of this
model is very similar to the one of the Random Subcubes Model (RSM) discussed in Sec.~\ref{sec:subcubes}.
We expect that adding quantum fluctuations should lead to an extremely complex spectrum, as it has been
discussed in Sec.~\ref{sec:qsubcubes} for the Quantum RSM (QRSM). 
In particular, we want to show that due to entropic effects, different
clusters evolve very differently under the action of the quantum term, leading to level crossings.

\subsubsection{Definition of the model}

The model is defined as follows. We consider $N$ Potts spins. Each of these
is a classical variable $\s_i \in {\cal X} \equiv \{1,\cdots,q\}$. To define the quantum model we introduce
for each spin the Hilbert space spanned by $|\s_i \rangle$ and the operators
\beq
\begin{split}
& \hT_i = \sum_{\s_i \neq \s_i'} | \s_i \rangle \langle \s_i' | \\
& \widehat{\D}_{ij} = \sum_{\s_i, \s_j} \d_{\s_i,\s_j} | \s_i \s_j \rangle \langle \s_i \s_j |
\end{split}
\eeq
where $\d_{\s,\s'}$ is the Kronecker delta.
We define the Hamiltonian as
\beq
\hH = \hH_P + \G\hH_Q = \sum_{\la i, j \ra} \widehat{\D}_{ij} - \G \sum_i \hT_i \ ,
\eeq
where the first sum runs on pairs of variables that are connected by a link of the random regular graph with fixed connectivity $c$, on which
the model is defined. In the classical case $\G=0$ it is easy to check that the model reduces to the classical
coloring problem (or equivalently to the antiferromagnetic Potts model) introduced at the beginning of Sec.~\ref{sec:classical_complexity}.
In the following we will call $m_x= \la \hT_i \ra$ the ``transverse magnetization'' by analogy with the Ising spin case.

We want to stress that the analysis of this model turns out to be extremely difficult for a combination of technical reasons.
First of all, the interesting regime is $q \geq 4$ (the case $q=3$ being special, because it displays a continuous transition at the classical level) 
and quite large connectivities $c$~\cite{col2}.
Hence, exact diagonalization is impossible, because the size of the Hilbert space grows as $q^N$, much faster
than for Ising spins, and moreover for large $c$ one needs to consider quite large systems to avoid finite size effects.
One therefore cannot have any direct information about the spectrum and needs to infer it from other techniques.
OQC methods are impossible to use, again because the size of the Hilbert space grows too quickly with the number of spins.

We mainly used the PIQC method, which however
is more difficult to apply to this model than to the XORSAT model. 
We found that large populations have to be considered to avoid finite population size effects: for the 1RSB computations reported below, 
we typically used $\Nint \sim$ 4000 populations made of $\Ntraj \sim$ 4000 trajectories, as in the XORSAT 
case (the coloring problem is also factorized). However,
for large $q$ and $c$ the PIQC solution algorithm is
slow (the running time grows roughly as $q^2 \, c$, see the discussion in Sec.~\ref{sec:methods}).
Moreover, in many cases we had to perform finite population size scalings to obtain reliable results.
We also used PIMC, which requires similar computational effort than in other models,
but suffers as usual from equilibration problems. 

Keeping these difficulties in mind, we now proceed to discuss briefly the structure of this model. We will keep the discussion 
short, and we will focus on the computationally simplest case that displays the phenomenology we are interested in, namely
$q=4$ and $c=9$.
A much more complete and detailed discussion of this problem can be found in~\cite{BSZ12}.

\subsubsection{Results}

For $q=4$ and $c=9$ the model is classically satisfiable (the ground state energy is zero and graphs are typically colorable). At $T=0$ the number of ground states (solutions) 
is exponentially large, and they are arranged in an exponential number of clusters~\cite{col2}. 
This corresponds to the region $c_{\rm d} < c < c_{\rm c}$ according to the discussion of Sec.~\ref{sec:classical_mean_field}.
In other words, $T_{\rm c}=0$ while $T_{\rm d} = 0.153(5)$ for the classical model at $\G=0$, 
which is described by the 1RSB cavity equations at Parisi parameter $m=1$, below $T_{\rm d}$ 
and down to $T=0$.

Let us briefly recall the expected behavior 
for small $\G>0$ and low temperature $T$, based on
the analysis of the QRSM (Sec.~\ref{sec:qsubcubes}).
At $\G=0$, clusters have internal entropies that are distributed according to a large deviation function (the complexity), 
$\NN(s) = \exp[N \Si(s)]$. 
Typical configurations are found in clusters that 
have a value of the entropy $s^*$ such that $\Si'(s^*)=-1$, and the corresponding complexity $\Si(s^*)$ is strictly positive.
However, many other clusters with larger and smaller entropies exist. 
When $\G\gtrsim 0$, each cluster $A$ of degenerate classical states transforms continuously into a set of quantum states, 
the lowest of which (the ``ground state of cluster $A$'' $|GS(A)\ra$) has an energy per spin
\beq\label{eq:e_COL}
\begin{split}
e(\G) &= e_{\rm cl}(A,\G) - \G m_x(A,\G) \ , \\
e_{\rm cl}(A,\G) &= \la GS(A) |  \hH_P | GS(A) \ra/N \ , \\
m_x(A,\G) &=  \la GS(A) |  \sum_i \hT_i | GS(A) \ra/N \ .
\end{split}\eeq
The crucial observation is that, as in the QRSM, we expect $m_x(A,\G)$ to be finite when $\G\to 0$,
$\lim_{\G\to 0} m_x(A,\G) = m_x^0(A)$, and we expect $m_x^0(A)$ to be positively
correlated with the classical entropy of the cluster.
At the same time, $e_{\rm cl}(A,\G) \propto \G^2$ at small $\G$.

Therefore, the energy of the ground state of a cluster is linear at small $\G$, $e(\G) \sim -\G m_x^0(A)$.
Largest clusters yield the greatest decrease in energy when quantum fluctuations are switched on, 
and they dominate at zero temperature as soon as $\G>0$.
Because these are the states with maximal entropy, they correspond to $\Si(s_{\rm max})=0$.
Hence as soon as $\G>0$, the zero temperature complexity drops abruptly to zero. 
In other words, we expect the system to condense into the largest cluster under an infinitesimal amount of quantum fluctuations. 
This in particular implies that a non-zero $T_{\rm c}(\G)$ should
emerge, and that $T_{\rm c} \propto \G$ for small $\G$, as in the QRSM.

\begin{figure}
\centering
\includegraphics[width = .48\textwidth]{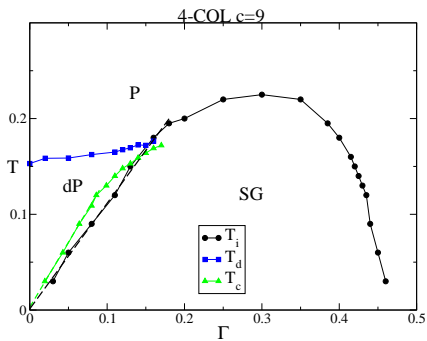}
\caption{
Phase diagram of the quantum coloring problem for $q=4$ and $c=9$.
The black solid line with circles represents the linear instability of the RS solution, $T_{\rm i}(\G)$ (a linear fit at small $\G$ is plotted with a black dashed line);
the blue line with squares is the clustering transition $T_{\rm d}(\G)$ separating the P and dP phases;
the green solid line with triangles is the condensation transition $T_{\rm c}(\G)$ separating the dP and 
SG phases (dashed green line is a linear fit of the small $\G$ behavior).
The classical problem at $T=0$
is in the SAT phase, with an exponential number of clusters.
It has therefore a finite $T_{\rm d} \sim 0.15$ and $T_{\rm c}=0$. For small $\G>0$, 
we observe a linear increase of $T_{\rm c}(\G) \sim \G$, as in the RSM, confirming that
the physics of the two models is very similar. At larger $\G$ the transition becomes a continuous transition
$T_{\rm i}(\G)$ that extends to a third order quantum 
critical point at $\G_{\rm i} \sim 0.47$.
}
\label{fig:col_4_9}
\end{figure}

This scenario is indeed confirmed by the PIQC results, see figure~\ref{fig:col_4_9}. 
We find that starting from finite temperature and increasing $\G$, the equilibrium 
complexity at Parisi parameter $m=1$ decreases from its (finite) classical value, until it vanishes at some $\G_{\rm c}(T)$. 
Inverting this function we find $T_{\rm c}(\G)$, which is indeed a linearly increasing function of $\G$ 
at small $\G$. This result is a strong indirect confirmation that the QRSM describes correctly the physics of complex models like the quantum coloring.
Next, we examine the evolution with $\G$ of $T_{\rm d}$, which is found by performing scans at constant $\G$ and $m=1$, 
starting at low $T$ and increasing $T$ until the non-trivial 1RSB solution is lost.
We find that also $T_{\rm d}$ increases slightly from its classical value for small $\G>0$. 
The behavior of $T_{\rm d}$ and $T_{\rm c}$ shows that, contrary to naive intuition, quantum fluctuations
promote glass formation, rather than make the glass unstable. 
This has been recently found in a series of different models of glasses by mean of different methods~\cite{FSZ10,FSZ11,MMBMRR10,OLG12}.

At larger $\G$, the two lines $T_{\rm d}(\Gamma)$ and $T_{\rm c}(\Gamma)$ approach each other and at some value of $\G$ they cross a third line $T_{\rm i}(\G)$
(see Fig.~\ref{fig:col_4_9}),
that corresponds to a linear instability of the RS solution of the cavity equations towards 1RSB 
(i.e. a continuous RSB transition).
Because this instability is a property of the RS solution, it is independent of $m$. Hence it has been detected by solving
the 1RSB cavity equations at $m=0$, initializing the population at $\G=0$ and $T>0$ in the RS solution 
(all external populations are identical), increasing $\G$ and finding the point where a
1RSB solution appears continuously. When the lines cross, 
the transition ceases to be a discontinuous 1RSB transition (i.e. a random first order transition) and becomes instead a continuous RSB transition.
We find that at larger $\G$ the transition remains continuous until $T_{\rm i}(\G)$ goes to zero around $\G_{\rm i} \sim 0.45$, 
leading to a second order quantum phase transition
(third-order in the thermodynamic sense), 
at variance with the XORSAT case where the transition is first order.
Correspondingly, in this case the paramagnetic phase is unique and we refer to it as P in Fig.~\ref{fig:col_4_9}.

\begin{figure}
\center
\includegraphics[width = .49\textwidth]{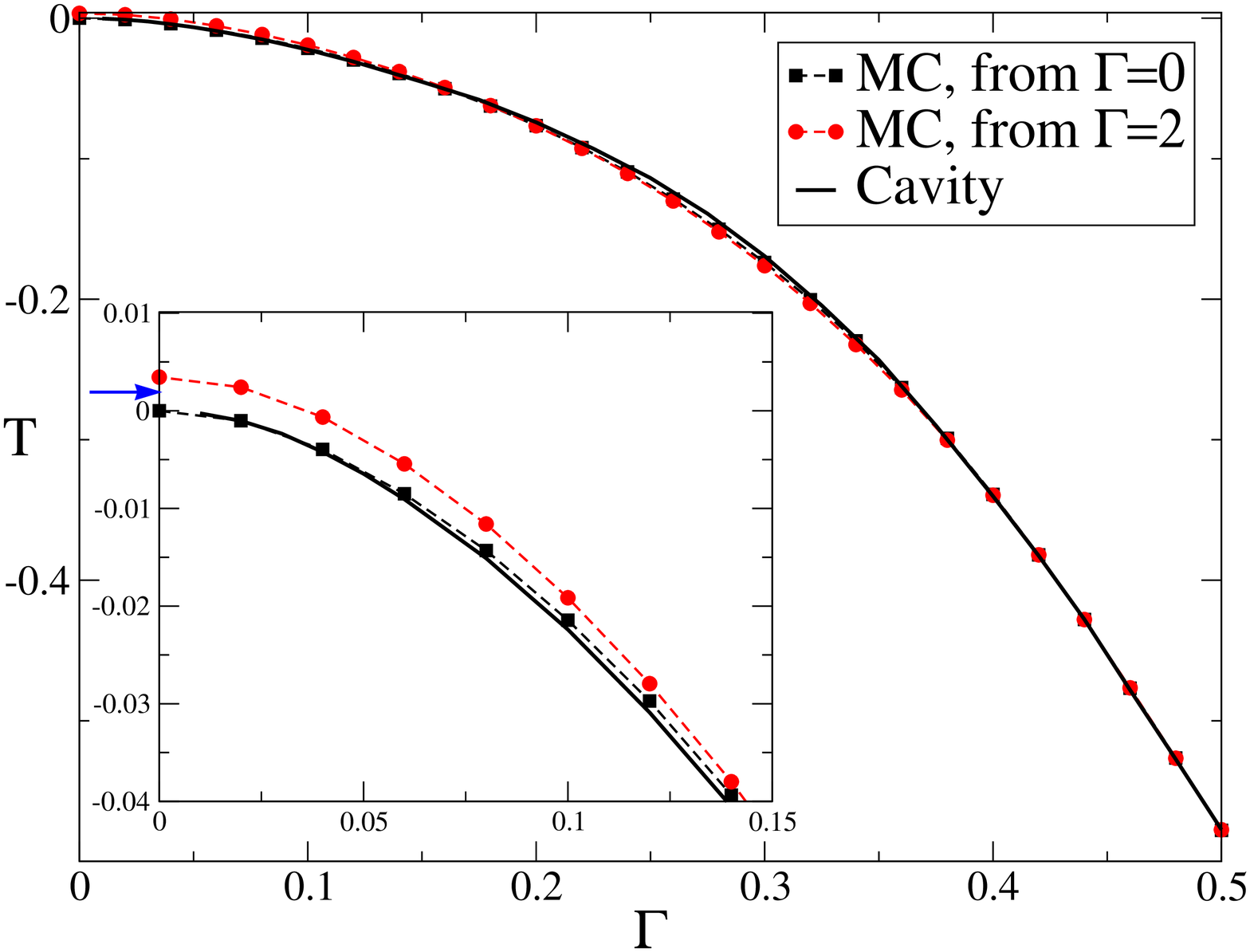}
\includegraphics[width = .49\textwidth]{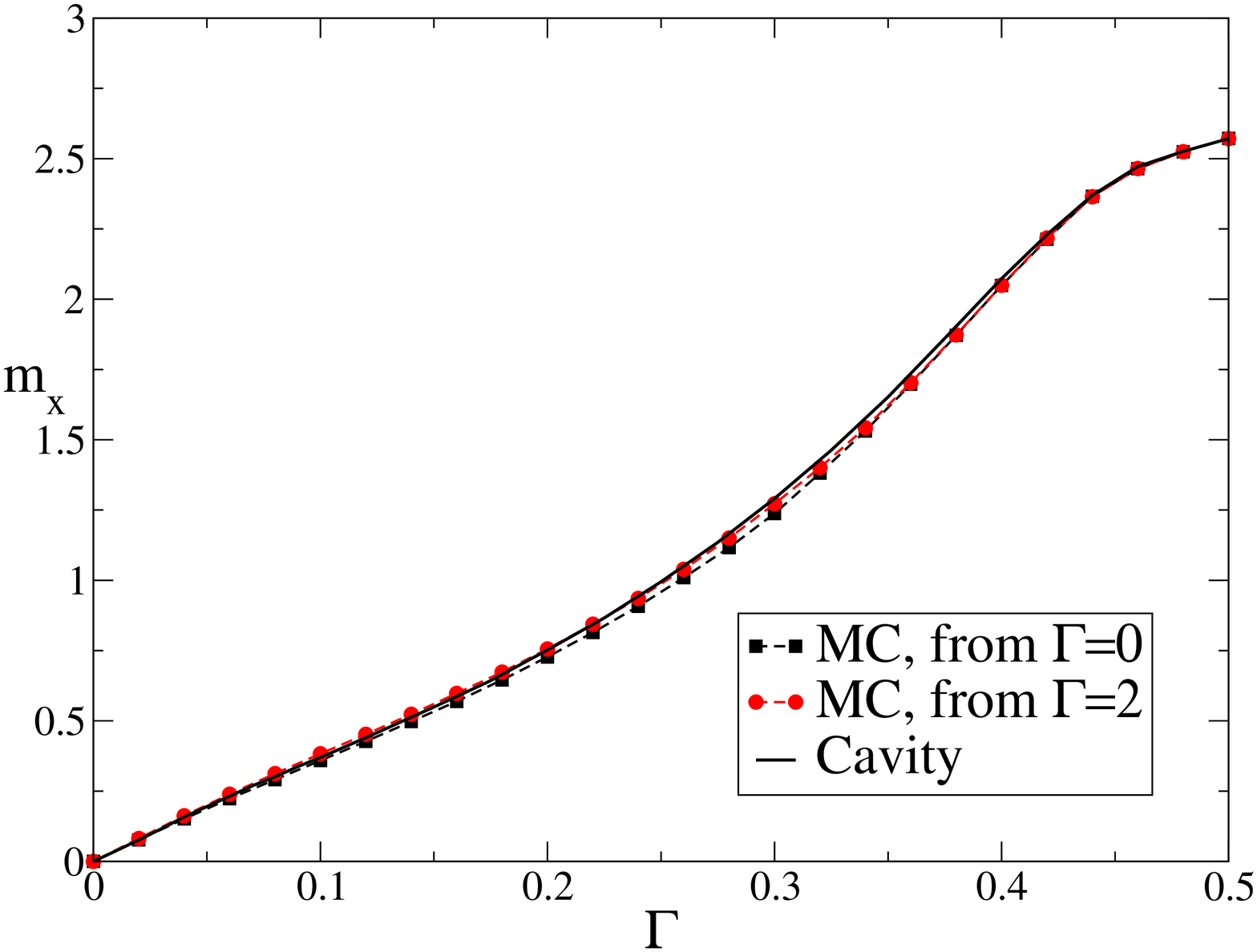}
\caption{
Energy {\it (left panel)} and transverse magnetization {\it (right panel)} as a function of $\G$ at fixed temperature $T=0.06$ for $q=4$ and $c=9$.
PIQC computations are reported as black solid lines.
PIMC simulations (for a single system of size $N=10000$) are reported as a 
dashed black line with squares 
for increasing $\G$ and as a dashed red line with circles for decreasing~$\G$.
{\it (Inset of left panel)} Zoom on the region of low $\G$ where hysteresis is observed in the PIMC.
The blue arrow indicates the value of classical energy (0.0019) that corresponds to a classical annealing starting from
$T_{\rm d}$.
}
\label{fig:mxe_col_4_9}
\end{figure}

In Fig.~\ref{fig:mxe_col_4_9} we show the energy and transverse magnetization as a function of $\Gamma$ at fixed temperature $T=0.06$,
as obtained from PIQC and PIMC.
For this temperature, the system undergoes first a condensation transition at $\G_{\rm c} \sim 0.0427$ (from dP to SG), then a second order
transition at $\G_{\rm i} \sim 0.45$ (from SG to P). 
The former transition is of second order, while the latter is of third
order. They both lead to weak singularities 
in the derivatives of $e$ and $m_x$ that are 
not visible in the figure.
The PIQC computations are 1RSB at $m=1$ for $\G<\G_{\rm c}$, 
while $m=m^*<1$ for $\G_{\rm c}<\G<\G_{\rm i}$.
Above $\G_{\rm i}$, the RS computation is correct.

PIMC simulations have been performed in two ways. 
In the first run, we prepared a typical graph together with one of its typical solutions via 
the ``quiet planting'' technique~\cite{QuietPlanting}; we initialized the PIMC at $\G=0$ in the solution, hence at zero energy,
and we then slowly increased $\G$.
In the second run, we initialized the PIMC at $\G=2$, we equilibrated in the paramagnetic phase, and we decreased $\G$ down to $\G=0$.
On the scales of the figure, no difference between the two PIMC runs is observed, however a closer look (inset of left panel in 
Fig.~\ref{fig:mxe_col_4_9})
reveals that decreasing
$\G$ one obtains a positive residual energy at $\G=0$. 
The latter is found to be larger than the residual energy after an infinitely slow
thermal annealing, which is obtained by preparing a quietly planted configuration at $T_{\rm d}$ and performing a slow classical annealing
down to $T=0$~\cite{QuietPlanting,0295-5075-90-6-66002}.
Although a direct comparison is not possible because the PIMC annealing has not been extrapolated in the infinitely slow limit,
this result suggests that an annealing of an imaginary time PIMC simulation (Sec.~\ref{sec:qmc_ann}) 
is not more efficient than a thermal annealing for this model,
and we believe that this is once again related to entropic level crossings inside the SG phase, as in the QRSM model.
A more detailed investigation of this point is one of the most important directions for future research.

\begin{figure}
\centering
\includegraphics[width = .7\textwidth]{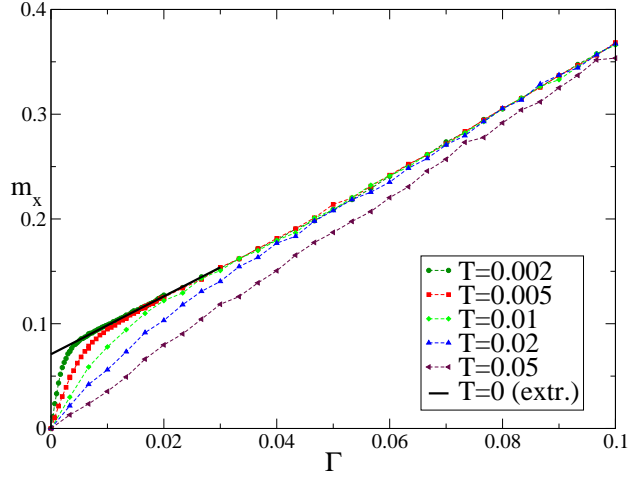}
\caption{
PIMC results for a single instance with $q=4$, $c=9$ and $N=1000$ spins, starting 
from a planted state at $T=0$ and increasing $\G$ at fixed $T$.
The thick black line is the extrapolation at $T=0$, that shows that $\lim_{\G\to 0} m_x(A,\G) = m_x^0 > 0$.
}
\label{fig:mx_col_4_9}
\end{figure}

Finally, we want to confirm numerically the assumptions in Eq.~(\ref{eq:e_COL}). Unfortunately, neither PIQC nor PIMC can
access directly the zero temperature properties of a cluster.
To solve the problem,
we performed several PIMC runs at different low temperatures,
starting from the same quietly planted zero-energy configuration at $\G=0$ and increasing $\G$.
In this way we assume that the PIMC follows the evolution of a single cluster, selected by the planted configuration, 
in $T$ and $\G$. Extrapolating the results to $T\to 0$ gives the ground state properties of the cluster.
In the QRSM, the transverse magnetization of a cluster is $m_x(A) = s(A) \tanh(\b \G)$, hence it vanishes linearly
in $\G$ at any finite temperature, but with a diverging slope that indicates a finite $m_x$ in the limit where $T\to 0$ first, 
and then $\G\to 0$. Note that also in the coloring problem one can show in perturbation theory that whenever 
floppy spins (that can be flipped without energy change) are present in a cluster, the slope of $m_x$ computed by perturbation theory for this
cluster diverges when $T\to 0$~\cite{BSZ12}.
The results of the PIMC runs, reported in Fig.~\ref{fig:mx_col_4_9}, are consistent with this expectation and the extrapolation
to $T\to 0$ clearly shows a finite $m_x$ at $\G=0$. In order to complete this investigation, one should show that larger clusters
have larger $m_x^0$; however generating configurations belonging to clusters of (appreciably) different size is not possible
using the quiet planting, leaving a direct investigation of this point as an open problem.

\subsubsection{Discussion}

Compared to the XORSAT model, the coloring problem shows a much more complex phase diagram, due to the non-trivial distribution of the energy and entropy densities of its pure states.
We studied the problem for $q=4$ and $c=9$, that classically is in the clustered phase with an exponential number of clusters.
By means of combined PIQC and PIMC computations, we showed that:
\begin{itemize}
\item The zero temperature transition from the spin glass to the paramagnetic phase is of third order, unlike in the XORSAT problem.
Although we suspect that the transition becomes first order for large enough $q$ and $c$, we cannot investigate the problem because of computational limitations.
\item The quantum spin glass phase has a complex structure similar to the one of the QRSM. Due to the finite entropy of clusters, the states are extremely delocalized even for $\G\to 0$,
leading at $T=0$ to a finite transverse magnetization $m_x^0$ for $\G\to 0$ and a 
linearly decreasing ground state energy $e(\G) \sim - \G m_x^0$. 
\item
We cannot give direct evidence that $m_x^0$ grows with the classical entropy of the cluster.
However, this is indirectly confirmed by the fact that $T_{\rm c}$ rises linearly from zero for small $\G$, as in the QRSM.
Because of this entropic phenomenon, quantum fluctuations promote the formation of the glass at low enough temperature and $\G$, 
as found in~\cite{FSZ10,FSZ11,MMBMRR10,OLG12}.
Hence, we expect that 
level crossings are induced in this model by an energy-entropy effect, exactly as in the QRSM. 
We believe that these crossings will lead to an everywhere exponentially
small gap in the spin glass phase.
\end{itemize}
We expect that similar results about level crossings hold for any other random optimization problem characterized by clusters of finite entropy.
However, the order of the zero temperature transition depends on the problem under investigation.

\section{Conclusions}
\label{sec:conclusions}

Let us now summarize the main messages we wanted to convey in this review,
and draw some perspectives for future work.
As the Quantum Adiabatic Algorithm (QAA) is a general purpose optimization procedure,
random constraint satisfaction problems provide natural benchmarks for its efficiency.
The tools of the statistical mechanics of disordered systems that have been developed for
their study, first in the classical case and more recently in a quantum context, have several
main outcomes.

Natural benchmarks for the QAA are random instances with a Unique Satisfying Assignment (USA),
because for these instances the minimal gap in unambiguously defined.
The generation process of USA instances
is enlightened by the detailed knowledge acquired on classical random constraint satisfaction problems; most of these
ensembles display a satisfiability transition that is strongly discontinuous in terms of the
number of their solutions, which jumps from being exponentially large to zero when the
control parameter is tuned across the transition. In consequence USA have an exponentially small
probability of existence even at the transition, so that alternative ensembles where USA exist
with finite probability should instead be used (Sec.~\ref{sec:generating_USA}).

The classical phase diagrams of random constraint satisfaction problems exhibits a rich 
phenomenology with several structural phase transitions, that are not all present in every model
(Sec.~\ref{sec:classical_mean_field}).
When quantum fluctuations are added zero temperature phase transitions as a function of the intensity
of the fluctuations generically appear. In some models they are of the first order type, thus 
accompanied by an exponentially small gap that implies an exponentially large running time for
the QAA. Other models exhibit higher order phase transitions, so that the
gap at the transition can be only polynomially small, 
yet in general their weakly quantum phase is
gapless because of the complex spin glass structure that induces a continuum of avoided level
crossings through the competition between the energy and the entropy of the classical pure states (Sec.~\ref{sec:results}).
The path integral quantum cavity method provides an analytic framework in which these transitions can
be quantitatively computed in the thermodynamic limit. Even if it contains an heavy numerical part, 
corresponding to the resolution of the cavity equations, it is of a different nature compared to 
quantum Monte Carlo methods that are plagued by equilibration problems when glassy features of
random constraint satisfaction problems come into play. 
It is also worth to note that the methods discussed in Sec.~\ref{sec:methods} are useful also for
condensed matter applications to bosonic 
or fermionic systems (see e.g.~\cite{STZ09,CTZ09,IM10,FSZ11}).

We would also like to emphasize that, as discussed for instance in Sec.~\ref{sec:thermalannealing},
there always exists an annealing (be it classical or quantum) path that allows to find a solution in polynomial time,
obtained by adding an external field in the direction of one solution of the problem.
The relevant question, in our opinion, is not about proving the existence of good annealing paths, but to 
determine whether an efficient annealing path can be constructed without assuming a detailed knowledge of the
ground states of the Hamiltonian to be minimized.
In this review we mainly considered annealing in a (possibly random) transverse field, but we believe that 
our results can be applied to more general Hamiltonians, at least those which do not have a sign problem, e.g.
Bosonic ones~\cite{STZ09,CTZ09,FSZ11}.

Even though the above statement sounds rather negative for the usefulness of the quantum adiabatic
algorithm, we want to emphasize that it might be more promising to consider it as an approximation
algorithm and not as an exact one. Classical computational complexity theory has indeed established
several hardness of approximation results and it would be interesting to study whether a quantum 
annealing performed on a timescale that does not respect the adiabaticity condition could lead to energies 
that, even if higher than the ground state one, would still be lower than the best approximation ratio
achievable by classical approximation algorithms. This question is obviously much more difficult to
tackle mathematically, as there are no general results as the adiabatic theorem to bound the residual
energy after a non-adiabatic evolution.

Finally, an important aspect that has not been touched upon at all in this review is the
description of actual quantum computers, that should necessarily take into account the
unavoidable coupling to the environment. Quantum adiabatic algorithm is claimed to be robust
with respect to a thermal coupling with the environment~\cite{CFP01,RC05,SL05}; 
we believe it would be worth investigating
the modification of the picture discussed in this review induced by an experimentally relevant
modelization of the coupling with the environment.

\section*{Acknowledgments}

We warmly thank G.~Biroli, E.~Farhi, D.~Gosset, J.~Kurchan, I.~Hen, C.~Laumann, A.~Rosso, S.~Sondhi, M.~Tarzia and P.~Young for many useful discussions and for collaborating with
some of us on the topics of this review.
We also wish to thank B.~Altshuler,  D.~Huse, S.~Mandr\`a, M.~M\'ezard, R.~Moessner, R.~Monasson, M.~Palassini, A.~Ramezanpour,
D.~Reichman, J.~Roland, G.~Santoro, A.~Scardicchio, L.~Zdeborov\'a, and R.~Zecchina for important discussions. 

Computations were performed in part at the MesoPSL computing center, 
with support from R\'egion Ile de France and ANR,
in part using HPC resources from GENCI-CCRT/TGCC (Grant 2012056924),
and in part using local computational resources that were bought thanks to the
PIR grant of ENS ``Optimization in complex system''.

We acknowledge funding from the MIT-France Seed Fund/MISTI Global Seed Fund ``Numerical Simulation and Quantum Adiabatic Algorithms".
FZ wishes to thank the Princeton Center for Theoretical Science for hospitality during part of this work.

\newpage

\bibliographystyle{elsarticle-num}
\bibliography{QOPT}

\begin{thebibliography}{100}
\expandafter\ifx\csname url\endcsname\relax
  \def\url#1{\texttt{#1}}\fi
\expandafter\ifx\csname urlprefix\endcsname\relax\def\urlprefix{URL }\fi
\expandafter\ifx\csname href\endcsname\relax
  \def\href#1#2{#2} \def\path#1{#1}\fi

\bibitem{GareyJohnson}
M.~R. Garey, D.~S. Johnson, Computers and Intractability, A Guide to the Theory
  of {NP}-Completeness, W.H. Freeman and Company, New York, 1979.

\bibitem{Papadimitriou94}
C.~H. Papadimitriou, Computational Complexity, Addison-Wesley, 1994.

\bibitem{Pa83}
C.~Papadimitriou, K.~Steiglitz, Combinatorial Optimization: Algorithms and
  Complexity, Dover, New York, 1998.

\bibitem{Fe82}
{R.P. Feynman}, Simulating physics with computers, Int. J. of Theor. Phys. 21
  (1982) 467.

\bibitem{Deutsch85}
D.~Deutsch, Quantum theory, the {C}hurch-{T}uring principle and the universal
  quantum computer, Proc. Royal Society of London A 400~(1818) (1985) 97.

\bibitem{NielsenChuang}
{M.A. Nielsen}, {I.L. Chuang}, Quantum Computation and Quantum Information,
  Cambridge University Press, Cambridge, 2000.

\bibitem{Mermin}
{N.D. Mermin}, Quantum Computer Science, Cambridge University Press, Cambridge,
  2007.

\bibitem{lectures_QI}
{D. Bru\ss}, {G. Leuchs} (Eds.), Lectures on Quantum Information, Wiley-VCH,
  Weinheim, 2007.

\bibitem{DeutschJozsa92}
{D. Deutsch}, {R. Jozsa}, Rapid solutions of problems by quantum computation,
  Proc. Royal Society of London A 439 (1992) 553.

\bibitem{Simon94}
D.~R. Simon, On the power of quantum computation, in: Foundations of Computer
  Science, 1994 Proceedings., 35th Annual Symposium, 1994, p. 116.

\bibitem{Shor94}
P.~Shor, Algorithms for quantum computation: discrete logarithms and factoring,
  in: Foundations of Computer Science, 1994 Proceedings., 35th Annual
  Symposium, 1994, p. 124.

\bibitem{Grover97}
L.~Grover, Quantum mechanics helps in searching for a needle in a haystack,
  Phys. Rev. Lett. 79~(2) (1997) 325.

\bibitem{BeVa97}
{E. Bernstein}, {U. Vazirani}, Quantum complexity theory, Siam J. Comput. 26
  (1997) 1411.

\bibitem{Wa00}
J.~Watrous, Succinct quantum proofs for properties of finite groups, in: Proc.
  41st Annual Symposium on Foundations of Computer Science, FOCS '00, IEEE
  Computer Society, Washington, 2000.

\bibitem{KiShVy02}
{A.Y. Kitaev}, {A.H. Shen}, {M.N. Vyalyi}, Classical and quantum computation,
  AMS, Providence, 2002.

\bibitem{qa_first}
B.~Apolloni, C.~Carvalho, D.~de~Falco, Quantum stochastic optimization, Stoc.
  Proc. Appl. 33 (1989) 233.

\bibitem{qa_second}
A.~Finnila, M.~Gomez, C.~Sebenik, C.~Stenson, J.~Doll, Quantum annealing: A new
  method for minimizing multidimensional functions, Chem. Phys. Lett. 219
  (1994) 343.

\bibitem{KaNi98}
T.~Kadowaki, H.~Nishimori, Quantum annealing in the transverse {I}sing model,
  Phys. Rev. E 58~(5) (1998) 5355.

\bibitem{Aeppli99}
J.~Brooke, D.~Bitko, T.~F., Rosenbaum, G.~Aeppli, Quantum annealing of a
  disordered magnet, Science 284~(5415) (1999) 779--781.

\bibitem{Fa01}
E.~Farhi, J.~Goldstone, S.~Gutmann, J.~Lapan, A.~Lundgren, D.~Preda, A quantum
  adiabatic evolution algorithm applied to random instances of an {NP}-complete
  problem, Science 292~(5516) (2001) 472--475.

\bibitem{qa_review_santoro}
G.~E. Santoro, E.~Tosatti, Optimization using quantum mechanics: quantum
  annealing through adiabatic evolution, J. Phys. A: Math. Gen. 39~(36) (2006)
  R393.

\bibitem{qa_book_das_chakrabarti}
A.~Das, {B.~K. Chakrabarti (Eds.)}, Quantum annealing and related optimization
  methods, Springer-Verlag, Berlin, 2005.

\bibitem{qa_rmp_das_chakrabarti}
A.~Das, B.~K. Chakrabarti, \textit{Colloquium} : Quantum annealing and analog
  quantum computation, Rev. Mod. Phys. 80 (2008) 1061--1081.

\bibitem{review_Nishimori}
S.~Morita, H.~Nishimori, Mathematical foundation of quantum annealing, J. Math.
  Phys. 49 (2008) 125210.

\bibitem{Messiah}
A.~Messiah, Quantum mechanics vol. 2, North-Holland, Amsterdam, 1962.

\bibitem{DaMoVa01}
{W. van Dam}, {M. Mosca}, {U. Vazirani}, How powerful is adiabatic quantum
  computation?, in Proc. 42nd FOCS (2001) 279.

\bibitem{DaVa01}
{W. van Dam}, {U. Vazirani}, Limits on quantum adiabatic optimization,
  unpublished.

\bibitem{farhi08}
E.~Farhi, J.~Goldstone, S.~Gutmann, D.~Nagaj, How to make the quantum adiabatic
  algorithm fail, Int. J. of Quantum Computation 6 (2008) 503.

\bibitem{ZnHo06}
{M. Znidaric}, {M. Horvat}, Exponential complexity of an adiabatic algorithm
  for an {NP}-complete problem, Phys. Rev. A 73 (2006) 022329.

\bibitem{Janson}
S.~Janson, T.~Luczak, A.~Rucinski, Random graphs, John Wiley and Sons, New
  York, 2000.

\bibitem{MitchellSelman92}
D.~G. Mitchell, B.~Selman, H.~J. Levesque, Hard and easy distributions for
  {SAT} problems, in: Proc. 10th AAAI, AAAI Press, Menlo Park, California,
  1992, pp. 459--465.

\bibitem{Beyond}
M.~M\'ezard, G.~Parisi, M.~Virasoro, Spin glass theory and beyond, World
  Scientific, Singapore, 1987.

\bibitem{sachdev2001}
S.~Sachdev, Quantum phase transitions, Cambridge University Press, 2001.

\bibitem{Go90}
Y.~Y. Goldschmidt, Solvable model of the quantum spin glass in a transverse
  field, Phys. Rev. B 41~(7) (1990) 4858--4861.

\bibitem{NR98}
T.~Nieuwenhuizen, F.~Ritort, {Quantum phase transition in spin glasses with
  multi-spin interactions}, Physica A 250~(1) (1998) 8--45.

\bibitem{BC01}
G.~Biroli, L.~F. Cugliandolo, Quantum {T}houless-{A}nderson-{P}almer equations
  for glassy systems, Phys. Rev. B 64~(1) (2001) 014206.

\bibitem{CGS01}
L.~F. Cugliandolo, D.~R. Grempel, C.~A. da~Silva~Santos, Imaginary-time replica
  formalism study of a quantum spherical $p$-spin-glass model, Phys. Rev. B
  64~(1) (2001) 014403.

\bibitem{JKKM08}
T.~J\"org, F.~Krzakala, J.~Kurchan, A.~C. Maggs, Simple glass models and their
  quantum annealing, Phys. Rev. Lett. 101~(14) (2008) 147204.

\bibitem{AC09}
M.~H.~S. Amin, V.~Choi, First-order quantum phase transition in adiabatic
  quantum computation, Phys. Rev. A 80~(6) (2009) 062326.

\bibitem{AKR10}
B.~Altshuler, H.~Krovi, J.~Roland, {Anderson localization makes adiabatic
  quantum optimization fail}, Proceedings of the National Academy of Sciences
  107~(28) (2010) 12446.

\bibitem{FGGGS10}
E.~Farhi, J.~Goldstone, D.~Gosset, S.~Gutmann, H.~B. Meyer, P.~W. Shor, Quantum
  adiabatic algorithms, small gaps, and different paths, Quantum Information
  {\&} Computation 11~(3{\&}4) (2011) 181--214.

\bibitem{Ch11}
{V. Choi}, Different adiabatic quantum optimization algorithms for the
  {NP}-complete exact cover and 3-{SAT} problems, Quantum Information \&
  Computation 11 (2011) 638.

\bibitem{DicAm11}
N.~G. Dickson, M.~H. Amin, Algorithmic approach to adiabatic quantum
  optimization, Phys. Rev. A 85 (2012) 032303.

\bibitem{vazirani2001}
V.~Vazirani, Approximation algorithms, Springer Verlag, 2001.

\bibitem{review_kz}
J.~Dziarmaga, Dynamics of a quantum phase transition and relaxation to a steady
  state, Advances in Physics 59~(6) (2010) 1063.

\bibitem{hastad01}
J.~Hastad, Some optimal inapproximability results, J. of the ACM 48 (2001) 798.

\bibitem{FSZ10}
L.~Foini, G.~Semerjian, F.~Zamponi, Solvable model of quantum random
  optimization problems, Phys. Rev. Lett. 105~(16) (2010) 167204.

\bibitem{JKSZ10}
T.~J\"org, F.~Krzakala, G.~Semerjian, F.~Zamponi, First-order transitions and
  the performance of quantum algorithms in random optimization problems, Phys.
  Rev. Lett. 104~(20) (2010) 207206.

\bibitem{BSZ12}
V.~Bapst, G.~Semerjian, F.~Zamponi, The effect of quantum fluctuations on the
  coloring of random graphs, in preparation.

\bibitem{review_Knysh}
V.~N. Smelyanskiy, E.~G. Rieffel, S.~I. Knysh, C.~P. Williams, M.~W. Johnson,
  M.~C. Thom, W.~G. Macready, K.~L. Pudenz, A near-term quantum computing
  approach for hard computational problems in space exploration (2012).
\newblock \href {http://arxiv.org/abs/1204.2821} {\path{arXiv:1204.2821}}.

\bibitem{review_Tanaka}
S.~Tanaka, R.~Tamura, Quantum annealing: from viewpoints of statistical
  physics, condensed matter physics, and computational physics (2012).
\newblock \href {http://arxiv.org/abs/1204.2907} {\path{arXiv:1204.2907}}.

\bibitem{review_Ohzeki}
M.~Ohzeki, Spin glass: A bridge between quantum computation and statistical
  mechanics (2012).
\newblock \href {http://arxiv.org/abs/1204.2865} {\path{arXiv:1204.2865}}.

\bibitem{RichardsonUrbanke}
{T. Richardson}, {R. Urbanke}, Modern Coding Theory, Cambridge University
  Press, Cambridge, 2008.

\bibitem{DiVicenzo_criteria}
D.~P. DiVincenzo, The physical implementation of quantum computation,
  Fortschritte der Physik 48~(9-11) (2000) 771--783.

\bibitem{LaMoScSo10}
{C.R. Laumann}, {R. Moessner}, {A. Scardicchio}, {S.L. Sondhi}, Statistical
  mechanics of classical and quantum computational complexity (2010).
\newblock \href {http://arxiv.org/abs/{\tt 1009.1635}} {\path{arXiv:{\tt
  1009.1635}}}.

\bibitem{DiVi95}
D.~P. DiVincenzo, Two-bit gates are universal for quantum computation, Phys.
  Rev. A 51 (1995) 1015--1022.

\bibitem{DeBaEk95}
D.~Deutsch, A.~Barenco, A.~Ekert, Universality in quantum computation,
  Proceedings: Math. and Phys. Sciences 449~(1937) (1995) 669.

\bibitem{Ll95}
S.~Lloyd, Almost any quantum logic gate is universal, Phys. Rev. Lett. 75
  (1995) 346--349.

\bibitem{BaBeChCletal95}
A.~Barenco, C.~H. Bennett, R.~Cleve, D.~P. DiVincenzo, N.~Margolus, P.~Shor,
  T.~Sleator, J.~A. Smolin, H.~Weinfurter, Elementary gates for quantum
  computation, Phys. Rev. A 52 (1995) 3457--3467.

\bibitem{ClEkMaMo98}
R.~Cleve, A.~Ekert, C.~Macchiavello, M.~Mosca, Quantum algorithms revisited,
  Proc. Royal Society of London A 454~(1969) (1998) 339--354.

\bibitem{RSA}
R.~L. Rivest, A.~Shamir, L.~Adleman, A method for obtaining digital signatures
  and public-key cryptosystems, Commun. ACM 21~(2) (1978) 120--126.

\bibitem{Shor_exp1}
{L. Vandersypen}, {M. Steffen}, {G. Breyta}, {C.S. Yannoni}, {M.H. Sherwood},
  {I.L. Chuang}, Experimental realization of {S}hor's quantum factoring
  algorithm using nuclear magnetic resonance, Nature 414 (2001) 883.

\bibitem{Shor_exp2}
C.-Y. Lu, D.~E. Browne, T.~Yang, J.-W. Pan, Demonstration of a compiled version
  of {S}hor's quantum factoring algorithm using photonic qubits, Phys. Rev.
  Lett. 99 (2007) 250504.

\bibitem{Shor_exp3}
B.~P. Lanyon, T.~J. Weinhold, N.~K. Langford, M.~Barbieri, D.~F.~V. James,
  A.~Gilchrist, A.~G. White, Experimental demonstration of a compiled version
  of {S}hor's algorithm with quantum entanglement, Phys. Rev. Lett. 99 (2007)
  250505.

\bibitem{Shor_exp4}
E.~Lucero, R.~Barends, Y.~Chen, J.~Kelly, M.~Mariantoni, A.~Megrant,
  P.~O'Malley, D.~Sank, A.~Vainsencher, J.~Wenner, T.~White, Y.~Yin, A.~N.
  Cleland, J.~M. Martinis, Computing prime factors with a {J}osephson phase
  qubit quantum processor (2012).
\newblock \href {http://arxiv.org/abs/{\tt1202.5707}}
  {\path{arXiv:{\tt1202.5707}}}.

\bibitem{AKS}
M.~Agrawal, N.~Kayal, N.~Saxena, P{RIMES} is in {P}, Ann. of Math. (2) 160~(2)
  (2004) 781.

\bibitem{Bennett97}
C.~H. Bennett, E.~Bernstein, G.~Brassard, U.~Vazirani, Strengths and weaknesses
  of quantum computing, SIAM J. Comput. 26~(5) (1997) 1510--1523.

\bibitem{Grover_exp}
A.~Dewes, R.~Lauro, F.~R. Ong, V.~Schmitt, P.~Milman, P.~Bertet, D.~Vion,
  D.~Esteve, Quantum speeding-up of computation demonstrated in a
  superconducting two-qubit processor, Phys. Rev. B 85 (2012) 140503.

\bibitem{qlineq_1}
A.~W. Harrow, A.~Hassidim, S.~Lloyd, Quantum algorithm for linear systems of
  equations, Phys. Rev. Lett. 103 (2009) 150502.

\bibitem{qlineq_2}
A.~Ambainis, Variable time amplitude amplification and quantum algorithms for
  linear algebra problems, Proc. of STACS 2012, C. D\"urr and T.~Wilke (eds)
  (2012) 636.

\bibitem{KeRe03}
J.~Kempe, O.~Regev, 3-local hamiltonian is {QMA}-complete, Quantum Inf. Comput.
  3 (2003) 258.

\bibitem{KeKiRe06}
J.~Kempe, A.~Kitaev, O.~Regev, The complexity of the local hamiltonian problem,
  SIAM J. on Computing 35~(5) (2006) 1070--1097.

\bibitem{AhGoIrKe09}
D.~Aharonov, D.~Gottesman, S.~Irani, J.~Kempe, The power of quantum systems on
  a line, Comm. Math. Phys. 287 (2009) 41--65.

\bibitem{Br06}
S.~Bravyi, Efficient algorithm for a quantum analogue of 2-{SAT} (2006).
\newblock \href {http://arxiv.org/abs/{\tt quant-ph/0602108}} {\path{arXiv:{\tt
  quant-ph/0602108}}}.

\bibitem{BrMoRu09}
S.~Bravyi, C.~Moore, A.~Russell, Bounds on the quantum satisfiability threshold
  (2009).
\newblock \href {http://arxiv.org/abs/0907.1297} {\path{arXiv:0907.1297}}.

\bibitem{AmKeSa10}
A.~Ambainis, J.~Kempe, O.~Sattath, A quantum {L}ovasz local lemma, in Proc.
  42nd STOC (2010) 151.

\bibitem{LaMoScSo10_qsat}
C.~Laumann, R.~Moessner, A.~Scardicchio, S.~Sondhi, Phase transitions and
  random quantum satisfiability, Quant. Inf. and Comp. 10 (2010) 1.

\bibitem{LaLaMoScSo10}
C.~R. Laumann, A.~M. L\"auchli, R.~Moessner, A.~Scardicchio, S.~L. Sondhi,
  Product, generic, and random generic quantum satisfiability, Phys. Rev. A 81
  (2010) 062345.

\bibitem{review_topological_qc}
C.~Nayak, S.~H. Simon, A.~Stern, M.~Freedman, S.~Das~Sarma, Non-abelian anyons
  and topological quantum computation, Rev. Mod. Phys. 80 (2008) 1083--1159.

\bibitem{oneway_qc}
R.~Raussendorf, H.~J. Briegel, A one-way quantum computer, Phys. Rev. Lett. 86
  (2001) 5188--5191.

\bibitem{qwalk_Farhi}
E.~Farhi, S.~Gutmann, Quantum computation and decision trees, Phys. Rev. A 58
  (1998) 915--928.

\bibitem{qwalk_Kempe}
J.~Kempe, Quantum random walks: An introductory overview, Contemporary Physics
  44~(4) (2003) 307--327.

\bibitem{qwalk_Ambainis}
A.~Ambainis, Quantum walks and their algorithmic applications, Int. J. of
  Quantum Information 1 (2003) 507.

\bibitem{qwalk_Childs}
A.~M. Childs, Universal computation by quantum walk, Phys. Rev. Lett. 102
  (2009) 180501.

\bibitem{qwalk_Reitzner}
{D. Reitzner}, {D. Nagaj}, {V. Buzek}, Quantum walks, Acta Physica Slovaca 61
  (2011) 603.

\bibitem{KirkpatrickGelatt83}
S.~Kirkpatrick, C.~D. {Gelatt Jr.}, M.~P. Vecchi, Optimization by simulated
  annealing, Science 220 (1983) 671--680.

\bibitem{StDaHoBrCh03}
M.~Steffen, W.~van Dam, T.~Hogg, G.~Breyta, I.~Chuang, Experimental
  implementation of an adiabatic quantum optimization algorithm, Phys. Rev.
  Lett. 90 (2003) 067903.

\bibitem{DWave12}
Z.~Bian, F.~Chudak, W.~G. Macready, L.~Clark, F.~Gaitan, Experimental
  determination of {R}amsey numbers with quantum annealing (2012).
\newblock \href {http://arxiv.org/abs/{\tt1201.1842}}
  {\path{arXiv:{\tt1201.1842}}}.

\bibitem{BoFo28}
M.~Born, V.~Fock, Beweis des {A}diabatensatzes, Zeit. Phys. A 51 (1928)
  165--180.

\bibitem{Kato50}
T.~Kato, On the adiabatic theorem of quantum mechanics, J. Phys. Soc. Jap.
  5~(6) (1950) 435--439.

\bibitem{ChHoWi11}
D.~Cheung, P.~Hoyer, N.~Wiebe, Improved error bounds for the adiabatic
  approximation, J. Phys. A 44 (2011) 415302.

\bibitem{ElHa12}
A.~Elgart, G.~Hagedorn, A note on the switching adiabatic theorem (2012).
\newblock \href {http://arxiv.org/abs/{\tt 1204.2318}} {\path{arXiv:{\tt
  1204.2318}}}.

\bibitem{FaGoGuSi00}
E.~Farhi, J.~Goldstone, S.~Gutmann, M.~Sipser, Quantum computation by adiabatic
  evolution (2000).
\newblock \href {http://arxiv.org/abs/{\tt quant-ph/0001106}} {\path{arXiv:{\tt
  quant-ph/0001106}}}.

\bibitem{RiOr11}
G.~Rigolin, G.~Ortiz, Adiabatic theorem for quantum systems with spectral
  degeneracy, Phys. Rev. A 85 (2012) 062111.

\bibitem{WiZe84}
F.~Wilczek, A.~Zee, Appearance of gauge structure in simple dynamical systems,
  Phys. Rev. Lett. 52 (1984) 2111--2114.

\bibitem{uniqueSAT}
{L.G. Valiant}, {V.V. Vazirani}, {NP} is as easy as detecting unique solutions,
  Th. Comp. Science 47 (1986) 85.

\bibitem{landau32}
L.~D. Landau, Zur {T}heorie der {E}nergieubertragung. {II}, Physics of the
  Soviet Union 2~(2) (1932) 46--51.

\bibitem{zener32}
C.~Zener, Non-adiabatic crossing of energy levels, Proc. Royal Society of
  London 137~(833) (1932) 696--702.

\bibitem{vitanov96}
N.~V. Vitanov, B.~M. Garraway, {L}andau-{Z}ener model: Effects of finite
  coupling duration, Phys. Rev. A 53~(6) (1996) 4288--4304.

\bibitem{vitanov99}
N.~V. Vitanov, Transition times in the {L}andau-{Z}ener model, Phys. Rev. A 59
  (1999) 988--994.

\bibitem{volkov06}
M.~V. Volkov, V.~N. Ostrovsky, Analytical results for state-to-state transition
  probabilities in the multistate {L}andau-{Z}ener model by nonstationary
  perturbation theory, Phys. Rev. A 75~(2) (2007) 022105.

\bibitem{SaMaToCa02}
G.~Santoro, R.~Marto{\v{n}}{\'a}k, E.~Tosatti, R.~Car, Theory of quantum
  annealing of an {I}sing spin glass, Science 295~(5564) (2002) 2427.

\bibitem{BaSe12}
V.~Bapst, G.~Semerjian, On quantum mean-field models and their quantum
  annealing, J. Stat. Mech. 2012~(06) (2012) P06007.

\bibitem{AhDaKeLaLlRe07}
D.~Aharonov, W.~van Dam, J.~Kempe, Z.~Landau, S.~Lloyd, O.~Regev, Adiabatic
  quantum computation is equivalent to standard quantum computation, SIAM J. on
  Computing 37~(1) (2007) 166.

\bibitem{zanardi10}
A.~T. Rezakhani, D.~F. Abasto, D.~A. Lidar, P.~Zanardi, Intrinsic geometry of
  quantum adiabatic evolution and quantum phase transitions, Phys. Rev. A 82
  (2010) 012321.

\bibitem{Roland}
J.~Roland, N.~J. Cerf, Quantum search by local adiabatic evolution, Phys. Rev.
  A 65~(4) (2002) 042308.

\bibitem{CaMuFaSa09}
T.~Caneva, M.~Murphy, T.~Calarco, R.~Fazio, S.~Montangero, V.~Giovannetti,
  G.~Santoro, Optimal control at the quantum speed limit, Phys. Rev. Lett.
  103~(24) (2009) 240501.

\bibitem{CaCaFaSaGiMo11}
T.~Caneva, T.~Calarco, R.~Fazio, G.~E. Santoro, S.~Montangero, Speeding up
  critical system dynamics through optimized evolution, Phys. Rev. A 84 (2011)
  012312.

\bibitem{NeMoEkSmFaCa11}
J.~Nehrkorn, S.~Montangero, A.~Ekert, A.~Smerzi, R.~Fazio, T.~Calarco, Staying
  adiabatic with unknown energy gap (2011).
\newblock \href {http://arxiv.org/abs/{\tt1105.1707}}
  {\path{arXiv:{\tt1105.1707}}}.

\bibitem{NiSe12}
Y.~Seki, H.~Nishimori, Quantum annealing with antiferromagnetic fluctuations,
  Phys. Rev. E 85 (2012) 051112.

\bibitem{SeNi12}
{B. Seoane}, {H. Nishimori}, Many-body transverse interactions in the quantum
  annealing of the p-spin ferromagnet, J. Phys. A 45 (2012) 435301.

\bibitem{ribeiro06}
P.~Ribeiro, R.~Mosseri, Adiabatic computation: A toy model, Phys. Rev. A 74
  (2006) 042333.

\bibitem{pcp}
S.~Arora, C.~Lund, R.~Motwani, M.~Sudan, M.~Szegedy, Proof verification and the
  hardness of approximation problems, J. of the ACM 45 (1998) 501--555.

\bibitem{qpcp_conjecture}
D.~Aharonov, I.~Arad, Z.~Landau, U.~Vazirani, The detectibility lemma and
  quantum gap amplification, Proc. 41st annual ACM symposium on Theory of
  computing 287 (2009) 417--426.

\bibitem{qpcp_hastings}
M.~Hastings, Trivial low energy states for commuting hamiltonians, and the
  quantum {{PCP}} conjecture (2012).
\newblock \href {http://arxiv.org/abs/1201.3387} {\path{arXiv:1201.3387}}.

\bibitem{qapprox_kempe}
S.~Gharibian, J.~Kempe, Approximation algorithms for {QMA}-complete problems,
  Proc. 26th CCC'11 (2011) 178.

\bibitem{qapprox_kempe2}
S.~Gharibian, J.~Kempe, Hardness of approximation for quantum problems (2012).
\newblock \href {http://arxiv.org/abs/1209.1055} {\path{arXiv:1209.1055}}.

\bibitem{CheesemanKanefsky91}
P.~Cheeseman, B.~Kanefsky, W.~M. Taylor, {Where the really hard problems are},
  in: Proc. 12th IJCAI, Morgan Kaufmann, San Mateo, CA, USA, 1991, pp.
  331--337.

\bibitem{Friedgut}
E.~Friedgut, Sharp thresholds of graph properties, and the k-sat problem, J. of
  the American Mathematical Society 12 (1999) 1017.

\bibitem{transition_lb}
J.~Franco, Results related to threshold phenomena research in satisfiability:
  lower bounds, Theor. Comput. Sci. 265 (2001) 147.

\bibitem{Achltcs}
D.~Achlioptas, {Lower bounds for random 3-SAT via differential equations.},
  Theor. Comput. Sci. 265~(1-2) (2001) 159--185.

\bibitem{transition_ub}
O.~Dubois, Upper bounds on the satisfiability threshold, Theor. Comput. Sci.
  265 (2001) 187.

\bibitem{transition_largek}
D.~Achlioptas, Y.~Peres, The threshold for random $k$-{SAT} is $2^k$ log $2 -
  {O}(k)$, J. American Math. Soc. 17 (2004) 947.

\bibitem{MoZe}
R.~Monasson, R.~Zecchina, Statistical mechanics of the random
  $k$-satisfiability model, Phys. Rev. E 56~(2) (1997) 1357--1370.

\bibitem{MeZe}
M.~M\'ezard, R.~Zecchina, Random $k$-satisfiability problem: From an analytic
  solution to an efficient algorithm, Phys. Rev. E 66~(5) (2002) 056126.

\bibitem{MezardParisi02}
M.~M{\'e}zard, G.~Parisi, R.~Zecchina, Analytic and algorithmic solution of
  random satisfiability problems, Science 297 (2002) 812--815.

\bibitem{MeMeZe}
S.~Mertens, M.~M\'ezard, R.~Zecchina, Threshold values of random $k$-{SAT} from
  the cavity method, Random Struct. Algorithms 28~(3) (2006) 340--373.

\bibitem{col1}
F.~Krzakala, A.~Pagnani, M.~Weigt, Threshold values, stability analysis, and
  high-$q$ asymptotics for the coloring problem on random graphs, Phys. Rev. E
  70~(4) (2004) 046705.

\bibitem{BiMoWe}
G.~Biroli, R.~Monasson, M.~Weigt, A variational description of the ground state
  structure in random satisfiability problems, Eur. Phys. J. B 14 (2000) 551.

\bibitem{KrMoRiSeZd}
F.~Krzakala, A.~Montanari, F.~Ricci-Tersenghi, G.~Semerjian, L.~Zdeborov{\'a},
  {Gibbs states and the set of solutions of random constraint satisfaction
  problems}, Proc. National Academy of Sciences 104~(25) (2007) 10318--10323.

\bibitem{FrLe}
S.~Franz, M.~Leone, {Replica bounds for optimization problems and diluted spin
  systems.}, J. Stat. Phys. 111~(3-4) (2003) 535--564.

\bibitem{PaTa}
D.~Panchenko, M.~Talagrand, {Bounds for diluted mean-fields spin glass
  models.}, Probab. Theory Relat. Fields 130~(3) (2004) 319--336.

\bibitem{MoraMezard05b}
H.~Daud\'e, T.~Mora, M.~M\'ezard, R.~Zecchina, Pairs of sat assignments and
  clustering in random boolean formulae, Th. Comp. Science 393 (2008) 260--279.

\bibitem{clus_rig_Fede}
D.~Achlioptas, F.~Ricci-Tersenghi, On the solution-space geometry of random
  constraint satisfaction problems, Proc. of the 38th annual ACM symposium on
  Theory of computing.

\bibitem{Coja11}
A.~Coja-Oghlan, On belief propagation guided decimation for random k-sat, Proc.
  22nd SODA (2011) 957.

\bibitem{Mo07}
R.~Monasson, Introduction to phase transitions in random optimization problems,
  in: {J.P. Bouchaud}, M.~M\'ezard, J.~Dalibard (Eds.), Complex Systems,
  Elsevier, Les Houches, France, 2007.

\bibitem{MM09}
M.~M\'ezard, A.~Montanari, Information, Physics and Computation, Oxford
  University Press, 2009.

\bibitem{BookCrisMoore}
C.~Moore, S.~Mertens, {The Nature of Computation}, Oxford University Press,
  Oxford, 2011.

\bibitem{An58}
P.~W. Anderson, Absence of diffusion in certain random lattices, Phys. Rev.
  109~(5) (1958) 1492.

\bibitem{De81}
B.~Derrida, Random-energy model: An exactly solvable model of disordered
  systems, Phys. Rev. B 24~(5) (1981) 2613--2626.

\bibitem{EdwardsAnderson75}
S.~F. Edwards, P.~W. Anderson, Theory of spin-glasses, J. Phys. F 5 (1975)
  965--974.

\bibitem{SK75}
D.~Sherrington, S.~Kirkpatrick, {Solvable Model of a Spin-Glass}, Phys. Rev.
  Lett. 35~(26) (1975) 1792--1796.

\bibitem{Pa80}
G.~Parisi, A sequence of approximated solutions to the {S-K} model for spin
  glasses, J. of Phys. A 13~(4) (1980) L115.

\bibitem{FH91}
K.~Fischer, J.~Hertz, {Spin Glasses}, Cambridge University Press, 1991.

\bibitem{Ta06}
M.~Talagrand, The {P}arisi formula, Annals of Mathematics 163 (2006) 221.

\bibitem{GuTo02}
F.~Guerra, F.~L. Toninelli, The thermodynamic limit in mean field spin glass
  models, Comm. Math. Phys. 230 (2002) 71--79.

\bibitem{GrossMezard84}
D.~Gross, M.~M\'ezard, The simplest spin glass, Nucl. Phys. B 240 (1984) 431.

\bibitem{VB85}
L.~Viana, A.~Bray, {Phase diagrams for dilute spin glasses}, J. Phys. C 18~(15)
  (1985) 3037--3051.

\bibitem{MezardParisi85}
M.~M{\'e}zard, G.~Parisi, Replicas and optimization, J. Physique 46 (1985)
  L771--L778.

\bibitem{FuAnderson86}
Y.~Fu, P.~W. Anderson, Application of statistical mechanics to {NP}-complete
  problems in combinatorial optimization, J. Phys. A 19 (1986) 1605--1620.

\bibitem{replica_diluted}
R.~Monasson, Optimization problems and replica symmetry breaking in finite
  connectivity spin glasses, J. Phys. A 31~(2) (1998) 513--529.

\bibitem{cavity}
M.~M\'ezard, G.~Parisi, The {B}ethe lattice spin glass revisited, Eur. Phys. J.
  B 20 (2001) 217.

\bibitem{cavity_T0}
M.~M\'ezard, G.~Parisi, {The cavity method at zero temperature.}, J. Stat.
  Phys. 111~(1-2) (2003) 1--34.

\bibitem{Ta03}
M.~Talagrand, {Spin Glasses: A Challenge for Mathematicians: Cavity and Mean
  Field Models}, Springer, 2003.

\bibitem{CC05}
T.~Castellani, A.~Cavagna, Spin-glass theory for pedestrians, J. Stat. Mech.
  2005~(05) (2005) P05012.

\bibitem{kirkpatrick:87}
T.~R. Kirkpatrick, P.~G. Wolynes, Stable and metastable states in mean-field
  {P}otts and structural glasses, Phys. Rev. B 36 (1987) 8552.

\bibitem{kirkpatrick:88}
T.~R. Kirkpatrick, D.~Thirumalai, Mean-field soft-spin {P}otts glass model:
  Statics and dynamics, Phys. Rev. B 37 (1988) 5342.

\bibitem{KW87}
T.~R. Kirkpatrick, P.~G. Wolynes, Connections between some kinetic and
  equilibrium theories of the glass transition, Phys. Rev. A 35~(7) (1987)
  3072--3080.

\bibitem{KTW89}
T.~R. Kirkpatrick, D.~Thirumalai, P.~G. Wolynes, Scaling concepts for the
  dynamics of viscous liquids near an ideal glassy state, Phys. Rev. A 40~(2)
  (1989) 1045--1054.

\bibitem{LW07}
V.~Lubchenko, P.~G. Wolynes, Theory of structural glasses and supercooled
  liquids, Annual Review of Physical Chemistry 58~(1) (2007) 235--266.

\bibitem{Cavagna09}
A.~Cavagna, Supercooled liquids for pedestrians, Physics Reports 476~(4–6)
  (2009) 51 -- 124.

\bibitem{BB09}
G.~Biroli, {J.P. Bouchaud}, {The Random First-Order Transition Theory of
  Glasses: a critical assessment}, in: P.~Wolynes, V.~Lubchenko (Eds.),
  Structural Glasses and Supercooled Liquids: Theory, Experiment and
  Applications, Wiley \& Sons, 2012.

\bibitem{BB11}
L.~Berthier, G.~Biroli, Theoretical perspective on the glass transition and
  amorphous materials, Rev. Mod. Phys. 83 (2011) 587--645.

\bibitem{gotzebook}
W.~G{\"o}tze, Complex dynamics of glass-forming liquids: A mode-coupling
  theory, Vol. 143, Oxford University Press, USA, 2009.

\bibitem{Kauzmann48}
W.~Kauzmann, The nature of the glassy state and the behavior of liquids at low
  temperatures, Chem. Rev. 43 (1948) 219.

\bibitem{Gardner85}
E.~Gardner, Spin glasses with $p$-spin interactions, Nuclear Physics B 257
  (1985) 747--765.

\bibitem{xor_1}
M.~M\'ezard, F.~Ricci-Tersenghi, R.~Zecchina, {Two solutions to diluted
  $p$-spin models and XORSAT problems.}, J. Stat. Phys. 111~(3-4) (2003)
  505--533.

\bibitem{xor_2}
S.~Cocco, O.~Dubois, J.~Mandler, R.~Monasson, Rigorous decimation-based
  construction of ground pure states for spin-glass models on random lattices,
  Phys. Rev. Lett. 90~(4) (2003) 047205.

\bibitem{MoSe}
A.~Montanari, G.~Semerjian, {On the dynamics of the glass transition on {B}ethe
  lattices}, J. Stat. Phys. 124 (2006) 103.

\bibitem{IKKM2011}
M.~Ibrahimi, Y.~Kanoria, M.~Kraning, A.~Montanari, The set of solutions of
  random {XORSAT} formulae, in: Proc. of the Twenty-Third Annual ACM-SIAM
  Symposium on Discrete Algorithms, SODA '12, SIAM, 2012, pp. 760--779.

\bibitem{achlioptas2011}
D.~Achlioptas, M.~Molloy, The solution space geometry of random linear
  equations (2011).
\newblock \href {http://arxiv.org/abs/{\tt 1107.5550}} {\path{arXiv:{\tt
  1107.5550}}}.

\bibitem{rcm}
T.~Mora, L.~Zdeborov{\'a}, Random subcubes as a toy model for constraint
  satisfaction problems, J. Stat. Phys. 131 (2008) 1121--1138.

\bibitem{col2}
L.~Zdeborov\'{a}, F.~Krzakala, Phase transitions in the coloring of random
  graphs, Phys. Rev. E 76~(3) (2007) 031131.

\bibitem{rearr_csp}
G.~Semerjian, On the freezing of variables in random constraint satisfaction
  problems, J. Stat.Phys. 130 (2008) 251.

\bibitem{AC08}
D.~Achlioptas, A.~Coja-Oghlan, Algorithmic barriers from phase transitions, in:
  Foundations of Computer Science, 2008. FOCS'08. IEEE 49th Annual IEEE
  Symposium on, IEEE, 2008, pp. 793--802.

\bibitem{circumspect}
M.~Alava, J.~Ardelius, E.~Aurell, P.~Kaski, S.~Krishnamurthy, P.~Orponen,
  S.~Seitz, Circumspect descent prevails in solving random constraint
  satisfaction problems, Proceedings of the National Academy of Sciences 105
  (2008) 15253.

\bibitem{PhysRevE.76.021122}
F.~Krzakala, J.~Kurchan, Landscape analysis of constraint satisfaction
  problems, Phys. Rev. E 76 (2007) 021122.

\bibitem{KZ08}
F.~Krzakala, L.~Zdeborov{\'a}, {Potts glass on random graphs}, Europhys. Lett.
  81 (2008) 57005.

\bibitem{MoSe2}
A.~Montanari, G.~Semerjian, {Rigorous inequalities between length and time
  scales in glassy systems.}, J. Stat. Phys. 125~(1) (2006) 23--54.

\bibitem{gross:85}
D.~J. Gross, I.~Kanter, H.~Sompolinsky, Mean-field theory of the {P}otts glass,
  Phys. Rev. Lett. 55 (1985) 304.

\bibitem{Geman}
S.~Geman, D.~Geman, Stochastic relaxation, {G}ibbs distributions, and the
  {B}ayesian restoration of images, IEEE Trans. Pattern Anal. Mach. Intell. 6
  (1984) 721.

\bibitem{0305-4470-29-14-012}
C.~Monthus, J.-P. Bouchaud, Models of traps and glass phenomenology, J. of
  Phys. A 29~(14) (1996) 3847.

\bibitem{springerlink:10.1007/s00220-008-0565-7}
G.~Arous, A.~Bovier, J.~Cerny, Universality of the {REM} for dynamics of
  mean-field spin glasses, Comm. Math. Phys. 282 (2008) 663--695.

\bibitem{BrMo87}
A.~J. Bray, M.~A. Moore, Chaotic nature of the spin-glass phase, Phys. Rev.
  Lett. 58~(1) (1987) 57--60.

\bibitem{FH88}
D.~S. Fisher, D.~A. Huse, Equilibrium behavior of the spin-glass ordered phase,
  Phys. Rev. B 38~(1) (1988) 386--411.

\bibitem{KrMa02}
F.~Krzakala, O.~Martin, Chaotic temperature dependence in a model of spin
  glasses, European Phys. J. B 28 (2002) 199--208.

\bibitem{0295-5075-90-6-66002}
F.~Krzakala, L.~Zdeborov{\'a}, Following {G}ibbs states adiabaticallyâ the
  energy landscape of mean-field glassy systems, Europhysics Lett. 90~(6)
  (2010) 66002.

\bibitem{ZK10}
L.~Zdeborov\'a, F.~Krzakala, Generalization of the cavity method for adiabatic
  evolution of {G}ibbs states, Phys. Rev. B 81~(22) (2010) 224205.

\bibitem{MourikSaad}
J.~van Mourik, D.~Saad, Random graph coloring: Statistical physics approach,
  Phys. Rev. E 66 (2002) 056120.

\bibitem{asat}
J.~Ardelius, E.~Aurell, Behavior of heuristics on large and hard satisfiability
  problems, Phys. Rev. E 74~(3) (2006) 037702.

\bibitem{ZM08}
L.~Zdeborov\'a, M.~M\'ezard, Locked constraint satisfaction problems, Phys.
  Rev. Lett. 101~(7) (2008) 078702.

\bibitem{LenkaThesis}
L.~Zdeborov{\'a}, Statistical physics of hard optimization problems, Acta
  Physica Slovaca 59 (2009) 169--303.

\bibitem{0305-4470-35-35-301}
A.~Braunstein, M.~Leone, F.~Ricci-Tersenghi, R.~Zecchina, Complexity
  transitions in global algorithms for sparse linear systems over finite
  fields, J. Phys. A 35~(35) (2002) 7559.

\bibitem{Jarvisalo}
H.~Haanp{\"a\"a}, M.~J{\"a}rvisalo, P.~Kaski, I.~Niemel{\"a}, Hard satisfiable
  clause sets for benchmarking equivalence reasoning techniques, J. on
  Satisfiability, Boolean Modeling and Computation 2 (2006) 27.

\bibitem{Ricci-Tersenghi17122010}
F.~Ricci-Tersenghi, Being glassy without being hard to solve, Science
  330~(6011) (2010) 1639--1640.

\bibitem{QuietPlanting}
L.~Zdeborov{\'a}, F.~Krzakala, Quiet planting in the locked constraint
  satisfaction problems, SIAM J. Discrete Math. 25 (2011) 750--770.

\bibitem{YKS10}
A.~P. Young, S.~Knysh, V.~N. Smelyanskiy, First-order phase transition in the
  quantum adiabatic algorithm, Phys. Rev. Lett. 104~(2) (2010) 020502.

\bibitem{FSZ11}
L.~Foini, G.~Semerjian, F.~Zamponi, Quantum {B}iroli-{M}\'ezard model: Glass
  transition and superfluidity in a quantum lattice glass model, Phys. Rev. B
  83 (2011) 094513.

\bibitem{BJ83}
R.~Botet, R.~Jullien, Large-size critical behavior of infinitely coordinated
  systems, Phys. Rev. B 28 (1983) 3955--3967.

\bibitem{DV04}
S.~Dusuel, J.~Vidal, Finite-size scaling exponents of the
  {L}ipkin-{M}eshkov-{G}lick model, Phys. Rev. Lett. 93 (2004) 237204.

\bibitem{DV05}
S.~Dusuel, J.~Vidal, Continuous unitary transformations and finite-size scaling
  exponents in the {L}ipkin-{M}eshkov-{G}lick model, Phys. Rev. B 71 (2005)
  224420.

\bibitem{DSSC06}
A.~Das, K.~Sengupta, D.~Sen, B.~K. Chakrabarti, Infinite-range {I}sing
  ferromagnet in a time-dependent transverse magnetic field: Quench and ac
  dynamics near the quantum critical point, Phys. Rev. B 74 (2006) 144423.

\bibitem{Fisher}
D.~S. Fisher, Critical behavior of random transverse-field {I}sing spin chains,
  Phys. Rev. B 51 (1995) 6411.

\bibitem{FiYo98}
D.~S. Fisher, A.~P. Young, Distributions of gaps and end-to-end correlations in
  random transverse-field {I}sing spin chains, Phys. Rev. B 58 (1998)
  9131--9141.

\bibitem{YoRi96}
A.~P. Young, H.~Rieger, Numerical study of the random transverse-field {I}sing
  spin chain, Phys. Rev. B 53 (1996) 8486--8498.

\bibitem{CaFaSa07}
T.~Caneva, R.~Fazio, G.~E. Santoro, Adiabatic quantum dynamics of a random
  {I}sing chain across its quantum critical point, Phys. Rev. B 76 (2007)
  144427.

\bibitem{BrMo80_qSK}
A.~J. Bray, M.~A. Moore, Replica theory of quantum spin glasses, J. Phys. C
  13~(24) (1980) L655.

\bibitem{AnMu_future}
{A. Andreanov}, {M. M\"uller}, Collective excitations and marginal stability of
  quantum {I}sing spin glasses (2012).
\newblock \href {http://arxiv.org/abs/{\tt 1204.4156}} {\path{arXiv:{\tt
  1204.4156}}}.

\bibitem{JKKMP10}
{T. J\"org}, {F. Krzakala}, {J. Kurchan}, {A.C. Maggs}, {J. Pujos}, Energy gaps
  in quantum first-order mean-field-like transitions: The problems that quantum
  annealing cannot solve, EPL 89~(4) (2010) 40004.

\bibitem{FiDuVi11}
M.~Filippone, S.~Dusuel, J.~Vidal, Quantum phase transitions in fully connected
  spin models: An entanglement perspective, Phys. Rev. A 83 (2011) 022327.

\bibitem{LaMoScSo12_2}
C.~Laumann, R.~Moessner, A.~Scardicchio, S.~Sondhi, The quantum adiabatic
  algorithm and scaling of gaps at first order quantum phase transitions
  (2012).
\newblock \href {http://arxiv.org/abs/{\tt1202.3646}}
  {\path{arXiv:{\tt1202.3646}}}.

\bibitem{DT90}
V.~Dobrosavljevic, D.~Thirumalai, $1/p$ expansion for a $p$-spin interaction
  spin-glass model in a transverse field, J. Phys. A 23 (1990) L767.

\bibitem{JKKM10}
T.~J\"org, F.~Krzakala, J.~Kurchan, A.~C. Maggs, Quantum annealing of hard
  problems, Progress of Theoretical Physics Supplement 184 (2010) 290--303.

\bibitem{BucDelScar11}
F.~Buccheri, A.~De~Luca, A.~Scardicchio, Structure of typical states of a
  disordered {R}ichardson model and many-body localization, Phys. Rev. B 84
  (2011) 094203.

\bibitem{Dic2011}
N.~G. Dickson, Elimination of perturbative crossings in adiabatic quantum
  optimization, New J. of Physics 13~(7) (2011) 073011.

\bibitem{KnySme10}
S.~Knysh, V.~Smelyanskiy, On the relevance of avoided crossings away from
  quantum critical point to the complexity of quantum adiabatic algorithm
  (2010).
\newblock \href {http://arxiv.org/abs/{\tt 1005.3011}} {\path{arXiv:{\tt
  1005.3011}}}.

\bibitem{Ch10}
V.~Choi, Adiabatic quantum algorithms for the {NP}-complete maximum-weight
  independent set, exact cover and 3-{SAT} problems (2010).
\newblock \href {http://arxiv.org/abs/{\tt 1004.2226}} {\path{arXiv:{\tt
  1004.2226}}}.

\bibitem{Ch10b}
V.~Choi, Avoid first order quantum phase transition by changing problem
  hamiltonians (2010).
\newblock \href {http://arxiv.org/abs/{\tt 1010.1220}} {\path{arXiv:{\tt
  1010.1220}}}.

\bibitem{DicAm10}
N.~G. Dickson, M.~H.~S. Amin, Does adiabatic quantum optimization fail for
  np-complete problems?, Phys. Rev. Lett. 106 (2011) 050502.

\bibitem{smallrank}
R.~C. Thompson, The behavior of eigenvalues and singular values under
  perturbations of restricted rank, Linear Algebra and its Applications
  13~(1-2) (1976) 69 -- 78.

\bibitem{MRS08}
A.~Montanari, F.~Ricci-Tersenghi, G.~Semerjian, Clusters of solutions and
  replica symmetry breaking in random $k$-satisfiability, J. Stat. Mech.
  2008~(04) (2008) P04004 (41pp).

\bibitem{DM10}
A.~Dembo, A.~Montanari, {{I}sing models on locally tree-like graphs}, Ann.
  Appl. Probab. 20 (2010) 565--592.

\bibitem{LSS08}
C.~Laumann, A.~Scardicchio, S.~L. Sondhi, Cavity method for quantum spin
  glasses on the {B}ethe lattice, Phys. Rev. B 78~(13) (2008) 134424.

\bibitem{KRSZ08}
F.~Krzakala, A.~Rosso, G.~Semerjian, F.~Zamponi, Path-integral representation
  for quantum spin models: Application to the quantum cavity method and monte
  carlo simulations, Phys. Rev. B 78~(13) (2008) 134428.

\bibitem{leifer2008}
M.~S. Leifer, D.~Poulin, Quantum graphical models and belief propagation,
  Annals of Physics 323~(8) (2008) 1899.

\bibitem{poulin2008}
D.~Poulin, E.~Bilgin, Belief propagation algorithm for computing correlation
  functions in finite-temperature quantum many-body systems on loopy graphs,
  Phys. Rev. A 77~(5) (2008) 052318.

\bibitem{Bil-Poul2010}
E.~Bilgin, D.~Poulin, Coarse-grained belief propagation for simulation of
  interacting quantum systems at all temperatures, Phys. Rev. B 81~(5) (2010)
  054106.

\bibitem{Poulin_Hasting2011}
D.~Poulin, M.~Hastings, Markov entropy decomposition: a variational dual for
  quantum belief propagation, Phys. Rev. Lett. 106~(8) (2011) 80403.

\bibitem{IM10}
L.~B. Ioffe, M.~M\'ezard, Disorder-driven quantum phase transitions in
  superconductors and magnets, Phys. Rev. Lett. 105~(3) (2010) 037001.

\bibitem{dimitrova2011}
O.~Dimitrova, M.~M\'ezard, The cavity method for quantum disordered systems:
  from transverse random field ferromagnets to directed polymers in random
  media, J. Stat. Mech. 2011 (2011) P01020.

\bibitem{ramezanpour2012}
A.~Ramezanpour, Cavity approach to variational quantum mechanics, Phys. Rev. B
  85 (2012) 125131.

\bibitem{STZ09}
G.~Semerjian, M.~Tarzia, F.~Zamponi, Exact solution of the {B}ose-{H}ubbard
  model on the {B}ethe lattice, Phys. Rev. B 80~(1) (2009) 014524.

\bibitem{factorgraph}
F.~Kschischang, B.~Frey, H.~Loeliger, Factor graphs and the sum-product
  algorithm, IEEE Transactions on Information Theory 47~(2) (2001) 498.

\bibitem{MolloyReed95}
{M. Molloy}, {B. Reed}, A critical point for random graphs with a given degree
  sequence, Random Struct. Alg. 6 (1995) 161.

\bibitem{Mo95}
R.~Monasson, Structural glass transition and the entropy of the metastable
  states, Phys. Rev. Lett. 75 (1995) 2847--2850.

\bibitem{CuKu93}
L.~Cugliandolo, J.~Kurchan, Analytical solution of the off-equilibrium dynamics
  of a long-range spin-glass model, Phys. Rev. Lett. 71~(1) (1993) 173.

\bibitem{abou1973}
R.~Abou-Chacra, D.~Thouless, P.~Anderson, A selfconsistent theory of
  localization, J. Phys. C 6 (1973) 1734.

\bibitem{particle_filters}
M.~Arulampalam, S.~Maskell, N.~Gordon, T.~Clapp, A tutorial on particle filters
  for online nonlinear/non-{G}aussian {B}ayesian tracking, IEEE Transactions on
  Signal Processing 50~(2) (2002) 174.

\bibitem{MoPaRi}
G.~P. Andrea~Montanari, F.~Ricci-Tersenghi, Instability of one-step
  replica-symmetry-broken phase in satisfiability problems, J. Phys. A 37
  (2004) 2073.

\bibitem{Gi68}
J.~Ginibre, Reduced density matrices of the anisotropic {H}eisenberg model,
  Commun. Math. Phys. 10 (1968) 140--154.

\bibitem{GMR68}
G.~Gallavotti, S.~Miracle-Sole, D.~Robinson, Analyticity properties of the
  anisotropic {H}eisenberg model, Commun. Math. Phys 10 (1968) 311--324.

\bibitem{FG92}
E.~Farhi, S.~Gutmann, The functional integral constructed directly from the
  {H}amiltonian, Annals of Physics 213 (1992) 182--203.

\bibitem{AiNa94}
{M.~Aizenman}, {B.~Nachtergaele}, Geometric aspects of quantum spin states,
  Commun. Math. Phys. 164 (1994) 17.

\bibitem{ChCrIoLe08}
L.~Chayes, N.~Crawford, D.~Ioffe, A.~Levit, The phase diagram of the quantum
  {C}urie-{W}eiss model, J. Stat. Phys. 133 (2008) 131--149.

\bibitem{Ioffe09}
D.~Ioffe, Stochastic geometry of classical and quantum {I}sing models, in:
  Methods of Contemporary Mathematical Statistical Physics, Vol. 1970 of
  Lecture Notes in Mathematics, Springer Berlin / Heidelberg, 2009, p.~87.

\bibitem{MaWo12}
F.~Martinelli, M.~Wouts, Glauber dynamics for the quantum {I}sing model in a
  transverse field on a regular tree, J. Stat. Phys. 146 (2012) 1059--1088.

\bibitem{hastings2007}
M.~Hastings, Quantum belief propagation: An algorithm for thermal quantum
  systems, Phys. Rev. B 76~(20) (2007) 201102.

\bibitem{FIM10}
M.~V. Feigel'man, L.~B. Ioffe, M.~M\'ezard, Superconductor-insulator transition
  and energy localization, Phys. Rev. B 82~(18) (2010) 184534.

\bibitem{Mu11}
M.~Mueller, Giant positive magnetoresistance and localization in bosonic
  insulators, arXiv:1109.0245.

\bibitem{DMFT}
A.~Georges, G.~Kotliar, W.~Krauth, M.~J. Rozenberg, Dynamical mean-field theory
  of strongly correlated fermion systems and the limit of infinite dimensions,
  Rev. Mod. Phys. 68 (1996) 13--125.

\bibitem{globerson}
A.~Globerson, T.~Jaakkola, Approximate inference using conditional entropy
  decompositions, in: Proc. of the 11th International Conference on Artificial
  Intelligence and Statistics, 2007.

\bibitem{Jastrow55}
R.~Jastrow, Many-body problem with strong forces, Phys. Rev. 98 (1955) 1479.

\bibitem{ramezanpour2012fermions}
{A. Ramezanpour}, {R. Zecchina}, Sign problem in the bethe approximation, Phys.
  Rev. B 86 (2012) 155147.

\bibitem{white1992}
S.~White, Density matrix formulation for quantum renormalization groups, Phys.
  Rev. Lett. 69~(19) (1992) 2863.

\bibitem{vidal2003}
G.~Vidal, Efficient classical simulation of slightly entangled quantum
  computations, Phys. Rev. Lett. 91~(14) (2003) 147902.

\bibitem{NFGSS08}
D.~Nagaj, E.~Farhi, J.~Goldstone, P.~Shor, I.~Sylvester, Quantum
  transverse-field {I}sing model on an infinite tree from matrix product
  states, Phys. Rev. B 77~(21) (2008) 214431.

\bibitem{Shi2006}
Y.-Y. Shi, L.-M. Duan, G.~Vidal, Classical simulation of quantum many-body
  systems with a tree tensor network, Phys. Rev. A 74 (2006) 022320.

\bibitem{Nagy12}
{\'A}.~Nagy, Simulating quantum systems on the {B}ethe lattice by
  translationally invariant infinite-tree tensor network, Annals of Physics 327
  (2012) 542.

\bibitem{Li12}
{W. Li}, {J. von Delft}, {T. Xiang}, Efficient simulation of infinite tree
  tensor network states on the bethe lattice (2012).
\newblock \href {http://arxiv.org/abs/1209.2387} {\path{arXiv:1209.2387}}.

\bibitem{Vidal07}
G.~Vidal, Classical simulation of infinite-size quantum lattice systems in one
  spatial dimension, Phys. Rev. Lett. 98~(7) (2007) 70201.

\bibitem{Lepetit}
M.-B. Lepetit, M.~Cousy, G.~Pastor, Density-matrix renormalization study of the
  {H}ubbard model on a {B}ethe lattice, European Phys. J. B 13 (2000) 421--427.

\bibitem{arpack}
{\tt http://www.caam.rice.edu/software/ARPACK/}.

\bibitem{hatano05}
N.~Hatano, M.~Suzuki, Finding exponential product formulas of higher orders,
  Optimization~(3) (2005) 22.

\bibitem{huyg90}
J.~Huyghebaert, H.~D. Raedt, Product formula methods for time-dependent
  schrodinger problems, J. Phys. A 23~(24) (1990) 5777.

\bibitem{poulin11}
D.~Poulin, A.~Quarry, R.~Somma, F.~Verstraete, Quantum simulation of
  time-dependent hamiltonians and the convenient illusion of hilbert space,
  Phys. Rev. Lett. 106~(17) (2011) 170501.

\bibitem{tdDMRG1}
M.~Cazalilla, J.~Marston, Time-dependent density-matrix renormalization group:
  A systematic method for the study of quantum many-body out-of-equilibrium
  systems, Phys. Rev. Lett. 88~(25) (2002) 256403.

\bibitem{tdDMRG2}
S.~White, A.~Feiguin, Real-time evolution using the density matrix
  renormalization group, Phys. Rev. Lett. 93~(7) (2004) 76401.

\bibitem{tdDMRG3}
A.~Daley, C.~Kollath, U.~Schollw{\"o}ck, G.~Vidal, Time-dependent
  density-matrix renormalization-group using adaptive effective hilbert spaces,
  J. Stat. Mech. 2004 (2004) P04005.

\bibitem{Ce95}
D.~M. Ceperley, Path integrals in the theory of condensed helium, Rev. Mod.
  Phys. 67 (1995) 279.

\bibitem{KTC91}
W.~Krauth, N.~Trivedi, D.~Ceperley, Superfluid-insulator transition in
  disordered boson systems, Phys. Rev. Lett. 67~(17) (1991) 2307.

\bibitem{RY94}
H.~Rieger, A.~P. Young, Zero-temperature quantum phase transition of a
  two-dimensional {I}sing spin glass, Phys. Rev. Lett. 72 (1994) 4141--4144.

\bibitem{BS96}
G.~Batrouni, R.~Scalettar, World line simulations of the bosonic {H}ubbard
  model in the ground state, Comp. Phys. Comm. 97~(1) (1996) 63--81.

\bibitem{BW96}
B.~B. Beard, U.-J. Wiese, Simulations of discrete quantum systems in continuous
  euclidean time, Phys. Rev. Lett. 77 (1996) 5130--5133.

\bibitem{PST98}
N.~V. Prokof'ev, B.~V. Svistunov, I.~S. Tupitsyn, "worm" algorithm in quantum
  monte carlo simulations, Physics Lett. A 238~(4-5) (1998) 253--257.

\bibitem{RK99}
H.~Rieger, N.~Kawashima, Application of a continuous time cluster algorithm to
  the two-dimensional random quantum {I}sing ferromagnet, European Phys. J. B
  9~(2) (1999) 233--236.

\bibitem{FGGM11}
E.~Farhi, J.~Goldstone, D.~Gosset, H.~Meyer, A quantum {M}onte {C}arlo method
  at fixed energy, Comp. Phys. Comm. 182 (2011) 1663--1673.

\bibitem{Sa99}
A.~W. Sandvik, Stochastic series expansion method with operator-loop update,
  Phys. Rev. B 59 (1999) R14157.

\bibitem{HY11}
I.~Hen, A.~P. Young, Exponential complexity of the quantum adiabatic algorithm
  for certain satisfiability problems, Phys. Rev. E 84 (2011) 061152.

\bibitem{YKS08}
A.~P. Young, S.~Knysh, V.~N. Smelyanskiy, Size dependence of the minimum
  excitation gap in the quantum adiabatic algorithm, Phys. Rev. Lett. 101
  (2008) 170503.

\bibitem{Hen2012}
I.~Hen, Excitation gap from optimized correlation functions in quantum monte
  carlo simulations, Phys. Rev. E 85 (2012) 036705.

\bibitem{MMBMRR10}
T.~E. Markland, J.~A. Morrone, B.~J. Berne, K.~Miyazaki, E.~Rabani, D.~R.
  Reichman, {The quantum liquid-glass transition}, Nature Physics 7 (2010)
  134--137.

\bibitem{GY11}
M.~Guidetti, A.~P. Young, Complexity of several constraint-satisfaction
  problems using the heuristic classical algorithm walksat, Phys. Rev. E 84
  (2011) 011102.

\bibitem{FGHSSYZ12}
E.~Farhi, D.~Gosset, I.~Hen, A.~Sandvik, P.~Shor, A.~Young, F.~Zamponi, The
  performance of the quantum adiabatic algorithm on random instances of two
  optimization problems on regular hypergraphs (2012).
\newblock \href {http://arxiv.org/abs/1208.3757} {\path{arXiv:1208.3757}}.

\bibitem{FLRZ01}
S.~Franz, M.~Leone, F.~Ricci-Tersenghi, R.~Zecchina, Exact solutions for
  diluted spin glasses and optimization problems, Phys. Rev. Lett. 87~(12)
  (2001) 127209.

\bibitem{FMRWZ01}
S.~Franz, M.~M\'ezard, F.~Ricci-Tersenghi, M.~Weigt, R.~Zecchina, A ferromagnet
  with a glass transition, Europhysics Lett. 55~(4) (2001) 465.

\bibitem{DPLL}
\href {http://arxiv.org/abs/{\tt http://code.google.com/p/relsat}}
  {\path{arXiv:{\tt http://code.google.com/p/relsat}}}.

\bibitem{max1}
\href {http://arxiv.org/abs/{\tt
  http://www.laria.u-picardie.fr/$\sim$cli/maxsatz2009.c}} {\path{arXiv:{\tt
  http://www.laria.u-picardie.fr/$\sim$cli/maxsatz2009.c}}}.

\bibitem{Gosset}
D.~Gosset, {PhD} Thesis, Case Studies in Quantum Adiabatic Optimization (2011).

\bibitem{OLG12}
B.~Olmos, I.~Lesanovsky, J.~Garrahan, Facilitated spin models of dissipative
  quantum glasses (2012).
\newblock \href {http://arxiv.org/abs/{\tt1203.6585}}
  {\path{arXiv:{\tt1203.6585}}}.

\bibitem{CTZ09}
G.~Carleo, M.~Tarzia, F.~Zamponi, {B}ose-{E}instein condensation in quantum
  glasses, Phys. Rev. Lett. 103~(21) (2009) 215302.

\bibitem{CFP01}
A.~M. Childs, E.~Farhi, J.~Preskill, Robustness of adiabatic quantum
  computation, Phys. Rev. A 65 (2001) 012322.

\bibitem{RC05}
J.~Roland, N.~J. Cerf, Noise resistance of adiabatic quantum computation using
  random matrix theory, Phys. Rev. A 71 (2005) 032330.

\bibitem{SL05}
M.~S. Sarandy, D.~A. Lidar, Adiabatic quantum computation in open systems,
  Phys. Rev. Lett. 95 (2005) 250503.

\end{thebibliography}

\end{document}